\documentclass[twoside,fleqn,floatflt]{ActaStyle}
\usepackage{times,cite}
\usepackage[english]{babel}
\usepackage{color}
\usepackage{latexsym}
\usepackage{amsmath}
\usepackage{amssymb}
\usepackage{graphicx}
\usepackage{epsfig}
\usepackage{fancyhdr}
\usepackage[colorlinks=true,pdfstartview=FitV,linkcolor=black,citecolor=black,urlcolor=black]{hyperref}

\def\tablename{Table }
\def\figurename{Fig.}

\def\sa{\omega}
\def\sbb{{\omega^{'}}}
\def\sc{{\omega^{''}}}
\def\sd{{\phi}}
\def\se{{\phi^{'}}}
\def\va{{\varrho}}
\def\vb{{\varrho^{'}}}
\def\vc{{\varrho^{''}}}
\def\vd{{\varrho^{'''}}}
\def\ve{{\varrho^{''''}}}

\def\mmt#1#2{\frac{m_{#1}^2m_{#2}^2}{(m_{#1}^2-t)(m_{#2}^2-t)}}
\def\mtt#1#2#3{\frac{m_{#1}^2m_{#2}^2m_{#3}^2}{(m_{#1}^2-t)(m_{#2}^2-t)
     (m_{#3}^2-t)}}
\def\mc#1{{m^2_#1}}
\def\zll#1#2#3#4{{\frac{#1-#2}{#3-#4}}}
\def\zlll#1#2#3#4#5#6#7#8{{\frac{(#1-#2)(#3-#4)}{(#5-#6)(#7-#8)}}}

\def\fff#1#2{(f^{(#1)}_{#2{NN}}/f_{#2})}

\def\mcsa{{\mc\sa}}
\def\mcsb{{\mc\sbb}}
\def\mcsd{{\mc\sd}}
\def\mcva{{\mc\va}}
\def\mcvb{{\mc\vb}}
\def\mcvc{{\mc\vc}}
\def\mcsc{{\mc\sc}}
\def\mcse{{\mc\se}}
\def\mcvd{{\mc\vd}}
\def\mcve{{\mc\ve}}

\renewcommand{\subsectionmark}[1]{}
\def\authorheading{S.Dubni\v cka}
\def\shorttitle{Analyticity in a phenomenology of electro-weak structure of hadrons}

\renewcommand{\thefigure}{\arabic{section}.\arabic{figure}}
\renewcommand{\thetable}{\arabic{section}.\arabic{table}}

\begin{document}

\pagerange{1}{153}
%
\title{ANALYTICITY IN A PHENOMENOLOGY OF ELECTRO-WEAK\\ STRUCTURE OF HADRONS}
\author{S.~Dubni\v cka\email{stanislav.dubnicka@savba.sk}}
{Institute of Physics, Slovak Academy of Sciences, Bratislava, Slovakia}
\author{A.Z.~Dubni\v ckov\'a\email{dubnickova@fmph.uniba.sk}}
{Department of Theoretical Physics, Comenius University, Bratislava, Slovakia}
\datumy{12 February 2010}{19 February 2010}
\abstract{The utility of an application of the analyticity in a phenomenology
of electro-weak structure of hadrons is demonstrated in a number
of obtained new and experimentally verifiable results. With this
aim first the problem of an inconsistency of the asymptotic
behavior of VMD model with the asymptotic behavior of form factors
of baryons and nuclei is solved generally and a general approach
for determination of the lowest normal and anomalous singularities
of form factors from the corresponding Feynman diagrams is
reviewed. Then many useful applications by making use of the
analytic properties of electro-weak form factors and amplitudes of
various electromagnetic processes of hadrons are carried out.}


\begin{minipage}{2.5cm}
\quad{\small {\sf KEYWORDS:}}
\end{minipage}
\begin{minipage}{10cm}
Electromagnetic Interactions, Polarization, Electromagnetic
Form\\ Factors, Deuteron, Strangeness, Analyticity, Sum Rules
\end{minipage}

\newpage
\tableofcontents

\setcounter{equation}{0} \setcounter{figure}{0} \setcounter{table}{0}\newpage
\section{Introduction}

    Up to the first half of fifties of 20th century all known elementary
particles were assumed to be structureless, i.e. point-like. The
latter property is reflected also in the principles of local
quantum field theory (QFT) (unifying consistently quantum theory,
the concept of the field and the relativistic invariance), which
is considered to be a dynamical theory of elementary particle
mutual interactions.

    Only Hofstadter experiments \cite{Hof} on elastic scattering of
electrons on protons at SLAC have clearly demonstrated
disagreement between theoretical expression for the cross-section
calculated in the framework of the quantum electro-dynamics (QED)
and the obtained experimental results. This phenomenon have
revealed the non-point-like nature of the proton, which later on
have been extended also to all other existing strongly interacting
particles.

    As a result at the calculation of the matrix element of the
elastic scattering of electrons on hadrons in one-photon-exchange
approximation one does not know explicitly (unlike the electron)
the electromagnetic current of considered hadron to be written
only symbolically and practically using various symmetries it is
decomposed into maximal number of linearly independent co-variants
constructed from momenta of incoming and outgoing hadrons and
their spin parameters. The corresponding coefficients are scalar
functions (the electromagnetic (EM) form factors (FFs)) of one
variable to be chosen the squared momentum $t=q^2=-Q^2$ transferred
by the virtual photon. Their number depends on the spin of the
considered hadron.

    Similarly can be introduced the weak FFs of hadrons,
representing the contribution of the weak structure of hadrons
into the dynamical quantities describing a weak interaction of
hadrons in various weak processes.

    A natural explanation of the electro-weak structure of hadrons
was obtained only after the discovery of quark-gluon structure of
strongly interacting observable particles.

    The behavior of the EM FFs in the whole interval of their
definition from $-\infty$ to $+\infty$ is expected to be
theoretically predicted by the quantum chromo-dynamics (QCD), the
gauge invariant local QFT describing mutual interactions of
colored quarks and gluons. However, as it is well known, merely at
sufficiently small distances (thanks to its asymptotic freedom)
QCD becomes a weakly coupled quark-gluon theory to be amenable to
a perturbative expansion in the running coupling constant
$\alpha_s$ and predicts  just the asymptotic behavior of FFs. In
low momentum transfer region, where $\alpha_s$ becomes large, the
quark-gluon perturbation theory breaks down and non-perturbative
methods in QCD are not well worked out to give interesting results
on the FFs of hadrons. The same is valid also for the low energy
time-like region where FFs are complex functions of their variable
and acquire the most complicated, resonant, behavior.

    In such situation a phenomenological approach based on the analyticity of
FFs starts to be very useful, which compensates above-mentioned
problems to some extent and renders possible to achieve a line of
interesting results.

    In an interpretation of experimental data on EM FFs appears to
be useful utilization of the analytic properties in the form of
integral (so-called dispersion) relations together with the
unitarity condition of FFs, which have brought the investigated
FFs into relations with other FFs and amplitudes of various
processes of strongly interacting particles. Such approach in the
case of nucleons have led to a prediction of an existence of
isoscalar and isovector vector mesons, and subsequently to the
vector-meson-dominance (VMD) model \cite{Sak}. The latter model is
based on the assumption (to be later on experimentally confirmed
in electron-positron annihilation into hadrons), that the
interaction of the virtual photon with hadron is realized by a
transformation of the photon to a vector meson with the same
quantum numbers and then this vector meson is strongly interacting
with considered hadron as in any other hadron collision.

    The VMD model was revealed before the discovery of the quark model
of hadrons. Despite of this fact the latter is consistent with VMD
model. Really, at the energy of the photon nearly to the mass of
the vector meson the latter is changed to the quark-antiquark
pair, which is as a result of the confinement effect immediately
bound into the vector meson with the photon quantum numbers.

    Though the VMD model from the point of view of the global
analysis of existing FF data has been in the past very frequently
applied, it suffers from a lot of shortcomings. It does not take
into account the instability of vector mesons, the unitarity
condition of FFs and also the analytic properties of FFs, which
could lead to a more realistic behavior of FFs in the time-like
resonant region. Other serious shortcoming is the same asymptotic
behavior for FFs of all hadrons, which is in contradiction with
the predictions of quark model for baryons and atomic nuclei.

    A solution from this situation is the universal
Unitary and Analytic $(U\&A)$ model of electro-weak FFs, which is
a unification of the experimental fact of a creation of unstable
vector-meson resonances in the electron-positron annihilation into
hadrons, the analytic properties of FFs in the complex $t$-plane
and the correct asymptotic behavior of FFs as predicted by the
quark model of hadrons. Its applications to many hadrons and
nuclei have led to a lot of interesting results to be verifiable
experimentally.

\setcounter{equation}{0} \setcounter{figure}{0} \setcounter{table}{0}\newpage
\section{Electromagnetic form factors of strongly interacting particles and their properties}

   In this section general properties of EM
FFs of strongly interacting particles, like number of FFs of the
hadron under consideration, their most usual parametrization and
its shortcomings, will be reviewed and subsequently resolved.

\medskip

      \subsection{Number of electromagnetic form factors of a considered
      hadron}\label{II1}

\medskip

   The elastic electron scattering $e^-h\to e^-h$ and the
annihilations $e^+e^-\leftrightarrow h\bar h$, where $h$ means an
arbitrary strongly interacting object (including also the atomic
nucleus), are the most usual processes, in which the concept of EM
FF appears. The cross-sections of these processes are proportional
to the absolute value squared of the corresponding scattering
amplitudes, which one knows formally to calculate in the framework
of the quantum electrodynamics (QED) perturbation expansion
according to the fine structure constant $\alpha\simeq 1/137$.
Since the value of the fine structure constant $\alpha\ll 1$,
those scattering amplitudes are considered practically in the
one-photon-exchange approximation as follows
\begin{equation}
M(e^-h\to e^-h)\simeq M^{(\gamma)}(s,t)=e^2\bar u(k_2)\gamma_{\mu}
u(k_1)\frac{g_{\mu\nu}}{q^2}\langle h|J_{\nu}^{EM}|h\rangle \label{sec21}
\end{equation}
and
\begin{equation}
M(e^+e^-\to \bar h h)\simeq M^{(\gamma)}(t,s)=e^2\bar
v(k_2)\gamma_{\mu} u(k_1)\frac{g_{\mu\nu}}{q^2}\langle
0|J_{\nu}^{EM}|h \bar h\rangle, \label{sec22}
\end{equation}
where $g_{\mu\nu}/q^2$ is the photon propagator and $\langle
h|J_{\nu}^{EM}|h\rangle$ (resp.$\langle 0|J_{\nu}^{EM}|h \bar
h\rangle$) is a matrix element of the hadron EM current, which,
however, due to the non-point-like nature of the hadron $h$, is
unknown. Therefore, in practice it is decomposed according to a
maximal number of linearly independent relativistic covariants
constructed from the four-momenta and spin parameters of $h$ as
follows
\begin{equation}
\langle h|J_{\mu}^{EM}|h\rangle = \sum_i R_{\mu}^i F_i(t)\label{sec23}
\end{equation}
or
\begin{equation}
\langle h \bar h|J_{\mu}^{EM}|0\rangle = \sum_i X_{\mu}^i F_i(t),\label{sec24}
\end{equation}
where the scalar coefficients $F_i(t)$ are the EM FFs of the
hadron $h$ as functions of one invariant variable $t$- the
momentum squared to be transferred by a virtual photon.

\medskip

The number of $F_i(t)$ depends essentially on the spin $S$ of $h$.

\medskip

    Let us consider the most topical cases.

\medskip

    We start with the nonet of pseudoscalar mesons
$\pi^+$, $\pi^0$, $\pi^-$, $K^+$, $K^0$, $\bar K^0$, $K^-$,
$\eta$, $\eta'$. Since they possess spin to be zero, for the
construction of covariants $(p_2-p_1)_\mu$ and $(p_2+p_1)_\mu$
only two four-momenta, $p_1$ and $p_2$, are available. By an
application of the gauge invariance of the EM interactions one
comes to the following final parametrization
\begin{equation}
\langle p_2|J_{\mu}(0)|p_1 \rangle =  F_P(t)(p_1+p_2)_{\mu} \label{sec25}
\end{equation}
only with one EM FF $F_P(t)$ completely describing the EM
structure of any member of the nonet of pseudoscalar mesons.
Moreover, making use of transformation properties of the EM
current operator $J_\mu(x)$ and the one-particle state vector with
regard to the all three discrete C, P, T transformations
simultaneously, one finds the relation between particle and
antiparticle FFs
\begin{equation}
F_P(t) = -F_{\bar P}(t)\label{cpt}
\end{equation}
from where it follows that for the true neutral pseudoscalar
mesons $\pi^0$, $\eta$, $\eta'$ the EM FFs are identically equal
zero for all values from $-\infty < t < +\infty$.

   The pseudoscalar meson EM FFs are normalized at $t = 0$ to the
charge of the considered meson.

   A consideration of the nonzero value of the isospin of the pion
does not enlarge a number of EM FFs and both charged pions are
described by the same FF.

   A completely different situation is with kaons. The $K^+$ and
$K^0$ belong to the same isomultiplet with $I = 1/2$. Therefore
instead of the positively charged and neutral kaons one can
introduce the EM current of the kaon and to investigate what
isotopic structure it has. One can show, that it splits on a sum
of isotopic scalar and isotopic vector. In connection with the
latter the isoscalar $F^s_K(t)$ and isovector $F^v_K(t)$ FFs of
the kaon are introduced to be expressed by $F_K^+(t)$ and
$F_K^0(t)$ as follows
\begin{equation}
F^s_K(t) = \frac{1}{2}[F_K^+(t) + F_K^0(t)]\label{ffKs}
\end{equation}
\begin{equation}
F^v_K(t) =\frac{1}{2}[F_K^+(t) - F_K^0(t)],\label{ffKv}
\end{equation}
from which immediately the normalization condition
\begin{equation}
F^s_K(0)=F^v_K(0)=\frac{1}{2}\label{normKff}
\end{equation}
is obtained.

   In principle, there is no problem of obtaining of the experimental
information on $|F_P(t)|$ in $t<0$ and $t>0$ regions as
$d\sigma/d\Omega\sim |F_P(t)|^2$. However, the data on nuclei (e.g
$He^4$, $C^{12}$, $O^{16}$) exist only for $t<0$ up to now and
there is no concept of the EM FF of nucleus for $t>0$ to be known
by nuclear physicists.

As a consequence of a compound nature of nuclei so-called
diffraction minima appear in $t<0$ region at the absolute value of
their charge FF $|F_c(t)|$ , which are interpreted as zeros of
$F_c(t)$ on the real axis of the complex $t$- plane. It is
observed, that if more compound nucleus is investigated more
diffraction minima emerge in the same range of momentum transfer
values.

\medskip

   In the case of the octet $1/2^+$ baryons $p$, $n$, $\Lambda$,
$\Sigma^+$, $\Sigma^0$, $\Sigma^-$, $\Xi^0$, $\Xi^-$ and e.g.
$He^3$, $H^3$ nuclei covariants $R_\mu(p_1,p_2)$ are constructed
by the four-momenta $p_1, p_2$, Dirac matrices and bispinors. The
final result for a parametrization of the matrix element of the EM
current of the octet $1/2^+$ baryons takes the form
\begin{equation}
\langle p_2|J_{\mu}(0)|p_1 \rangle = \frac{1}{2\pi^3}\bar
u(p_2)\{\gamma_{\mu} F_{1B}(t) +
\frac{1}{2m_B}\sigma_{\mu\nu}(p_2-p_1)_{\nu}F_{2B}(t)\}u(p_1),\label{sec26}
\end{equation}
where $F_{1B}(t)$ and $F_{2B}(t)$ are  Dirac and Pauli FFs,
respectively and $m_B$ is the baryon mass.
   From the practical point of view it is more suitable to describe the
EM structure of the octet $1/2^+$ baryons by means of the Sachs
electric $G_{EB}(t)$ and magnetic $G_{MB}(t)$ FFs, defined by the
following expressions
\begin{equation}
G_{EB}(t)=F_{1B}(t) +\frac{t}{4m_B^2}F_{2B}(t);\quad
G_{MB}(t)=F_{1B}(t) +F_{2B}(t). \label{sec27}
\end{equation}
There is a special coordinate system (the Breit reference frame),
in which $G_{EB}(t)$ and $G_{MB}(t)$ describe a distribution of
the charge and magnetic moment of the baryon. Hence they are
called the electric and magnetic FFs to be normalized to the
charge and magnetic moment of the baryon, respectively, for $t=0$.

   Similarly to kaons, one can consider instead of the EM current
of every member of the octet $1/2^+$ baryons the EM current of the
corresponding isomultiplets and to look for their splitting into
isoscalar and isovector parts. As a result one finds the following
decomposition of the nucleon and $\Lambda-$, $\Sigma-$ and $\Xi-$
hyperon electric and magnetic FFs into isoscalar and isovector
parts of the Dirac and Pauli FFs
\begin{eqnarray}
G_{\rm Ep}(t)&=&[F_{1N}^{\rm s}(t)+F_{1N}^{\rm v}(t)]+
\frac{t}{4m^2_p}[F_{2N}^{\rm s}(t)+F_{2N}^{\rm v}(t)]  \label{sec28}\\
G_{\rm Mp}(t)&=&[F_{1N}^{\rm s}(t)+F_{1N}^{\rm v}(t)]+
[F_{2N}^{\rm s}(t)+F_{2N}^{\rm v}(t)] \nonumber \\
G_{\rm En}(t)&=&[F_{1N}^{\rm s}(t)-F_{1N}^{\rm v}(t)]+
\frac{t}{4m^2_n}[F_{2N}^{\rm s}(t)-F_{2N}^{\rm v}(t)]  \label{sec29}\\
G_{\rm Mn}(t)&=&[F_{1N}^{\rm s}(t)-F_{1N}^{\rm v}(t)]+
[F_{2N}^{\rm s}(t)-F_{2N}^{\rm v}(t)] \nonumber
\end{eqnarray}
\begin{eqnarray}
G_{\rm E\Lambda}(t)&=&F_{1\Lambda}^{\rm s}(t)+
\frac{t}{4m^2_{\Lambda}}F_{2\Lambda}^{\rm s}(t) \nonumber \\
G_{\rm M\Lambda}(t)&=&F_{1\Lambda}^{\rm s}(t)+ F_{2\Lambda}^{\rm
s}(t)\label{sec210}
\end{eqnarray}
\begin{eqnarray}
G_{\rm E\Sigma^+}(t)&=&[F_{1\Sigma}^{\rm s}(t)+F_{1\Sigma}^{\rm
v}(t)]+
\frac{t}{4m^2_{\Sigma^+}}[F_{2\Sigma}^{\rm s}(t)+F_{2\Sigma}^{\rm v}(t)] \nonumber \\
G_{\rm M\Sigma^+}(t)&=&[F_{1\Sigma}^{\rm s}(t)+F_{1\Sigma}^{\rm
v}(t)]+
[F_{2\Sigma}^{\rm s}(t)+F_{2\Sigma}^{\rm v}(t)] \nonumber \\
G_{\rm E\Sigma^0}(t)&=&F_{1\Sigma}^{\rm s}(t)+
\frac{t}{4m^2_{\Sigma^0}}F_{2\Sigma}^{\rm s}(t) \nonumber \\
G_{\rm M\Sigma^0}(t)&=&F_{1\Sigma}^{\rm s}(t)+
F_{2\Sigma}^{\rm s}(t) \\
\label{sec211} G_{\rm E\Sigma^-}(t)&=&[F_{1\Sigma}^{\rm
s}(t)-F_{1\Sigma}^{\rm v}(t)]+
\frac{t}{4m^2_{\Sigma^-}}[F_{2\Sigma}^{\rm s}(t)-F_{2\Sigma}^{\rm v}(t)] \nonumber \\
G_{\rm M\Sigma^-}(t)&=&[F_{1\Sigma}^{\rm s}(t)-F_{1\Sigma}^{\rm
v}(t)]+ [F_{2\Sigma}^{\rm s}(t)-F_{2\Sigma}^{\rm v}(t)] \nonumber
\end{eqnarray}
\begin{eqnarray}
G_{\rm E\Xi^0}(t)&=&[F_{1\Xi}^{\rm s}(t)+F_{1\Xi}^{\rm v}(t)]+
\frac{t}{4m^2_{\Xi^0}}[F_{2\Xi}^{\rm s}(t)+F_{2\Xi}^{\rm v}(t)] \nonumber \\
G_{\rm M\Xi^0}(t)&=&[F_{1\Xi}^{\rm s}(t)+F_{1\Xi}^{\rm v}(t)]+
[F_{2\Xi}^{\rm s}(t)+F_{2\Xi}^{\rm v}(t)]  \\
\label{sec212} G_{\rm E\Xi^-}(t)&=&[F_{1\Xi}^{\rm
s}(t)-F_{1\Xi}^{\rm v}(t)]+
\frac{t}{4m^2_{\Xi^-}}[F_{2\Xi}^{\rm s}(t)-F_{2\Xi}^{\rm v}(t)] \nonumber \\
G_{\rm M\Xi^-}(t)&=&[F_{1\Xi}^{\rm s}(t)-F_{1\Xi}^{\rm v}(t)]+
[F_{2\Xi}^{\rm s}(t)-F_{2\Xi}^{\rm v}(t)]. \nonumber
\end{eqnarray}

\medskip

 The experimental information on $G_{EB}(t)$, $G_{MB}(t)$ in $t<0$ region can be easily
determined from parameters of the straight-line (so-called
Rosenbluth) plot of
\begin{eqnarray}
&&\frac{d\sigma(e^- B\to e^- B)}{d\Omega}/\left \{
\frac{\alpha^2\cos^2(\vartheta/2)}{4E^2\sin^4(\vartheta/2)[1+(2E/m_B)\sin^2(\vartheta/2)]}\right \}= \nonumber \\
&=&A(t)+B(t)\tan^2(\theta/2),\label{sec213}
\end{eqnarray}
where
\begin{eqnarray}
 A(t)&=&\frac{G^2_{EB}(t)-\frac{t}{4m_B^2}G^2_{MB}(t)}{1-\frac{t}{4m_B^2}},\nonumber \\
 B(t)&=&-2\frac{t}{4m_B^2}G^2_{MB}(t)\label{sec214}
\end{eqnarray}
in the laboratory system versus $\tan^2(\vartheta/2)$ at fixed $t$
and for nuclei again diffraction minima appear.

   Till now all existing data on $G_{EB}(t)$, $G_{MB}(t)$ in  $t>0$
region are obtained  from $\sigma_{tot}(e^+e^-\leftrightarrow
B\bar B)$ under the assumption that $|G_{EB}(t)|=|G_{MB}(t)|$.

\smallskip

   The covariants $R_\mu(p_1, p_2)$ for EM FFs of the nonet of
vector mesons, $\rho^+$, $\rho^0$, $\rho^-$, $K^{*+}$, $K^{*0}$,
$\bar K^{*0}$, $K^{*-}$, $\omega$, $\Phi$ and also of the deuteron
are constructed by the four-momenta $p_1$, $p_2$ and polarization
vectors. Then a parametrization of the matrix element of the EM
current of vector-particles takes the following form
\begin{equation}
\langle p_2|J_{\mu}^{EM}|p_1\rangle = F_1(t)(\xi'^*\cdot
\xi)d_{\mu} + F_2(t)[\xi_{\mu}(\xi'^*\cdot q)-
\xi_{\mu}'^*(\xi\cdot q)]-F_3(t)\frac{(\xi\cdot q)(\xi'^*\cdot
q)}{2m_V^2}d_{\mu}, \label{sec215}
\end{equation}
where $\xi$ and $\xi'$ are polarization vectors for incoming and
outgoing particles of four-momenta $p_1$ and $p_2$, respectively
\begin{align*}
&\xi'\cdot p_2=0; \quad \xi\cdot p_1=0; \quad \xi'^2=-1; \quad \xi^2=-1;\\
&d_{\mu}=(p_2+p_1)_{\mu}; \quad q_{\mu}=(p_2-p_1)_{\mu}.
\end{align*}

   Practically, it is convenient to describe the EM structure of
vector particles by an analogue of the Sachs FFs of nucleons

\begin{eqnarray}
G_C(t)&=&F_1(t)-\frac{t}{6m_H^2}G_Q(t);\quad G_M(t)=F_2(t);\nonumber \\
G_Q(t)&=&F_1(t)-F_2(t)+(1-\frac{t}{4m_H^2})F_3(t)\label{sec216}
\end{eqnarray}
the names of which, the charge $G_C(t)$, the magnetic $G_M(t)$ and
the quadrupole $G_Q(t)$ FFs, are derived from the fact that their
static values correspond to the charge, magnetic and quadrupole
moment of the vector particles.

   One can determine all $G_C(t)$, $G_M(t)$, $G_Q(t)$ FFs from
$d\sigma/d\Omega$ of the $e^- V\to e^- V$ process provided that
polarized particles are used in the corresponding experiments.
Otherwise only the elastic structure functions $A(t)$ and $B(t)$
can be drawn out from
\begin{equation}\label{sigdeutlab}
\frac{d\sigma}{d\Omega}=\frac{\alpha^2E'\cos^2(\vartheta/2)}{4E^3\sin^4(\vartheta/2)}[A(t)
+B(t)\tan^2(\vartheta/2)],
\end{equation}
where
\begin{eqnarray}
A(t)&=&-\frac{t}{6m_V^2}(1-\frac{t}{4m_V^2})G_M^2(t)+G_C^2(t)+ \frac{t^2}{18m_V^4}G_Q^2(t)\nonumber \\
B(t)&=& -\frac{t}{3m_V^2}(1-\frac{t}{4m_V^2})^2G_M^2(t)\label{deutelsf}.
\end{eqnarray}

On the other hand, from
\begin{equation}\label{sec218}
\sigma_{tot}(e^+e^-\to V\bar V
)=\frac{\pi\alpha^2\beta^3_V}{3t}\left \{3|G_C(t)|^2 +
\frac{t}{m_V^2}\left[|G_M(t)|^2+
\frac{1}{6m_V^2}|G_Q(t)|^2\right]\right \}
\end{equation}
one can see immediately that it is not a single task to obtain any
experimental information on the corresponding EM FFs in $t>0$
region.

\medskip

For strongly interacting particles $h$ with $S >1$ a situation is
even more complicated and generally it is not solved up to now.

\medskip

     \subsection{Properties of electromagnetic form factors of hadrons}\label{II2}

\medskip

   Summarizing our knowledge about the experimental behavior of EM
FFs we come to a conclusion, that all of them have a similar
behavior in the shape. But they differ in the asymptotic behavior,
normalization, number of bumps corresponding to vector-meson
resonances and also in the shape and height of those bumps.

   A behavior of EM FFs is a matter of predictions of a strong
interaction dynamical theory. However, there is no such theory
able to predict a correct behavior of $|F_h(t)|$ for $-\infty <t
<+\infty $ up to now and only partial successes were achieved in
this direction.

   The great discovery in the elementary particle physics was a
revelation of the quark-gluon structure of hadrons and its direct
relation \cite{Matv,Brod1} to the asymptotic behavior of EM FFs to
be determined by a number of constituent quarks $n_q$ of the
hadron $h$ as follows
\begin{equation}
F_h(t)_{|t|\to\infty} \sim t^{1-n_q},\label{asym}
\end{equation}
which is in a qualitative agreement with existing experimental
data.

   On the other hand, it is well known that on the role of a true
dynamical theory of strong interactions QCD, the gauge-invariant
local quantum field theory of interactions of quarks and gluons,
is pretending. But, as a consequence of the asymptotic freedom of
QCD, in the framework of the perturbation theory, the latter is
able to reproduce \cite{Lep,Far,Efr} just the asymptotic behavior
(\ref{asym}) up to logarithmic corrections.

   Not even the nonperturbative QCD sum rules \cite{Shif} by means of
which a prediction \cite{Nest,Iof} of a behavior of EM FFs in a
restricted $t<0$ region is achieved, solve the problem of a
reconstruction of EM FFs in the framework of QCD completely.

   For a completeness we mention  also the chiral perturbation
approach \cite{Gass}, in the framework of which a correct behavior
of EM FFs of hadrons around the point $t=0$ is predicted. This is
very important to be mentioned as the chiral perturbation approach
is equivalent to QCD at low energies where the running coupling
constant $\alpha_s(t)$ takes large values and the PQCD is
nonapplicable.

   Summarizing, QCD (not even its equivalent form) gives no
quantitative predictions in the most important part of the
time-like ($4m^2_{\pi}<t\leq 4GeV^2$) region, where EM FFs are
already complex functions of $t$ and the $e^+e^-$ annihilation
experiments exhibit a nontrivial behavior of measured
cross-sections caused by a creation of various unstable
vector-meson states.

   Therefore for the present an appropriate phenomenological approach
based on the synthesis of the experimental fact of a creation of
vector-mesons in $e^+e^-$- annihilation processes into hadrons,
the asymptotic behavior (\ref{asym}) and the well-established
analytic properties, leading to the U\&A model of EM structure of
strongly interacting particles, is still the most successful way
in a global theoretical reconstruction of EM FFs of hadrons.

   The vector-meson creation in $e^+e^-$- annihilation processes
into hadrons is taken  into account by means of the
vector-meson-dominance (VMD) model given for isoscalar and
isovector parts of EM FFs by the relation
\begin{equation}
F_h^{s,v}(t)=\sum_{V=1}^n \frac{m^2_V}{(m^2_V - t)}(f_{Vh\bar
h}/f_V)\label{vmd}
\end{equation}
where $f_{Vh\bar h}$ and $f_V$ are the vector-meson-hadron and the
universal vector-meson coupling constants, respectively, and $m_V$
is the vector-meson mass.

   The analyticity consists in a hypothesis that all EM FFs are analytic
functions in the whole complex $t$- plane besides infinite number
of branch points on the positive real axis corresponding to normal
and anomalous thresholds.

\subsection{Vector-meson-dominance model for form factors of hadrons}\label{II3}

\medskip

   There are experimentally confirmed neutral vector-mesons \cite{Rpp} with
quantum numbers to be identical with photon. They have the isospin
either 0 or 1.

   On the other hand the EM current of hadrons is by a rotation in
the isospin-space transformed like the sum of isotopic scalar and
the third component of isotopic vector. The latter transformation
properties reflect well known fact of the non-conservation of the
isospin in the EM interactions and lead automatically to a
phenomenon, observed experimentally, that at the absorption and
creation of virtual photon by hadron the isospin value can be
changed by 1. Therefore the photon can be considered to be a
superposition of states with isospin value 0 (the isoscalar
photon) and isospin value 1 (the isovector photon). Then neutral
vector-mesons with quantum numbers $J^{PC}=1^{- -}$ differ from
the photon only by the mass and there is no obstacle to assume
that there are between virtual photons and vector-mesons
transitions with a definite probability determined by the coupling
constant
\begin{equation}
g_{\gamma {\rm V}}= e\frac{m^2_{\rm V}}{f_{\rm V}}, \label{gmmV}
\end{equation}
where $e$ is the electric charge , $m_{\rm V}$ is the mass of
vector-meson and $f_V$ is the so-called universal vector-meson
coupling constant.

\begin{figure}[t]
    \centering
    \scalebox{1.5}{\includegraphics{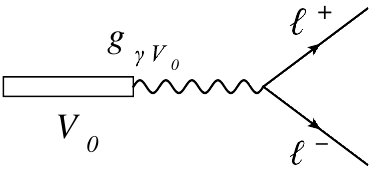}}
    \caption{\small{Decay of neutral vector meson into lepton pair.}}
    \label{fig1}
\end{figure}

   The transition of virtual photons to neutral vector mesons is
confirmed practically by the neutral vector-meson decay into
lepton-antilepton pair. Really, if we assume the transition $\gamma
\leftrightarrow V^0$, then the $V^0 \to \ell^+\ell^-$ decay can be
explained as a result of a change of $V^0$ into virtual photon
$\gamma^*$ which subsequently is creating the lepton-antilepton pair
(Fig.~\ref{fig1}).

   It seems to be natural a generalization of this mechanism to any
process of an interaction of photon with hadron, in which first
the photon is changed to vector-meson and then the latter is
interacting with the hadron like in other hadron collisions by
strong interactions.

\begin{figure}[tb]
    \centering
    \scalebox{0.6}{\includegraphics{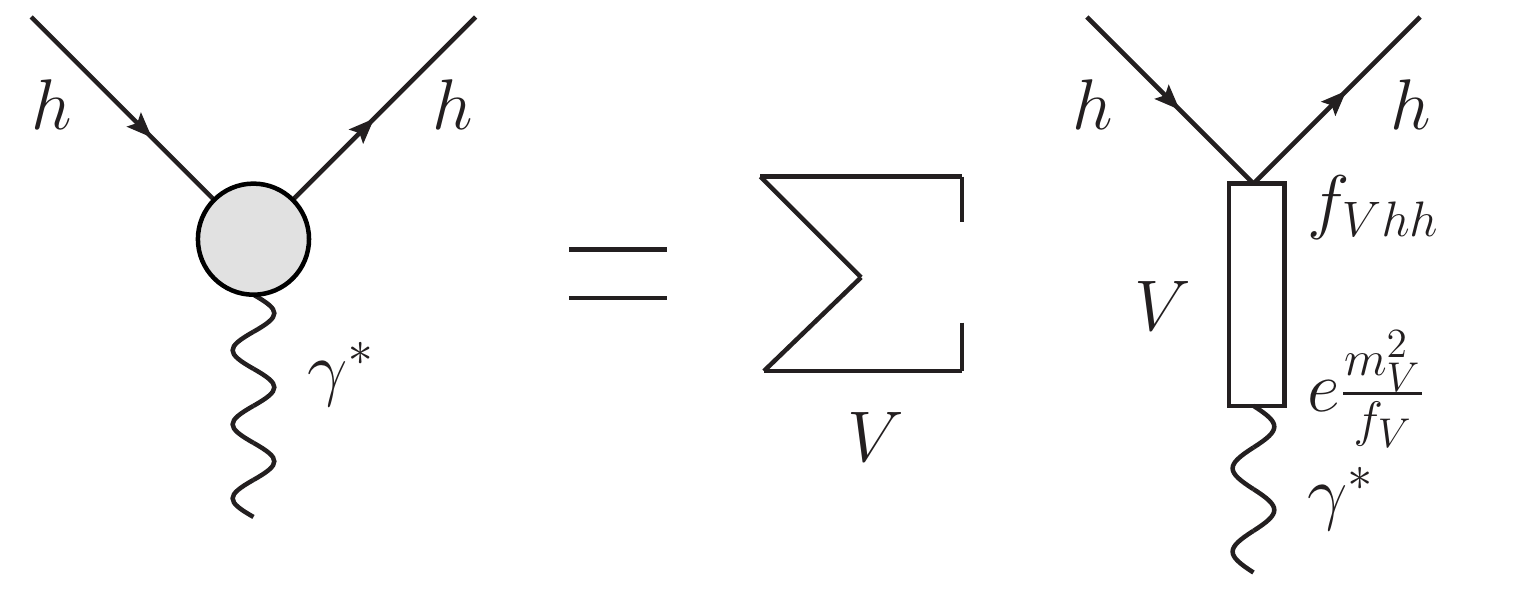}}
    \caption{\small{Approximation of hadron FF by a sum of VMD terms.}}
    \label{fig2}
\end{figure}

   In conformity with the idea of VMD model any EM FF of hadron in
the first approximation can be represented by a sum of Feynman
diagrams with and exchange of vector-mesons $V^0$ (Fig.~\ref{fig2}).

   By an application of standard methods of quantum field theory to
an explicit calculation of contributions of this sum of Feynman
diagrams one obtains
\begin{equation}
eF_h(t)= \sum_{\rm V} \frac{(g_{\gamma {\rm V}}.f_{{\rm V} h\bar
h})}{m^2_{\rm V}-t} \label{sec219}
\end{equation}
the VMD parametrization (in the zero width, i.e. $\Gamma_{\rm V}
=0$ approximation) of the hadron EM FF, which is, taking into
account also the relation (\ref{gmmV}), normalized for $t=0$ in
the form
\begin{equation}
F_h(0)= \sum_{\rm V} (f_{{\rm V}h\bar h}/f_{\rm V}) \label{sec220}
\end{equation}
and it possess the asymptotic behavior
\begin{equation}
F_h(t) \sim t^{-1}_{|t|\to\infty} \label{asymvmd}
\end{equation}
to be the same for EM FFs of all strongly interacting particles.

   The relation (\ref{gmmV}) can be derived by taking, first the
expression (\ref{sec219}) for the pion FF, the isovector part of
the kaon FF and the isovector part of the Dirac nucleon FF to be
saturated by only the lowest vector-meson, the $\rho$-meson. As a
result one can write the following three independent expressions

\begin{equation}
eF_\pi(t)= \frac{g_{\gamma \rho}.f_{\rho \pi \pi}}{m^2_\rho - t}\label{sec221}
\end{equation}
\begin{equation}
eF^v_K(t)= \frac{g_{\gamma \rho}.f_{\rho K\bar K}}{m^2_\rho - t}\label{sec222}
\end{equation}
\begin{equation}
eF^v_{1N}(t)= \frac{g_{\gamma \rho}.f^1_{\rho N\bar N}}{m^2_\rho -
t}\label{sec223}
\end{equation}
in which the normalization condition at $t = 0$ gives
\begin{equation}
e=\frac{g_{\gamma \rho}.f_{\rho \pi \pi}}{m^2_\rho}\label{sec224}
\end{equation}
\begin{equation}
\frac{e}{2}=\frac{g_{\gamma \rho}.f_{\rho K\bar K}}{m^2_\rho}\label{sec225}
\end{equation}
\begin{equation}
\frac{e}{2}=\frac{g_{\gamma \rho}.f^1_{\rho N\bar N}}{m^2_\rho}.\label{sec226}
\end{equation}
By a multiplication of the last two equations by 2, one can obtain
the following universal interaction of the $\rho$-meson to be
expressed by the relations
\begin{equation}
f_{\rho \pi \pi}=2f_{\rho K\bar K}=2f^1_{\rho N\bar N}=f_\rho,\label{sec227}
\end{equation}
where the coupling constant $f_\rho$ is named to be the universal
coupling constant of the $\rho$-meson and the coupling constant of
the $\rho$-meson with the virtual photon $g_{\gamma \rho}$ can be
written as follows
\begin{equation}
g_{\gamma \rho}=e\frac{m^2_\rho}{f_\rho}.\label{sec228}
\end{equation}

   Generalization of the previous relation to any other vector
meson V with quantum numbers of the photon leads to the form of
(\ref{gmmV}).

\medskip

          \subsection{Asymptotic conditions for form factors represented by VMD
          model} \label{II4}

\medskip

   As we have noticed above, from one side the quark model of hadrons
predicts (\ref{asym}) the asymptotic behavior of EM FF of hadron
to be dependent, in conformity with existing experimental data, on
the number of constituent quarks of considered hadron.

   On the other hand, the VMD model for EM FFs (\ref{II3}) of
strongly interacting particles gives the same asymptotic behavior
(\ref{asymvmd}) independently on the number of constituent quarks.

   This conflict is solved in this paragraph  \cite{Adam1} by
a derivation of the so-called asymptotic conditions. However,
there are known in practice two seemingly different asymptotic
conditions. Further we clearly demonstrate their equivalence.

   Generally, let us assume that the FF in (\ref{vmd}) is saturated by $n$
different vector meson pole terms and it has to have the
asymptotic behavior
\begin{equation}
F_h(t)_{|t|\to\infty} \sim t^{-m}, \label{asym2}
\end{equation}
where $m\le n$.

Transforming the VMD pole representation (\ref{vmd}) into a common
denominator one obtains FF in the form of a rational function with a
polynomial of $(m-1)$ degree
\begin{equation}
P_{n-1}(t)= A_0 + A_1\cdot t + A_2\cdot t^2+\cdots + A_{n-1}\cdot
t^{n-1} \label{polyn}
\end{equation}
in the numerator, where
\begin{eqnarray*}
A_{n-1}&=&(-1)^{n-1}\sum_{j=1}^nm_j^2a_j \nonumber \\
A_{n-2}&=&(-1)^{n-2}\sum_{i=1\atop i\ne j}^n m_i^2\sum_{j=1}^n m_j^2a_j \nonumber \\
A_{n-3}&=&(-1)^{n-3}\sum_{i_1,i_2=1\atop i_1< i_2,i_r\ne j}^n m_{i_1}^2m_{i_2}^2\sum_{j=1}^n m_j^2a_j \nonumber \\
A_{n-4}&=&(-1)^{n-4}\sum_{i_1,i_2,i_3=1\atop i_1< i_2<i_3,i_r\ne
j}^n m_{i_1}^2m_{i_2}^2m_{i_3}^2\sum_{j=1}^n m_j^2a_j \nonumber
\end{eqnarray*}
\begin{center}
\ldots \ldots \ldots
\end{center}
\begin{eqnarray}
A_{n-(m-1)}&=&(-1)^{n-m+1}\sum_{i_1,i_2,\cdots i_{m-2}=1\atop i_1< i_2\cdots <i_{m-2},i_r\ne j}^n m_{i_1}^2m_{i_2}^2\cdots m_{i_{m-2}}^2\sum_{j=1}^n m_j^2a_j \label{z5}\\
A_{n-m}&=&(-1)^{n-m}\sum_{i_1,i_2,\cdots i_{m-1}=1\atop i_1<
i_2\cdots <i_{m-1},i_r\ne j}^n m_{i_1}^2m_{i_2}^2\cdots
m_{i_{m-2}}^2m_{i_{m-1}}^2\sum_{j=1}^n m_j^2a_j \nonumber
\end{eqnarray}
\begin{center}
\ldots\ldots\ldots
\end{center}
\begin{eqnarray*}
A_2&=&(-1)^2\sum_{i_1,i_2,\cdots i_{n-3}=1\atop i_1< i_2\cdots <i_{n-3},i_r\ne j}^n m_{i_1}^2m_{i_2}^2\cdots m_{i_{n-3}}^2\sum_{j=1}^n m_j^2a_j   \\
A_1&=&(-1)\sum_{i_1,i_2,\cdots i_{n-2}=1\atop i_1< i_2\cdots <i_{n-2},i_r\ne j}^n m_{i_1}^2m_{i_2}^2\cdots m_{i_{n-3}}^2m_{i_{n-2}}^2\sum_{j=1}^n m_j^2a_j     \\
A_0&=&\sum_{i_1,i_2,\cdots i_{n-1}=1\atop i_1< i_2\cdots
<i_{n-1},i_r\ne j}^n m_{i_1}^2m_{i_2}^2\cdots
m_{i_{n-2}}^2m_{i_{n-1}}^2\sum_{j=1}^n m_j^2a_j
\end{eqnarray*}
and $a_j=(f_{jhh}/f_j)$.

In order to achieve the assumed asymptotic behavior (\ref{asym2})
one requires in (\ref{polyn}) the first $(m-1)$ coefficients from the
highest powers of $t$ to be zero and as a result the following
first system of linear homogeneous algebraic equations for the
coupling constant ratios is obtained
\begin{eqnarray}
& \sum\limits_{j=1}^n&\hspace{-0.8cm}m_j^2a_j=0 \nonumber \\
& \sum\limits_{i=1\atop i\ne j}^n&\hspace{-0.8cm}m_i^2\sum_{j=1}^nm_j^2a_j=0\label{z6}\nonumber  \\
& \sum\limits_{i_1,i_2=1\atop i_1<i_2, i_r\ne j}^n&\hspace{-0.8cm}m_{i_1}^2m_{i_2}^2\sum_{j=1}^nm_j^2a_j=0\\
& \sum\limits_{i_1,i_2,i_3=1\atop i_1<i_2<i_3, i_r\ne
j}^n&\hspace{-0.8cm}m_{i_1}^2m_{i_2}^2m_{i_3}^2\sum_{j=1}^nm_j^2a_j=0\nonumber
\end{eqnarray}

\hspace{3cm}\ldots\ldots\ldots

\begin{eqnarray}\nonumber \hspace{-0.3cm}
&\sum\limits_{i_1,i_2,\cdots i_{m-2}=1\atop i_1<i_2\cdots <i_{m-2}, i_r\ne j}^n&\hspace{-1.2cm}m_{i_1}^2m_{i_2}^2\cdots m_{i_{m-2}}^2\sum\limits_{j=1}^nm_j^2a_j=0.
\end{eqnarray}
As one can see from (\ref{z6}) with increased $m$ the coefficients
become sums of more and more complicated products of squared
vector-meson masses.

\medskip

   For a derivation of the second system we employ the assumed
analytic properties of EM FFs of hadrons, consisting of infinite
number of branch points on the positive real axis, i.e. cuts. The
first cut extends from the lowest branch point $t_0$ to $+\infty$.
Then one can apply the Cauchy theorem to FF in $t$- plane
\begin{equation}
\frac{1}{2\pi i}\oint F_h(t)dt=0 \label{z7}
\end{equation}
where the closed integration path consists of the circle $C_R$ of
the radius $R\to \infty$ and the path avoiding the cut on the
positive real axis. As a result (\ref{z7}) can be rewritten into a
sum of the following four integrals
\begin{equation}
\frac{1}{2\pi
i}\left\{\int_{C_R}F_h(t)dt+\int_{+\infty}^{t_0}F_h(t-i\epsilon)dt+
\int_{C_r/2}F_h(t)dt+\int^{+\infty}_{t_0}F_h(t+i\epsilon)dt
\right\}=0 \label{z8}
\end{equation}
where $\epsilon \ll 1$ and $C_r/2$ is the half-circle joining the
upper boundary of the cut with the lower-boundary of the cut
around the lowest branch point $t_0$. The contribution of the
first integral in (\ref{z8}) is zero as $F_h(t)$ for $R\to \infty$
is vanishing. One can prove also that the third integral in
(\ref{z8}) for $r\to 0$ is zero. As a result one gets
\begin{equation}
\frac{1}{2\pi i}\int_{t_0}^{\infty}[F_h(t+i\epsilon)-
F_h(t-i\epsilon)]dt=0. \label{z9}
\end{equation}
Then, taking into account the reality condition of FF
\begin{equation}
F_h^*(t)= F_h(t^*) \label{z10}
\end{equation}
following from the general Schwarz reflection principle in the
theory of analytic functions, one arrives at the integral
superconvergent sum rule
\begin{equation}
\frac{1}{\pi}\int_{t_0}^{\infty} Im F_h(t)dt=0 \label{z11}
\end{equation}
for the imaginary part of the FF under consideration.

   Repeating the same procedure for the functions $t F_h(t)$, $t^2
F_h(t)$, $\cdots$, $t^{m-2} F_h(t)$ which possess the same
analytic properties in the complex $t$-plane as $F_h(t)$, one gets
another $(m-2)$ superconvergent sum rules
\begin{eqnarray}
\frac{1}{\pi}\int_{t_0}^{\infty} t\cdot Im F_h(t)dt &=&0 \nonumber \\
\frac{1}{\pi}\int_{t_0}^{\infty} t^2\cdot Im F_h(t)dt &=&0
\label{z12}
\end{eqnarray}
$$
\ldots\ldots\ldots
$$
\begin{equation*}
\frac{1}{\pi}\int_{t_0}^{\infty} t^{m-2}\cdot Im F_h(t)dt =0.
\end{equation*}
Now, approximating the FF imaginary part by $\delta$- function in
the following form
\begin{equation}
ImF(t)=\pi \sum_i^n a_i \delta(t-m_i^2)m_i^2 \label{z13}
\end{equation}
and substituting it into (\ref{z11}) and (\ref{z12}) one obtains
the second system of $(m-1)$ linear homogeneous algebraic
equations for coupling constant ratios $a_i=(f_{ihh}/f_i)$
\begin{eqnarray}
& & \sum_{i=1}^n m_i^2a_i =0 \nonumber \\
& & \sum_{i=1}^n m_i^4a_i =0 \nonumber \\
& & \sum_{i=1}^n m_i^6a_i =0 \label{z14}
\end{eqnarray}

\hspace{2cm}\ldots\ldots\ldots

\begin{eqnarray*}
& & \sum_{i=1}^n m_i^{2(m-2)}a_i =0 \\
& & \sum_{i=1}^n m_i^{2(m-1)}a_i =0,
\end{eqnarray*}
where coefficients are simply even powers of the vector-meson
masses.

   Further we demonstrate explicitly that both systems of the algebraic
equations, (\ref{z6}) and (\ref{z14}), are equivalent, despite the
fact that they have been derived starting from different
properties of the EM FF, and thus they appear to be different.

   We start with the equations (\ref{z6}). From a direct comparison
of systems (\ref{z6})  and (\ref{z14}) one can see immediately the
identity of the first equations in them.

   The second equation in (\ref{z6}) can be written explicitly as
follows
\begin{eqnarray}
\nonumber
&\phantom{+}&(m_2^2+m_3^2+\cdots +m_n^2)m_1^2a_1+ (m_1^2+m_3^2+\cdots
+m_n^2)m_2^2a_2+\cdots +\\
&+&(m_1^2+m_2^2+\cdots +m_{n-1}^2)m_n^2a_n=0.
\label{z15}
\end{eqnarray}
Adding and subtracting $m_1^4a_1$ to the first term of the sum,
$m_2^4a_2$ to the second term of the sum $\cdots$ etc. and finally
$m_n^4a_n$ to the last term of the sum, the equation  (\ref{z15})
can be modified into the form
\begin{equation}
\sum_{i=1}^nm_i^2\sum_{j=1}^nm_j^2a_j- \sum_{j=1}^nm_j^4a_j=0,
\label{z16}
\end{equation}
from where one can see immediately that the second equation in
(\ref{z14}) is fulfilled
\begin{equation}
\sum_{j=1}^nm_j^4 a_j=0 \label{z17}
\end{equation}
as $\sum_{j=1}^nm_j^2 a_j=0$ is just the first equation in
(\ref{z6}) and (\ref{z14}) as well.

   The third equation in (\ref{z6}) can be written explicitly as
follows
\begin{eqnarray}
& &(m^2_2m_3^2+ m^2_2m_4^2+\cdots + m^2_2m_n^2+ m^2_3m_4^2+ m^2_3m_5^2+\cdots  \nonumber \\
&+& m^2_3m_n^2+\cdots +m^2_{n-1}m_n^2)m_1^2a_1+ \nonumber \\
&+&(m^2_1m_3^2+ m^2_1m_4^2+\cdots + m^2_1m_n^2+ m^2_3m_4^2+ m^2_3m_5^2+\cdots  \nonumber \\
&+& m^2_3m_n^2+\cdots +m^2_{n-1}m_n^2)m_2^2a_2+\nonumber \\
&+&(m^2_1m_2^2+ m^2_1m_4^2+\cdots + m^2_1m_n^2+ m^2_2m_4^2+ m^2_2m_5^2+\cdots  \nonumber \\
&+& m^2_2m_n^2+\cdots +m^2_{n-1}m_n^2)m_3^2a_3+
\nonumber
\end{eqnarray}

\hspace{2cm}\ldots \ldots \ldots

\begin{eqnarray}
&+&(m^2_1m_2^2+ m^2_1m_3^2+\cdots + m^2_1m_{n-2}^2+ m^2_1m_n^2+m^2_2m_3^2+\cdots  \nonumber \\
&+& m^2_2m_{n-2}^2+m^2_2m_n^2\cdots +\nonumber \\
&+& m^2_{n-2}m_n^2)m_{n-1}^2a_{n-1}+ \label{z18} \\
&+&(m^2_1m_2^2+ m^2_1m_3^2+\cdots + m^2_1m_{n-2}^2+ m^2_1m_{n-1}^2+ m^2_2m_3^2+\cdots  \nonumber \\
&+& m^2_2m_{n-2}^2+m^2_2m_{n-1}^2+\nonumber\\
&+&\cdots +m^2_{n-2}m_{n-1}^2)m_n^2a_n =0. \nonumber
\end{eqnarray}
Now adding and subtracting all missing terms in (\ref{z18}) from
$\sum_{{i_1},{i_2}=1\atop {i_1}<{i_2}}^nm_{i_1}^2m_{i_2}^2
\sum_{j=1}^nm_j^2a_j$ which in the substraction form can be
written explicitly as follows
\begin{eqnarray}
&-&(m_2^2+m_3^2+m_4^2+\cdots m_n^2)m_1^4a_1- \nonumber \\
&-&(m_1^2+m_3^2+m_4^2+\cdots m_n^2)m_2^4a_2- \nonumber \\
&-&(m_1^2+m_2^2+m_4^2+\cdots m_n^2)m_3^4a_3- \label{z19}
\end{eqnarray}

\hspace{2cm}\ldots \ldots \ldots

\begin{eqnarray*}
&-&(m_1^2+m_2^2+\cdots +m_{n-2}^2+ m_n^2)m_{n-1}^4a_{n-1}-  \\
&-&(m_1^2+m_2^2+\cdots +m_{n-2}^2+ m_{n-1}^2)m_n^4a_n
\end{eqnarray*}
and again subtracting and adding  $m_1^6a_1$ in the first line of
(\ref{z19}), $m_2^6a_2$ in the second line of (\ref{z19})...etc.,
and finally $m_n^6a_n$ in the last line of (\ref{z19}), one can
rewrite (\ref{z18}) into the form
\begin{equation}
\sum_{i_1,i_2=1\atop i_1<i_2}^n
m_{i_1}^2m_{i_2}^2\sum_{j=1}^nm_j^2a_j-\sum_{i=1}^nm_i^2\sum_{j=1}^nm_j^4a_j+\sum_{j=1}^nm_j^6a_j=0.
\label{z20}
\end{equation}
From this expression, taking into account the first two equations
in (\ref{z14}), the third equation of (\ref{z14})
\begin{equation}
\sum_j^n m_j^6a_j=0 \label{z21}
\end{equation}
follows.

   The fourth equation in (\ref{z6}) takes the following explicit
form
\begin{eqnarray}
& & (m^2_2m_3^2m_4^2+\cdots + m^2_2m_3^2m_n^2 + m^2_2m_4^2m_5^2+\cdots + m^2_2m_4^2m_n^2+\cdots  + \nonumber\\
&+&  m_{n-2}^2m_{n-1}^2m_n^2)m_1^2a_1+ \nonumber \\
&+& (m^2_1m_3^2m_4^2+\cdots + m^2_1m_3^2m_n^2 + m^2_1m_4^2m_5^2+\cdots + m^2_1m_4^2m_n^2+\cdots  + \nonumber\\
&+& m_{n-2}^2m_{n-1}^2m_n^2)m_2^2a_2+ \label{z22} \\
&+& (m^2_1m_2^2m_4^2+\cdots + m^2_1m_2^2m_n^2 + m^2_1m_4^2m_5^2+\cdots + m^2_1m_4^2m_n^2+\cdots  + \nonumber \\
&+& m_{n-2}^2m_{n-1}^2m_n^2)m_3^2a_3+ \nonumber
\end{eqnarray}

\hspace{2cm}\ldots \ldots\ldots

\begin{eqnarray*}
 &+&(m^2_1m_2^2m_3^2+\cdots + m_1^2m_2^2m_n^2 + m_1^2m_3^2m_4^2+\cdots + m^2_1m_3^2m_n^2+\cdots + \\
&+&m_{n-3}^2m_{n-2}^2m_n^2)m_{n-1}^2a_{n-1}+  \\
&+&(m^2_1m_2^2m_3^2+\cdots + m^2_1m_2^2m_{n-1}^2 + m^2_1m_3^2m_4^2+\cdots + m^2_1m_3^2m_{n-1}^2+\cdots  +  \\
&+&m_{n-3}^2m_{n-2}^2m_{n-1}^2)m_2^na_n=0.
\end{eqnarray*}

   First, adding and subtracting all missing terms in (\ref{z22})
from
$$\sum_{{i_1},{i_2},{i_3}=1\atop i_1< i_2< i_3}^nm_{i_1}^2m_{i_2}^2 m_{i_3}^2\sum_{j=1}^nm_j^2a_j,$$
the equation (\ref{z22}) takes the form
\begin{eqnarray}
& &\sum_{i_1,i_2,i_3=1\atop i_1<i_2<i_3}^n m_{i_1}^2m_{i_2}^2m_{i_3}^2\sum_{j=1}^nm_j^2a_j- \nonumber \\
&-&\sum_{i_1,i_2=1\atop i_1<i_2, i_r \ne
j}^nm_{i_1}^2m_{i_2}^2\sum_{j=1}^nm_j^4a_j=0. \label{z23}
\end{eqnarray}

   Second, subtracting and adding of all the missing terms in
(\ref{z23}) from $\sum_{{i_1},{i_2}=1\atop i_1<
i_2}^nm_{i_1}^2m_{i_2}^2$ $\sum_{j=1}^nm_j^4a_j$ one gets the
equation
\begin{eqnarray}
& &\sum_{i_1,i_2,i_3=1\atop i_1<i_2<i_3}^n m_{i_1}^2m_{i_2}^2m_{i_3}^2\sum_{j=1}^nm_j^2a_j- \label{z24} \\
&-&\sum_{i_1,i_2=1\atop
i_1<i_2}^nm_{i_1}^2m_{i_2}^2\sum_{j=1}^nm_j^4a_j+\sum_{i=1\atop
i\ne j}^nm_{i}^2\sum_{j=1}^nm_j^6a_j=0. \nonumber
\end{eqnarray}
Finally, additions and substractions of all missing terms in
(\ref{z24}) from $\sum_{i=1}^nm_i^2\sum_{j=1}^nm_j^6a_j$ lead to
the definitive form of the fourth equation in (\ref{z6})
\begin{eqnarray}
& &\sum_{i_1,i_2,i_3=1\atop i_1<i_2<i_3}^n m_{i_1}^2m_{i_2}^2m_{i_3}^2\sum_{j=1}^nm_j^2a_j- \label{z25} \\
&-&\sum_{i_1,i_2=1\atop
i_1<i_2}^nm_{i_1}^2m_{i_2}^2\sum_{j=1}^nm_j^4a_j+\sum_{i=1}^n
m_i^2\sum_{j=1}^n m_j^6a_j-\sum_{j=1}^nm_j^8a_j=0. \nonumber
\end{eqnarray}
From here, taking into account the first three equations in
(\ref{z14}), the fourth equation in (\ref{z14})
\begin{equation}
\sum_{j=1}^nm_j^8a_j=0 \label{z26}
\end{equation}
follows.

   It is now easy to give a straightforward generalization of the
above procedures
\begin{itemize}
\item[i)] the q-th equation in (\ref{z6}) can be decomposed into q-terms
(see (\ref{z16}), (\ref{z20}) and (\ref{z25})) consisting of the
product of two parts, where the first part is just the sum of
decreasing numbers of products of different vector-meson masses
squared, starting from (q-1) coefficients and ending with the
constant 1. The second term takes the form
$\sum_{j=1}^nm_j^{\alpha}a_j$ with increasing even power $\alpha$
starting from $\alpha=2$ up to 2q;
\item[ii)]
there is an alternating sign in front of every term in that
decomposition, while the first term is always positive.
\end{itemize}

Now, in order to carry out a general proof of the equivalence of
the two systems of algebraic equations under consideration, let us
assume an equivalence of $(m-2)$ equations in (\ref{z6}) and
(\ref{z14}). Then, taking into account a generalization of our
procedure defined by rules $i)$ and $ii)$ above, one can decompose
the $(m-1)$-equation in (\ref{z6}) into the following form
\begin{eqnarray}
& & \sum_{i_1,i_2,i_3,\cdots ,i_{m-2} =1\atop i_1<i_2<i_3<\cdots <i_{m-2}}^n m_{i_1}^2m_{i_2}^2\cdots m_{i_{m-2}}^2\sum_j^n m_j^2a_j- \nonumber \\
&-&\sum_{i_1,i_2,i_3,\cdots ,i_{m-3} =1\atop i_1<i_2<i_3<\cdots <i_{m-3}}m_{i_1}^2m_{i_2}^2\cdots m_{i_{m-3}}^2\sum_{j=1}^nm_j^4a_j+ \label{z27} \\
&+& \sum_{i_1,i_2,i_3,\cdots ,i_{m-4} =1\atop i_1<i_2<i_3<\cdots <i_{m-4}}^n m_{i_1}^2m_{i_2}^2\cdots m_{i_{m-4}}^2\sum_j^n m_j^6a_j+\cdots +\nonumber \\
&+&(-1)^{m-3} \sum_{i=1}^nm_i^2\sum_{j=1}^nm_j^{2(m-2)}a_j+
(-1)^{m-2} \sum_{j=1}^nm_j^{2(m-1)}a_j=0,\nonumber
\end{eqnarray}
from where one can see immediately that the $(m-1)$ equation in
(\ref{z14}) is satisfied
\begin{equation}
\sum_{j=1}^nm_j^{2(m-1)}a_j=0 \label{z28}
\end{equation}
as  $\sum_{j=1}^nm_j^2a_j=0$,  $\sum_{j=1}^nm_j^4a_j=0$, $\cdots$,
$\sum_{j=1}^nm_j^{2(m-2)}a_j=0$ are just the first $(m-2)$
equations in (\ref{z14}) assumed to be valid.

At the end we would like to draw an attention to the proof of the
equivalence of the systems of algebraic equations (\ref{z6}) and
(\ref{z14})  from the other point of view.

If the sums $\sum_{j=1}^nm_j^2a_j$,  $\sum_{j=1}^nm_j^4a_j$,
$\sum_{j=1}^nm_j^6a_j$, $\cdots$,  $\sum_{j=1}^nm_j^{2(m-3)}a_j$,
$\sum_{j=1}^nm_j^{2(m-2)}$ $a_j$, $\sum_{j=1}^nm_j^{2(m-1)}a_j$ are
considered to be independent variables, then the first equation in
(\ref{z6}) together with the modified forms (\ref{z16}),
(\ref{z20}), (\ref{z25}),..,(\ref{z27}) form a system of $(m-1)$
homogeneous algebraic equations for these variables and the
equations (\ref{z14}) are just its trivial solutions.

\medskip

       \subsection{General solution of asymptotic conditions} \label{II5}

\medskip

   In the previous paragraph we have derived two different systems
of $(m-1)$ linear homogenous algebraic equations for coupling
constants ratios, starting from different properties of EM FF
$F_h(t)$ of strongly interacting particles.

   In this paragraph, with regard to the proof of equivalence of
both systems, we shall be interested in (\ref{z14}) (the simpler
one of them), though it is derived in our opinion by incorrect way
by means of the superconvergent sum rules for the imaginary part
of the EM FF to be multiplied by the powers of the momentum
transfer squared. The coefficients in this system  are even powers
of the vector-meson masses. We find a general solution
\cite{Dubgs} of it, finally producing the VMD representation of
$F_h(t)$ with the required asymptotic behavior.

 First, we  look for a general solution of the asymptotic conditions to
be combined with the FF norm  when FF is saturated by more
vector-meson resonances than the  power determining the FF
asymptotics.

   If we assume that EM FF of any strongly interacting particle is
well approximated by a finite number $n$ of vector-meson exchange
tree Feynman diagrams (see Fig.~\ref{fig2}),  one finds the VMD pole
parametrization (\ref{vmd}) and its asymptotics (\ref{asym2}) is
required to be determined by the power $m$. The normalization of
(\ref{vmd}) at $t=0$ is
\begin{equation}
F_h(0)=F_0. \label{ro3}
\end{equation}

   The requirement for the conditions (\ref{ro3}) and (\ref{asym2}) to
be satisfied by (\ref{vmd}) (including also the results of ref.
\cite{Adam1}) leads to the following system of $m$ linear algebraic
equations
\begin{eqnarray}
\sum_{i=1}^n a_i&=&F_0  \label{ro4} \\
\sum_{i=1}^n m_i^{2r}a_i&=&0,  \quad\quad\quad r=1,2,...,m-1 \nonumber
\end{eqnarray}
for $n$ coupling constant ratios $a_i=(f_{ihh}/f_i)$. Therefore,
a solution of (\ref{ro4}) will be looked for  $m$ unknowns
$a_1$,...,$a_m$ and $a_{m+1}$,...,$a_n$ will be considered as free
parameters of the model. Then, the system (\ref{ro4}) can be
rewritten in the matrix form
\begin{equation}
\mathbf{M}\mathbf{a}=\mathbf{b}, \label{ro5}
\end{equation}
with the $m\times m$ Vandermonde matrix $\mathbf{M}$
\begin{equation}
\mathbf{M}=\begin{pmatrix}
1 & 1 & \hdots & 1\\
m_1^2 & m_2^2 & \hdots & m_m^2 \\
m_1^4 & m_2^4 & \hdots & m_m^4 \\
\hdotsfor 4 \\
m_1^{2(m-1)} & m_2^{2(m-1)} & \hdots &  m_m^{2(m-1)} \\
\end{pmatrix}\qquad                                        \label{ro6}
\end{equation}
and the column vectors
\begin{eqnarray}
\mathbf{a}=\begin{pmatrix} a_1\\a_2 \\a_3 \\
\vdots\\a_m\\\end{pmatrix},\qquad \;\;\;
\mathbf{b}=\begin{pmatrix}
F_0-\sum_{k=m+1}^na_k\\-\sum_{k=m+1}^nm^2_k a_k \\
-\sum_{k=m+1}^nm^4_k a_k
\\ \hdotsfor 1\\ -\sum_{k=m+1}^nm^{2(m-1)}_k a_k\\
\end{pmatrix}. \qquad
\label{ro7}
\end{eqnarray}
The Vandermonde determinant of the matrix (\ref{ro6}) is different
from zero
\begin{equation}
det\mathbf{M}=\prod^m_{\substack{j,l=1,\\ j<l }} (m_l^2-m_j^2).
\label{ro8}
\end{equation}
This has been proved explicitly by reducing the matrix (\ref{ro6})
to the triangular form and then taking into account the fact  that
the determinant of a triangular matrix is the product of its main
diagonal elements.

   As a consequence of (\ref{ro8}) a nontrivial solution of
(\ref{ro5}) exists. To find  the latter we use Cramer's Rule
despite  the fact that computationally Cramer's Rule for $m>3$
offers no advantages over the Gaussian elimination method.
However, in our case (as one can see further) all calculations are
for the most part reduced  to a calculation of the Vandermonde
type determinants,  and there is no problem to come to the
explicit solutions.

   So, the corresponding solutions of (\ref{ro5}) for $i=1,...,m$ are
\begin{equation}
a_i=\frac{det\mathbf{M_i}}{det\mathbf{M}} \label{ro9}
\end{equation}
where the matrix $\mathbf{M_i}$ takes the following form
\begin{equation}
\mathbf{M_i}=\begin{pmatrix} 1   \hdots & 1 & F_0- \sum_{\substack
{k=m+1}}^na_k & 1 & \hdots & 1\\ m_1^2   \hdots & m_{i-1}^2 & 0-
\sum_{\substack {k=m+1}}^n m_k^2a_k & m_{i+1}^2 & \hdots & m_m^2
\\ m_1^4  \hdots & m_{i-1}^4 & 0-
\sum_{\substack {k=m+1}}^n m_k^4a_k& m_{i+1}^4  & \hdots & m_m^4 \\ \hdotsfor 6\\
m_1^{2(m-2)}  \hdots  & m_{i-1}^{2(m-2)} &  0- \sum_{\substack {k=m+1}}^nm_k^{2(m-2)}a_k & m_{i+1}^{2(m-2)} & \hdots & m_m^{2(m-2)} \\
m_1^{2(m-1)}   \hdots & m_{i-1}^4  &  0-
\sum_{\substack {k=m+1}}^nm_k^{2(m-1)}a_k & m_{i+1}^{2(m-1)} & \hdots &  m_m^{2(m-1)} \\
\end{pmatrix}.\qquad           \label{ro10}
\end{equation}
Since any determinant is an additive function of each column,
 for each scalar $C$ we have
 \begin{equation*}
det(A_1,...,CA_i,...A_n)=Cdet(A_1,...A_i,...A_n)
\end{equation*}
 and
\begin{equation*}
det(A_1,..,A_{i-1},\sum_k x_kA_k,A_{i+1},...,A_n)=\sum_kx_k
det(A_1,...,A_{i-1},A_k,A_{i+1},...,A_n).\end{equation*}
 As a result, for a
determinant of the matrix $\mathbf{M_i}$ one can write the
decomposition
\begin{eqnarray}
& &det\mathbf{M_i}= \begin{vmatrix}
1 & 1 & \hdots & F_0  & \hdots & 1 \\
m_1^2 & m_2^2 & \hdots & 0 &\hdots & m_m^2 \\
m_1^4 & m_2^4 & \hdots & 0 & \hdots & m_m^4 \\
\hdotsfor 6 \\
m_1^{2(m-2)} & m_2^{2(m-2)} & \hdots & 0 & \hdots &  m_m^{2(m-2)} \\
m_1^{2(m-1)} & m_2^{2(m-1)} & \hdots & 0 & \hdots & m_m^{2(m-1)} \\
\end{vmatrix}\qquad \label{ro11}\\
& & - \sum_{k=m+1}^n a_k
\begin{vmatrix}
1 & 1 & \hdots & 1 & \hdots & 1 \\
m_1^2 & m_2^2 & \hdots & m_k^2 &\hdots & m_m^2 \\
m_1^4 & m_2^4 & \hdots & m_k^4 & \hdots & m_m^4 \\
\hdotsfor 6 \\
m_1^{2(m-2)} & m_2^{2(m-2)} & \hdots & m_k^{2(m-2)} & \hdots &  m_m^{2(m-2)} \\
m_1^{2(m-1)} & m_2^{2(m-1)} & \hdots & m_k^{2(m-1)} & \hdots & m_m^{2(m-1)} \\
\end{vmatrix}\qquad \nonumber
\end{eqnarray}
from where, if in the first determinant the Laplace expansion by
the entries of the column $i$ is used, the explicit form is
obtained
\begin{eqnarray}
det\mathbf{M_i}&=& F_0 (-1)^{1+i}\prod^m_{\substack{j=1\\j\neq
i}}m_j^2\prod^m_{\substack{j,l=1\\j<l,j,l\neq i}}(m_l^2-m_j^2)- \label{ro12} \\
&-&(-1)^{i-1}\prod^m_{\substack{j,l=1\\j<l, j,l\neq
i}}(m_l^2-m_j^2)\sum_{k=m+1}^na_k\prod^m_{\substack{j=1\\j\neq
i}}(m_j^2-m_k^2).\nonumber
\end{eqnarray}
Now, substituting (\ref{ro8}) and (\ref{ro12}) into (\ref{ro9}),
one gets the solutions of (\ref{ro5}) as follows
\begin{eqnarray}
a_i&=&\frac{F_0(-1)^{1+i}\prod^m_{\substack{j=1\\j\neq
i}}m_j^2\prod^m_{\substack{j,l=1\\j<l,j,l\neq
i}}(m_l^2-m_j^2)}{\prod^m_{\substack{j,l=1\\j<l}}(m_l^2-m_j^2)}-  \label{ro13} \\
&-&\frac{(-1)^{i-1}\prod^m_{\substack{j,l=1\\j<l,j,l\neq
i}}(m_l^2-m_j^2)\sum_{k=m+1}^na_k\prod^m_{\substack{j=1\\j\neq
i}}(m_j^2-m_k^2)}{\prod^m_{\substack{j,l=1\\j<l}}(m_l^2-m_j^2)}.
\nonumber
\end{eqnarray}

   In order to find, by means of (\ref{ro13}), an explicit form of
$F_h(t)$ to be automatically normalized  with the required
asymptotic behavior, let us separate the sum in (\ref{vmd}) into two
parts with the subsequent transformation of the first one into a
common denominator as follows
\begin{eqnarray}
F_h(t)&=& \sum_{i=1}^m\frac{m_i^2a_i}{m_i^2-t} +
\sum_{k=m+1}^n\frac{m_k^2a_k}{m_k^2-t}= \label{ro14} \\
&=& \frac{\sum_{i=1}^m\prod^m_{\substack{j=1\\j\neq
i}}(m_j^2-t)m_i^2a_i}{\prod^m_{j=1}(m_j^2-t)} +
\sum^n_{k=m+1}\frac{m_k^2a_k}{m_k^2-t}. \nonumber
\end{eqnarray}

Then (\ref{ro13}) together with (\ref{ro14}) gives
\begin{eqnarray}
&&F_h(t)=F_0\frac{\sum_{i=1}^m(-1)^{1+i}m_i^2\prod^m_{\substack{j=1\\j\neq
i}}m_j^2\prod^m_{\substack{j=1\\j\neq
i}}(m_j^2-t)\prod^m_{\substack{j,l=1\\j<l, j,l\neq
i}}(m_l^2-m_j^2)}{\prod^m_{j=1}(m_j^2-t)\prod^m_{\substack{j,l=1\\j<l}}(m_l^2-m_j^2)}-\nonumber
\\
&-& \frac{\sum_{i=1}^m(-1)^{i-1}m_i^2\prod^m_{\substack{j=1\\j\neq
i}}(m_j^2-t)\prod^m_{\substack{j,l=1\\j<l, j,l\neq
i}}(m_l^2-m_j^2)\sum_{k=m+1}^na_k\prod^m_{\substack{j=1\\j\neq
i}}(m_j^2-m_k^2)}{\prod^m_{j=1}(m_j^2-t)\prod^m_{\substack{j,l=1\\j<l}}(m_l^2-m_j^2)}+
\nonumber \\
&+& \sum_{k=m+1}^n\frac{m_k^2a_k}{m_k^2-t}. \label{ro15}
\end{eqnarray}
The first term in (\ref{ro15}) can be rearranged into the form
\begin{equation}
F_0\frac{\prod^m_{j=1}m_j^2}{\prod^m_{j=1}(m_j^2-t)}\frac{\sum_{i=1}^m(-1)^{1+i}
\prod^m_{\substack{j=1\\j\neq
i}}(m_j^2-t)\prod^m_{\substack{j,l=1\\j<l, j,l\neq
i}}(m_l^2-m_j^2)}{\prod^m_{\substack{j,l=1\\j<l}}(m_l^2-m_j^2)}\label{ro16}
\end{equation}
in which one can prove explicitly the identity
\begin{equation}
\sum_{i=1}^m(-1)^{1+i}\prod^m_{\substack{j=1\\j\neq
i}}(m_j^2-t)\prod^m_{\substack{j,l=1\\j<l, j,l\neq
i}}(m_l^2-m_j^2)\equiv
\prod^m_{\substack{j,l=1\\j<l}}(m_l^2-m_j^2) \label{ro17}
\end{equation}
leading to remarkable simplification of the term under
consideration as follows
\begin{equation}
F_0\frac{\prod^m_{j=1}m_j^2}{\prod^m_{j=1}(m_j^2-t)}. \label{ro18}
\end{equation}

One could prove (\ref{ro17}) by rewriting its left-hand side into
the following form
\begin{eqnarray}
& &\hspace{-0.6cm}\sum_{i=1}^m(-1)^{1+i}\times \nonumber \\
\hspace{-0.8cm}& &\hspace{-0.6cm}\begin{vmatrix}
(m_1^2-t)   \hdots &(m_{i-1}^2-t)  & (m_{i+1}^2-t) \hdots & (m_m^2-t) \\
m_1^2(m_1^2-t)   \hdots & m_{i-1}^2(m_{i-1}^2-t) & m_{i+1}^2(m_{i+1}^2-t) \hdots & m_m^2(m_m^2-t) \\
m_{1}^4(m_{1}^2-t)   \hdots & m_{i-1}^4(m_{i-1}^2-t) & m_{i+1}^4(m_{i+1}^2-t) \hdots & m_m^4(m_{m}^2-t)\\
\hdotsfor 4 \\
m_1^{2(m-3)}(m_{1}^2-t)   \hdots & m_{i-1}^{2(m-3)}(m_{i-1}^2-t) &m_{i+1}^{2(m-3)}(m_{i+1}^2-t) \hdots &  m_m^{2(m-3)}(m_{m}^2-t) \\
m_1^{2(m-2)}(m_{1}^2-t)   \hdots & m_{i-1}^{2(m-2)}(m_{i-1}^2-t) & m_{i+1}^{2(m-2)}(m_{i+1}^2-t) \hdots & m_m^{2(m-2)}(m_{m}^2-t) \\
\end{vmatrix}\qquad \nonumber \hspace{-0.7cm}\\
&& \label{ro19}
\end{eqnarray}
and then by using various basic properties of the determinants
decomposing it into the sum of large number of various
determinants of the same order with their subsequent explicit
calculations. Since this procedure seems to be, from the
calculational point of view, not simple, with the aim of a proving
(\ref{ro17}) let us define  the new matrix
\begin{equation}
\mathbf{D(t)}= \begin{pmatrix}
1& 1  \hdots & 1& \hdots & 1\\
(m_1^2-t) & (m_2^2-t)   \hdots &(m_{i}^2-t)  & \hdots & (m_m^2-t) \\
(m_1^2-t)^2  & (m_2^2-t)^2 \hdots & (m_{i}^2-t)^2 & \hdots & (m_m^2-t)^2 \\
(m_{1}^2-t)^3  & (m_2^2-t)^3\hdots & (m_{i}^2-t)^3 & \hdots & (m_{m}^2-t)^3\\
\hdotsfor 5 \\
(m_{1}^2-t^{m-2} & (m_2^2-t)^{m-2} \hdots & (m_{i}^2-t)^{m-2} & \hdots & (m_{m}^2-t)^{m-2} \\
(m_{1}^2-t)^{m-1} & (m_2^2-t)^{m-1} \hdots & (m_{i}^2-t)^{m-1} &\hdots & (m_{m}^2-t)^{m-1} \\
\end{pmatrix}.\qquad\label{ro20}
\end{equation}
Denoting $(m_i^2-t)$=$x_i$ one gets the Vandermonde matrix
\begin{equation}
\mathbf{D(t)}= \begin{pmatrix}
1& 1  \hdots & 1 & \hdots & 1\\
x_1 & x_2   \hdots &x_{i}  & \hdots & x_m \\
x_1^2  & x_2^2 \hdots & x_{i}^2 & \hdots & x_m^2 \\
x_{1}^3  & x_2^3\hdots & x_{i}^3 & \hdots & x_m^3\\
\hdotsfor 5 \\
x_{1}^{m-2} & x_2^{m-2} \hdots & x_{i}^{m-2} & \hdots & x_{m}^{m-2} \\
x_{1}^{m-1} & x_2^{m-1} \hdots & x_{i}^{m-1} &\hdots & x_{m}^{m-1} \\
\end{pmatrix}\qquad \label{ro21}
\end{equation}
the determinant of which is equal just to the right-hand side of
(\ref{ro17})
\begin{equation}
det\mathbf{D(t)}=\prod^m_{\substack{j,l=1\\j<l}}(x_l-x_j)\equiv
\prod^m_{\substack{j,l=1\\j<l}}(m_l^2-t-m_j^2+t)=\prod^m_{\substack{j,l=1\\j<l}}(m_l^2-m_j^2).
\label{ro22}
\end{equation}

   On the other hand, if in the determinant of the matrix
(\ref{ro20}) the Laplace expansion by the entries of the first row
with a subsequent pulling out of common factors in all columns of
the subdeterminants is carried out,  one gets the expression
\begin{eqnarray}
&&det\mathbf{D(t)}=\sum_{i=1}^m(-1)^{1+i}\prod^m_{\substack{j=1\\j\neq
i}}(m_j^2-t)\times  \label{ro23}\\
& &\begin{vmatrix}
 1  &\hdots & 1 & 1  \hdots & 1\\
(m_1^2-t) &   \hdots &(m_{i-1}^2-t)  & (m_{i+1}^2-t)\hdots & (m_m^2-t) \\
(m_1^2-t)^2  &  \hdots & (m_{i-1}^2-t)^2 &(m_{i+1}^2-t)^2 \hdots & (m_m^2-t)^2 \\
(m_{1}^2-t)^3  & \hdots & (m_{i-1}^2-t)^3 &(m_{i+1}^2-t)^3 \hdots & (m_{m}^2-t)^3\\
\hdotsfor 5 \\
(m_{1}^2-t)^{m-3} &  \hdots & (m_{i-1}^2-t)^{m-3} & (m_{i+1}^2-t)^{m-3}\hdots & (m_{m}^2-t)^{m-3} \\
(m_{1}^2-t)^{m-2} &  \hdots & (m_{i-1}^2-t)^{m-2} &(m_{i+1}^2-t)^{m-2}\hdots & (m_{m}^2-t)^{m-2} \\
\end{vmatrix}.\qquad \nonumber
\end{eqnarray}
Then calculating explicitly the determinant in (\ref{ro23})  by
using again the denotation $x_k=(m_k^2-t)$ for
$k=1,...,i-1,i+1,..m$,  one finally obtains
\begin{equation}
det\mathbf{D(t)}=\sum_{i=1}^{m}(-1)^{1+i}\prod^m_{\substack{j=1\\j\neq
i}}(m_j^2-t)\prod^m_{\substack{j,l=1\\j<l,j,l\neq i}}(m_l^2-m_j^2)
\label{ro24}
\end{equation}
just the left-hand side of (\ref{ro17}) and in this way the
identity under consideration is clearly proved.

   The second and third term in (\ref{ro15}), transforming them to a
common denominator, can be unified into one following term:
\begin{eqnarray}
& & \sum_{k=m+1}^n\biggl \{
\frac{m_k^2\prod^m_{j=1}(m_j^2-t)\prod^m_{\substack{j,l=1\\j<l}}(m_l^2-m_j^2)}
{(m_k^2-t)\prod_{j=1}^m(m_j^2-t)\prod_{\substack{j,l=1\\j<l}}^m(m_l^2-m_j^2)}+\label{ro25}
\\
&+&\frac{(m_k^2-t)\sum^m_{i=1}(-1)^im_i^2\prod^m_{\substack{j=1\\j\neq
i}}(m_j^2-m_k^2)\prod^m_{\substack{j,l=1\\j<l,j,l\neq
i}}(m_l^2-m_j^2)\prod^m_{\substack{j=1\\j\neq
i}}(m_j^2-t)}{(m_k^2-t)\prod_{j=1}^m(m_j^2-t)\prod^m_{\substack{j,l=1\\j<l}}(m_l^2-m_j^2)}\biggr
 \}a_k \nonumber
\end{eqnarray}
the numerator of which is exactly the Laplace  expansion by the
entries of the first row of the determinant of the matrix of the
$(m+1)$ order
\begin{equation}
\mathbf{N(t)}= \begin{pmatrix}
m_k^2 & m_1^2  & \hdots  & m_m^2\\
(m_k^2-t) & (m_1^2-t)      & \hdots & (m_m^2-t) \\
(m_k^2-t)^2  & (m_1^1-t)^2   & \hdots & (m_m^2-t)^2 \\
(m_k^2-t)^3  & (m_1^2-t)^3  & \hdots & (m_{m}^2-t)^3\\
\hdotsfor 4 \\
(m_k^2-t)^{m-1} & (m_1^2-t)^{m-1}   & \hdots & (m_{m}^2-t)^{m-1} \\
(m_k^2-t)^{m} & (m_1^2-t)^{m}  &\hdots & (m_{m}^2-t)^{m} \\
\end{pmatrix}.\qquad \label{ro26}
\end{equation}
If we define the new matrix of the $(m+1)$ order
\begin{equation}
\mathbf{R(t)}= \begin{pmatrix}
1& 1 & \hdots  & 1\\
(m_k^2-t) & (m_1^2-t)    & \hdots & (m_m^2-t) \\
(m_k^2-t)^2  & (m_1^2-t)^2  & \hdots & (m_m^2-t)^2 \\
(m_k^2-t)^3  & (m_1^2-t)^3  & \hdots & (m_{m}^2-t)^3\\
\hdotsfor 4 \\
(m_k^2-t)^{m-1} & (m_1^2-t)^{m-1}   & \hdots & (m_{m}^2-t)^{m-1} \\
(m_k^2-t)^{m} & (m_1^2-t)^{m}  &\hdots & (m_{m}^2-t)^{m} \\
\end{pmatrix},\qquad \label{ro27}
\end{equation}
then for the determinant of both matrices, (\ref{ro26}) and
(\ref{ro27}), the equation
\begin{equation}
det\mathbf{N(t)}- t\cdot det\mathbf{R(t)}\equiv det
\mathbf{S(t)}=0 \label{ro28}
\end{equation}
is fulfilled under the assumption  that $det\mathbf{S(t)}$ is
obtained by   multiplication of the first row of
$det\mathbf{R(t)}$ by $t$, and the substraction of the resultant
determinant from $\mathbf{N(t)}$ is carried out explicitly.

There is valid also a relation
\begin{equation}
det\mathbf{N(0)}=0 \label{ro29}
\end{equation}
as in $det\mathbf{N(0)}$ (like in $det\mathbf{S(t)}$) the first
two rows are identical.

   Now, in order to arrange the numerator of (\ref{ro25})
conveniently, we write  $det\mathbf{N(t)}$ in the form
\begin{equation}
det\mathbf{N(t)}=t\cdot det\mathbf{R(0)}-det\mathbf{N(0)},
\label{ro30}
\end{equation}
taking into account (\ref{ro28}),
(\ref{ro29}) and the identity
\begin{equation} det\mathbf{R(t)}\equiv det \mathbf{R(0)}.
\label{ro31}
\end{equation}
In (\ref{ro30}) we apply the Laplace expansion by entries of the
first row to $det\mathbf{R(0)}$ \\ and $det\mathbf{N(0)}$,
separately.
As a result, one gets
\begin{eqnarray}
& & det\mathbf{N(t)}=t\cdot \begin{vmatrix}
m_1^2 & m_2^2 &  \hdots & m_m^2 \\
m_1^4 & m_2^4  & \hdots & m_m^4 \\
\hdotsfor 4 \\
m_1^{2(m-1)} & m_2^{2(m-1)} & \hdots &   m_m^{2(m-1)} \\
m_1^{2m} & m_2^{2m} & \hdots  & m_m^{2m} \\
\end{vmatrix}\qquad + \label{ro32}\\
&+&\sum_{i=1}^m(-1)^it\begin{vmatrix}
m_k^2 & m_1^2  \hdots & m_{i-1}^2& m_{i+1}^2  & \hdots & m_m^2 \\
m_k^4 & m_1^4  \hdots & m_{i-1}^4 & m_{i+1}^4 &\hdots & m_m^4 \\
\hdotsfor 6 \\
m_k^{2(m-1)} & m_1^{2(m-1)}   \hdots & m_{i-1}^{2(m-1)} & m_{i+1}^{2(m-1)} & \hdots &  m_m^{2(m-1)} \\
m_k^{2m} & m_1^{2m}  \hdots &m_{i-1}^{2m} & m_{i+1}^{2m}  & \hdots & m_m^{2m} \\
\end{vmatrix}\qquad -\nonumber \\
&-& m_k^2 \begin{vmatrix}
m_1^2 & m_2^2 &  \hdots & m_m^2 \\
m_1^4 & m_2^4   &\hdots & m_m^4 \\
\hdotsfor 4 \\
m_1^{2(m-1)} & m_2^{2(m-1)} & \hdots &   m_m^{2(m-1)} \\
m_1^{2m} & m_2^{2m} & \hdots  & m_m^{2m} \\
\end{vmatrix}\qquad -\nonumber \\
&-&\sum_{i=1}^m(-1)^im_i^2\begin{vmatrix}
m_k^2 & m_1^2  \hdots & m_{i-1}^2& m_{i+1}^2  & \hdots & m_m^2 \\
m_k^4 & m_1^4  \hdots & m_{i-1}^4 & m_{i+1}^4 &\hdots & m_m^4 \\
\hdotsfor 6 \\
m_k^{2(m-1)} & m_1^{2(m-1)}   \hdots & m_{i-1}^{2(m-1)} & m_{i+1}^{2(m-1)} & \hdots &  m_m^{2(m-1)} \\
m_k^{2m} & m_1^{2m}  \hdots &m_{i-1}^{2m} & m_{i+1}^{2m}  & \hdots & m_m^{2m} \\
\end{vmatrix}\qquad \nonumber
\end{eqnarray}
or calculating explicitly the corresponding subdeterminants
\begin{eqnarray}
& &det \mathbf{N(t)}=
(t-m_k^2)\prod_{j=1}^mm_j^2\prod^m_{\substack{j,l=1\\j<l}}(m_l^2-m_j^2)+
\label{ro33} \\
&+& \sum_{i=1}^m(-1)^i(t-m_i^2)m_k^2\prod^m_{\substack{j=1\\
j\neq i}}m_j^2\prod^m_{\substack{j=1\\j\neq
i}}(m_j^2-m_k^2)\prod^m_{\substack{j,l=1\\j<l,j,l\neq
i}}(m_l^2-m_j^2).\nonumber
\end{eqnarray}
Substituting the latter into (\ref{ro25}) one obtains
\begin{equation}
\sum_{k=m+1}^n\biggl \{ -\frac{\prod_{j=1}^m
m_j^2}{\prod_{j=1}^m(m_j^2-t)}
+\sum_{i=1}^m\frac{m_k^2}{(m_k^2-t)}\frac{\prod_{\substack{j=1\\j\neq
i}}^mm_j^2}{\prod^m_{\substack{j=1\\j\neq
i}}(m_j^2-t)}\frac{\prod^m_{\substack{j=1\\j\neq
i}}(m_j^2-m_k^2)}{\prod^m_{\substack{j=1\\j\neq
i}}(m_j^2-m_i^2)}\biggr \}a_k \label{ro34}
\end{equation}
and combining  this result with (\ref{ro18}), one gets the form
factor $F_h(t)$ to be saturated by $n$-vector mesons $(n>m)$ in
the form suitable for the unitarization
\begin{eqnarray}
& &F_h(t)=F_0\frac{\prod^m_{j=1}m_j^2}{\prod^m_{j=1}(m_j^2-t)}+ \label{ff35} \\
&+& \sum_{k=m+1}^n\biggl\{\sum_{i=1}^m\frac{m_k^2}{(m_k^2-t)}
\frac{\prod_{\substack{j=1\\j\neq
i}}^mm_j^2}{\prod_{\substack{j=1\\j\neq
i}}^m(m_j^2-t)}\frac{\prod^m_{\substack{j=1\\j\neq
i}}(m_j^2-m_k^2)}{\prod^m_{\substack{j=1\\j\neq
i}}(m_j^2-m_i^2)}-\frac{\prod_{j=1}^mm_j^2}{\prod_{j=1}^m(m_j^2-t)}\biggr
\}a_k \nonumber
\end{eqnarray}
for which the asymptotic behavior (\ref{asym2}) and for $t=0$ the
normalization (\ref{ro3}) are fulfilled automatically.

The asymptotic behavior in (\ref{ff35}) is transparent. However, for
the normalization (\ref{ro3}) the following identity
\begin{equation}
\sum_{i=1}^m\frac{\prod^m_{\substack{j=1\\j\neq
i}}(m_j^2-m_k^2)}{\prod^m_{\substack{j=1\\j\neq
i}}(m_j^2-m_i^2)}=1 \label{ro36}
\end{equation}
has to be valid in the second term of (\ref{ff35}) generally.

   For $m=2,3,4,5$ it can be proved explicitly. And for an arbitrary
finite $m$ it follows directly from (\ref{ro25}), the numerator of
which is exactly the Laplace expansion by the entries of the first
row of the  determinant of the matrix (\ref{ro26}). Then, just
relation (\ref{ro29}) causes the term (\ref{ro25}) and also
(\ref{ro34}) at $t=0$ for arbitrary nonzero values of $a_k$ to be
zero. Hence, every term in the wave-brackets of (\ref{ro34}) for
$t=0$ has to be zero and this is true if and only if  identity
(\ref{ro36}) is fulfilled.

\medskip

   Now we consider the case of equations (\ref{ro4}) for $n=m$. Then
it can  also be rewritten into the matrix form (\ref{ro5}) with
the  $m\times m$ Vandermonde matrix $\mathbf{M}$ (\ref{ro6}) and
the same column vector $\mathbf{a}$, but with the $\mathbf{b}$
vector of the following form
\begin{equation}
\mathbf{b}=\begin{pmatrix} F_0\\0 \\0 \\
\vdots\\ 0 \\0 \\ \end{pmatrix}.\qquad \label{ro37}
\end{equation}
So, the corresponding solutions are again looked for in the form
$$
a_i=\frac{det\mathbf{M_i}}{det\mathbf{M}},
$$
with the matrix $\mathbf{M_i}$
\begin{equation}
\mathbf{M_i}=\begin{pmatrix} 1 &  \hdots & 1 & F_0  & 1 & \hdots &
1\\ m_1^2 &  \hdots & m_{i-1}^2 & 0 & m_{i+1}^2 & \hdots & m_m^2
\\ m_1^4 & \hdots & m_{i-1}^4 & 0 & m_{i+1}^4  & \hdots & m_m^4 \\ \hdotsfor 7\\
m_1^{2(m-2)} &  \hdots  & m_{i-1}^{2(m-2)} &  0 & m_{i+1}^{2(m-2)} & \hdots & m_m^{2(m-2)} \\
m_1^{2(m-1)} &  \hdots & m_{i-1}^{2(m-1)} & 0 & m_{i+1}^{2(m-1)} & \hdots &  m_m^{2(m-1)} \\
\end{pmatrix},\qquad           \label{ro38}
\end{equation}
and its determinant to be
\begin{equation}
det\mathbf{M_i}=F_0(-1)^{1+i}\prod^m_{\substack{j=1\\j\neq
i}}m_j^2\prod^m_{\substack{j,l=1\\j<l, j,l\neq i}}(m_l^2-m_j^2).
\label{ro39}
\end{equation}

   The solutions
\begin{eqnarray}
a_i&=&F_0 \frac{(-1)^{1+i}\prod^m_{\substack{j=1\\j\neq
i}}m_j^2\prod^m_{\substack{j,l=1\\j<l, j,l\neq
i}}(m_l^2-m_j^2)}{\prod^m_{\substack{j,l=1\\j<l}}(m_l^2-m_j^2)}=\nonumber
\\
&=& F_0\frac{\prod^m_{\substack{j=1\\j\neq
i}}m_j^2(-1)^{1+i}}{\prod^m_{\substack{j=1\\j\neq
i}}(m_j^2-m_i^2)(-1)^{i-1}}\label{ro40}
\end{eqnarray}
are then completely expressed through only the masses of $m$
vector-mesons  by means of which the considered FF is saturated.

   Substituting the solutions (\ref{ro40}) into the VMD
parametrization (\ref{vmd}) of the EM FF one comes to the following
representation
\begin{equation}
   F_h(t) = F_0 \frac{\prod^m_{\substack
   {j=1}}m_j^2}{\prod^m_{\substack{j=1}}(m_j^2-t)}\label{simpl}
\end{equation}
dependent only on the considered vector-meson masses and the
required asymptotic behavior (\ref{asym2}) is transparent to be
fulfilled automatically.

\medskip

   The third case with the $(m-1)$ linear homogeneous algebraic
equations for the $n$ $\quad\quad $ $(n>m)$ coupling constant ratios
without any normalization of FF appears naturally in the
determination of the so-called strange FF behaviors of a strongly
interacting particles with the spin $s>0$ from the isoscalar parts
of the corresponding EM FFs, as we shall see later on.

   Then, we have only the equations \begin{equation}
\sum_{i=1}^nm_i^{2r}a_i = 0, \quad\quad\quad r=1,2,...m-1 \label{ro41}
\end{equation}
which can be rewritten in the matrix form (\ref{ro5}) with the
$(m-1)\times(m-1)$ matrix $\mathbf{M}$
\begin{equation}
\mathbf{M}=\begin{pmatrix}
m_1^2 & m_2^2 & \hdots & m_{m-1}^2 \\
m_1^4 & m_2^4 & \hdots & m_{m-1}^4 \\
\hdotsfor 4 \\
m_1^{2(m-1)} & m_2^{2(m-1)} & \hdots &  m_{m-1}^{2(m-1)} \\
\end{pmatrix}\qquad                                        \label{ro42}
\end{equation}
and the column vectors
\begin{eqnarray}
\mathbf{a}=\begin{pmatrix} a_1\\a_2 \\a_3 \\
\vdots\\a_{m-1}\\\end{pmatrix},\qquad \;\;\;
\mathbf{b}=\begin{pmatrix}
-\sum_{k=m}^nm_k^2a_k\\-\sum_{k=m}^nm^4_k a_k \\
-\sum_{k=m}^nm^6_k a_k
\\ \hdotsfor 1\\ -\sum_{k=m}^nm^{2(m-1)}_k a_k\\
\end{pmatrix}. \qquad
\label{ro43}
\end{eqnarray}
The determinant of the matrix $\mathbf{M}$
\begin{equation}
det\mathbf{M}=\prod^{m-1}_{j=1}
m_j^2\prod^{m-1}_{\substack{j,l=1\\j<l}}(m_l^2-m_j^2) \label{ro44}
\end{equation}
is different from zero, and thus, a nontrivial solution of
(\ref{ro41}) exists
\begin{equation*}
a_i=\frac{det\mathbf{M_i}}{det\mathbf{M}}
\end{equation*}
where the matrix $\mathbf{M_i}$ takes the form
\begin{equation}
\mathbf{M_i}=\begin{pmatrix}
 m_1^2 & \hdots & m_{i-1}^2 &
-\sum_{k=m}^nm_k^2a_k & m_{i+1}^2 & \hdots & m_{m-1}^2
\\ m_1^4 & \hdots & m_{i-1}^4 & -\sum_{k=m}^nm_k^4a_k& m_{i+1}^4  & \hdots & m_{m-1}^4 \\ \hdotsfor 7\\
 m_1^{2(m-1)} &  \hdots & m_{i-1}^{2(m-1)}  & -\sum_{k=m}^nm_k^{2(m-1)}a_k & m_{i+1}^{2(m-1)} & \hdots &  m_m^{2(m-1)} \\
\end{pmatrix}.\qquad           \label{ro45}
\end{equation}
Then employing the basic properties of the determinants one gets
\begin{eqnarray}
&&det\mathbf{M_i}=\nonumber \\
&=&-\sum_{k=m}^na_k
\begin{vmatrix}
 m_1^2 & \hdots & m_{i-1}^2 &
m_k^2 & m_{i+1}^2 & \hdots & m_{m-1}^2\\
 m_1^4 & \hdots & m_{i-1}^4 & m_k^4 & m_{i+1}^4  & \hdots & m_{m-1}^4 \\
 m_1^6 & \hdots & m_{i-1}^6 & m_k^6 & m_{i+1}^6  & \hdots & m_{m-1}^6 \\
\hdotsfor 7\\
 m_1^{2(m-1)} &  \hdots & m_{i-1}^{2(m-1)}  & m_k^{2(m-1)} & m_{i+1}^{2(m-1)} & \hdots &  m_{m-1}^{2(m-1)} \\
\end{vmatrix}=\qquad           \nonumber \\
&=&-\sum_{k=m}^na_km_k^2\times\nonumber \\
&\times&\prod^{m-1}_{\substack{j=1\\j\neq i}}m_j^2
\begin{vmatrix} 1& \hdots& 1 & 1& 1&\hdots & 1\\
 m_1^2 & \hdots & m_{i-1}^2 &
m_k^2 & m_{i+1}^2 & \hdots & m_{m-1}^2
\\ m_1^4 & \hdots & m_{i-1}^4 & m_k^4 & m_{i+1}^4  & \hdots & m_{m-1}^4 \\
\hdotsfor 7\\
 m_1^{2(m-2)} &  \hdots & m_{i-1}^{2(m-2)}  & m_k^{2(m-2)} & m_{i+1}^{2(m-2)} & \hdots &  m_m^{2(m-2)} \\
\end{vmatrix}=\qquad          \nonumber \\
&=&-\sum_{k=m}^n m_k^2a_k\prod^{m-1}_{\substack{j=1\\j\neq
i}}m_j^2 \prod^{m-1}_{\substack{j,l=1\\j<l, j,l\neq
i}}(m_l^2-m_j^2)(-1)^{i-1} \prod^{m-1}_{\substack{j=1\\j\neq
i}}(m_j^2-m_k^2).\label{ro46}
\end{eqnarray}
As a result,
\begin{equation}
a_i=\frac{-\sum^n_{k=m} m_k^2a_k\prod^{m-1}_{\substack{j=1\\j \neq
i}}m_j^2\prod^{m-1}_{\substack{j,l=1\\j<l, j,l\neq
i}}(m_l^2-m_j^2)(-1)^{i-1}\prod^{m-1}_{\substack{j=1\\j\neq
i}}(m_j^2-m_k^2)}{\prod^{m-1}_{j=1}m_j^2\prod^{m-1}_{\substack{j,l=1\\j<l}}(m_j^2-m_l^2)}
\label{ro47}
\end{equation}
or finally,
\begin{equation}
a_i=-\sum_{k=m}^n\frac{m_k^2}{m_i^2}\frac{\prod^{m-1}_{\substack{j=1\\j
\neq i}}(m_j^2-m_k^2)}{\prod^{m-1}_{\substack{j=1\\ j\neq
i}}(m_j^2-m_i^2)}a_k, \;\;\; i=1,2...,m-1. \label{ro48}
\end{equation}
Now  substituting (\ref{ro48}) into
\begin{equation}
F_h(t)=\frac{\sum^{m-1}_{i=1}\prod^{m-1}_{\substack{j=1\\j\neq
i}}(m_j^2-t)m^2_ia_i}{\prod_{j=1}^{m-1}(m_j^2-t)} +
\sum_{k=m}^n\frac{m_k^2a_k}{m_k^2-t} \label{ro49}
\end{equation}
and transforming both terms into a common denominator one gets the
relation
\begin{eqnarray}
&&F_h(t)=\sum_{k=m}^n\biggl\{\frac{\prod^{m-1}_{j=1}(m_j^2-t)\prod^{m-1}_{\substack{j,l=1\\j<l}}(m^2_l-m^2_j)}
{\prod^{m-1}_{j=1}(m_j^2-t)\prod^{m-1}_{\substack{j,l=1\\j<l}}(m^2_l-m^2_j)}+
\label{ro50} \\
&+&\frac{(m_k^2-t)\sum_{i=1}^{m-1}(-1)^i\prod^{m-1}_{\substack{j=1\\j\neq
i}}(m_j^2-t)\prod^{m-1}_{\substack{j,l=1\\j<l, j,l\neq
i}}(m_l^2-m_j^2)\prod^{m-1}_{\substack{j=1\\j\neq
i}}(m_j^2-m_k^2)}{\prod^{m-1}_{j=1}(m_j^2-t)\prod^{m-1}_{\substack{j,l=1\\j<l}}(m^2_l-m^2_j)}\biggr\}\times\nonumber
\\
& &\times\frac{m_k^2}{m_k^2-t}a_k, \nonumber
\end{eqnarray}
in which the numerator of the first term under the sum is just the
Laplace expansion by the entries of the first row of the
determinant of the matrix $\mathbf{T(t)}$ of the  $m$ order
\begin{equation}
\mathbf{T(t)}= \begin{pmatrix}
1& 1   &\hdots  & 1\\
(m_k^2-t) & (m_1^2-t)   &  \hdots   &  (m_{m-1}^2-t) \\
(m_k^2-t)^2  & (m_1^2-t)^2   & \hdots & (m_{m-1}^2-t)^2 \\
\hdotsfor 4 \\
(m_{k}^2-t)^{m-1} & (m_1^2-t)^{m-1} & \hdots  & (m_{m-1}^2-t)^{m-1} \\
\end{pmatrix}.\qquad\label{ro51}
\end{equation}
One can  immediately prove that
\begin{equation}
det \mathbf{T(t)}\equiv det \mathbf{T(0)}. \label{ro52}
\end{equation}
Then calculating $det\mathbf{T(0)}$ explicitly
\begin{equation}
det\mathbf{T(0)}=\prod^{m-1}_{j=1}(m_j^2-m_k^2)\prod^{m-1}_{\substack{j,l=1\\j<l}}(m_l^2-m_j^2)
\label{ro53}
\end{equation}
and substituting the result into (\ref{ro50}) instead of the
numerator of the term in the wave-brackets, one finally obtains
the parametrization
\begin{equation}
F_h(t)=\sum_{k=m}^n\frac{\prod^{m-1}_{j=1}(m_j^2-m_k^2)}{\prod_{j=1}^{m-1}m_j^2}\frac{\prod^{m-1}_{j=1}m_j^2}
{\prod_{j=1}^{m-1}(m_j^2-t)}\frac{m_k^2}{m_k^2-t}a_k  \label{ro54}
\end{equation}
for which the asymptotic behavior (\ref{asym2}) is fulfilled
automatically.

   \subsection{Analytic properties of electromagnetic form factors}\label{II6}

   In principle there are two sources on the analytic properties
of FFs of strongly interacting particles.

   The first one resides in the exact proof of the analytic
properties of FFs starting form the first principles of the local
QFT. In this way the analytic properties of the pion FF were
proven \cite{Schw}, though at present days in connection with
quark-gluon structure of hadrons this pretentious proof is
possible to accept in such approximation, in which the considered
hadron can be brought into compatibility with the local quantum
field. Unfortunately, there are no exact proofs of the analytic
properties of FFs of any other hadrons, though they are utilized
practically for many years.

   There is a general belief that all EM FFs are analytic functions in
the complex plane of the momentum transfer squared $t$ besides a
cut from the lowest branch point $t_0$ on the real axis to
$+\infty$. The positions of the branch points are found by an
investigation of the analytic properties of Feynman diagrams
\cite{Land,Eden,Dub75} representing separate terms of a formal
series of EM FFs obtained in the framework of a quantum field
perturbation theory.

   As a consequence of the hadron EM current to be Hermitian, all EM
FFs are real on the real axis for $t < t_0$. Then by an
application of the Schwarz reflection principle to EM FFs one
finds the so-called reality condition
\begin{equation}
F_h^*(t)=F_h(t^*) \label{sec233}
\end{equation}
reflecting the reality of the EM FFs on the real axis below $t_0$
and the relation of values of EM FFs on the upper and lower
boundary of the cut
\begin{equation}
F_h^*(t+i\epsilon)=F_h(t^*-i\epsilon), \quad\quad\quad \epsilon \ll 1
\label{sec234}
\end{equation}
automatically.

   The discontinuity across the cut is given by the unitarity
condition
\begin{equation}
\frac{1}{2i}\{ \langle h\bar h|J_{\mu}(0)|0\rangle -\langle
0|J_{\mu}(0)|h\bar h \rangle^*\}=\sum_n\langle h\bar
h|T^+|n\rangle\langle n|J_{\mu}(0)|0\rangle, \label{sec235}
\end{equation}
where the sum in (\ref{sec235}) is carried out over a complete
set of intermediate states allowed by various conservation laws
and $T^+$ means Hermitian conjugate amplitudes.

   Moreover, just from the unitarity condition (\ref{sec235}) it follows that
there is an infinite number of branch points on the positive real
axis between $t=t_0$ and $+\infty$, which always correspond to a
new allowed intermediate state $n$ in (\ref{sec235}). In order to
fulfil the reality condition (\ref{sec233}), the cuts associated
with these branch points are chosen to be extended to $+\infty$
along the real axis.

   Further we demonstrate the main principles of investigation of
the analytic properties of Feynman diagrams.

   Individual terms of the formal series of the perturbation
theory possess the analytic properties which are in no
contradiction with first principles of the local QFT. An arbitrary
Feynman diagram can be represented by the integral
\begin{equation}
I_{\varepsilon}(t)=\int
d^{4}k_{1}...d^{4}k_{l}\frac{N}{\prod_{j=1}^{n}(q_{j}^2+m_{j}^2+i\varepsilon)},
\hspace{0.5cm} \varepsilon>0, \hspace{0.5cm} \varepsilon\ll1,
\label{Feyin}
\end{equation}

\smallskip

where $q_{j}$ is four-momentum of $j$ propagator at the diagram
and $m_{j}$ is the corresponding mass. $N$ at the numerator is
representing a spin structure of Feynman diagram and it does not
contribute to an appearance of any singularity. The variables
$k_{j}$ $(j =1,...,l)$ are four-momenta related to $l$ independent
loops to be chosen arbitrary, nevertheless with regard to the
conservation law of four-momenta.

   At the investigation of the analytic properties of Feynman diagrams
appears to be the most suitable the so-called $\alpha$-approach.
By denotation $q_{j}^2+m_{j}^2=u_{j}$ and an application of the
Feynman relation

\begin{equation}
\frac{1}{u_{1}...u_{n}}=(n-1)!\int_{0}^{1}
d\alpha_{1}...d\alpha_{n}\frac{\delta(1-\sum\alpha_{j})}
{[\sum_{j=1}^{n}\alpha_{j}u_{j}]^n}
\end{equation}

\smallskip

the integral (\ref{Feyin}) can be transformed into the following
form

\begin{equation}
I_{\varepsilon}(t)=(n-1)!\int_{0}^{1}
d\alpha_{1}...d\alpha_{n}\int
d^{4}k_{1}...d^{4}k_{l}\frac{N\delta(1-\sum\alpha_{j})}
{[\sum_{j=1}^{n}\alpha_{j}u_{j}+i\varepsilon]^n}.\label{ieps}
\end{equation}

   Now, to find singularities of an individual term of the
perturbation expansion means to scrutinize the analytic properties
of integrals of the type $I_{\varepsilon}(t)$ in (\ref{ieps}).

   The problem of a search of singularities of the general integral
(\ref{ieps}) has been reduced by Landau \cite{Land} and Cutkosky
\cite{Cutk} into a series of necessary conditions (so-called
Landau-Cutkosky equations) to be formulated in the following way.
The integral $I_{\varepsilon}(t)$ possesses a singularity:

\medskip

\begin{itemize}
\item
 if either $\alpha_{j}=0$ or $q_{j}^2=-m_{j}^2$ for every $j$ inner line;
\item
 $\sum_{i=1}^{n}\alpha_{i}q_{i}=0$ for every independent loop.
\end{itemize}
If one is restricted to a search of singularities only on the
physical sheet of the Riemann surface, then to the above-mentioned
conditions another one is joined in
\begin{itemize}
\item
 $\alpha_{j}$ have to be real positive numbers.
\end{itemize}

   From the first condition it follows, that at the diagram
generating a singularity either every inner four-momentum is on
the mass shell or the corresponding $\alpha_{j}$ is equal zero. In
the last case the four-momentum $q_{j}$ does not appear in the
second condition, so a presence of the inner line at the diagram
does not influence the created singularity. Or in other words, the
same singularity can be found from the Feynman diagram, in which
the inner line belonging to $\alpha_{j}$ at the primary diagram,
is substituted by point.

   Diagrams, obtained from the primary diagram by removing one or
more inner lines (the corresponding vertices are joined into one
point) are called to be reduced diagrams.

\begin{figure}[tb]
    \centering
    \scalebox{0.99}{\includegraphics{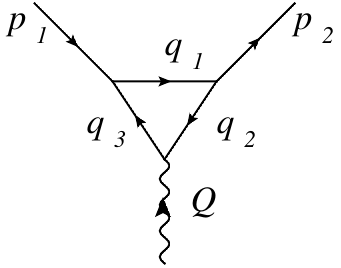}}
    \caption{Triangle diagram as reduced from primary diagram}
    \label{fig3}
\end{figure}
\begin{figure}[ht]
    \centering
    \scalebox{0.99}{\includegraphics{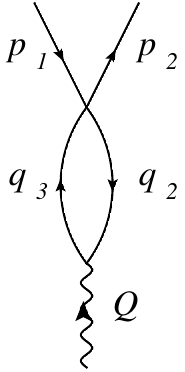}}
    \caption{\small{Two-point diagram as reduced from primary diagram}}
    \label{fig4}
\end{figure}

   Further there are described two types of Feynman diagrams, a triangle
presented in Fig.~\ref{fig3} and  a two-point diagram presented in
Fig.~\ref{fig4}. Both of them are creating branch points, the
triangle the anomalous and the two-point the normal (physical)
threshold. The two-point diagram can in another time appear to be
joined with some triangle or more-component diagram.

\medskip

   At the determination of the singularity one can start from the
two-point diagram (see Fig.~\ref{fig4}), from which it follows

\begin{equation}
q_{2}^{2}=-m_{2}^{2}, \hspace{1cm} q_{3}^{2}=-m_{3}^{2},
\hspace{1cm} \alpha_{2}q_{2}+\alpha_{3}q_{3}=0.
\end{equation}
By a multiplication of the last equation by $q_{2}$ and $q_{3}$
one after the other one obtains the system of algebraic equations

\begin{align}
-\alpha_{2}m_{2}^{2}+\alpha_{3}q_{2}q_{3}&=0 \nonumber \\
\alpha_{2}q_{2}q_{3}-\alpha_{3}m_{3}^{2}&=0,
\end{align}

\smallskip

which possesses nontrivial solution for $\alpha_{j}$ at that time
and only at that time when the determinant of the system is equal
zero, i.e.
\begin{equation}\label{sec236}
\det
\begin{pmatrix}
  -m_{2}^{2} & q_{2}q_{3} \\
  q_{2}q_{3} & -m_{3}^{2}
\end{pmatrix}
=0
\end{equation}

\smallskip

or $m_{2}^{2}m_{3}^{2}-(q_{2}q_{3})^2=0$.

   By utilization of the conservation law of the four-momenta
 $Q=p_{2}-p_{1}=q_{3}-q_{2}$ and  $t=-Q^{2}$ one gets
the expression

\begin{equation*}
q_{2}q_{3}=\frac{1}{2}[t-(m_{2}^{2}+m_{3}^{2})], \nonumber
\end{equation*}

\smallskip

which together with the relation for the determinant gives the
following quadratic equation

\begin{equation}
t^{2}-2(m_{2}^{2}+m_{3}^{2})t+(m_{2}^{2}-m_{3}^{2})^2=0.\label{sec237}
\end{equation}

\smallskip

Its solution gives

\begin{equation}
t=(m_{2}\pm m_{3})^2,\label{sec238}
\end{equation}

\smallskip

in which only the root with the positive sign fulfills the third
condition. As a result the two-point diagram in Fig.~\ref{fig4} creates the
branch point on the physical sheet of the Riemann surface at
$t=(m_{2}+ m_{3})^2$ to be known as the normal threshold.

\medskip

   Further we are interested in the triangle diagram in Fig.~\ref{fig3}, where

\begin{equation*}
q_{1}^{2}=-m_{1}^{2},\hspace{1cm} q_{2}^{2}=-m_{2}^{2},
\hspace{1cm} q_{3}^{2}=-m_{3}^{2}
\end{equation*}
and
\begin{equation}
\alpha_{1}q_{1}+\alpha_{2}q_{2}+\alpha_{3}q_{3}=0. \label{k116}
\end{equation}

   From the last equation by a subsequent multiplication by
$q_{1}$, $q_{2}$ a $q_{3}$ one obtains the system of three
algebraic equations

\begin{eqnarray}
-\alpha_{1}m_{1}^{2}+\alpha_{2}q_{1}q_{2}+\alpha_{3}q_{1}q_{3}&=&0\nonumber\\
\alpha_{1}q_{1}q_{2}-\alpha_{2}m_{2}^{2}+\alpha_{3}q_{2}q_{3}&=&0 \label{k117} \\
\alpha_{1}q_{1}q_{3}+\alpha_{2}q_{2}q_{3}-\alpha_{3}m_{3}^{2}&=&0,\nonumber
\end{eqnarray}

\smallskip

possessing a nontrivial solution only in the case of the zero
determinant
\begin{equation}
\det
\begin{pmatrix}
  -m_{1}^{2} & q_{1}q_{2} & q_{1}q_{3} \\
  q_{1}q_{2} & -m_{2}^{2} & q_{2}q_{3} \\
  q_{1}q_{3} & q_{2}q_{3} & -m_{3}^{2}\label{det0}
\end{pmatrix}
=0,
\end{equation}

\smallskip

or
\begin{equation}
-m_{1}^{2}m_{2}^{2}m_{3}^{2}+2(q_{1}q_{2})(q_{1}q_{3})(q_{2}q_{3})+
m_{1}^{2}(q_{2}q_{3})^{2}+m_{2}^{2}(q_{1}q_{3})^{2}+m_{3}^{2}(q_{1}q_{2})^{2}=0.
\nonumber
\end{equation}

   By utilization of the four-momenta conservation laws at the
corresponding vertices at the triangle diagram

\begin{align}
p_{2}-p_{1}&=q_{3}-q_{2}\nonumber \\
p_{1}&=q_{1}-q_{3} \label{k122}\\
p_{2}&=q_{1}-q_{2}\nonumber
\end{align}

\smallskip

and the relation $p_{1}^{2}=p_{2}^{2}=-M^{2}$ one obtains for
scalar products of four-momenta appearing at the determinant

\begin{align}
(q_{2}q_{3})&=\frac{1}{2}[t-(m_{2}^{2}+m_{3}^{2})] \nonumber \\
(q_{1}q_{3})&=\frac{1}{2}[M^{2}-(m_{1}^{2}+m_{3}^{2})] \nonumber \\
(q_{1}q_{2})&=\frac{1}{2}[M^{2}-(m_{1}^{2}+m_{2}^{2})], \nonumber
\end{align}

\smallskip

on the base of which the condition (\ref{det0}) leads to the
quadratic equation
\begin{eqnarray}
&&m_{1}^{2}t^{2}+\Big\{[M^{2}-(m_{1}^{2}+m_{2}^{2})][M^{2}-(m_{1}^{2}+m_{3}^{2})]
-2m_{1}^{2}(m_{2}^{2}+m_{3}^{2})\Big\}t+ \nonumber \\
&+&m_{1}^{2}(m_{2}^{2}-m_{3}^{2})^{2}+m_{2}^{2}[M^{2}-(m_{1}^{2}+m_{3}^{2})]^{2}+
m_{3}^{2}[M^{2}-(m_{1}^{2}+m_{2}^{2})]^{2}-\label{mmm} \\
&-&(m_{2}^{2}+m_{3}^{2})[M^{2}-(m_{1}^{2}+m_{2}^{2})][M^{2}-(m_{1}^{2}+m_{3}^{2})]=0.
\nonumber
\end{eqnarray}

Its solution takes the form

\begin{equation*}
t=(m_{2}^{2}+m_{3}^{2})-\frac{1}{2m_{1}^{2}}[M^{2}-(m_{1}^{2}+m_{2}^{2})]
[M^{2}-(m_{1}^{2}+m_{3}^{2})]
\end{equation*}
\begin{equation}
\pm
\frac{1}{2m_{1}^{2}}\sqrt{4m_{1}^{2}m_{2}^{2}-[M^{2}-(m_{1}^{2}+m_{2}^{2})]^{2}}
\sqrt{4m_{1}^{2}m_{3}^{2}-[M^{2}-(m_{1}^{2}+m_{3}^{2})]^{2}},\label{k127}
\end{equation}

\smallskip

determining singularities of EM FFs from the triangle diagram to
be called the anomalous thresholds. The latter, unlike the normal
thresholds, do not correspond to some physical processes.

   The anomalous thresholds are found on the physical sheet of the
Riemann surface only in the case of the real positive solutions
$\alpha_{j}$ of the system of equations (\ref{k117}). This
condition is fulfilled by the solution (\ref{k127}) with the
positive sign.

\begin{figure}[th]
\centering
    \scalebox{0.99}{\includegraphics{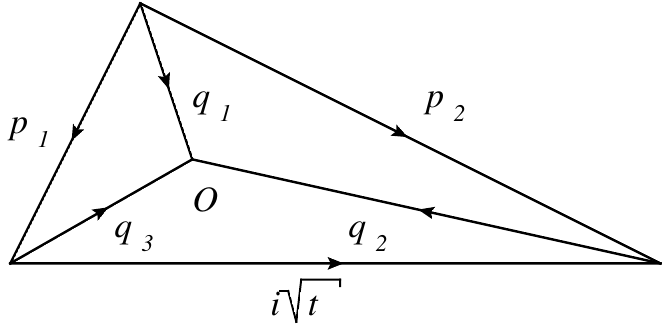}}
    \caption{The dual diagram to triangle  Feynman diagram in Fig.~\ref{fig3}}
    \label{fig5}
\end{figure}

   The most effective approach for a determination of the position of the
anomalous threshold of the triangle diagram seems to be a
geometrical way of a solution of the Landau-Cutkosky equations. It
consists in the following. If  $Q$ is expressed as
$i\sqrt{t}$, then from the four-momenta conservation law
(\ref{k122}) follows

\begin{equation*}
p_{1}+i\sqrt{t}=p_{2},
\end{equation*}
\\
to be pictured graphically in Fig.~\ref{fig5} and called a dual diagram to
Fig.~\ref{fig3}.

   Every inner triangle on Fig.~\ref{fig5} represents the four-momenta
conservation law at the vertices of the diagram on Fig.~\ref{fig3}.
From its form one can reveal wether the singularity is placed on the
physical sheet or on one of the un-physical sheets of the Riemann
surface.

   The Landau-Cutkosky equation (\ref{k116}), expressing the linear
dependence among $q_{1}, q_{2}, q_{3},$, requires all three vectors
to be in one plane. The lengths of $q_{j}$ are equal to the
corresponding masses $m_{j}$ and the lengths of $p_{1}$ a $p_{2}$ to
the mass M. From the latter there is unambiguous boundary on the
form of the outer triangle in Fig.~\ref{fig5}.

   The anomalous threshold is determined by an expression for
$t$ in Fig.~\ref{fig5} to be found by the methods of elementary
geometry.

   From the form of the dual diagram one can conclude, whether the
anomalous threshold is located on the physical sheet or on the one
of the un-physical sheets of the Riemann surface.

   The condition of a positivity of the parameters $\alpha_{j}$,
securing for the singularity to be on the physical sheet, leads the
point $0$ to be inside of the outer triangle in Fig.~\ref{fig5}. If
the point $0$ is outside of the outer triangle, the corresponding
anomalous singularity is placed on some un-physical sheet.

   So, the position of the anomalous threshold of EM FF on
the physical sheet of the Riemann surface, generating by the
triangle diagram, is numerically determined by (\ref{k127}) with
the positive sign.

\medskip

   \subsection{Unitary and Analytic model of electromagnetic structure of
   hadrons}\label{II7}

\medskip

   In the previous paragraphs we have solved the conflict between
the uniform asymptotic behavior of the VMD model (\ref{asymvmd}) and the
asymptotic behavior (\ref{asym}) of EM FFs as predicted by the quark
model of strongly interacting particles by finding the three VMD
parametrizations  (\ref{ff35}), (\ref{simpl}) and (\ref{ro54})
fulfilling the quark model asymptotic behavior (\ref{asym})
automatically. Every of them are compound of the products of the
resonant terms
\begin{equation}
\frac{m^2_r}{m^2_r-t} \label{pole}
\end{equation}
which further will be unitarized by an incorporation of the well
known analytic properties of EM FFs consisting of an infinite
number of branch points on the positive real axis. They are branch
points corresponding to normal (given by the unitarity condition)
and anomalous (given by allowed triangle diagrams) thresholds
generating many-sheeted Riemann surface. The first sheet is called
physical, all other sheets of the Riemann surface are unphysical.

   Practically, further we are restricting ourselves to a
two-square-root-cut approximation of the analytic properties
generating the four-sheeted Riemann surface. Then any resonant
stay is always associated with a pair of complex conjugate poles
on unphysical sheets to be generated by the branch points on the
positive real axis of the t-plane.

   In order to transform (\ref{pole}) into one analytic function
\begin{itemize}
\item[i\raisebox{-1mm}{)}]
with two square-root branch points on the positive real axis,
\item[ii\raisebox{-1mm}{)}]
with two pairs of complex conjugate poles on unphysical sheets
corresponding to the resonance $r$,
\end{itemize}
one proceeds as follows
\begin{itemize}
\item[{\normalfont\bfseries \textendash}]
first, the nonlinear transformation
\begin{equation}
t = t_0 + \frac{4(t_{in}-t_0)}{[1/W(t)-W(t)]^2} \label{texp}
\end{equation}
with $t_0$~-~the square-root branch point corresponding to the
lowest threshold and $t_{in}$~-~an effective square-root branch
point simulating contributions of all other relevant thresholds
given by the unitarity condition is applied, which automatically
generates the relations
\begin{equation}
m^2_r = t_0 + \frac{4(t_{in}-t_0)}{[1/W_{r0}-W_{r0}]^2}
\label{m2exp}
\end{equation}
and
\begin{equation}
0 = t_0 + \frac{4(t_{in}-t_0)}{[1/W_N-W_N]^2} \label{0exp}
\end{equation}
\item[{\normalfont\bfseries \textendash}]
then relations between $W_{r0}$ and $W^*_{r0}$ are utilized
\item[{\normalfont\bfseries \textendash}]
and finally, the instability of the resonance is introduced by its
non-zero width $\Gamma_r \neq 0$.
\end{itemize}

   The application of the (\ref{texp})-(\ref{0exp}) to (\ref{pole})
leads to the following factorized form
\begin{equation}\label{basefact}
\frac{m^2_r}{m^2_r-t} =\left(\frac{1-W^2}{1-W^2_N}\right)^2
\frac{(W_N-W_{r0})(W_N+W_{r0})(W_N-1/W_{r0})(W_N+1/W_{r0})}
{(W-W_{r0})(W+W_{r0})(W-1/W_{r0})(W+1/W_{r0})}
\end{equation}
with asymptotic term (the first term, completely determining the
asymptotic behavior of (\ref{pole})) and on the so-called
finite-energy term (for $|t|\to\infty$ it turns out to be a real
constant) giving a resonant behavior around $t=m^2_r$.

   One can prove
\begin{eqnarray}
{\rm a\raisebox{-1mm}{)}~~~if~~} m^2_r-\Gamma^2_r/4 < t_{in} &\Rightarrow& W_{r0} = - W^*_{r0} \nonumber \\
 & & \\
{\rm b\raisebox{-1mm}{)}~~~if~~} m^2_r-\Gamma^2_r/4 > t_{in}
&\Rightarrow&  W_{r0} =  1/W^*_{r0} \nonumber
\end{eqnarray}
which lead the eq. (\ref{basefact}) in the case
a\raisebox{-1mm}{)} to the expression
\begin{equation}\label{afact}
\frac{m^2_r}{m^2_r-t} =\left(\frac{1-W^2}{1-W^2_N}\right)^2
\frac{(W_N-W_{r0})(W_N-W^*_{r0})(W_N-1/W_{r0})(W_N-1/W^*_{r0})}
{(W-W_{r0})(W-W^*_{r0})(W-1/W_{r0})(W-1/W^*_{r0})}
\end{equation}
and in the case b\raisebox{-1mm}{)} to the following expression
\begin{equation} \label{bfact}
\frac{m^2_r}{m^2_r-t}=\left(\frac{1-W^2}{1-W^2_N}\right)^2
\frac{(W_N-W_{r0})(W_N-W^*_{r0})(W_N+W_{r0})(W_N+W^*_{r0})}
{(W-W_{r0})(W-W^*_{r0})(W+W_{r0})(W+W^*_{r0})}.
\end{equation}

   Lastly, introducing the non-zero width of the resonance by a
substitution
\begin{equation}
m^2_r\to (m_r-\Gamma_r/2)^2
\end{equation}
i.e. simply one has to rid of "0" in sub-indices of (\ref{afact})
and
(\ref{bfact}), one gets:
in a\raisebox{-1mm}{)} case
\begin{eqnarray} \label{aafact}
\frac{m^2_r}{m^2_r-t} &\to&
\left(\frac{1-W^2}{1-W^2_N}\right)^2
\frac{(W_N-W_r)(W_N-W^*_r)(W_N-1/W_r)(W_N-1/W^*_r)}
{(W-W_r)(W-W^*_r)(W-1/W_r)(W-1/W^*_r)} =  \nonumber\\
& =& \left(\frac{1-W^2}{1-W^2_N}\right)^2 L(W_r)
\end{eqnarray}
and in b\raisebox{-1mm}{)} case
\begin{eqnarray} \label{bbfact}
\frac{m^2_r}{m^2_r-t} &\to&
\left(\frac{1-W^2}{1-W^2_N}\right)^2
\frac{(W_N-W_r)(W_N-W^*_r)(W_N+W_r)(W_N+W^*_r)}
{(W-W_r)(W-W^*_r)(W+W_r)(W+W^*_r)} = \nonumber\\
& =& \left(\frac{1-W^2}{1-W^2_N}\right)^2 H(W_r)
\end{eqnarray}
where no more equality can be used in (\ref{aafact}) and
(\ref{bbfact}) between the pole-term and the transformed
expressions.

   The latter are then analytic functions defined on four-sheeted
Riemann surface, what can be seen explicitly by the inverse
transformation to (\ref{texp})
\begin{equation}
W(t) = i\frac{\sqrt{\left(\frac{t_{in}-t_0}{t_0}\right)^{1/2} +
                         \left(\frac{t-t_0}{t_0}\right)^{1/2}} -
           \sqrt{\left(\frac{t_{in}-t_0}{t_0}\right)^{1/2} -
                         \left(\frac{t-t_0}{t_0}\right)^{1/2}}}
                  {\sqrt{\left(\frac{t_{in}-t_0}{t_0}\right)^{1/2} +
                         \left(\frac{t-t_0}{t_0}\right)^{1/2}} +
           \sqrt{\left(\frac{t_{in}-t_0}{t_0}\right)^{1/2} -
                         \left(\frac{t-t_0}{t_0}\right)^{1/2}}}.\label{confmap}
\end{equation}
These expressions have two pairs of complex conjugate poles on
\begin{itemize}
\item[a\raisebox{-1mm}{)}]
case (\ref{aafact})~-~the second and fourth sheets
\item[b\raisebox{-1mm}{)}]
case (\ref{bbfact})~-~the third and fourth sheets,
\end{itemize}
describing always one resonance under consideration, and the
asymptotic behavior $\sim 1/t$, to be completely given by the
asymptotic term $\left(\frac{1-W^2}{1-W^2_N}\right)^2$, more
specifically, by its power "2".

   Here we would like also to note, that expressions (\ref{aafact})
or (\ref{bbfact}) (it depends on the fact if the resonance is
under the threshold $t_{in}$ or above it) are more sophisticated
analogue of the Breit-Wigner form and they can be applied to
determine resonance parameters $m_r$, $\Gamma_r$ from experimental
data on  EM FFs in the region of the resonance
under consideration.

\setcounter{equation}{0} \setcounter{figure}{0} \setcounter{table}{0}\newpage
       \section{Electro-weak structure of nonet of
       pseudoscalar mesons}\label{III}

   There is only one scalar function of the momentum transfer
squared $t$ completely describing the electro-weak structure of
every member of the nonet of pseudoscalar mesons.

   The EM FFs are identically equal to zero for
 $\pi^0$, $\eta$ and $\eta'$ for all $t$ from $-\infty$ to
$+\infty$, if one takes into account the relation (\ref{cpt}).
Then practically one has to consider only three EM FFs for psedoscalar meson nonet
 corresponding to the charged pions, charged
kaons and neutral kaons, respectively.

   The pseudoscalar mesons are the bound states of $q\bar q$ pairs and
as a result of (\ref{asym}) the asymptotic behavior of their FFs
is
\begin{equation}
F_P(t)_{|t|\to\infty} \sim \frac{1}{t}\label{mesasym}
\end{equation}
to be identical with the VMD model (\ref{vmd}) asymptotics.

   Consequently, the pseudoscalar meson FF in the framework of the $U\&A$ model can be
represented by a sum of $i$-terms (\ref{aafact}) of resonances
below the threshold $t_{in}$ and $j$-terms (\ref{bbfact}) of
resonances above the threshold $t_{in}$ as follows
\begin{eqnarray}
F_P[W(t)] &=& \left(\frac{1-W^2}{1-W^2_N}\right)^2\times\label{ffgen}\\
& \times &\left\{
\sum_i\frac{(W_N-W_i)(W_N-W^*_i)(W_N-1/W_i)(W_N-1/W^*_i)}
{(W-W_i)(W-W^*_i)(W-1/W_i)(W-1/W^*_i)}(f_{iPP}/f_i) + \right. \nonumber \\
& +&\left.  \sum_j\frac{(W_N-W_j)(W_N-W^*_j)(W_N+W_j)(W_N+W^*_j)}
{(W-W_j)(W-W^*_j)(W+W_j)(W+W^*_j)}(f_{jPP}/f_j) \right\},\nonumber
\end{eqnarray}
which takes into account just $n=i+j$ vector-meson resonances with
quantum numbers of the photon. It is analytic in the whole complex
$t$-plane except for two cuts on the positive real axis.

\medskip

   \subsection{Electromagnetic form factor of charged pions}\label{III1}

\medskip

   The $\pi$- meson is the lightest hadron with spin $S=0$ and isospin
$I=1$, i.e. three charge (positive, negative and neutral) states
of the pion do exist. We consider the pion to have an internal
(commonly quark-gluon) structure that for the time being we do not
know to reproduce theoretically but that can be parametrized in
some rather general way.

\begin{figure}[tb]
    \centering
    \scalebox{0.8}{\includegraphics{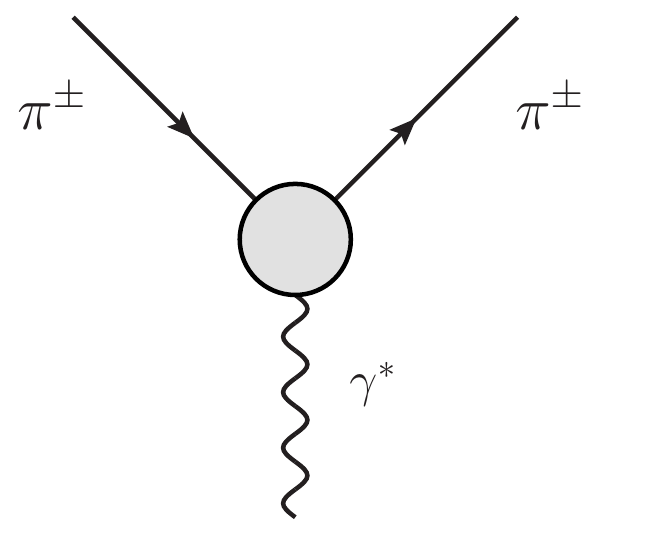}}
    \caption{\small{The virtual photon-pion vertex}}
    \label{fig6}
\end{figure}

   The virtual photon-pion vertex with the internal structure of the
pion is graphically presented in Fig.~\ref{fig6}.

 The blob means various
virtual processes caused by the strong interactions and
theoretically it is expressed by a matrix element $\langle
p_2|J_{\mu}^{EM}(0)|p_1\rangle$ of the electromagnetic current
$J_{\mu}^{EM}(x)$ of the pion, that can be decomposed into a sum
of products of invariant coefficients and linearly independent
covariants constructed now only from the four-momenta $p_1, p_2$
as the spin of the pion is zero.

   The corresponding coefficients, so-called invariant FFs, are
functions of only one invariant variable.

   Really, from the conservation of four-momenta at the virtual
photon-pion vertex
\begin{equation}
p_1 +q= p_2 \label{3cons}
\end{equation}
and by a multiplication of both sides of the latter by one after
the other four-momenta $q$, $p_1$ and $p_2$, one gets a system of
the following three algebraic equations
\begin{eqnarray}
(qp_1) + q^2 &=& (qp_2) \nonumber \\\label{3alge}
p_1^2 + (qp_1) &=& (p_1p_2)\\
(p_1p_2) + (qp_2)
&=&p_2^2 \nonumber
\end{eqnarray}
for four unknown invariant variables, $(qp_1)$, $(qp_2)$, $q^2$,
and $(p_1p_2)$, provided $p_1^2=p_2^2=-m_{\pi}^2$, where $m_{\pi}$
is the pion mass.

   By a solution of (\ref{3alge}) one can express three invariant
variables by means of an arbitrary value of the fourth one. The
latter is usually chosen to be just the four-momentum transfer
squared $t=q^2=-Q^2$.

   A decomposition of the matrix element of the pion electromagnetic
current looks as follows
\begin{equation}
\langle p_2|J_{\mu}^{EM}(0)|p_1\rangle =e[F_\pi^E(t)(p_1 +
p_2)_{\mu} + G_\pi^E(t)(p_2-p_1)_{\mu}] \label{dpiec}
\end{equation}
where instead of two independent four-momenta $p_1$, $p_2$ we have
used their suitable combinations, $(p_1+p_2)_{\mu}$ and
$(p_2-p_1)_{\mu}=q_{\mu}$ and $F_\pi^E(t)$, $G_\pi^E(t)$ are
invariant FFs.

   The gauge invariance of the EM interactions implies the current
conservation
\begin{equation}
\partial_{\mu} J_{\mu}^{EM}(x) = 0
\end{equation}
from where the following condition
\begin{equation}
q_{\mu}\langle p_2|J_{\mu}^{EM}(0)|p_1\rangle = 0 \label{gicon}
\end{equation}
has to be fulfilled. Since $q_{\mu}(p_1+p_2)_{\mu}=p_2^2-p_1^2=0$,
the condition (\ref{gicon}) will be valid if in (\ref{dpiec})
$G_\pi^E(t)=0$. Then the matrix element of the pion EM current takes
finally the well-known form
\begin{equation}
 \langle p_2|J_{\mu}^{EM}(0)|p_1\rangle =eF_\pi^E(t)(p_1 +
 p_2)_{\mu},\label{piffdef}
\end{equation}
where $F_{\pi}^E(t)$ is the so-called pion EM FF.

   On the other hand, making use of transformation properties of the
EM current operator $J_{\mu}^{EM}(x)$ and the one-particle state
vectors of pions, $|\pi^+\rangle$, $|\pi^-\rangle$ and
$|\pi^0\rangle$, with regard to the all three discrete $C$, $P$,
$T$ transformations simultaneously, one can derive relations as
follows
\begin{equation}
\langle\pi^-|J_{\mu}^{EM}(0)|\pi^-\rangle =
-\langle\pi^+|J_{\mu}^{EM}(0)|\pi^+\rangle \label{cptcr}
\end{equation}
and
\begin{equation}
\langle\pi^0|J_{\mu}^{EM}(0)|\pi^0\rangle =
-\langle\pi^0|J_{\mu}^{EM}(0)|\pi^0\rangle. \label{cptnr}
\end{equation}
Then a substitution of the parametrization (\ref{piffdef}) into
(\ref{cptcr}) and (\ref{cptnr}) provides for the EM FFs of charged
and neutral pions relations
\begin{equation}
F_{\pi^-}^E(t) = - F_{\pi^+}^E(t)\label{chffspirel}
\end{equation}
and
\begin{equation}
F_{\pi^0}^E(t) \equiv 0,\label{nffpizer}
\end{equation}
respectively, for all values of the four-momentum transfer squared
$-\infty < t <+\infty$.

   Since the charge $e$ appears as a prefactor in (\ref{piffdef}), the pion EM FF
at $t=0$ is normalized to one
\begin{equation}
F_{\pi}^E(0) = 1.\label{normff}
\end{equation}

   In the framework of the axiomatic quantum field theory one can
prove \cite{Schw} that $F_{\pi^\pm}(t)$ is an analytic function in
the whole complex $t$- plane, besides a cut from $4m_{\pi}^2$ to
$+\infty$. The same analytic properties can be derived within the
framework of QCD \cite{Oeh}. Further, as a consequence of the
pion EM current $J_{\mu}^E(x)$  to be Hermitian, the pion EM FF is
real on the real axis for $t< 4m^2_{\pi}$. Then by an application
of the Schwarz reflection principle to the pion EM FF one finds
the so-called reality condition
\begin{equation}
[F_{\pi}^E(t)]^* =F_{\pi}^E (t^*)\label{recon}
\end{equation}
reflecting the reality of the pion EM FF on the real axis below
$4m^2_{\pi}$ and the relation of values of the pion EM FF on the
upper and lower boundary of the cut
\begin{equation}
[F_{\pi}^E(t + i\epsilon)]^* =F_{\pi}^E (t - i\epsilon),
\quad\quad\quad\quad\quad\quad\quad\quad \epsilon \ll 1\label{reconeps}
\end{equation}
automatically.

   The discontinuity across the cut is given by the unitarity
condition
\begin{equation}\label{unicon}
\frac{1}{2i}\left \{ \langle\pi^+\pi^-|J_{\mu}^{EM}(0)|0 \rangle -
\langle 0|J_{\mu}^{EM}(0)|\pi^+\pi^- \rangle^* \right \}
= \sum_n \langle \pi^+\pi^- |T^+|n \rangle \langle
n|J_{\mu}^{EM}(0)|0 \rangle ,
\end{equation}
where
\begin{equation}
\langle \pi^+\pi^- |J_{\mu}^{EM}(0)|0 \rangle = e (k_{\pi^+} -
k_{\pi^-})_{\mu}F_{\pi}^E(t)
\end{equation}
is a matrix element of the pion EM current defining the pion EM FF
in the time-like ($t > 4m^2_{\pi}$) region. The sum in
(\ref{unicon}) is carried out over a complete set of intermediate
states allowed by various conservation laws and $T^+$ means
Hermitian conjugate amplitudes.

   Moreover, the unitarity condition (\ref{unicon}) tells us that there is an
infinite number of branch points on the positive real axis between
$t = 4m^2_{\pi}$ and $ + \infty$, which always correspond to a new
allowed intermediate state $n$ in (\ref{unicon}). In order to
fulfil the reality condition (\ref{recon}), the cuts associated
with these branch points are chosen to extend to $+\infty$ along
the real axis.

   We would like to note that one can specify \cite{Land,Eden,Dub75} the
same singularities of the pion EM FF by means of an investigation
of the analytic properties of the corresponding Feynman diagrams
of a formal pion EM FF perturbation series.

   Consequently, the first branch point of the pion EM FF is at $t =
4m^2_{\pi}$, the second one at $t = 16m^2_{\pi}$, then at $t=
(m_{\pi^0} + m_{\omega})^2$, $t=4m_K^2$, etc.

   If we restrict ourselves in (\ref{unicon}) only to $|n\rangle =
|\pi^+\pi^-\rangle$, then the elastic unitarity condition of the
pion EM FF is obtained
\begin{equation}
\frac{1}{2i}\left \{ F_{\pi}^E(t +i\epsilon) - [F_{\pi}^E(t
+i\epsilon)]^*\right \} = [A^1_1(t +i\epsilon)]^* F_{\pi}^E(t
+i\epsilon)\label{eunicon}
\end{equation}
where $A^1_1 (t+ i\epsilon)$ is the $P$- wave isovector $\pi\pi$-
scattering amplitude.

   If the complete set of intermediate states in (\ref{unicon}) is taken into account
then the pion EM FF unitarity condition can be written in the form
\begin{equation}
\frac{1}{2i}\left \{ F_{\pi}^E(t +i\epsilon) - [F_{\pi}^E(t
+i\epsilon)]^*\right \} = [A^1_1(t +i\epsilon)]^* F_{\pi}^E(t
+i\epsilon) +\sigma(t+i\epsilon)\label{inlunicon}
\end{equation}
where $\sigma(t+i\epsilon)$ represents all higher contributions.

   Now we prove that the lowest singularity of the pion EM FF at  $t
= 4m^2_{\pi}$ is a square-root type branch point.

   The idea consists in the analytic continuation of the pion EM FF
through the upper and lower boundary of the cut between $t =
4m^2_{\pi}$ and $t = 16m^2_{\pi}$ on the next sheet of the Riemann
surface by means of the elastic unitarity condition (\ref{eunicon})
and in a subsequent comparison of obtained functional expressions.

   By using the reality condition of the pion EM FF (\ref{recon}) and similar
condition \cite{Mart}
\begin{equation}
[A^1_1(t+i\epsilon)]^*=- A^1_1(t-i\epsilon)\label{reconam}
\end{equation}
for the $P$- wave isovector $\pi\pi$ - scattering amplitude, one
can rewrite (\ref{eunicon}) into the form as follows
\begin{equation}
F_{\pi}^E(t+i\epsilon)=\frac{F_{\pi}^E(t-i\epsilon)}{1 + 2i
A^1_1(t-i\epsilon)}\label{unshff}.
\end{equation}

   If we denote by $[F_{\pi}^E(t-i\epsilon)]^+$ and
$[F_{\pi}^E(t+i\epsilon)]^-$ the analytic continuation of the pion
EM FF on the Riemann sheets achieved through the upper and lower
boundary of the elastic cut, respectively, then as a consequence
of a continuity of the pion EM FF we have
\begin{equation}
[F_{\pi}^{E}(t-i\epsilon )]^+ \equiv F_{\pi}^E(t+i\epsilon
)\label{acon+}
\end{equation}
\begin{equation}
[F_{\pi}^{E}(t+i\epsilon )]^- \equiv F_{\pi}^E(t-i\epsilon
)\label{acon-}.
\end{equation}
Then by a substitution of (\ref{acon+}) into (\ref{unshff}) one
gets
\begin{equation}
[F_{\pi}^E(t-i\epsilon)]^+=\frac{F_{\pi}(t-i\epsilon)}{1 + 2i
A^1_1(t-i\epsilon)}\label{anlcon+}.
\end{equation}

   On the other hand, by a complex conjugation of (\ref{eunicon}) and the reality
conditions (\ref{reconeps}) and (\ref{reconam}), the unitarity
condition of the pion EM FF on the lower boundary of the elastic
cut is obtained
\begin{equation*}
\frac{1}{2i}\left \{ F_{\pi}^E(t -i\epsilon) - [F_{\pi}^E(t-
i\epsilon)]^*\right \} = \{ A^1_1(t -i\epsilon) \}^* F_{\pi}^E(t
-i\epsilon),
\end{equation*}
from which the relation follows
\begin{equation*}
F_{\pi}^E(t -i\epsilon)\{ 1 + 2iA^1_1(t+i\epsilon)\} = F_{\pi}^E(t
+i\epsilon)
\end{equation*}
or by using (\ref{acon-}), finally one gets
\begin{equation}
[F_{\pi}^E(t+i\epsilon)]^-=\frac{F_{\pi}^E(t+i\epsilon)}{1 + 2i
A^1_1(t+i\epsilon)}\label{anlcon-}.
\end{equation}

   Comparing (\ref{anlcon-}) with (\ref{anlcon+}) we see that by the
analytic continuation of the pion EM FF through upper and lower
boundary of the elastic cut one gets the identical functional
expressions. The latter convinces us that one has come on the same
sheet of the Riemann surface, which further will be called the
second one and denoted by II. It follows just from here that the
branch point $t=4 m^2_{\pi}$ is of a square-root type, i.e. it
generates just two sheets of the Riemann surface, on which the
pion EM FF is defined.

   Moreover, the analytic  continuation of (\ref{anlcon+}) and (\ref{anlcon-}) leads to the
same expression of the pion EM FF on the second Riemann sheet as
follows
\begin{equation}
[F_{\pi}^E(t)]^{II}=\frac{F_{\pi}^E(t)}{1 + 2i
A^1_1(t)}\label{ffsecsh}.
\end{equation}

   One can read out from (\ref{ffsecsh}), that the pion EM FF on the
second Riemann sheet has all the branch points of the pion EM FF
on the first sheet and besides also the branch points of the $P$-
wave isovector $\pi\pi$- scattering  partial amplitude $A^1_1(t)$.

   It is well known \cite{Mart} that the analytic properties of
$A^1_1(t)$ consist of the right-hand unitary cut for $4m^2_{\pi}
<t <\infty$ and the left-hand dynamical cut for $-\infty < t < 0$.
From the latter and (\ref{ffsecsh}) it follows that the pion EM
FF on the second Riemann sheet has the left hand cut too.

   The unitarity condition in the language of the $S^1_1(t)$ matrix,
connected with the amplitude $A^1_1(t)$ in (\ref{ffsecsh}) by the
relation
\begin{equation}
S^1_1(t) = 1+ 2i A^1_1(t)\label{smatr},
\end{equation}
has the form
 \begin{equation}
S^1_1(t)[S^1_1(t)]^* =1\label{uconsmat}
\end{equation}
from where one can parametrize
\begin{equation}
S^1_1(t)= e^{2i\delta_1^1(t)}\label{smatpar}
\end{equation}
with $\delta_1^1(t)$ to be the $P$- wave isovector
$\pi\pi$-scattering phase shift. By means of a substitution of
(\ref{smatr}) into (\ref{uconsmat}) one gets the unitarity
condition for $A^1_1(t)$ in the form
\begin{equation}
Im A^1_1(t) = |A^1_1(t)|^2\label{ampluncon}
\end{equation}
which is exactly valid for $4m_{\pi}^2 < t < 16m_{\pi}^2$. However
it follows just from data \cite{Esta,Hyam,Prot} on $A^1_1(t)$ that
it works approximately up to 1${\rm GeV}^2$.

   On the other hand by a combination of the relations (\ref{smatpar}) and
(\ref{smatr}), the following parame\-trization of $A^1_1(t)$ is
obtained
\begin{equation}
 A_1^1(t)= \frac{1}{2i}\left ( e^{2i\delta^1_1(t)} -1\right )=
 e^{i\delta^1_1(t)}\sin\delta_1^1(t)\label{amplpar}
\end{equation}
 which is consistent with the unitarity condition (\ref{ampluncon}) automatically.

   We note, that the partial wave amplitude $A^1_1(t)$ has a resonant
behavior as the derivative $d\delta^1_1(t)/dt$ has a pronounced
maximum to be not caused by an opening of some new threshold.
Since $A^1_1(t)$ is an analytic function, the rapid variation of
$\delta^1_1(t)$ can be described by a pole in the second Riemann
sheet, which corresponds just to the well known $\rho$(770)
resonance. Farther, starting only from general considerations we
show that the latter pole has to appear also on the second Riemann
sheet of the pion EM FF.

   Really, starting from the unitarity condition (\ref{ampluncon}), analogically
to the pion EM FF, one can prove that the branch point
$4m_{\pi}^2$ of the amplitude $A_1^1(t)$ is a square root type and
also one can find an expression of the amplitude $A^1_1$ on the
second Riemann sheet to be
\begin{equation}
[A_1^1(t)]^{II}= -\frac{A^1_1(t)}{1+ 2i
A^1_1(t)}\label{amplsecsh}.
\end{equation}

   By a comparison of (\ref{amplsecsh}) and (\ref{ffsecsh}) we see
they have an identical denominator. So, if $[A_1^1(t)]^{II}$ has
the $\rho$(770)-meson pole created by a zero of the denominator
$1+ 2i A^1_1(t)$, then the same zero has to be present in the
denominator of $[F_{\pi}(t)]^{II}$, and as a consequence, the
$\rho$(770)- meson pole has to be created on the second Riemann
sheet of the pion EM FF.

   Since such a pole is shifted (due to the instability of
$\rho$(770)) from the real axis into the complex plane, as a
result of the reality condition (\ref{recon}), there must exist
also the complex conjugate pole to the latter.

\begin{figure}[tb]
    \centering
    \scalebox{1.1}{\includegraphics{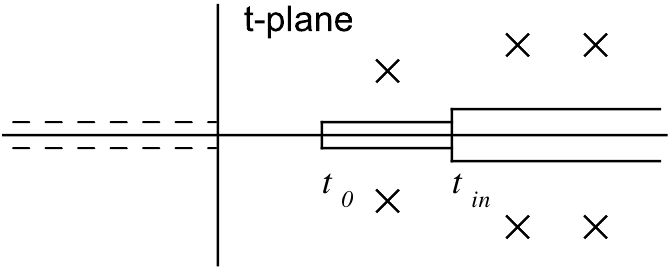}}
    \caption{\small{Analytic properties of the pion EM FF}}
    \label{fig7}
\end{figure}

   We note, there are radial excitations \cite{Rpp} of the $\rho$(770)-meson.
The corresponding poles are also placed on the unphysical sheets of
the Riemann surface on which the pion EM FF is defined. Then all
information on the analytic properties of the pion EM FF looks like
it is presented in Fig.~\ref{fig7}.

   Defining a phase of the pion EM FF for $t >4m_{\pi}^2$ by the
relation
\begin{equation}
F_{\pi}^E(t) =|F_{\pi}^E(t)| e^{i\delta_{\pi}(t)}\label{defffphas}
\end{equation}
and substituting the latter, together with (\ref{amplpar}), into
the unitarity condition (\ref{eunicon}) one finds the identity
\begin{equation}
\delta_{\pi}(t)\equiv \delta_1^1(t)\label{phasiden}
\end{equation}
valid in the whole elastic region.

\medskip

   The next property of the pion EM FF follows from the threshold
behavior of the $P$-wave isovector $\pi\pi$- scattering partial
wave amplitude
\begin{equation}
A^1_1(t)_{|_{q\to 0}} \sim a^1_1 q^3 \label{amplthrbeh}
\end{equation}
where $a^1_1$ is the $P$- wave isovector $\pi\pi$- scattering
length and $q$ is the absolute value of the pion three momentum in
the center of mass (c.m.) system, which is connected with the
momentum transfer squared by the following  relation
\begin{equation}
t = 4(q^2 +m_{\pi}^2).
\end{equation}

   Taking into account (\ref{amplthrbeh}) and the parametrization
(\ref{amplpar}) for $q\to 0$ one gets the threshold behavior of
the phase $\delta_1^1(t)_{q\to 0}\sim a_1^1q^3$ which by means of
the (\ref{phasiden}) gives the threshold behavior of the pion EM
FF phase
\begin{equation}
\delta_{\pi}(t)_{|_{q\to 0}} \sim a_1^1q^3.\label{ffphthrbeh}
\end{equation}

\medskip

   One of the basic properties of the EM FFs of strongly interacting
particles is their asymptotic behavior. However, until the
discovery of the quark-gluon structure of hadrons the latter was
unknown theoretically and only polynomial bounds were always
assumed to be fulfilled.

   According to the quark model, the large momentum transfer behavior
of the hadron EM FF is related to the number of constituent quarks
of the considered hadron (\ref{asym}). Thus for the pion $(n_q
=2)$ one gets (\ref{mesasym}), which is in a qualitative agreement
with existing data. This is a prediction of a dimensional counting
rule and some simple assumptions which are based on the case of an
underlying scale-invariant theory \cite{Matv,Brod1}.

\begin{figure}[tb]
    \centering
    \scalebox{1.4}{\includegraphics{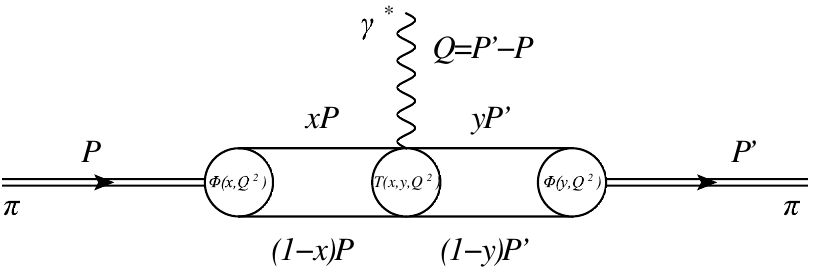}}
    \caption{\small{Representation of the pion FF in QCD.}}
    \label{fig8}
\end{figure}

   The asymptotic behavior (\ref{mesasym}) of the pion EM FF was proved
\cite{Lep,Far,Efr} (up to the logarithmic correction) in the
framework of the perturbative QCD. Here the pion EM FF takes the
following form \cite{Lep} (see Fig.~\ref{fig8})

\begin{equation}
F_{\pi}^E(Q^2)=\int_0^1dx\int_0^1dy
\Phi^+(y,Q^2)T(x,y,Q^2)\Phi(x,Q^2)
\end{equation}
where $\Phi(x,Q^2)$ is a wave function of the pion, which
represents a probability to find the quark to carry the fractional
momentum $x$ of the total pion momentum and $T(x,y,Q^2)$ is the
amplitude of an interaction of the virtual photon with quarks and
gluons. For the latter one can write the following perturbative
expansion
\begin{equation}
T(x,y,Q^2) =\alpha_s(Q^2)T_B(x,y,Q^2)\left [ 1 +
\alpha_s(Q^2)T_2(x,y,Q^2) +....\right ]
\end{equation}
where $\alpha_s(Q^2)$ is the effective constant of the quark-gluon
interactions defined (in the lowest order) by the expression
\begin{equation}
\alpha_s(Q^2) =\frac{4\pi}{(11 -
2/3n_f)\ln{Q^2/\Lambda^2}}\label{alfqcd}
\end{equation}
with $n_f$ to be a quark number flavor and $\Lambda$ as a QCD
scale parameter.

\begin{figure}[tb]
    \centering
    \scalebox{1.2}{\includegraphics{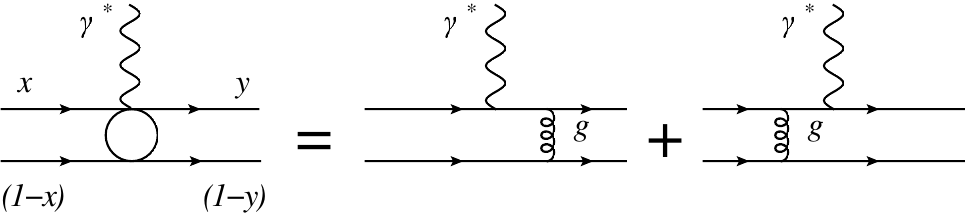}}
    \caption{\small{Dominating Feynman diagrams in the lowest order according to $\alpha_s(Q^2)$.}}
    \label{fig9}
\end{figure}

   In the lowest order according to $\alpha_s(Q^2)$ there are
contributing only two Feynman diagrams (see Fig.~\ref{fig9}) into
$T(x,y,Q^2)$.

   By means of a calculation of the latter one gets an explicit form
for the Born approximation $T_B$ of the amplitude $T(x,y,Q^2)$ to
be
\begin{equation*}
T_B(x,y,Q^2)=\frac{16\pi C_F}{Q^2(1 -x)(1- y)}
\end{equation*}
where $C_F=3/4$ is the so-called colour factor.

   Now one has to find $\Phi(x,Q^2)$. It satisfies the differential
equation (see e.g.\cite{Brod2,Chos})
\begin{equation}
\frac{d\Phi(x,Q^2)}{d\kappa}=\int_0^1du V_{q\bar q\to q\bar
q}(u,x)\Phi(u,Q^2)\label{qcdeq}
\end{equation}
in which
\begin{equation*}
\kappa
=2/\beta_0\ln{\frac{\alpha_s(Q^2_0)}{\alpha_s(Q^2)}},\quad\quad\quad\quad\quad
\beta= 11 - 2/3n_f
\end{equation*}
and the kernel takes the form \cite{Lep} as follows
\begin{equation}
V_{q\bar q\to q\bar q}(u,x)= 1/2C_F\left \{ \frac{1- u}{1- x} (1 +
\frac{1}{u-x})_+\Theta(u-
x)+\frac{u}{x}(1+\frac{1}{x-u})_+\Theta(x-u)\right \}.
\end{equation}

   If the function $\Phi(x,Q^2)$ is known at $Q^2_0$, then by using
the equation (\ref{qcdeq}) one can calculate it for arbitrary
value $Q^2$. In principle $Q^2_0$ can be chosen to be arbitrary as
soon as the condition
\begin{equation}
\frac{\alpha_s(Q^2_0)}{4\pi} < 1
\end{equation}
is fulfilled.

   The equation (\ref{qcdeq}) can be solved easily if the property
\cite{Lep}
\begin{equation}
\int_0^1dz C_n^{(3/2)}(2z -1)V_{q\bar q\to q\bar q}(z,x)=1/2
A_n^{NS}C_n^{3/2}(x)
\end{equation}
is used, where
\begin{equation}
A_n^{NS} =C_F \left [ -\frac{1}{2} +\frac{1}{(n+1)(n+2)}
-2\sum_{j=2}^{n+1}\frac{1}{j}\right ]
\end{equation}
are the so-called non-singlet anomalous dimensions and
$C_n^{3/2}(x)$ are the Gegenbauer polynomials of the order 3/2.

   If one decomposes $\Phi(x,Q^2)$ into a series according to the
polynomials $C_n^{(3/2)}(x)$ as follows
\begin{equation}
\Phi(x,Q^2) =
x(1-x)\sum_{n=0}^{\infty}\Phi_n(Q^2)C_n^{(3/2)}(2x-1)
\end{equation}
and then substitute it into (\ref{qcdeq}), one gets the equation
for the coefficients of the previous expansion in the form
\begin{equation}
\frac{d\Phi_n(Q^2)}{d\kappa}= A_n^{NS} \Phi_n(Q^2).
\end{equation}

   By means of a solution of the latter one obtains
\begin{equation}
\Phi_n(Q^2) =  exp(\kappa A^{NS}_n)\Phi_n(Q^2_0)\equiv \left \{
\frac{\alpha_s(Q^2)}{\alpha_s(Q_0^2)}\right \}^{d_n}\Phi_n(Q_0^2)
\end{equation}
where
\begin{equation}
d_n = -2 A_n^{NS}/\beta_0.
\end{equation}

   Now collecting all partial results we come to the expression of
the pion EM FF
\begin{equation}
F_{\pi}^E(Q^2)= 4\pi(\frac{4}{3})\frac{\alpha_s(Q^2)}{Q^2}\left
|\sum_{n=0}^{\infty}\Phi_n(Q^2_0)\left \{
\frac{\alpha_s(Q^2)}{\alpha_s(Q_0^2)}\right \}^{d_n}\right |
\end{equation}
from which the relation
\begin{equation}
F_{\pi}^E(Q^2)= 4\pi(\frac{4}{3})\frac{\alpha_s(Q^2)}{Q^2}\left [
|\Phi_0(Q^2_0)|^2 +2 Re\quad (\Phi_0(Q^2_0)\Phi^*_1(Q^2_0))\left \{
\frac{\alpha_s(Q^2)}{\alpha_s(Q^2_0)}\right \}^{d_1}+...\right
]\label{ffqcdexp}
\end{equation}
follows with $d_1$=0.427 for $n_f$=4.

   If one restricts  in the previous result only to the first leading
term, then
\begin{equation*}
F_{\pi}^E(Q^2)=
4\pi(\frac{4}{3})\frac{\alpha_s(Q^2)}{Q^2}|\Phi_0(Q_0^2)|^2.
\end{equation*}
In this case Farrar and Jackson have found \cite{Far} the relation
\begin{equation}
\Phi_0(Q_0^2) ={\sqrt{3}} f_{\pi},\label{phi}
\end{equation}
where $f_{\pi} =92.4\pm 0.2 \,{\rm MeV}$ is the constant of the weak pion
decay into $\mu$- meson and antineutrino. As a result one comes to
the expression for the asymptotic behavior of the pion EM FF
\begin{equation}
F_{\pi}^E(Q^2)_{Q^2\to \infty} \sim \frac{16\pi
f_{\pi}^2\alpha_s(Q^2)}{Q^2}\label{31}
\end{equation}
or by using (\ref{alfqcd})
\begin{equation}
F_{\pi}^E(Q^2)_{Q^2\to \infty} \sim \frac{64\pi^2
f_{\pi}^2}{(11-2/3n_f)Q^2\ln{Q^2/\Lambda}}\label{qcdasym}
\end{equation}
which confirms (up to the logarithmic correction) the quark
counting rule prediction (\ref{mesasym}).

   Higher orders in (\ref{ffqcdexp}) depend on the models chosen for the pion
wave function.

\medskip

   Till now we have considered only the invariant pion EM FF in the
momentum representation to be sufficient for a description of
observable phenomena.

   Nevertheless, there is a special Breit system in which the
space-component of the pion EM current is equal to zero  and as a
consequence one can write down there the Fourier transform of the
pion EM FF giving just the static charge distribution density as
follows
\begin{equation}
\rho(\vec{r})=\frac{1}{(2\pi)^3}\int
F_{\pi}^E(\vec{Q}^2)e^{i\vec{Q}\vec{r}}d^3Q.\label{chardis}
\end{equation}

   The inverse Fourier transform to (\ref{chardis}) has the following form
\begin{equation}
F_{\pi}^E(\vec{Q}^2)=\int\rho(\vec{r})e^{-i\vec{Q}\vec{r}}d^3r.\label{fourtransf}
\end{equation}
If $\rho(\vec{r})$ is a spherically symmetric distribution, then
one can rewrite (\ref{fourtransf}) into the spherical coordinates
and by an integration over $\theta$ and $\varphi$ angles one gets
\begin{equation*}
F_{\pi}^E(\vec{Q}^2)=4\pi\int_0^{\infty}\frac{\sin Qr}{Q
r}\rho(\vec{r})r^2dr.
\end{equation*}
For the case of $Qr\ll 1$
\begin{equation*}
\sin Qr \simeq Qr - \frac{(Qr)^3}{6} +......
\end{equation*}
and
\begin{equation*}
F_{\pi}^E(\vec{Q}^2)=4\pi\int_0^{\infty}r^2\rho(\vec{r})dr
-\frac{4\pi Q^2}{6}\int_0^{\infty}r^4\rho(\vec{r})dr +...  \cdot
\end{equation*}

   Now, taking into account the charge distribution density
normalization
\begin{equation*}
\int_0^{\infty}\rho(\vec{r})4\pi r^2dr= 1
\end{equation*}
and a definition of the mean square charge radius
\begin{equation*}
\langle r^2_{\pi}\rangle = \int_0^{\infty}r^2\rho(\vec{r})4\pi
r^2dr
\end{equation*}
one gets finally
\begin{equation*}
F_{\pi}^E(\vec{Q}^2)= 1- \frac{Q^2}{6}\langle r^2_{\pi}\rangle
+...
\end{equation*}
from where the well known rule for a calculation of the mean
square pion charge radius
\begin{equation}
\langle r^2_{\pi}\rangle =
6\frac{dF_{\pi}^E(t)}{dt}_{|_{t=0}}\label{ffrad}
\end{equation}
is obtained.

   From the relations (\ref{chffspirel}), (\ref{nffpizer}) and
(\ref{ffrad}) it is transparent to see that the mean square charge
radius of the $\pi^+$- meson takes a positive value, the mean
square charge radius of the $\pi^-$- meson takes a negative value
and the mean square charge radius of the $\pi^0$- meson is
identically to be zero.

\medskip

   \subsection{Experimental information on the absolute value of charged pions form factor} \label{III2}

\medskip

   The pion EM FF is the simplest one of all others EM FFs of hadrons, the absolute
value of which can be measured (directly or indirectly) everywhere
on the real axis of the complex $t$-plane, on which the pion EM FF
is defined, by using the following five different processes
\begin{itemize}
\item
the electroproduction of pions on nucleons ($e^-N\to e^-\pi N$);
\item
the scattering of charged pions on atomic electrons;
($\pi^-e^-\to\pi^-e^-$)
\item
the inverse electroproduction processes ($\pi^- p\to e^+e^- n$);
\item
the electron-positron annihilation into two charged pions;
($e^+e^-\to\pi^+\pi^-$)
\item
$J/\Psi$ decay into two charged pions ($J/\Psi\to\pi^+\pi^-$).
\end{itemize}

   Up to the present time there are more than 25 independent
experiments realized in which an information on  $|F^E_{\pi}(t)|$
in the space-like ($t <0$) and in the time-like ($t >0$) regions,
for the range of momenta $-10 {\rm GeV}^2 < t < 10 {\rm GeV}^2$, was obtained.

   The crucial (though not very precise) experiment was carried out
in Novosibirsk \cite{Ausl}, where the cross-section on the
$e^+e^-\to\pi^+\pi^-$ process was measured and an information on
$|F^E_{\pi}(t)|$ at the region of the $\rho$(770)- resonance was
extracted. As a result, a creation of the $\rho(770)$- meson in
the electron-positron annihilation into hadrons was confirmed
experimentally for the first time. Later on this resonant region
was re-measured with a higher precision in ORSAY \cite{Benks} and
the isospin violating $\omega(782)\to \pi^+\pi^-$ decay
contribution to $e^+e^-\to \pi^+\pi^-$ process, the so-called
$\rho-\omega$ interference effect, was experimentally revealed.

   Afterwards, during the next four decades many new experiments with
the aim of obtaining of the experimental information on the pion
EM FF in the space-like and time-like regions were proposed and
realized.

   Since the pion is an unstable particle, one can not make a fixed
pion target experiment of the elastic electron scattering on
pions, like on the protons, in order to investigate the pion EM
structure in the space-like region. Therefore, for very small
values of $t$, the scattering of charged pions on atomic
electrons, $\pi^- e^-\to \pi^- e^-$, was employed
\cite{Dally,Amend1} and for higher negative values of $t$ the
electro-production process, $e^- N\to e^-\pi N$, was used
\cite{Bebek,Volmer,Horn}.

   In the time-like region, for $0 <t < 4m^2_{\pi}$, an information
on $|F_{\pi}^E(t)|$ was obtained \cite{Berzh1,Berzh2,Alidze} from
data on the inverse electro-production process $\pi^- p\to e^-e^+
n$. Also the process $\bar p n\to \pi^-\ell^+\ell^-$ is a suitable
candidate \cite{Dubnaz} for the latter, provided that all
corresponding nucleon EM FFs are known in this region.

   On the other hand, as the process $e^+e^-\to \pi^+\pi^-$ is
of the EM nature, one can treat the latter in the
one-photon-exchange approximation and as a result, there are no
model ingredients in an extraction of $|F_{\pi}^E(t)|$ from the
measured cross-section of the process under consideration.
Therefore, the most reliable up to now information on the
$|F_{\pi}^E(t)|$ was obtained in a systematic investigations of
the $e^+e^-\to\pi^+\pi^-$ process in ORSAY
\cite{Quen,Cosme,Bisell}, Frascati \cite{Bolli,Espo,Alois},
Novosibirsk \cite{Bark,Akhm,Achas} and CERN \cite{Amend2}.

\begin{figure}[t,h]
    \centering
    \scalebox{.6}{\includegraphics{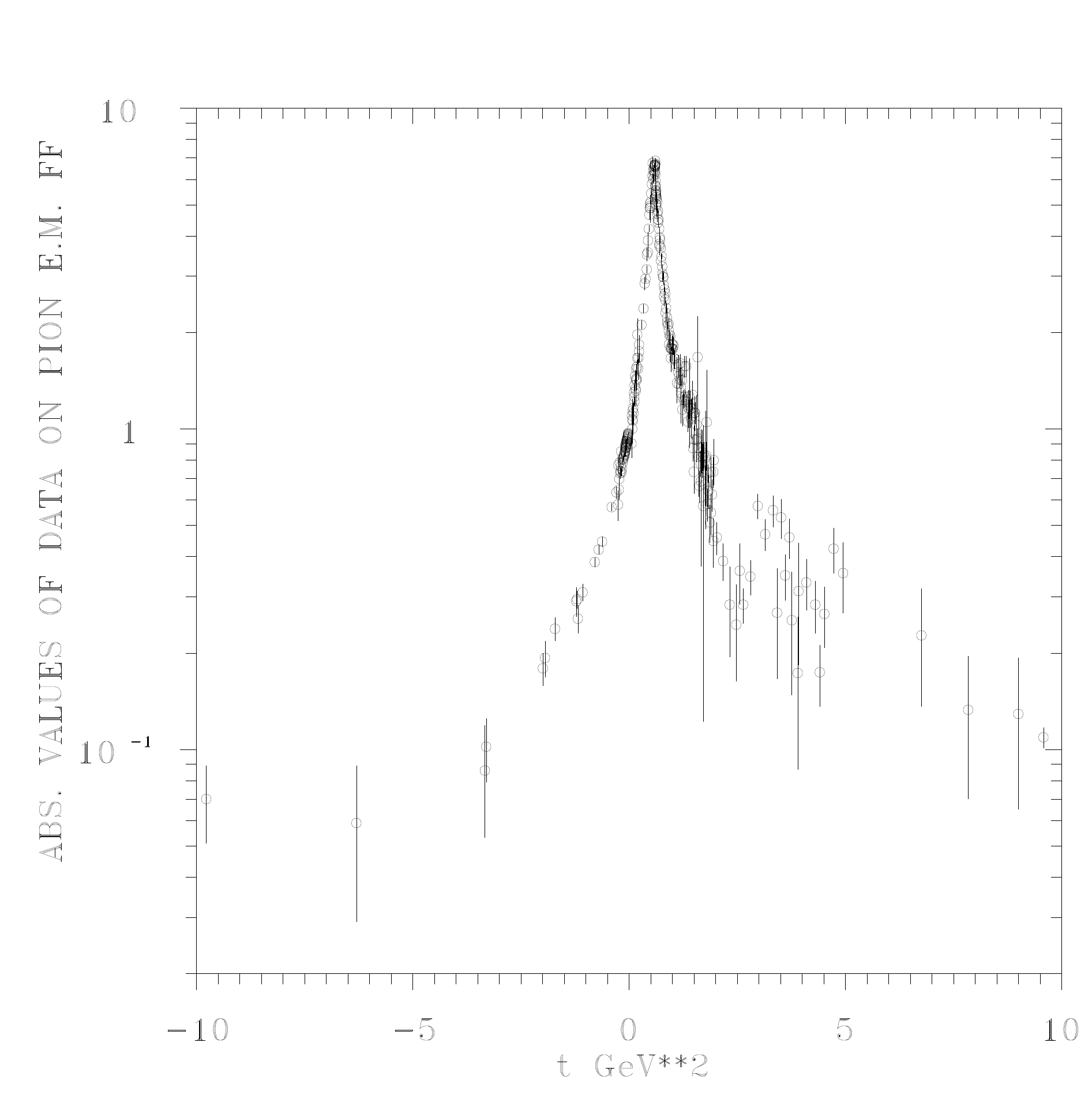}}
    \caption{\small{Experimental data on $|F_{\pi}^E(t)|$ }}
    \label{fig10}
\end{figure}

The experimental point on $|F_{\pi}^E(t)|$ at the highest value of
$t$, $t$= 9.579 ${\rm GeV}^2$, was obtained \cite{Baltr} from the
$J/\Psi \to\pi^+\pi^-$ decay, which must be electromagnetic (i.e.
it is realized through one virtual photon) provided that the $G$-
parity is conserved by strong interactions.

As a result there are more than 300 experimental points on
$|F_{\pi}^E(t)|$ (see Fig.~\ref{fig10}) for the range of momenta $-10 {\rm GeV}^2
< t < +10 {\rm GeV}^2$. However, they do not seem to be all mutually
consistent and as a result too high values of the $\chi^2/{\rm NDF}$ is
in the fitting procedure by $U\&A$ pion EM FF models found.

\medskip

   \subsection{Unitary and analytic model of charged pions electromagnetic
   structure}\label{III3}

\medskip

    Further we present a construction of the model which reflects all known properties
 of $F_{\pi}^E(t)$, including also the threshold behavior of $Im F_{\pi}^E(t)$
 and describes all existing data.
 It represents just the most accomplished pion electromagnetic structure
 model to be constructed up to now.

 However, first very briefly all known properties of the pion EM FF are reviewed.
 \begin{itemize}
 \item[1.]
 The function $F_{\pi}^E(t)$ is normalized (\ref{normff}) at $t=0$.

 \item[2.]
 $F_{\pi}^E(t)$ can be considered as a complex function of a complex
 variable $t$ and it is analytic function in the whole complex $t$-plane
 besides the cut from $4m^2_{\pi}$ to $+\infty$.

 \item[3.]
 A discontinuity across the latter cut is given by the unitarity
 condition (\ref{unicon}), in which the first three allowed intermediate
 states are $|\pi^+\pi^- \rangle$  $|\pi^+\pi^-\pi^+\pi^- \rangle$
 and $|\pi^0\omega\rangle$.

 \item[4.]
 To every intermediate state  $|n \rangle$ in the unitarity
 condition (\ref{unicon})
 a branch point corresponds at the value
 of $t$ to be equal to the squared sum of masses of the corresponding particles,
 i.e. they are at $t=(2m_{\pi})^2$ , $(4m_{\pi})^2$,  $(m_{\pi^0}+m_{\omega})^2$, etc..

 \item[5.]
 From the Hermitivity of $J_{\mu}^{EM}$ i.e.
 \begin{equation}
 (J_{\mu}^{EM})^+ = J_{\mu}^{EM}
 \label{chap16}
 \end{equation}
 the reality of $F_{\pi}^E(t)$ on the real axis from $-\infty$ to
 $4m_{\pi}^2$ and the reality condition (\ref{recon}), as well as
 a relation (\ref{reconeps}) between values of $F_{\pi}^E(t)$
 on the upper $(t+i\epsilon)$ and lower $(t-i\epsilon)$ boundary of the cut ( $t >
 4m^2_{\pi}$)
 follow.

 \item[6.]
 By using of (\ref{eunicon}) one can continue analytically $F_{\pi}^E(t)$ through the
 upper and lower boundary of the cut at the interval $4m^2_{\pi} < t < 16m^2_{\pi}$
 and simultaneously one can prove in a such way that the first branch point at
 $t=4m_{\pi}^2$ is a square root type.

 \item[7.]
 As a by-product of the latter one obtains the expression
 (\ref{ffsecsh})
 for the pion EM FF on the second sheet of the Riemann surface on which
  $F_{\pi}^E(t)$ is completely defined as an unambiguous function of $t$.

 \item[8.]
 One can read out from (\ref{ffsecsh}) the analytic properties of $F_{\pi}^E(t)$
 on the second Riemann sheet. As the analytic properties of $A^1_1$ on
 the first (physical) Riemann sheet consist of the right-hand unitary cut for
 $4m_{\pi}^2< t < +\infty$ and the left hand dynamical cut for
 $-\infty < t < 0$, the pion EM FF on the second Riemann sheet has the left
 hand cut for $-\infty < t < 0$ too. In order to obtain a correct reproduction of
 the pion EM FF phase, a contribution of the latter can not be neglected as it
 was done in many phenomenological models in the past.

 \item[9.]
 As the $\pi\pi$ scattering amplitude is dominated by the $\rho$(770) resonance
 to be placed in the form of the pole on the second sheet and $[A^1_1]^{II}(t)$
 has identical denominator with (\ref{ffsecsh}), then $F_{\pi}^E(t)$ has also
 the pole on the second Riemann sheet corresponding just to the $\rho$(770) resonance.
 All excited states of $\rho$(770) are also placed on some unphysical sheets
 of the constructed Riemann surface.

 \item[10.]
 As a consequence of the reality condition (\ref{recon}) to every resonance
 pole also a complex conjugate pole has to exist .

 \item[11.]
 The identity (\ref{phasiden}) is practically valid for $4m_{\pi}^2 < t \le 1 {\rm {\rm GeV}}^2$.

 \item[12.]
 Directly from (\ref{defffphas}) the imaginary part of the pion EM FF takes the form
 \begin{equation}
 Im F_{\pi}^E(t)=|F_{\pi}^E(t)|\sin\delta_{\pi}(t)
 \label{chap117}
 \end{equation}
 from where by using (\ref{ffphthrbeh}) one gets the threshold
 behavior of the pion EM FF imaginary part
 \begin{equation}
 Im F_{\pi}^E(t)_{|_{q=0}}=|F_{\pi}^E(t)|a^1_1 q^3\label{imffthrbeh}
 \end{equation}
 which can be transformed into the following three conditions
 \begin{equation}
 Im F_{\pi}^E(t)_{|_{q=0}}=\frac{d ImF_{\pi}^E(t)}{d q}_{|_{q=0}}= \frac{d^2 ImF_{\pi}^E(t)}{d q^2}_{|_{q=0}} =0.
 \label{chap119}
 \end{equation}

 \item[13.]
 The $\pi\pi$ scattering length $a^1_1$ in (\ref{amplthrbeh}) can be calculated from the pion
 EM FF by means of the expression
 \begin{equation}
 a^1_1=\frac{ \frac{d^3 Im F_{\pi}^E(t)}{dq^3}}{6|F_{\pi}^E(t)|}_{|_{q=0}}.
 \label{chap120}
 \end{equation}

 \item[14.]
 The asymptotic behavior (\ref{qcdasym}) of the pion EM FF obtained in the framework of the
 pQCD is equivalent (up to the logarithmic correction) to the quark
 counting rule prediction (\ref{mesasym}).
 \end{itemize}

 To our knowledge there are no other properties of $F_{\pi}^E(t)$
 to be known for the time being.

\medskip

   Starting from the general parametrization (\ref{ffgen}) one
obtains for the pion EM FF the expression

\begin{eqnarray}
 F_{\pi}^E[W(t)]&=&\left (\frac{1-W^2}{1-W_N^2}\right )^2\times \label{chap127} \\
 &\times&\left [ \frac{(W_N-W_{\rho})(W_N-W_{\rho}^*)(W_N-1/W_{\rho})(W_N-1/W_{\rho}^*)}
 {(W-W_{\rho})(W-W_{\rho}^*)(W-1/W_{\rho})(W-1/W_{\rho}^*)}
 (\frac{f_{{\rho}\pi\pi}}{f_{\rho}})\right.+ \nonumber \\
 &+&\left. \sum_{v=\rho',\rho''}\frac{(W_N-W_v)(W_N-W_v^*)(W_N+W_v)(W_N+W_v^*)}
    {(W-W_v)(W-W_v^*)(W+W_v)(W+W_v^*)}
    (\frac{f_{v\pi\pi}}{f_v})\right ] \nonumber
 \end{eqnarray}
 to be defined on the four-sheeted Riemann surface reflecting all
 above mentioned properties, besides the left-hand cut contribution
 and the threshold behavior (\ref{imffthrbeh}) of the $Im F_{\pi}^E(t)$.

   The left hand cut contribution manifests itself mainly about the
first $4m^2_{\pi}$ threshold as it provides a possibility to
achieve a correct ratio of $Im F_{\pi}^E(t)$ to $Re F_{\pi}^E(t)$
at the expression
 \begin{equation}
 \tan\delta_{\pi}(t)=\frac{Im F_{\pi}^E(t)}{Re F_{\pi}^E(t)}
 \label{chap128}
 \end{equation}
 and in a such way as a result of (\ref{phasiden}) also a  reproduction of the experimental
 data on $\delta^1_1(t)$.

    From literature it is well known, that the contribution of a cut of an arbitrary analytic
 function can be in the framework of the Pad\`e approximations represented
 by alternating zeros and poles on the place of that cut.

    This method was employed in \cite{Dubn79,Dubn81} to demonstrate explicitly that
 experimental data on $|F_{\pi}^E(t)|$ and $\delta^1_1(t)$ can be
 described consistently only by the pion EM FF model including the correct
 left-hand cut contribution.

    Really, in \cite{Dubn79} $\tan\delta_1^1(t)$ was approximated by a rational
 function in $q$- variable with a minimal number of coefficients to be
 determined from a fit of data on $\delta^1_1(t)$. Then by means of a
 dispersion integral a rational function for $F_{\pi}^E(t)$ was obtained  in which
 one zero and one pole just on the place of the cut from the second Riemann
 sheet were found. The latter result clearly demonstrates that the left-hand cut contribution
 can not be neglected in any analytic pion EM FF model.

    The same result was confirmed in \cite{Dubn81} by a different way. Here, first, contributions
 of the isospin violating $\omega({\rm 783})\to \pi^+\pi^-$ decay from the data
 on $\sigma_{tot}(e^+e^-\to\pi^+\pi^-)$ were separated. Then the resultant
 data were combined with data on $\delta^1_1(t)$ in order to get the data on
 $Re F_{\pi}^E(t)$ and $Im F_{\pi}^E(t)$. The latter were described by various
 Pad\`e-type approximants in $q$-variable, in which repeatedly one zero
 and one pole were identified on the place of the left-hand cut from the second Riemann
 sheet. Moreover, the found positions are almost identical with those from \cite{Dubn79}.

    By means of the conformal mapping (\ref{confmap})  the left-hand
 cut from the second Riemann sheet is mapped into the interval $0 < W < +1$.
 So, on the base of the results obtained in \cite{Dubn79,Dubn81} the left-hand cut
 contribution is in our model approximated by one normalized zero and one normalized pole as follows
 \begin{equation}
 \frac{(W-W_z)(W_N-W_p)}{(W_N-W_z)(W-W_p)}
 \label{chap129}
 \end{equation}
 where $W_z$ and $W_p$ are left to be free parameters.

    The most natural incorporation of (\ref{chap129}) into our model of $F_{\pi}^E(t)$
 could be as a multiplicative factor to the $\rho$(770)-meson term in
 (\ref{chap127}) as the left-hand cut from the second Riemann sheet is near-by to the
elastic region $4m^2_{\pi} <t< 1 {\rm GeV}^2$.

    However, just for the sake of a consistent incorporation of the threshold
 behavior of $Im F_{\pi}^E(t)$, given by (\ref{imffthrbeh}), we incorporate
 (\ref{chap129}) as a common factor to all $\rho$-meson resonances contributing
 to $F_{\pi}^E(t)$ as follows
 \begin{eqnarray}
 F_{\pi}^E[W(t)]&=&\left (\frac{1-W^2}{1-W_N^2}\right )^2\frac{(W-W_z)(W_N-W_p)}{(W_N-W_z)(W-W_p)}\times \label{chap130} \\
 & \times &\left \{ \frac{(W_N-W_{\rho})(W_N-W_{\rho}^*)(W_N-1/W_{\rho})(W_N-1/W_{\rho}^*)}
 {(W-W_{\rho})(W-W_{\rho}^*)(W-1/W_{\rho})(W-1/W_{\rho}^*)}
 (\frac{f_{{\rho}\pi\pi}}{f_{\rho}})\right.+ \nonumber \\
 &+&\left. \sum_{v=\rho',\rho''}\frac{(W_N-W_v)(W_N-W_v^*)(W_N+W_v)(W_N+W_v^*)}
    {(W-W_v)(W-W_v^*)(W+W_v)(W+W_v^*)}
    (\frac{f_{v\pi\pi}}{f_v})\right \}. \nonumber
 \end{eqnarray}
 Then all three terms in the wave-brackets give a nonzero contribution to the
 $Im F_{\pi}^E(t)$, though $\rho'$- and $\rho''$-term (unlike the $\rho$-meson term)
 are pure real in the interval $4m^2_{\pi} <t< t_{in}$. Thus the left hand cut
 contribution in a consideration of the threshold behavior of $Im F_{\pi}^E(t)$
 in the $U\&A$ pion EM FF model is crucial.

    Now, calculating the $Im F_{\pi}^E(t)$ from  (\ref{chap130}) explicitly, one can immediately verify that \-
 $Im F_{\pi}^E(t)_{|_{q=0}}$=$0$. A requirement of the first derivative of
 $Im F_{\pi}^E(t)$ at $q=0$ to be zero gives the following condition on the coupling
 constant ratios
 \begin{eqnarray}
 &&[(W_z-W_p)+2W_z\cdot W_p Re W_{\rho}(1+|W_{\rho}|^{-2})]N_{\rho}(f_{\rho\pi\pi}/f_{\rho})+ \label{chap131} \\
 &+&\sum_{v=\rho',\rho''}\frac{W_z-W_p}{|W_v|^4}N_v(f_{v \pi\pi}/f_{v})=0, \nonumber
 \end{eqnarray}
 where
\begin{equation*}
 N_{\rho}=(W_N-W_{\rho})(W_N-W_{\rho}^*)(W_N-1/W_{\rho})(W_N-1/W_{\rho}^*)
 \end{equation*}
 and
\begin{equation*}
 N_{v}=(W_N-W_{v})(W_N-W_{v}^*)(W_N+W_{v})(W_N+W_{v}^*).
 \end{equation*}
 If condition (\ref{chap131}) is fulfilled, then the second derivative of
 $Im F_{\pi}^E(t)$ at $q=0$ is automatically zero. The condition (\ref{chap131}),
 together with the relation
 \begin{equation}
 \sum_{v=\rho,\rho',\rho''}(f_{v \pi\pi}/f_{v})=1,
 \label{chap132}
 \end{equation}
 following from the normalization  (\ref{normff}) and (\ref{chap130}), give
 a system of two algebraic equations with three unknown coupling constant
 ratios. A solution of the latter can be written, e.g. in the form as follows

 \begin{eqnarray}
 (f_{\rho'\pi\pi}/f_{\rho'})&=&\frac{\frac{N_{\rho''}}{|W_{\rho''}|^4}}
 {\frac{N_{\rho'}}{|W_{\rho'}|^4}-
 \frac{N_{\rho''}}{|W_{\rho''}|^4}}-\label{chap133} \\
 &-&\frac{\frac{N_{\rho''}}{|W_{\rho''}|^4}+
 \left (1+2\frac{W_z.W_p}{W_z-W_p}.Re \left [W_{\rho}(1+|W_{\rho}|^{-2})\right ]\right )N_{\rho}} {\frac{N_{\rho'}}{|W_{\rho'}|^4}-
 \frac{N_{\rho''}}{|W_{\rho''}|^4}}\cdot (f_{\rho\pi\pi}/f_{\rho})
 \nonumber \\
 (f_{\rho''\pi\pi}/f_{\rho''})&=&1-\frac{\frac{N_{\rho''}}{|W_{\rho''}|^4}}
 {\frac{N_{\rho'}}{|W_{\rho'}|^4}-
 \frac{N_{\rho''}}{|W_{\rho''}|^4}}+\label{chap134} \\
&+&\left [\frac{\frac{N_{\rho''}}{|W_{\rho''}|^4}
 (1+2\frac{W_zW_p}{W_z-W_p}.Re [W_{\rho}(1+|W_{\rho}|^{-2})])N_{\rho}}{\frac{N_{\rho'}}{|W_{\rho'}|^4}-
 \frac{N_{\rho''}}{|W_{\rho''}|^4}}-1\right ].(f_{\rho\pi\pi}/f_{\rho})\nonumber
 \end{eqnarray}

    Then the expression (\ref{chap130}), together with relations (\ref{chap133})
 and (\ref{chap134}), represents the most accomplished $U\&A$ pion EM FF model, which
 is defined on four sheeted Riemann surface with complex conjugate poles (corresponding
 to unstable $\rho$-meson resonances) on unphysical sheets and reflecting
 all properties (including also the threshold behavior (\ref{imffthrbeh}) of the
 $Im F_{\pi}^E(t)$) briefly reviewed at the beginning of this paragraph. It depends
 on 10 physically interpretable free parameters, $t_{in}$, $m_{\rho}$, $\Gamma_{\rho}$,
 $(f_{\rho,\pi\pi}/f_{\rho})$, $m_{\rho'}$, $\Gamma_{\rho'}$, $m_{\rho''}$, $\Gamma_{\rho''}$,
 $W_z$ and $W_p$. They are determined from the fit of existing reliable
 experimental points on  $|F_{\pi}^E(t)|$, containing, however, also a contribution
 of the isospin violating $\omega({\rm 783})\to \pi^+\pi^-$ decay, leading to the so-called
 $\rho-\omega$ interference effect, which can not be excluded by experimentalists
 in a measurement of $\sigma_{tot}(e^+e^-\to\pi^+\pi^-)$. In  order to take into
 account the latter effect we have carried out the fit of data by
 \begin{equation}
 |F_{\pi}^E[W(t)] + Re^{i\phi}\frac{m^2_{\omega}}{ m^2_{\omega}-t-im_{\omega}\Gamma_{\omega}}|
 \label{chap135}
 \end{equation}
 where
 \begin{equation}
 \phi=\arctan{\frac{m_{\rho}\Gamma_{\rho}}{m_{\rho}^2-m_{\omega}^2}}
 \label{chap136}
 \end{equation}
 is the $\rho-\omega$ interference phase and $R$ is the corresponding amplitude to be left as an
 additional, eleventh, free parameter.
\begin{figure}[tb]
    \centering\vspace{-0.4cm}
    \scalebox{.45}{\includegraphics{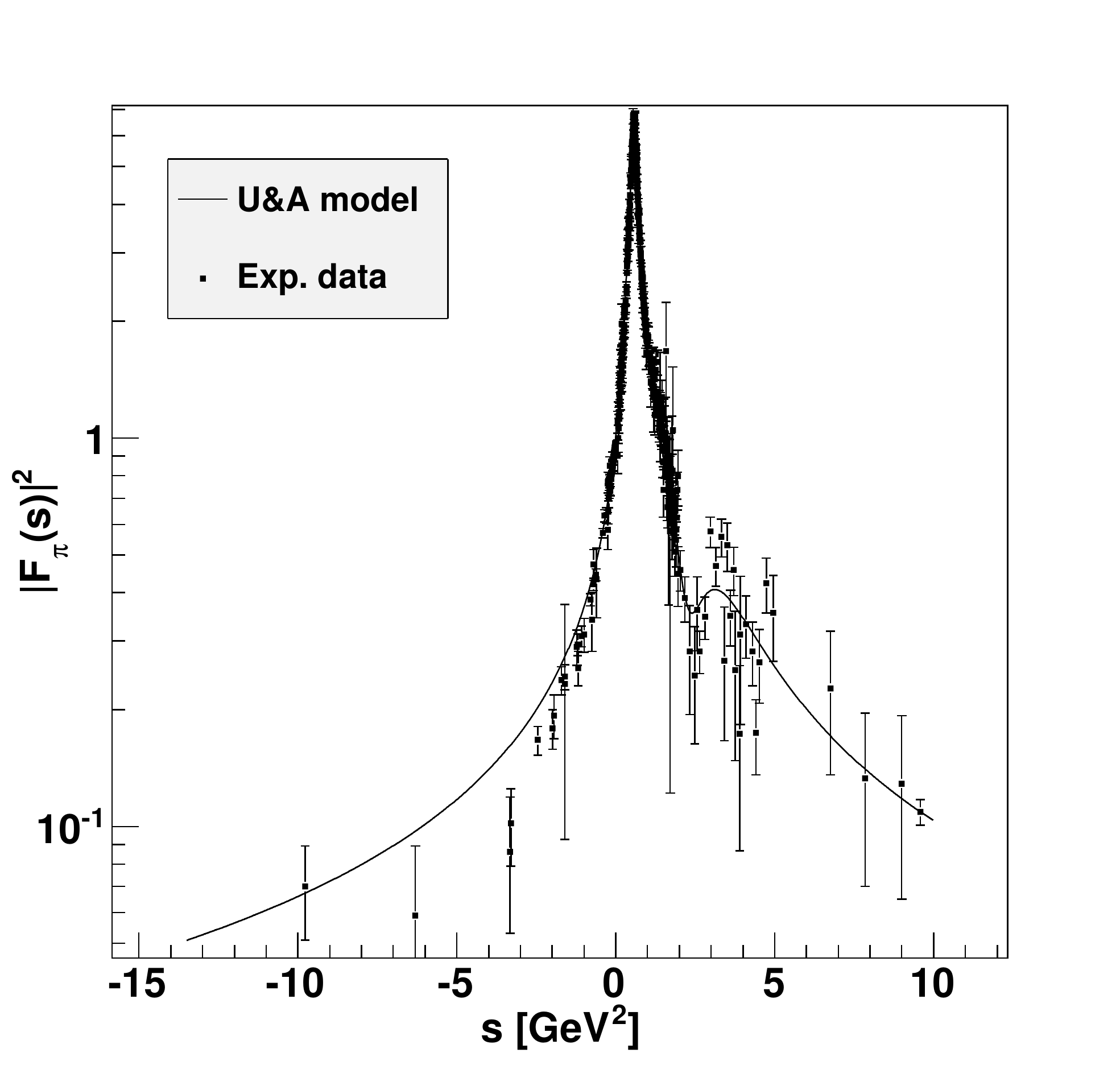}}
    \caption{\small{Description of pion FF data by three resonance $U\&A$ model}}
    \label{figpion}
\end{figure}

\begin{figure}[t]
    \centering
    \scalebox{0.6}{\includegraphics[clip]{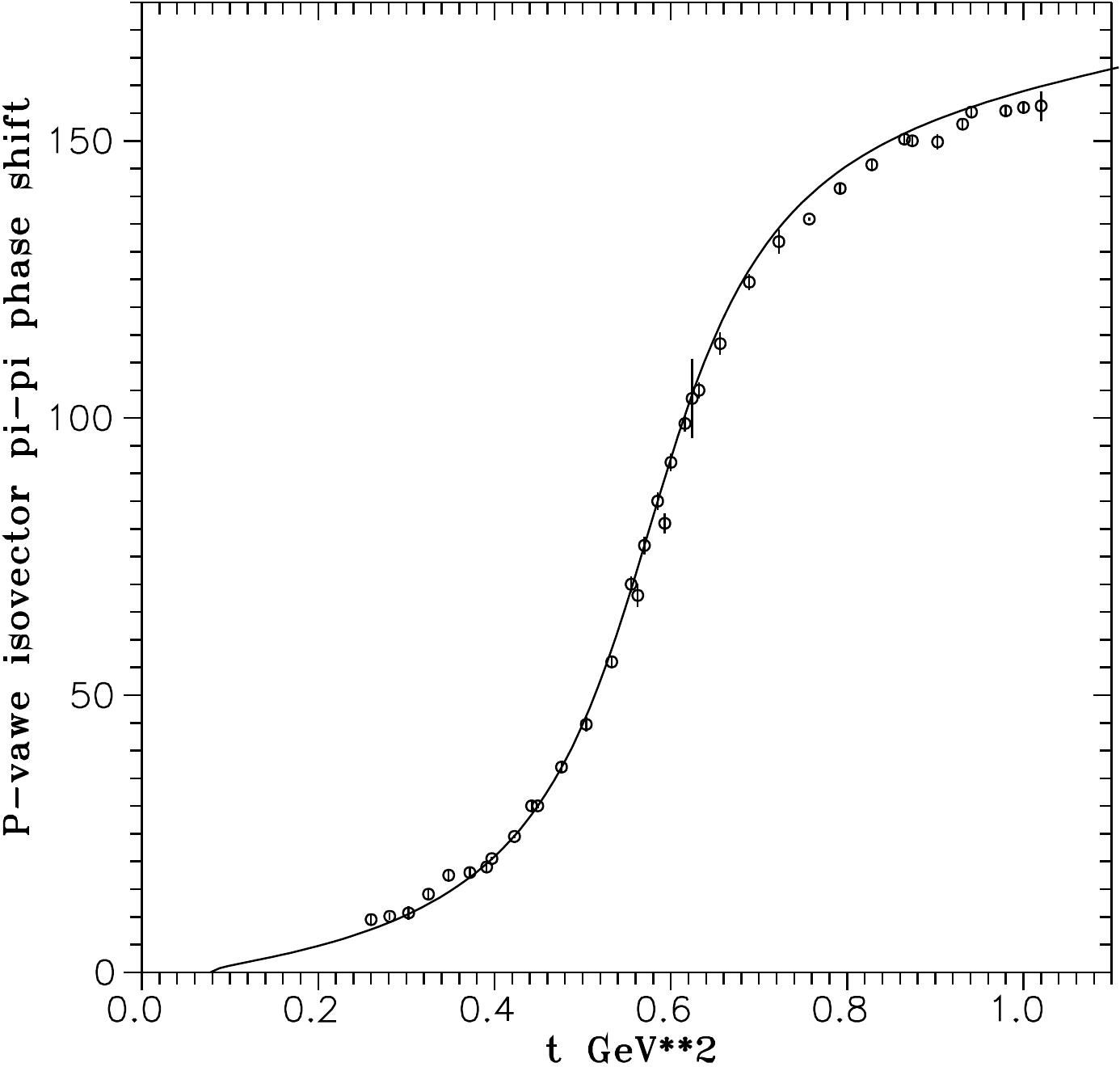}}
    \caption{\small{P-wave isovector $\pi\pi$ phase shift.}}
    \label{figdeltael}
\end{figure}

    A description of all existing data by means of the most accomplished up to now
 $U\&A$ pion EM FF model (\ref{chap130}) with (\ref{chap133}) and
 (\ref{chap134}), and the found values of parameters
 \vspace{-0.1cm}\begin{equation}
 \begin{array}{lcl}
 t_{in}=(1.296\pm 0.011){\rm GeV}^2 && R=0.0123\pm 0.0032\\ \nonumber
 W_z=0.3722\pm 0.0008 && W_p=0.5518\pm 0.0003  \\ \nonumber
  m_{\rho}=(759.26\pm 0.04)\,{\rm MeV}  && \Gamma_{\rho}=(141.90\pm 0.13)\,{\rm MeV}\label{Achap137}\\
  m_{\rho'}=(1395.9\pm 54.3)\,{\rm MeV} && \Gamma_{\rho'}=(490.9\pm 118.8)\,{\rm MeV}\\  \nonumber
  m_{\rho''}=(1711.5\pm 63.6)\,{\rm MeV} && \Gamma_{\rho''}=(369.5\pm 112.7)\,{\rm MeV}\\  \nonumber
 (f_{\rho\pi\pi}/f_{\rho})=1.0063\pm 0.0024 && \chi^2/{\rm NDF}=1.58\\ \nonumber
 \end{array}
 \end{equation}\vspace{-0.1cm}
 is presented in Fig.~\ref{figpion}. A prediction of $\delta_{\pi}(t)$ behavior by this $U\&A$ model and its
 comparison with data on $\delta^1_1(t)$ is shown in Fig.~\ref{figdeltael}.

\begin{figure}[t,h]
    \centering
    \scalebox{0.6}{\includegraphics[clip]{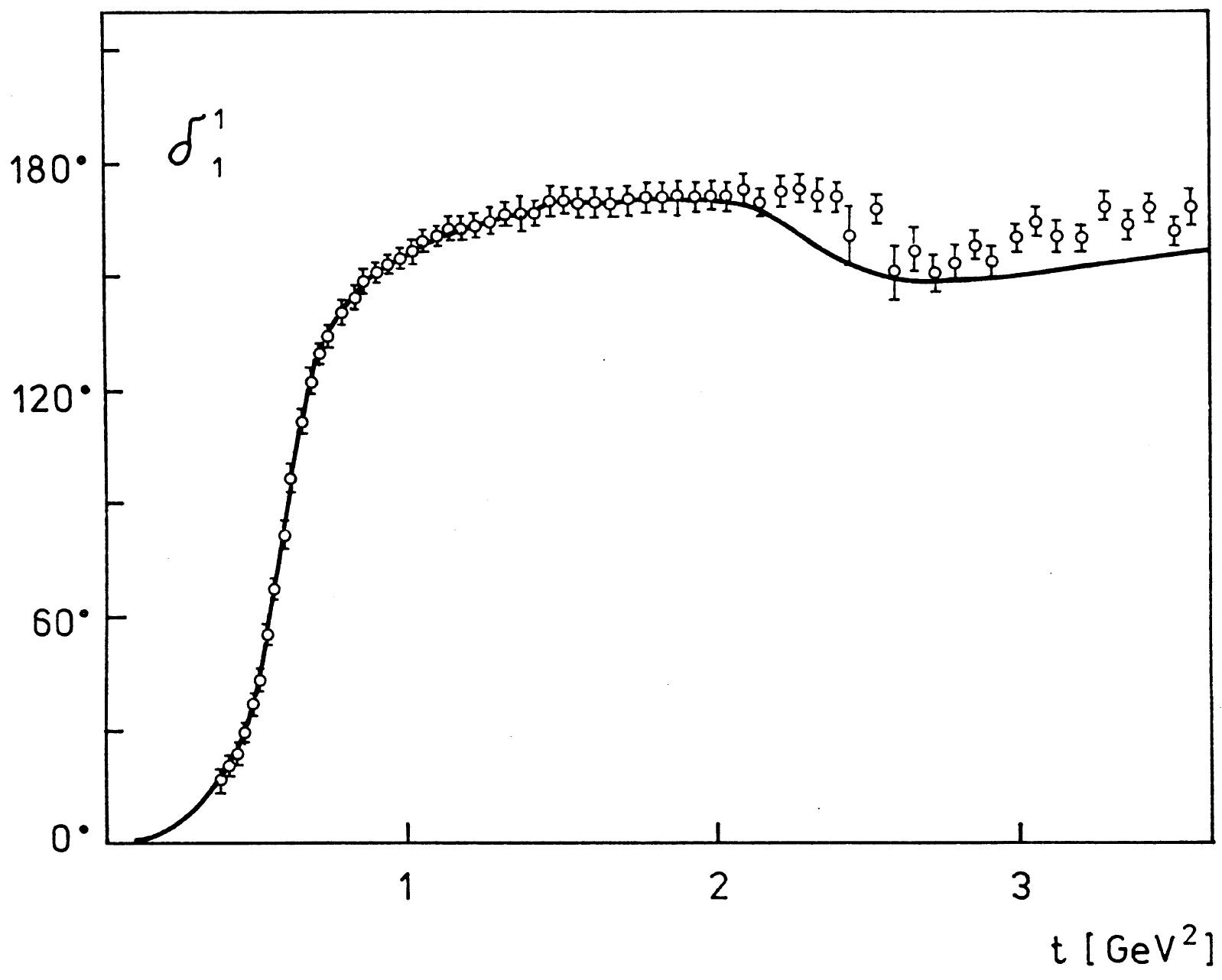}}
    \caption{\small{P-wave isovector $\pi\pi$ phase shift predicted by U\&A model }}
    \label{figdeltainel}
\end{figure}

    Just the coincidence of the predicted behavior of $\delta_{\pi}(t)$ with
existing data on $\delta^1_1(t)$ confirms that the $U\&A$ pion EM
FF model (\ref{chap130}) with (\ref{chap133}) and (\ref{chap134})
fulfills the unitarity condition (\ref{eunicon}), at least from
$4m^2_{\pi}$ to aproximatelly $1 {{\rm GeV}}^2$.

 \medskip

   \subsection{Prediction of P-wave isovector  $\pi \pi$  phase shift and
   inelasticity above inelastic threshold from $e^+e^- \to \pi^+
   \pi^-$ process}\label{III4}

 \medskip

\begin{figure}[tb]
    \centering
    \scalebox{0.6}{\includegraphics[clip]{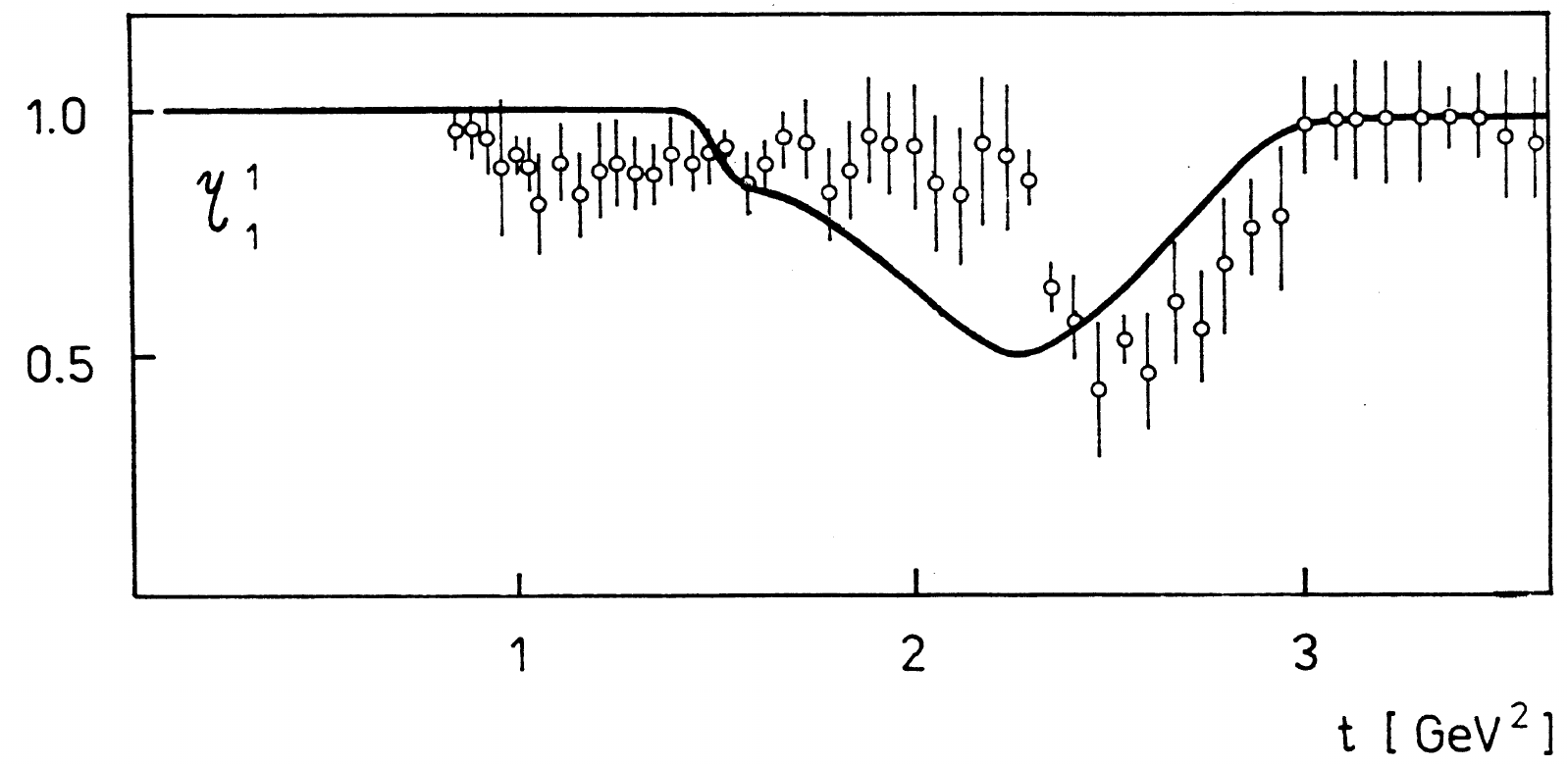}}
    \caption{\small{Predicted behavior of inelasticity of the P- wave}}
    \label{figeta}
\end{figure}

    The unitary and analytic model (\ref{chap130}) of $F^E_\pi[W(t)]$ has one elastic
cut $4m^2_\pi<s<+\infty$ and one effective cut $t_{in}\approx
1{\rm {{\rm GeV}}}^2<t<+\infty$.

   The elastic unitarity condition (\ref{eunicon}) can be utilized
for the analytic continuation of $F^E_\pi[W(t)]$ through the
elastic cut on the second Riemann sheet and as a result one
obtains (\ref{ffsecsh}), the FF on the II. sheet to be expressed
by FF on the I. sheet and the $\pi\pi$ partial wave amplitude
$A^1_1(t)$ on the I. sheet, from where one obtains
\begin{equation}
\left[A^1_1(t)\right]^{\rm I} = \frac{\left[F^E_\pi(t)\right]^{\rm
I} - \left[F^E_\pi(t)\right]^{\rm
II}}{2i\left[F^E_\pi(t)\right]^{\rm II}} \label{A}
\end{equation}
to be valid in the whole complex $t$-plane.

   Now, substituting a standard parametrization of I=J=1 $\pi\pi$
scattering amplitude at the physical region
\begin{equation}
A^1_1(t+i\varepsilon) = \frac{\eta^1_1(t+i\varepsilon)\quad {\rm
e}^{2i\delta^1_1(t+i\varepsilon)}-1}{2i}
\end{equation}
into (\ref{A}), one gets
\begin{equation}
\eta^1_1(t+i\varepsilon){\rm e}^{2i\delta^1_1(t+i\varepsilon)} =
\frac{\left[F^E_\pi[W(t+i\varepsilon)]\right]^{\rm
I\hphantom{I}}}{\left[F^E_\pi[W(t+i\varepsilon)]\right]^{\rm II}}
\end{equation}
from where it is straightforward to find
\begin{equation}
\delta^1_1(t+i\varepsilon)=\frac{1}{2}{\rm arctg} \frac{{\rm
Im}\quad\frac{\left[F^E_\pi[W(t+i\varepsilon)]\right]^{\rm
I\hphantom{I}}}{\left[F^E_\pi[W(t+i\varepsilon)]\right]^{\rm II}}}
{{\rm Re}\quad\frac{\left[F^E_\pi[W(t+i\varepsilon)]\right]^{\rm
I\hphantom{I}}}{\left[F^E_\pi[W(t+i\varepsilon)]\right]^{\rm
II}}}; \quad\quad\quad
\eta^1_1(t+i\varepsilon)=\left|\frac{\left[F^E_\pi[W(t+i\varepsilon)]\right]^{\rm
I\hphantom{I}}}{\left[F^E_\pi[W(t+i\varepsilon)]\right]^{\rm
II}}\right|. \label{deltaeta}
\end{equation}

   Substituting the $U\&A$ model (\ref{chap130}) of the pion EM FF with (\ref{chap133}),
(\ref{chap134}), (\ref{ffsecsh}) and (\ref{A}) into (\ref{deltaeta}) one
predicts \cite{Dubnlnc} behavior of $\delta^1_1(t+i\varepsilon)$ and
$\eta^1_1(t+i\varepsilon)$ in a perfect agreement with existing data
(see Figs.~\ref{figdeltainel} and ~\ref{figeta}) also in the
inelastic region, i.e. above 1~{{\rm GeV}}$^2$.

   The latter is clear demonstration of the analyticity as one of the
powerful means in the elementary particle physics phenomenology.

\medskip

   \subsection{Excited states of the $\rho$(770)-meson}\label{III5}

\medskip

   Here we demonstrate another example of the utilization of the
analyticity as the powerful means in the elementary particle
physics phenomenology.

   From Fig.~\ref{fig10}, where the data on the pion EM FF are collected, one
can see dominating role of the $\rho(770)$-meson. Besides the
latter one can notice in the data also another explicit resonance
around the energy $s=2.9$~{{\rm GeV}}$^2$. But it does not mean that there
are no more other hidden $\rho$-resonances in the pion EM FF to be
cowered in shadow of some other resonances and the background. In
such case it is difficult to identify the resonance by the
Breit-Wiegner form. Here the $U\&A$ pion EM FF model, providing
one analytic function in the whole interval of definition, is
starting to be very suitable. So, the correct approach in an
investigation of the latter problem is then to take the expression
(\ref{chap130}), first, with two resonances, then with three
resonances etc. and to look always for minimal value of $\chi^2$
in experimental data fit.

   Such program has been practically realized. Considering only two resonances in
(\ref{chap130}), $\chi^2/{\rm NDF}$ = 539/279 was achieved and
$\rho(770)$ with $\rho''(1700)$ were identified.

   However, if three resonances were taken into account in
(\ref{chap130}), $\chi^2/{\rm NDF}$ = 382/276 was found and two excited
states, $\rho'(1450)$ and $\rho''(1700)$, were identified
\cite{rho3}.
\begin{figure}[tb]
    \centering
    \scalebox{0.6}{\includegraphics[clip]{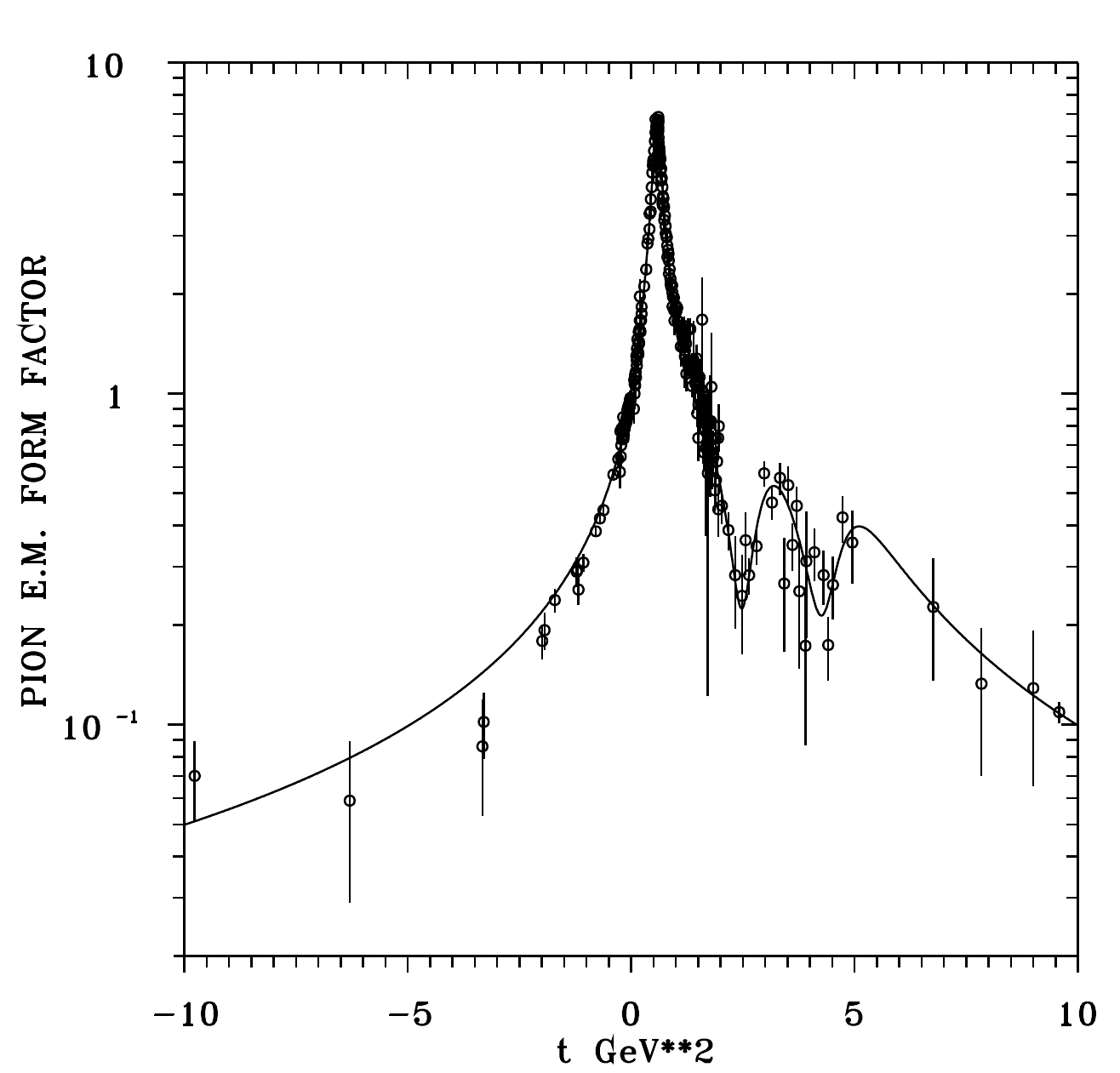}}
    \caption{\small{Description of pion FF data by four resonance $U\&A$ model}}
    \label{fig4pion}
\end{figure}

   Finally, by a consideration of four resonances in (\ref{chap130}),
$\chi^2/{\rm NDF}$ = 343/273 was reached and in addition to
$\rho'(1450)$ and $\rho''(1700)$, also the third excited state
$\rho'''(2150)$ of the $\rho(770)$-meson was revealed \cite{rho4}
(see Fig.~\ref{fig4pion}).

   \subsection{Unitary and analytic model for kaon electromagnetic structure} \label{III6}

   Unlike the pion EM FF, little has been done theoretically for the kaon
EM FFs $F^E_{K^+}(t)$ and $F^E_{K^0}(t)$ up to now. The main reason
was, first, the existence of a rather broad unphysical region
$0< t$ $ <4m^2_K$ unattainable experimentally, in which the dominating
resonances $\rho(770)$, $\omega(783)$ and $\phi(1020)$ of the kaon
EM FFs are spread out, and secondly, shortage of reliable and
compact experimental information outside the internal
$0 <t$ $<4m_K^2$.

   For instance, there are no data on the neutral kaon in the
space-like ($t<0$) region, and for the charge kaon only 25
experimental points in the very narrow interval of momenta
$-0.1145 \quad{{\rm GeV}}^2$ $ \leq t\leq -0.0175 \quad{{\rm GeV}}^2$ exist
\cite{Dallykmes,Amendkmes}. Slightly better situation is in the
time-like ($t>4m^2_K$) region, where the data by means of the
$e^+e^-\to K^+K^-$ and $e^+e^-\to K_LK_S$ processes have been
obtained up to  $t =4.5 \,{{\rm GeV}}^2$.

   The most successful description of this experimental information
on the kaon EM FFs was achieved \cite{rho4,Dubnkmes} in the
framework of the $U\&A$ model (\ref{ffgen}) to be applied to the
isoscalar and isovector parts of the kaon EM FFs
\begin{eqnarray}
F_K^s(t)&=&\left(\frac{1-V^2}{1-V_N^2}\right)^2 \times \nonumber\\
&\times & \left[ \sum_{s=\omega,\phi}
         \frac{(V_N-V_s)(V_N-V_s^*)(V_N-1/V_s)(V_N-1/V_s^*)}
         {(V-V_s)(V+V_s)(V-1/V_s)(V+1/V_s)}
         (f_{sK\bar K}/f_s) +\right.\nonumber \\
&+ &\left.
\frac{(V_N-V_{\phi'})(V_N-V_{\phi}^*)(V_N+V_{\phi'})(V_N+V_{\phi'}^*)}
{(V-V_{\phi'})(V-V_{\phi'}^*)(V+V_{\phi'})(V+V_{\phi'}^*)}
(f_{\phi'K\bar K}/f_{\phi'})\right]\nonumber \\
F_K^v(t)&=&\left(\frac{1-W^2}{1-W_N^2}\right)^2 \times \label{chap35} \\
&\times & \left[ \sum_{v=\rho,\rho'}
\frac{(W_N-W_v)(W_N-W_v^*)(W_N-1/W_v)(W_N-1/W_v^*)}
{(W-W_v)(W-W_v^*)(W-1/W_v)(W-1/W_v^*)}
(f_{vK\bar K}/f_v)+\right.\nonumber
\end{eqnarray}
\begin{eqnarray}
& +&\left.
\frac{(W_N-W_{\rho''})(W_N-W_{\rho''}^*)(W_N+W_{\rho''})(W_N+W_{\rho''}^*)}
{(W-W_{\rho''})(W-W_{\rho''}^*)(W+W_{\rho''})(W+W_{\rho''}^*)}
(f_{\rho''K\bar K}/f_{\rho''})\right],\nonumber
\end{eqnarray}
which are related to the charge and neutral kaon EM FFs by the
relations
\begin{equation}
  F^E_{K^+}(t) = F^s_K(t) + F^v_K(t)\label{Kffch}
\end{equation}
\begin{equation}
  F^E_{K^0}(t) = F^s_K(t) - F^v_K(t)\label{Kff0}
\end{equation}
as it follows directly from (\ref{ffKs}) and (\ref{ffKv}).

   The expressions (\ref{chap35}) are analytic functions in
the whole $t$-plane besides two cuts on the positive real axis
starting, in the case of $F_K^s(t)$ from $t_0^s= 9m^2_{\pi}$ and
$t_{in}^s$ and in the case of the $F_K^v(t)$ from
$t_0^v=4m_{\pi}^2$ and $t_{in}^v$, to be real on the whole
negative real axis up to positive values  $t_0^s= 9 m^2_{\pi}$ and
$t_0^v= 4 m^2_{\pi}$, respectively, automatically normalized
(\ref{normKff}) if
\begin{equation}
\sum_s(f_{s K\bar K}/f_s) = \sum_v(f_{vK\bar K}/f_v) =1/2,
\label{chap36}
\end{equation}
with $ImF_K^s(t)\neq 0$ starting from $9m^2_{\pi}$ up to $+\infty$
and $ImF_K^v(t)\neq 0$ starting from $4m^2_{\pi}$ up to $+\infty$,
with poles corresponding to vector-mesons placed in the complex
conjugate pairs on the unphysical sheets of the Riemann surface,
on which both EM FFs $F_K^s(t)$ and $F_K^v(t)$ are defined and
governing the asymptotic behavior (\ref{asym}) with $n_q=2$ as
predicted by the quark model of hadrons.

\begin{figure}[tb]
\begin{center}
\includegraphics[scale=.4,clip]{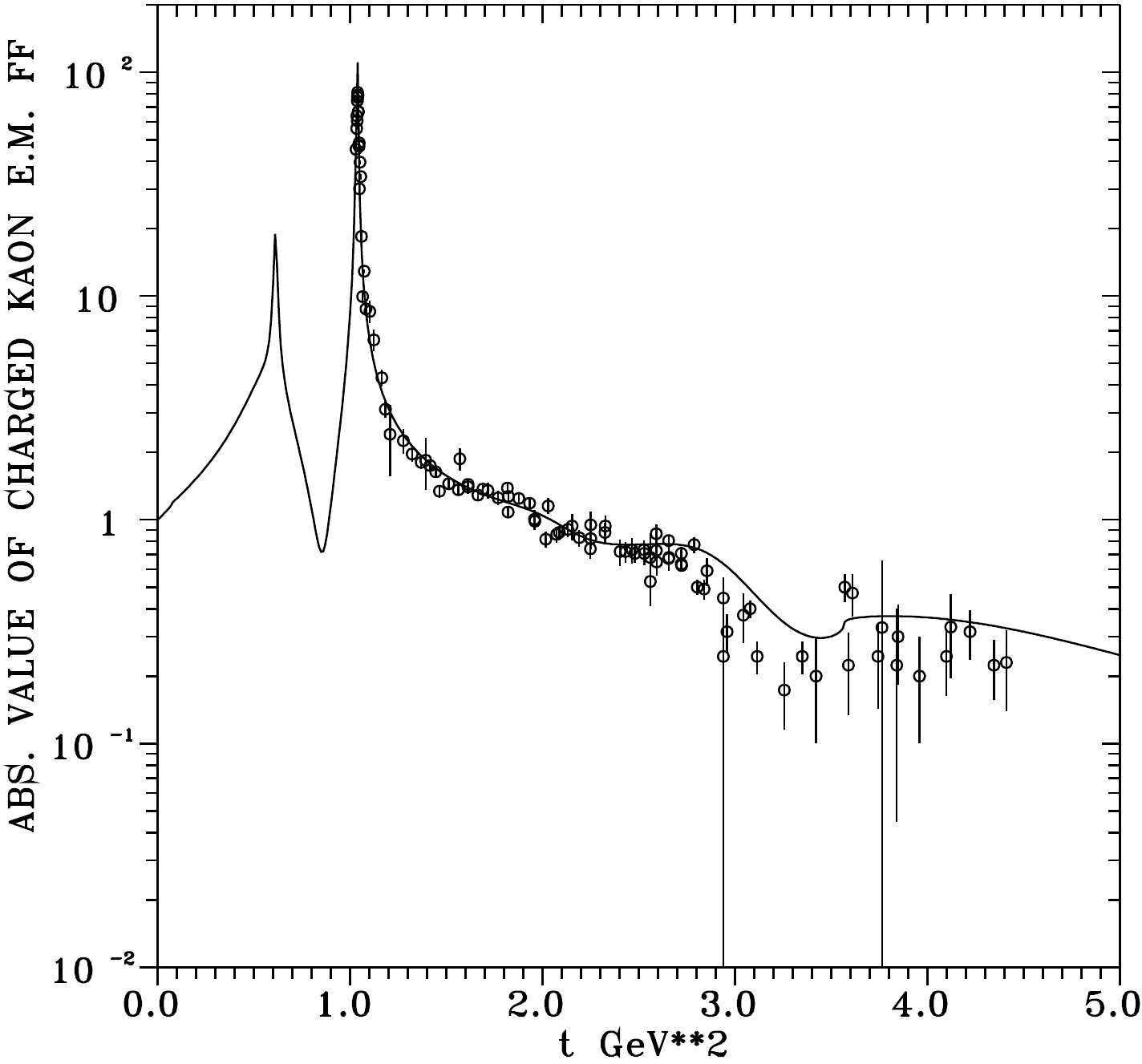}
   \hspace{.5cm}
\includegraphics[scale=.4,clip]{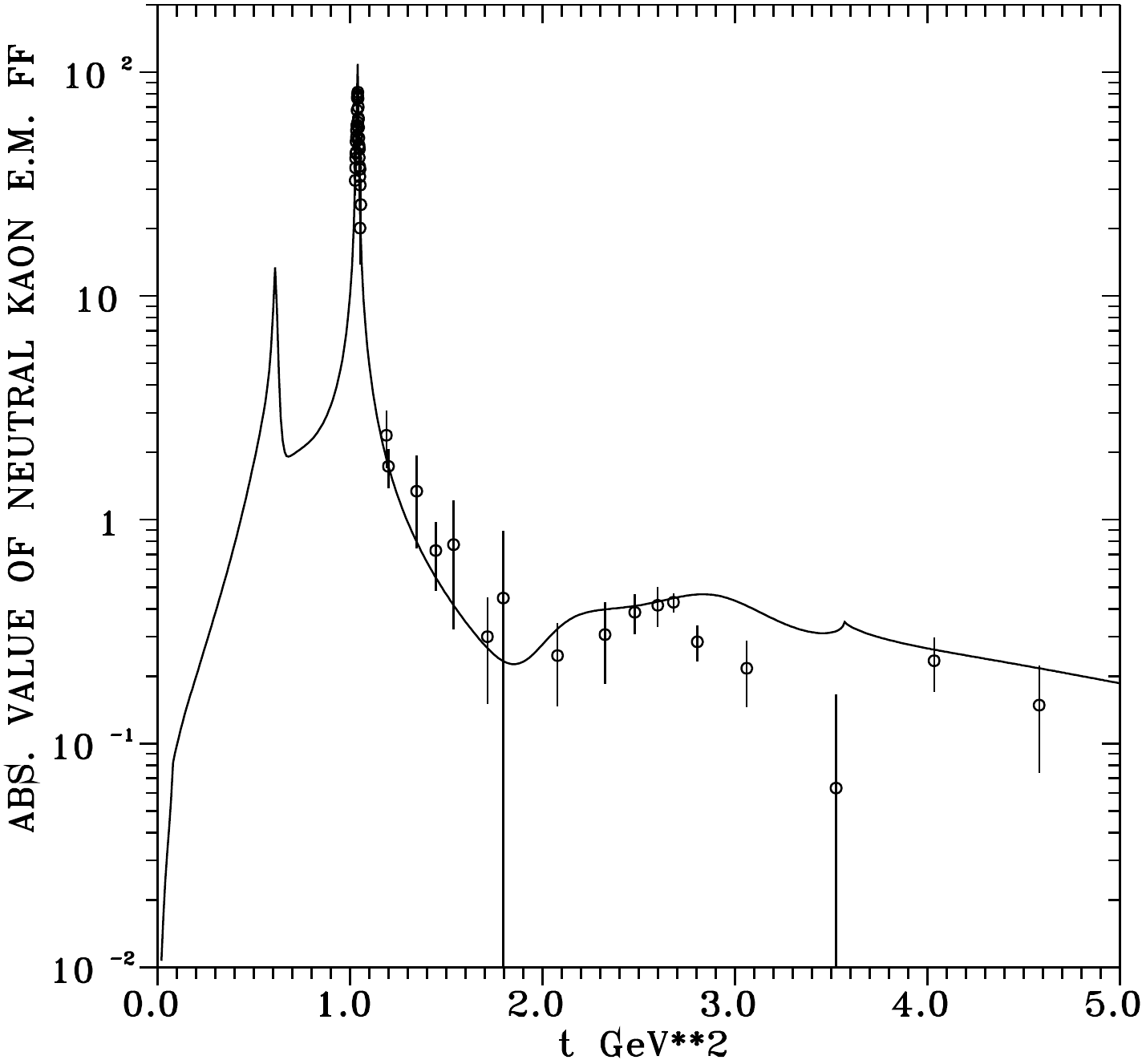}
\end{center}
  \caption{Description of the charged and  neutral kaon FF data by $U\&A$ model.}
  \label{fig16-17}
\end{figure}

   The reproduction of all existing data on $|F_{K^+}(t)|$ and
$|F_{K^0}(t)|$ by means of (\ref{chap35}), (\ref{Kffch}),
(\ref{Kff0}) and the free parameters of the model (the values of
parameters $m_{\rho}$, $\Gamma_{\rho}$, $m_{\omega}$,
$\Gamma_{\omega}$ are fixed at the world averaged values)
\begin{equation}
\begin{array}{lcl}
q_{in}^s=\sqrt{(t_{in}^s-9)/9}= 2.2326 [m_{\pi}] && q_{in}^v=\sqrt{(t_{in}^v-4)/4}= 6.6721 [m_{\pi}]\\
(f_{\omega K\bar K}/f_{\omega})=0.14194 && (f_{\rho K\bar K}/f_{\rho})=0.5615\\
m_{\phi}=7.2815 [m_{\pi}]  &&  m_{\rho'}=10.3940 [m_{\pi}] \\
\Gamma_{\phi}= 0.03733 [m_{\pi}] && \Gamma_{\rho'}=1.6284 [m_{\pi}]\\
(f_{\phi K\bar K}/f_{\phi})=0.4002  && (f_{\rho'K\bar K}/f_{\rho'})=-0.3262\\
m_{\phi'}=11.8700 [m_{\pi}]  &&  m_{\rho''}=13.5650 [m_{\pi}] \\
\Gamma_{\phi'}= 1.3834 [m_{\pi}] && \Gamma_{\rho''}=3.4313 [m_{\pi}]\\
(f_{\phi'K\bar K}/f_{\phi'})=-.04214 && (f_{\rho''K\bar K}/f_{\rho''})=-.02888\\
\chi^2/{\rm NDF} =181/166 &&
\end{array}
\label{chap37}
\end{equation}

are presented in Fig.~\ref{fig16-17}.

   From the obtained results one observes that the contribution of
the $\rho'''(2150)$ resonance to the $e^+e^-\to K\bar K$ processes
is favored prior to the $\rho''(1700)$ one by the existing data in
the charge and neutral kaon EM FFs.

   In the Review of Particle Properties \cite{Rpp} are besides
$\omega(783)$ meson also its higher excitation states,
$\omega(1420)$ and $\omega(1600)$, presented. Therefore it could
be interesting to investigate their contributions to the
$e^+e^-\to K\bar K$ processes by means of the $U\&A$ model of the
kaon Em FFs.

   However, the latter problem can be considered seriously only in
the case that more abundant and more precise experimental
information on the kaon EM FFs will appear.

\medskip

   \subsection{Conserved-vector-current (CVC) hypothesis and the $\bar
   \nu_e e \to M^-M^0$ and $\tau^- \to \nu_\tau M^-M^0$ weak
   processes}\label{III7}

\medskip

   At the end of fifties, by Gerstein and Zeldowich \cite{Gerstzeld}, and independently
by Feynman and Gell-Mann \cite{Feyngelm}, the
conserved-vector-current (CVC) hypothesis, as a theoretical ground
for an explanation of an approximate numerical equality of the muon
decay constant $G_{\mu}$ and the neutron decay vector constant
$G_{\rm V}$, was postulated in the framework of the V-A weak
interaction theory.

The latter hypothesis, besides others, provides a relation between a
matrix element of the vector part of the weak charged hadronic
current and a corresponding matrix element of the EM current to be
taken between two pion states. As a result of the foregoing, a
probability of the $\pi^+$-meson beta-decay
$\pi^+\to\pi^0+e^++{\nu}_e$ was predicted \cite{Gerstzeld,Feyngelm}
theoretically. Its agreement with experimental results
was presented \cite{Nambu} as a demonstration of the general validity of the CVC
hypothesis in the weak interaction theory.

Further, based on the CVC hypothesis and $U\&A$
model of the pseudoscalar meson EM FF's formulated above, we
predict \cite{Dubnazrek1},\cite{Dubnazrek2} a behavior
of
 $\sigma_{\rm tot}(E^{\rm lab}_{\nu})$
and d$\sigma$/d$E^{\rm lab}_{\pi}$ of the weak $\bar{\nu}_{\rm
e}e^-\to M^-M^0$ processes for the first time and the ratio of
total probabilities  $\Gamma(\tau^-\to\bar{\nu}_{\tau}
M^-M^0)/\Gamma_e$ are determined  and compared with
experimental values and theoretical values of other estimates.

\begin{figure}[tb]
\centerline{\includegraphics[scale=.7]{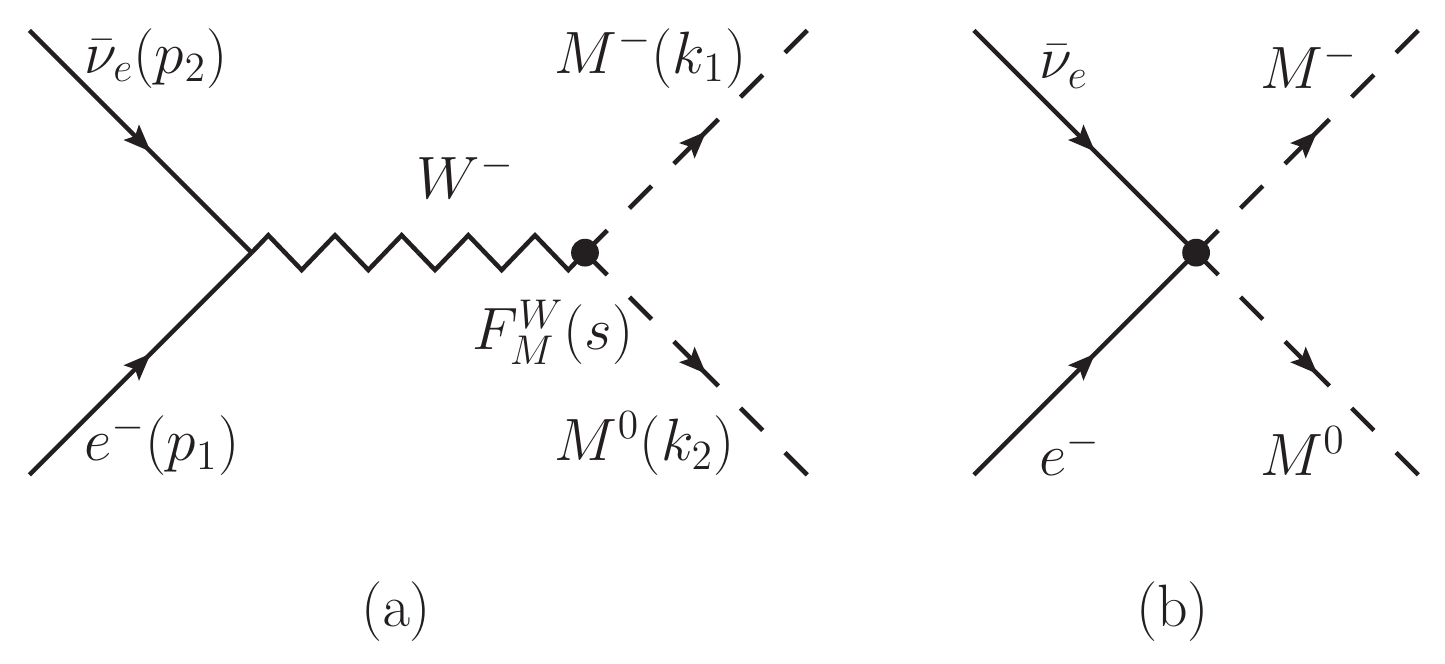}}
\caption{\small{Reduction of the Feynman diagram to contact one}}
\label{fignee}
\end{figure}

   In general, the differential cross-section of the weak $\bar{\nu}_e e^-\to M^-M^0$
reactions in the c.m. system is given by the following expression
\begin{equation}
\frac{{\rm d}\sigma}{{\rm d}{\mit
\Omega}}=\frac{1}{(2\mbox{s}_{\bar{\nu}}+1)(2\mbox{s}_{\rm e}+1)}
\frac{1}{64\pi^2s}\frac{k^{\rm c.m.}}{p^{\rm c.m}}
\sum_{\mbox{s}_{\bar{\nu}},\mbox{s}_e} |{\cal{M}}|^2,
\label{chap71}
\end{equation}
where $s\ge 4m_\pi^2$ is the c.m. energy squared, $k^{\rm
c.m.}=[(s-4m_\pi^2)/4]^{1/2}$ is a length of a 3-dimensional
momentum of produced pseudoscalar mesons, $p^{\rm
c.m.}=\sqrt{s/4}$ is a length of a 3-dimensional neutrino-momentum
(assuming $m_{\nu_{\rm e}}=0$) and $s_{\bar{\nu}}$ and $s_{\rm e}$
are spins of the antineutrino and electron, respectively. The
matrix element $M$ in the lowest order of a perturbation expansion
can be calculated from the Feynman diagram presented in Fig.~\ref{fignee}a
(the $F_M^{\rm W}(s)$ is the weak FF of a charged $W^-$-boson
transition $(W^-)^*\to M^-M^0$), which for $s\ll m_{{\rm W}^-}^2$ is
reduced to a contact diagram shown in Fig.~\ref{fignee}b.

In the standard electro-weak theory an effective Hamiltonian
describing the $\bar{\nu}_{\rm e}e^-\to M^-M^0$ process is
\begin{equation}
{\cal H}_{\rm I}=\frac{G}{\sqrt{2}}\left[\bar{\nu}_{\rm
e}\gamma_{\mu}\left(1+\gamma_5\right)e\right]J_{\mu}^W+h.c.,
\label{chap72}
\end{equation}
where $G=1.1663\times 10^{-5}\mbox{\,{{\rm GeV}}}^{-2}$ is the Fermi
constant of the weak interactions and $J_\mu^W$ is a pseudoscalar
meson weak current. Then the matrix element corresponding to the
diagrams in Fig.~\ref{fignee} takes the form
\begin{equation}
{\cal M}=\frac{G}{\sqrt{2}}\bar{\nu}_{\rm
e}(p_2)\gamma_{\mu}\left(1+\gamma_5\right)
e(p_1)\left(k_1-k_2\right)^\mu F_M^{\rm W}(s) \label{chap73}
\end{equation}
and as a result, the differential cross-section (\ref{chap71})
 can be calculated explicitly
\begin{equation}
\frac{{\rm d}\sigma}{{\rm d}{\mit
\Omega}}=\frac{G^2}{128\pi^2}\cdot s\beta^3_M |F^{\rm
W}_M(s)|^2\sin^2{\vartheta}, \label{chap74}
\end{equation}
where $\beta_M=\sqrt{1-4m^2_M/s}$ is the velocity of produced
pions and $\vartheta$ is the scattering angle in the c.m. system.

Now considering an explicit form of  d$\mit\Omega$=$\sin\vartheta$
d$\vartheta$ d$\varphi$ and then integrating over angles
$\vartheta$ and $\varphi$ in (\ref{chap74}), one gets the total
cross-section
\begin{equation}
\sigma_{\rm tot}(s)=\frac{G^2}{48\pi} s\cdot\beta_M^3 |F_M^{\rm
W}(s)|^2 \label{chap75}
\end{equation}
of the weak $\bar{\nu}_{\rm e}e^-\to M^-M^0$ process.

In order to predict the behavior of (\ref{chap74}) or
(\ref{chap75}) as dependent on $s$, one is in need of a knowledge
of the weak pion FF $F_M^{\rm W}(s)$ as a function of $s$. Since,
there are neither data on the latter, nor accomplished dynamical
theory of strong interactions able to predict the $F_M^{\rm W}(s)$
behavior for $s\ge 4m_M^2$, not even a phenomenological approach
like it is in the case of the EM FF's, one has to use the
CVC-hypothesis \cite{Gerstzeld,Feyngelm} providing a relation of the weak
$F_M^{\rm W}(s)$ just with the pure isovector part of EM FF's of
the pseudoscalar mesons.

In order to derive the  latter relations we start with the
standard expression of the pseudoscalar meson weak current
\begin{equation}
J^{\rm W}_\mu=V_\mu+A_\mu \label{chap76}
\end{equation}
where $V_\mu$ and  $A_\mu$ are the vector and axial-vector,
respectively.

On the other hand, the EM current of hadrons is composed of the
sum
\begin{equation} J^{\rm E}_\mu=(J_3)_\mu+(J_{\rm S})_\mu,
\label{chap77}
\end{equation}
where $(J_3)_\mu$ is a third component of an isotopic vector
current $\vec{J}^\mu(J_1^\mu,J_2^\mu,J_3^\mu)$ and $(J_{\rm
S})0_\mu$ is an isoscalar current.

In the second-half of the fifties a very predictable postulation
was introduced
\begin{equation}
V^\mu=J_1^\mu-i J_2^\mu, \label{chap78}
\end{equation}
i.e. that the charged weak vector current $V^\mu$ and the
isovector part $J_3^\mu$ of the EM current are components of the
same isotopic vector $\vec{J}^\mu$. Since, strong interactions are
invariant with respect to the isotopic SU(2) group, the isotopic
vector current $\vec{J}^\mu$ obeys the relation
\begin{equation}
\partial_\mu J_i^\mu=0,
\label{chap79}
\end{equation}
which directly results in the formalism of the
conserved-vector-current (CVC) hypothesis \begin{equation}
\partial_\mu V^\mu=0.
\label{chap780}
\end{equation}

Further, in order to derive a relation between $F_M^{\rm W}(s)$
and $F_M^{\rm E,I=1}(s)$, we start with a commutation relation
\begin{equation}
\left [T_i,J^\mu_j\right ]=i\varepsilon_{ijk} J_k^\mu,
\label{chap781}
\end{equation}
with $T_i$ being an isospin operator. The relation (\ref{chap781})
is fulfilled automatically if $\vec{J}^\mu$ is transformed in the
isospin space like a vector. Now defining
\begin{eqnarray}T_-=T_1-i\,T_2\nonumber\\ \makebox[1cm][l]{and}\label{chap782}\\
J^\mu_-=J_1^\mu-i\,J_2^\mu,\nonumber\end{eqnarray}
 one can prove the following relation
\begin{equation} \left[T_-,J_3^\mu\right]=J_-^\mu\,\equiv\,V^\mu
\label{chap783}
\end{equation}
by using expressions (\ref{chap782}).

If we multiply (\ref{chap783}) from the right-hand side by a state
vector of $M^+$ pseudoscalar meson $|M^+\rangle$ and from the
left-hand side by a state vector of $M^0$ pseudoscalar meson
$\langle M^0|$ of the same isomultiplet, and simultaneously we use
the relations
\begin{eqnarray}
 T_-|T,T_3\rangle&=&\sqrt{(T+ T_3)(T- T_3+1)}\quad|T,T_3-1\rangle\label{chap784}\\
\langle T,T_3|T_-&=&\sqrt{(T- T_3)(T+ T_3+1)}\quad\langle
T,T_3+1|\nonumber
\end{eqnarray}
we get
\begin{eqnarray}
& &\langle M^0|V^\mu|M^+\rangle=\langle M^0|\left [T_-,J^\mu_3\right ]|M^+\rangle= \label{chap785} \\
& &=\sqrt{(T- T_3)(T+ T_3+1)}\quad\langle M^+|J^\mu_3|M^-\rangle
-\sqrt{(T+ T_3)(T- T_3+1)}\quad\langle
M^0|J^\mu_3|M^0\rangle.\nonumber
\end{eqnarray}
Then from (\ref{chap785}) for pions one gets the relations as
follows
\begin{equation}
\langle\pi^0|V^\mu|\pi^+\rangle=\sqrt{2}\langle\pi^+|J^\mu_3|\pi^+\rangle
-\sqrt{2}\langle\pi^0|J^\mu_3|\pi^0\rangle, \label{chap786}
\end{equation}
in which the second term is zero due to the charge conjugation.
Really, if $U_{\rm C}$ is a unitary charge conjugation operator,
and
\begin{equation}
U_{\rm C}|\pi^0\rangle=|\pi^0\rangle,\quad\quad\quad U_{\rm C}J^\mu_3U_{\rm
C}^{-1}=-J^\mu_3, \label{chap787}
\end{equation}
then
\begin{equation}
\langle\pi^0|J_3^\mu|\pi^0\rangle=\langle\pi^0|U_{\rm
C}^{-1}U_{\rm C}J^\mu_3U_{\rm C}^{-1}U_{\rm
C}|\pi^0\rangle=-\langle\pi^0|J^\mu_3|\pi^0\rangle\equiv 0.
\label{chap788}
\end{equation}
So, finally one gets the relation
\begin{equation}
\langle\pi^0|V^\mu|\pi^+\rangle=\sqrt{2}\langle\pi^+|J^\mu_3|\pi^+\rangle.
\label{chap789}
\end{equation}

On the other hand, if we multiply (\ref{chap77}) from the
right-hand side and left-hand side by a state vector of $\pi^+$
meson, we get
\begin{equation}
\langle\pi^+|J^\mu_{\rm
E}|\pi^+\rangle=\langle\pi^+|J^\mu_3|\pi^+\rangle\,+\,\langle\pi^+|J^\mu_S|\pi^+\rangle.
\label{chap790}
\end{equation}
Because strong interactions are invariant under the
G-transformation, that is a combination of the charge conjugation
and an isotopic rotation for an angle  180$^{\circ}$ around the
second axis in the isospin space, the second term in
(\ref{chap790}) is equal to zero. Really, considering that
\begin{equation}
U_{\rm G}|\pi^+\rangle=-|\pi^+\rangle,\makebox[3cm]{ }U_{\rm
G}J^\mu_{\rm S}U_{\rm G}^{-1}=-J^\mu_{\rm S} \label{chap791}
\end{equation} one has
\begin{equation}
\langle\pi^+|J_{\rm S}^\mu|\pi^+\rangle=\langle\pi^+|U_{\rm
G}^{-1}U_{\rm G}J^\mu_SU_{\rm G}^{-1}U_{\rm
G}|\pi^+\rangle=-\langle\pi^+|J^\mu_{\rm S}|\pi^+\rangle\equiv 0
\label{chap792}
\end{equation}
and as a result,
\begin{equation} \langle\pi^+|J_{\rm E}^\mu|\pi^+\rangle=\langle\pi^+|J^\mu_3|\pi^+\rangle.\label{chap793} \end{equation}
Comparison of (\ref{chap793}) with (\ref{chap789}) leads to
\begin{equation}
\langle\pi^0|V^\mu|\pi^+\rangle=\sqrt{2}\langle\pi^+|J^\mu_{\rm
E}|\pi^+\rangle. \label{chap794}
\end{equation}
Then parametrizing the matrix elements in (\ref{chap794}) through
corresponding FF's one gets finally the  relation between the pure
isovector part of the pion EM FF  and the weak pion FF in the
following form
\begin{equation} F_\pi^{\rm W}(s)=\sqrt{2}F_\pi^{\rm E,I=1}(s).\label{chap795}\end{equation}

For kaons the expression (\ref{chap785}) gives the following
relation
\begin{equation}
\langle K^0|V^\mu|K^+\rangle=\langle
K^+|J^\mu_3|K^+\rangle-\langle K^0|J^\mu_3|K^0\rangle.
\label{chap796}
\end{equation}
By using the relation (\ref{chap77}) one can write down for the
difference
\begin{eqnarray}
& &\langle K^+|J^{\mu}_E|K^+\rangle- \langle K^0|J^{\mu}_E|K^0\rangle= \label{chap797}\\
&=&\langle K^+|J^{\mu}_3|K^+\rangle- \langle
K^0|J^{\mu}_3|K^0\rangle +\langle K^+|J^{\mu}_S|K^+\rangle-
\langle K^0|J^{\mu}_S|K^0\rangle.\nonumber
\end{eqnarray}
Now applying the relations
\begin{eqnarray}
 T_+|T,T_3\rangle&=&\sqrt{(T- T_3)(T+ T_3+1)}\quad|T,T_3+1\rangle\label{chap798}\\
\langle T,T_3|T_+&=&\sqrt{(T+ T_3)(T- T_3+1)}\quad\langle
T,T_3-1|\nonumber
\end{eqnarray}
to kaons as follows
\begin{equation}
T_+|K^0\rangle =|K^+\rangle; \quad\quad\quad \langle K^+|T_+=\langle K^0|
\label{chap799}
\end{equation}
one gets
\begin{equation}
\langle K^+|J^\mu_S|K^+\rangle=\langle
K^+|J^\mu_ST_+|K^0\rangle=\langle K^+|T_+J^\mu_S|K^0\rangle=
\langle K^0|J^\mu_S|K^0\rangle \label{chap7100}
\end{equation}
which in (\ref{chap797}) leads to
\begin{equation}
\langle K^+|J^\mu_E|K^+\rangle -\langle K^0|J^\mu_E|K^0\rangle=
\langle K^+|J^\mu_3|K^+\rangle -\langle K^0|J^\mu_3|K^0\rangle.
\label{chap7101}
\end{equation}
A comparison of (\ref{chap796}) and (\ref{chap7101}) gives finally
\begin{equation}
\langle K^0|V^\mu|K^+\rangle= \langle K^+|J^\mu_E|K^+\rangle
-\langle K^0|J^\mu_E|K^0\rangle. \label{chap7102}
\end{equation}
Then parametrizing the matrix elements in (\ref{chap7101}) through
corresponding FF's, we obtain the relation between the isovector
part of the kaon EM FF's and the weak kaon FF in the following
form
\begin{equation}
F_K^W(s)= F_{K^+}(s)-
F_{K^0}(s)=F_K^S(s)+F_K^V(s)-F_K^S(s)+F_K^V(s)=2 F_K^V(s).
\label{chap7103}
\end{equation}
Substituting the relations  (\ref{chap795}) and (\ref{chap7103})
into (\ref{chap74}) and (\ref{chap75}) and using the unitary and
analytic model of the pion and kaon EM structure formulated in
we get the differential and total cross-sections
\begin{equation}
\frac{{\rm d}\sigma(\bar\nu_e e^-\to\pi^-\pi^0)}{{\rm
d}\Omega}=\frac{G^2}{64\pi^2}\beta^3_{\pi}\cdot  s |F^{\rm
E,I=1}_{\pi}(s)|^2\sin^2{\vartheta} \label{chap7104}
\end{equation}
\begin{equation}
\sigma_{\rm tot}(\bar\nu_e e^-\to\pi^-\pi^0) =
\frac{G^2}{24\pi}\beta^3_{\pi} \cdot s |F_\pi^{\rm E,I=1}(s)|^2
\label{chap7105}
\end{equation}
and
\begin{equation}
\frac{{\rm d}\sigma(\bar\nu_e e^-\to K^-K^0)}{{\rm
d}\Omega}=\frac{G^2}{32\pi^2}\beta^3_K \cdot s
|F^V_K(s)|^2\sin^2{\vartheta} \label{chap7106}
\end{equation}
\begin{equation}
\sigma_{\rm tot}(\bar\nu_e e^-\to K^-K^0) =
\frac{G^2}{12\pi}\beta^3_K \cdot s |F_K^V(s)|^2 \label{chap7107}
\end{equation}
of the weak annihilation processes $\bar{\nu}_{\rm e}e^-\to
\pi^-\pi^0$ and $\bar{\nu}_{\rm e}e^-\to K^-K^0$ respectively, for
the first time \cite{Dubnazrek1,Dubnazrek2}. The comparison of $\sigma_{\rm
tot}(\bar{\nu}_{\rm e}e^-\to \pi^-\pi^0)$ with $\sigma_{\rm
tot}(e^+e^-\to \pi^+\pi^-)$,  and comparison of $\sigma_{\rm
tot}(\bar{\nu}_{\rm e}e^-\to K^-K^0)$ with $\sigma_{\rm
tot}(e^+e^-\to K^+K^-)$ and $\sigma_{\rm
tot}(e^+e^-\to K^0_SK^0_L)$ is presented in Fig.~{\ref{sigpipi} and Fig.~\ref{sigkk},
respectively.

\begin{figure}[tb]
\centerline{\includegraphics[scale=.7]{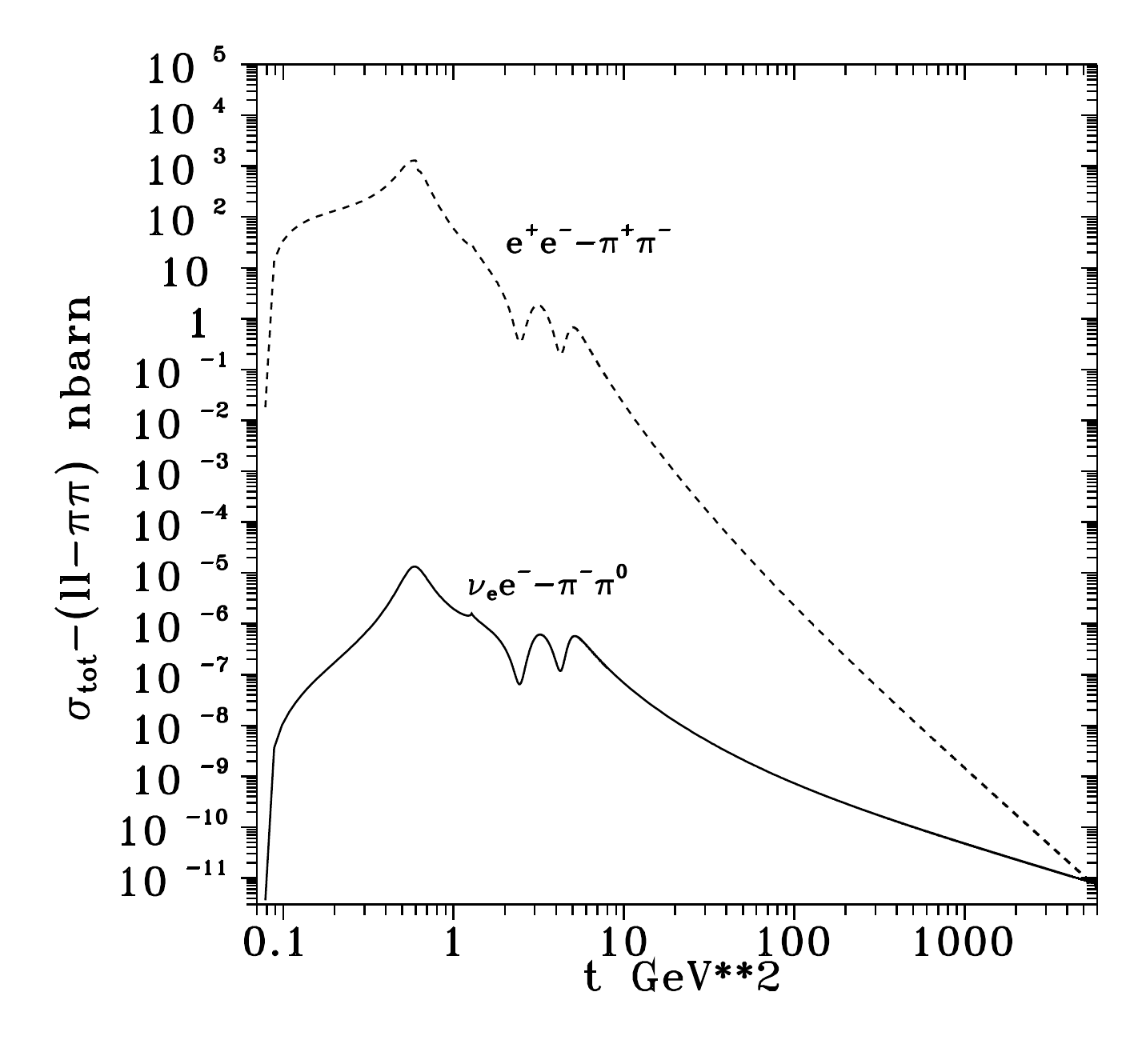}}
\caption{\small{Cross-section comparison of electromagnetic and weak process.}}
\label{sigpipi}
\end{figure}

\begin{figure}[tb]
\centerline{\includegraphics[scale=.75]{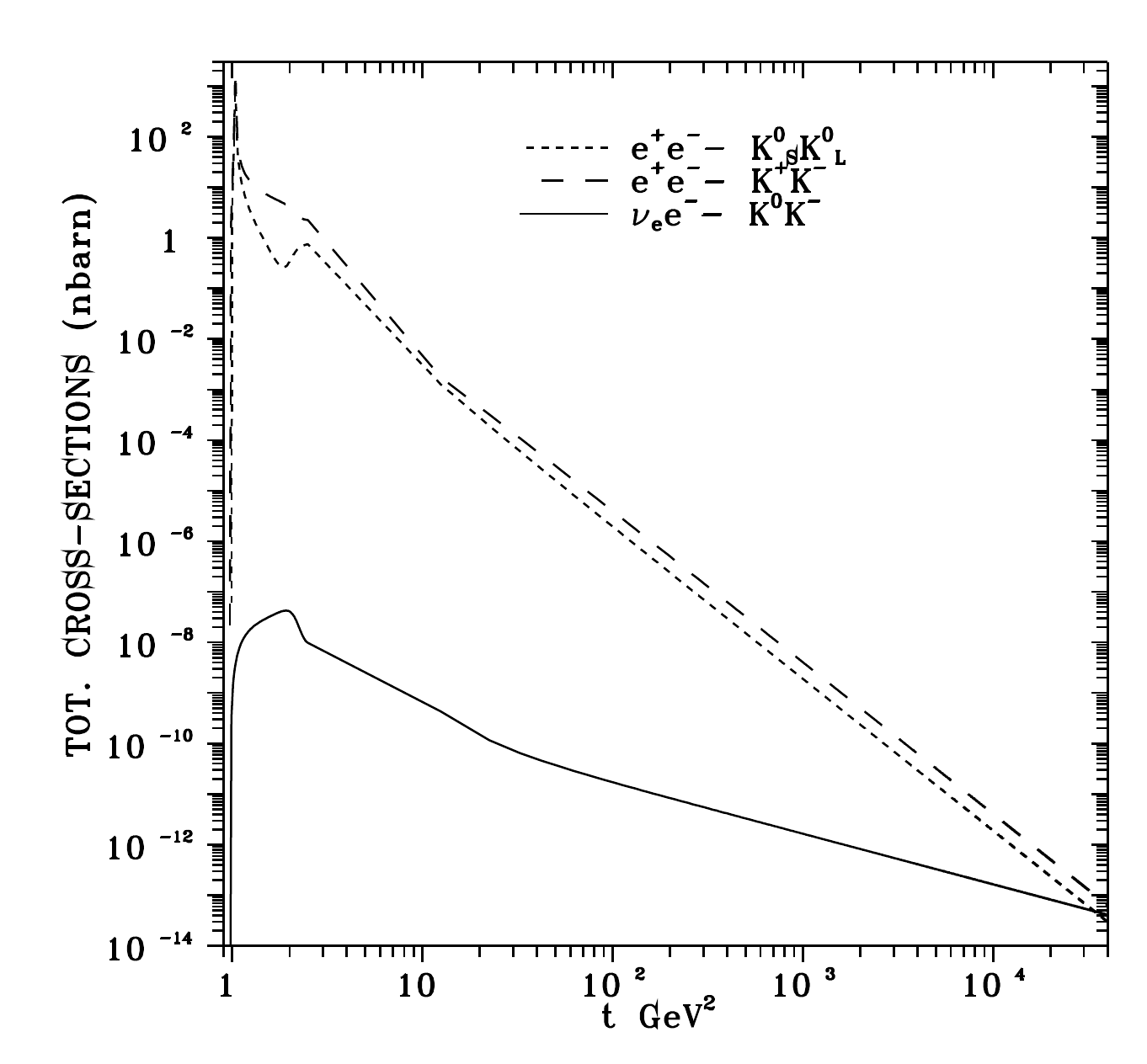}}
  \caption{\small{Cross-section comparison of electromagnetic and weak process.}}
  \label{sigkk}
\end{figure}

However, the results in Fig.~\ref{sigpipi} and Fig.~\ref{sigkk} are not very interesting
for experimentalists because the weak  annihilation
process $\bar{\nu}_{\rm e}e^-\to M^-M^0$  can be in principle
measured only in an interaction of antineutrino beams with atomic
electrons in the laboratory system.

The corresponding total cross-sections  in the laboratory system
can be found by using the following relation
\begin{equation}
s=m^2_{\rm e}+2m_{\rm e}\; E_\nu^{\rm lab} \label{chap7108}
\end{equation}
which leads to the expressions
\begin{equation}
\sigma_{\rm tot}^{lab}(\bar{\nu}_{\rm e}e^-\to
\pi^-\pi^0)=\frac{G^2}{24\pi} (m_{\rm e}^2+2m_{\rm e}E_\nu^{\rm
lab})\beta^3_{\pi} |F_\pi^{\rm E,I=1}(E_\nu^{\rm lab})|^2
\label{chap7109}
\end{equation}
and
\begin{equation}
\sigma_{\rm tot}^{lab}(\bar{\nu}_{\rm e}e^-\to
K^-K^0)=\frac{G^2}{12\pi} (m_{\rm e}^2+2m_{\rm e}E_\nu^{\rm
lab})\beta^3_K |F_K^V(E_\nu^{\rm lab})|^2, \label{chap7110}
\end{equation}
respectively.

It is also interesting to have a prediction of the energy
distribution of the pseudoscalar mesons  created in the final
state of the weak  $\bar{\nu}_{\rm e}e^-\to M^-M^0$  process given
by the differential cross-section  d$\sigma/{\rm d}E_M^{\rm lab}$
in the laboratory system. With this aim we  use in
(\ref{chap7104}) and (\ref{chap7106}) the explicit form of
d${\mit\Omega}=-{\rm d}\cos\vartheta\quad{\rm d}\varphi$ and
integrate the corresponding expressions over the angle $\varphi$.
Then we insert
\begin{eqnarray}
{\rm d}\cos\vartheta&=&-\frac{m_{\rm e}}{k^{\rm c.m.} p^{\rm
c.m.}}{\rm d}E_M^{\rm lab}; \quad\quad\quad\quad
s\approx 2m_{\rm e}E_\nu^{\rm lab};\label{chap7111} \\
\sin^2\vartheta&=&1-\frac{(E_M^{\rm c.m.}E_{\rm e}^{\rm
c.m.}-m_{\rm e}E_M^{\rm lab})^2}{(k^{\rm c.m.})^2 (p^{\rm
c.m.})^2},
\end{eqnarray}
where
\begin{equation}
E_M^{\rm c.m}\approx E_{\rm e}^{\rm c.m.}\approx p^{\rm
c.m}\approx \sqrt{{m_{\rm e}\over 2}E_\nu^{\rm lab}};
\quad\quad\quad\quad\mbox{and}\quad\quad\quad k^{\rm c.m.}\approx \sqrt{{m_{\rm e}\over
2}(E_\nu^{\rm lab}-E_\nu^{(0)})} \label{chap7112}
\end{equation}
with the threshold energy $E_\nu^{(0)}=(4m_M^2-m_e^2)/2m_e$ of the
antineutrino beam to be $E_{\nu \pi}^{(0)}=76.7 \,{{\rm GeV}}$ for pions and
$E_{\nu K}^{(0)}=962 \,{{\rm GeV}}$ for kaons. As a result, the following
expressions for the energy distributions of pions and kaons
\begin{equation}
\frac{{\rm d}\sigma(\bar{\nu}_{\rm e}e^-\to \pi^-\pi^0)}{{\rm
d}E_\pi^{\rm lab}}= \frac{m_{\rm
e}G^2}{8\pi}\left(\frac{E_\nu^{\rm
lab}-E_{\nu\pi}^{(0)}}{E_\nu^{\rm lab}}\right)
\left\{1-\frac{(E_\nu^{\rm lab}-2E_\pi^{\rm lab})^2}{E_\nu^{\rm
lab}(E_\nu^{\rm lab}-E_{\nu\pi}^{(0)})} \right\}|F_\pi^{\rm
E,I=1}(E_\nu^{\rm lab})|^2 \label{chap7113}
\end{equation}
and
\begin{equation}
\frac{{\rm d}\sigma(\bar{\nu}_{\rm e}e^-\to K^-K^0)}{{\rm
d}E_K^{\rm lab}}= \frac{m_{\rm e}G^2}{4\pi}\left(\frac{E_\nu^{\rm
lab}-E_{\nu K}^{(0)}}{E_\nu^{\rm lab}}\right)
\left\{1-\frac{(E_\nu^{\rm lab}-2E_K^{\rm lab})^2}{E_\nu^{\rm
lab}(E_\nu^{\rm lab}-E_{\nu K}^{(0)})} \right\}|F_K^V(E_\nu^{\rm
lab})|^2 \label{chap7114}
\end{equation}
are obtained, respectively.

\medskip

   In the last few years a considerable improvement in the detection of small
exclusive hadronic tau-decay processes has been achieved. As a
result, also experimental values on the branching ratio
$BR(\tau^-\to \nu_{\tau} M^- M^0)$ were obtained recently, where
$M$ means a pseudoscalar meson.

In order to predict it theoretically, we start with a general
expression for a probability of $\tau^-(k_1)\to \nu_{\tau}(k_2)
M^-(p_1) M^0(p_2)$ decay process
\begin{equation}
d\Gamma=(2\pi)^4\int\frac{\bar |{\cal
M}|^2}{2m_{\tau}}\delta(k_1-k_2-p_1-p_2)\frac{d^3k_2}{(2\pi)^32E_{\nu}}\frac{d^3p_1}{(2\pi)^32E_1}\frac{d^3p_2}{(2\pi)^32E_2},
\label{chap7115}
\end{equation}
where the corresponding matrix element takes the following form
\begin{equation}
{\cal M}=\frac{G_F \cos{\theta_C}}{\sqrt{2}}\bar
u(k_2)\gamma_{\mu}(1+\gamma_5)u(k_1) J_{\mu}^W \label{chap7116}
\end{equation}

and $\theta_C$ is the Cabibbo angle. In the expression
(\ref{chap7116}) the $W$-boson exchange mechanism is assumed to be
realized. However, since $m_{\tau}^2/m^2_W<<1$, all effects due
to the $W$-boson exchange can be neglected. $G_F$ means the Fermi
constant of weak interactions and $J_{\mu}^W$ is a matrix element
of the weak charge vector current responsible for the $W^-\to
K^-K^0$ transition.

Then the absolute value squared of the matrix element
(\ref{chap7116}) takes the form as follows
\begin{equation}
|\bar{\cal M}|^2 =\frac{G_F^2 \cos^2{\theta_C}}{2}[\ell_{\mu\nu}
+r_{\mu\nu}]J_{\mu}^W (J_{\mu}^W)^* \label{chap7117}
\end{equation}
with
\begin{equation}
\ell_{\mu\nu}= 4[k_{1\mu}k_{2\nu}+ k_{1\nu}k_{2\mu}
-g_{\mu\nu}(k_{1}\cdot k_{2}] \label{chap7118}
\end{equation}
\begin{equation*}
r_{\mu\nu}=4i\epsilon_{\mu\nu\alpha\beta}s_{\alpha}k_{2\beta}
-4m_{\tau}[s_{\mu}k_{2\nu}+ s_{\nu}k_{2\mu} -g_{\mu\nu}(s\cdot
k_{2})],
\end{equation*}
where $s$ is the $\tau$-lepton spin four-vector, $s\cdot k_1=0$
and the bar (\ref{chap7117}) means an average over the spin of the
$\tau$-lepton and a summation over the spin states of the neutrino
$\nu_{\tau}$.

It is straightforward to rewrite (\ref{chap7115}) into the
following three equivalent forms
\begin{eqnarray}
d\Gamma&=&\frac{|{\cal M}|^2}{64\pi^3}\frac{dEdE_{\nu}}{m_{\tau}},\label{chap7119}\\
d\Gamma&=&\frac{|{\cal M}|^2}{128\pi^3}\frac{dEdk^2}{m_{\tau}^2},\label{chap7120}\\
d\Gamma&=&\frac{|{\cal
M}|^2}{256\pi^3}m_{\tau}dxdy,\label{chap7121}
\end{eqnarray}
where $E$ is the energy of the $M^-$ pseudoscalar meson at the
$\tau$-lepton rest reference frame,
$k^2=(k_1-k_2)^2=m_{\tau}(m_{\tau}-2E_{\nu})$ is the effective
mass squared of the $M^-M^0$ system, $x=k^2/m_{\tau}^2$ and
$y=2E/m_{\tau}$.

If we define $Q=4m^2_M/m_{\tau}$, then
\begin{equation}
R\leq x \leq 1; \quad\quad\quad{\rm and}\quad\quad  \sqrt{R}\leq y\leq 1.
\end{equation}

The matrix element of the weak vector current of the $W^-\to
M^-M^0$ transition can be parametrized in the following way
\begin{equation}
J_{\mu}^W=(p_1-p_2)_{\mu} F_M^W(k^2), \label{chap7122}
\end{equation}
where $F_M^W(k^2)$ is the weak FF of the corresponding
pseudoscalar mesons. By using   (\ref{chap7122}),
(\ref{chap7117})-(\ref{chap7118}) and an integration over the
energy of the $M^-$ pseudoscalar meson in (\ref{chap7118}) one
gets the relation as follows
\begin{equation}
d\Gamma=\frac{G_F^2
\cos^2{\theta_C}}{768\pi^3m_{\tau}^3}dk^2(m_{\tau}^2-k^2)^2(m_{\tau}^2+2k^2)(1-\frac{4m_M^2}{k^2})^{3/2}
|F_M^W(k^2)|^2. \label{chap7123}
\end{equation}
Now, if instead of the weak pseudoscalar meson FF in
(\ref{chap7123}) the EM FFs by (\ref{chap795}) and (\ref{chap7103})
are substituted, one gets the decay widths
\begin{equation}
\Gamma(\tau^-\to \nu_{\tau}\pi^-\pi^0)
=\frac{G_F^2\cos^2{\theta_C}}{384\pi^3m_{\tau}^3}\int_{4m_{\pi}^2}^{m_{\tau}^2}dk^2(m_{\tau}^2-k^2)^2(m_{\tau}^2+2k^2)(1-\frac{4m_{\pi}^2}{k^2})^{3/2}
|F_\pi^{I=1,E}(k^2)|^2 \label{chap7124}
\end{equation}
\begin{equation}
\Gamma(\tau^-\to \nu_{\tau}K^-K^0)
=\frac{G_F^2\cos^2{\theta_C}}{192\pi^3m_{\tau}^3}\int_{4m_K^2}^{m_{\tau}^2}dk^2(m_{\tau}^2-k^2)^2(m_{\tau}^2+2k^2)(1-\frac{4m_K^2}{k^2})^{3/2}
|F_K^v(k^2)|^2, \label{chap7125}
\end{equation}
respectively. By using in (\ref{chap7124}) the pion EM FF $U\&A$ model
(\ref{chap130}) with the corresponding parameter values  and in
(\ref{chap7125}) the isovector part of kaon EM FF $U\&A$ model
 (\ref{chap35}) with the corresponding parameter values  (\ref{chap37}), and simultaneously
$G_F=0.2278\times 10^{-6}[m^{-2}_{\pi}]$, $\sin\theta_ C=0.22$  we
finally get the decay widths
\begin{equation}
\Gamma(\tau^-\to\nu_{\tau}\pi^-\pi^0) = 0.3898\times 10^{-11}
[m_{\pi}] \label{chap7126}
\end{equation}
and
\begin{equation}
\Gamma(\tau^-\to\nu_{\tau}K^-K^0) =0.2574\times 10^{-13}
[m_{\pi}], \label{chap7127}
\end{equation}
respectively. Dividing the latter values by the total width of the
$\tau$-lepton to be
\begin{equation}
\Gamma^{tot} = 0.159 \times 10^{-10} [m_{\pi}] \label{chap7128}
\end{equation}
one gets the corresponding branching ratios
\begin{equation*}
BR(\tau^-\to\nu_{\tau}\pi^-\pi^0) =24.52 \%
\end{equation*}
and
\begin{equation*}
BR(\tau^-\to\nu_{\tau}K^-K^0) =0.16 \%
\end{equation*}
respectively. They can be compared with the recent experimental
values
\begin{equation*}
BR^{exp}(\tau^-\to\nu_{\tau}\pi^-\pi^0) =(25.30\pm 0.20)\%
\end{equation*}
and
\begin{equation*}
BR^{exp}(\tau^-\to\nu_{\tau}K^-K^0) =(0.26\pm 0.09) \%
\end{equation*}
obtained by ALEPH Collaboration \cite{Buskulic96} in CERN. There is very
good agreement of our theoretical predictions with experiment.

\medskip

   \subsection{Parameter differences of the charged and neutral
   $\rho$-meson family} \label{III8}

\medskip

  Isotopic spin is to a very good approximation the conserved
quantum number in strong interactions. A breaking of the
corresponding symmetry occurs as a consequence of EM interactions
and the mass difference of the up and down quarks. Practically, it
is demonstrated in nature by a splitting of hadrons into
isomultiplets.

   In this paragraph we are concerned with $\rho$-meson resonances.
The cleanest determination of their parameters comes from the
${e}^+{e}^-$ annihilation and $\tau$-lepton decay. However, as it
is declared by the Review of Particle Physics \cite{Rpp},
experimental accuracy is not yet sufficient for unambiguous
conclusions. The difference for $\rho^0$ and $\rho^\pm$ are
presented in averaged to be $m_{\rho^0}-m_{\rho^\pm}=-0.7\pm0.8
{MeV}$ and $\Gamma_{\rho^0}-\Gamma_{\rho^\pm}=0.3\pm1.3 {MeV}$,
respectively.

   Nowadays the situation is changed in two aspects.

   On one hand, new very accurate KLOE data \cite{Alois} on the
pion EM form factor (FF) at the energy range $0.35{GeV}^2\leq
t\leq 0.95 {GeV}^2$ were obtained by the radiative return method
in Frascati. Also corrected CMD-2 \cite{Akhm} and SND \cite{Achas}
Novosibirsk ${e}^+{e}^-\to\pi^+\pi^-$ data have appeared recently.

   On the other hand the weak pion FF accurate data
\cite{Fujikawa} from the measurement of the
$\tau^-\to\pi^-\pi^0\nu_{\tau}$ decay by Belle (KEK) experiment
were published recently.

   The $U\&A$ model (\ref{chap130}) of the pion
EM FF, to be represented by one analytic function for
$-\infty<t<+\infty$ was elaborated, which is always successfully
applied for a description of existing data on the pion EM FF from
${e}^+{e}^-\to\pi^+\pi^-$ and due to the CVC hypothesis
\cite{Dubnazrek1} equally well also for a description of existing
data on the weak pion FF from $\tau^-\to\pi^-\pi^0\nu_{\tau}$
decay.

   As a result more sophisticated evaluation of a difference of
the $\rho$-meson families parameters can be achieved \cite{BDDFH}.

   For the pion EM FF there is almost continuous
interval of 381 experimental points for\\
 $-9.77 {GeV}^2\leq t\leq 13.48 {GeV}^2$, which all are described by the pion EM FF model
(\ref{chap130}) in the space-like and the time-like regions
simultaneously. The most important from them are accurate KLOE
data \cite{Alois} at the energy range $0.35 {GeV}^2\leq t\leq 0.95
{GeV}^2$ obtained in Frascati by  the radiative return method and
also the corrected Novosibirsk CMD-2 data \cite{Akhm} at the range
$0.36{GeV}^2\leq t\leq 0.9409 {GeV}^2$ and SND data \cite{Achas}
at the range $0.1521 {GeV}^2\leq t\leq 0.9409 {GeV}^2$, which can
influence the finite results substantially. They are supplemented
at the interval $-9.77 {GeV}^2\leq t\leq 0.3364 {GeV}^2$ and
$0.9557 {GeV}^2\leq t\leq 13.48 {GeV}^2$ by other existing data.

   An application of the pion EM FF $U\&A$ model (\ref{chap130})
with three lowest resonances, in order to take into account the
fact that the mass of the $\tau$ lepton in decay
$\tau^-\to\pi^-\pi^0\nu_{\tau}$ allows to reach not more than the
energy corresponding just to the second radially excited state
$\rho''(1700)$ meson, leads to the best description of all
existing 381 experimental points with the values of the parameters
presented in the second column of the Tab.~\ref{table:1}.

   The weak decay $\tau^-\to\pi^-\pi^0\nu_{\tau}$ is, like the $e^+e^- \to \pi^+\pi^-$
process, dominated by the $\rho^-$-meson family resonances and
thus it can be used to extract information on the charged
$\rho$-meson family properties.

   From the conservation of vector current (CVC) theorem
it follows (\ref{chap795}), i.e. the $\pi^-\pi^0$ mass spectrum in
the $\tau^-\to\pi^-\pi^0\nu_{\tau}$ decay can be related to the
total cross section of the ${e}^+{e}^-\to\pi^+\pi^-$ process. As a
result the same $U\&A$ pion EM FF model (\ref{chap130}) can be
applied to a description of the weak pion FF data, which can be
drawn from the measured normalized invariant mass-spectrum.

   Though there are measurements of $\tau^-\to\pi^-\pi^0\nu_{\tau}$
decay to be carried out previously by ALEPH \cite{Schael} and CLEO
\cite{Anderson}, here we are concentrated only on the
high-statistics measurement \cite{Fujikawa} of the weak pion FF
from $\tau^-\to\pi^-\pi^0\nu_{\tau}$ decay with the Belle detector
at the KEK-B asymmetric-energy ${e}^+{e}^-$ collider as they are
charged by the lowest total errors.

   An application of the pion EM FF $U\&A$ model
(\ref{chap130}) to the best description of 62 experimental points
\cite{Fujikawa} on the weak pion FF leads to the values of the
parameters also presented in Tab.~\ref{table:1}, however, in the
third column.

   There may be a question about the fact that the obtained $\rho$-meson
family parameters are differing from those presented at the Review
of the Particle Physics \cite{Rpp}, especially of the $\rho(770)$-
and $\rho'(1450)$-mesons. As it is well known, the resonance
parameters depend on the parametrization used in a fit of data.
However, in this contribution it plays no crucial role as finally
we are interested in a difference of the corresponding parameters.

   Moreover, in a determination of the parameters we exploit the
$U\&A$ model of the pion FF, in which any resonance is defined as
a pole on the unphysical sheets of the Riemann surface to be
considered as the most sophisticated approach in a description of
resonant states.

   The difference of the $\rho$-meson family resonance parameters
is presented in Tab.~\ref{table:1} in the fourth column.

   On the base of these results one can declare that the masses of
the neutral $\rho$-meson family are lower than the masses of the
charged $\rho$-meson family.

   Their widths are just reversed.

   Considering the evaluated errors, one can confidently affirm only
in the case of the $\rho(770)$-meson parameters and also in the
case of the $\rho'(1450)$-meson widths, that for the charged and
the neutral states they are different.

   For other $\rho$-meson family parameters one can say nothing
definitely and one has to wait for even more precise experimental
data.
\begin{table}[!tb]
\caption{The values of the fitting parameters for the fit of pion
FF. The values are shown for two cases, the result of fitting
${e}^+{e}^-$ data to the U\&A model of pion EM FF (second column),
the result of fitting $\tau^-$ data to the $U\&A$ model of weak pion
FF (third column). The differences of the values for both cases
are presented in fourth column.} \label{table:1}
\renewcommand{\tabcolsep}{0.5pc} 
\renewcommand{\arraystretch}{1.2} 
\centering
\begin{tabular}{@{}lccc}
\\[2pt]
\hline
Parameter & $\rho^0$ & $\rho^\pm$ & $\Delta$ ($\rho^\pm-\rho^0$) \\
\hline
$t_{in}$ [Gev$^2$] & $1.3646\pm 0.0198$ & $1.2432\pm 0.0157$  & \\%
$W_Z$     & $0.1857\pm 0.0004$           & $0.4078\pm 0.0013$            & \\%
$W_P$     & $0.2335\pm 0.0005$           & $0.6197\pm 0.0007$            & \\%
$m_{\rho}$ [ MeV] & $758.2260\pm 0.4620$ & $761.6000\pm 0.9520$  &  $3.3740\pm 1.0582$ \\%
$m_{\rho'}$ [ MeV]  & $1342.3060\pm 46.6200$ & $1373.8340\pm 11.3680$  &  $31.5280\pm 47.9860$ \\%
$m_{\rho''}$ [ MeV]  & $1718.5000\pm 65.4360$ & $1766.8000\pm 52.3600$  &  $48.3000\pm 83.8060$ \\%
$\Gamma_{\rho}$ [ MeV]  & $144.5640\pm 0.7980$ & $139.9020\pm 0.4620$  &  $-4.6620\pm 0.8502$ \\%
$\Gamma_{\rho'}$ [ MeV] & $492.1700\pm 138.3760$ & $340.8720\pm 23.8420$  &  $-151.2980\pm 140.4150$ \\%
$\Gamma_{\rho''}$ [ MeV]  & $489.5800\pm 16.9540$ & $414.7080\pm 119.4760$  & $-74.8720\pm 120.6729$ \\%
$f_{\rho\pi\pi}/f_{\rho}$  & $1.0009\pm 0.0001$ & $0.9998\pm 0.0002$  & \\%
$\chi^2/{\rm NDF}$  & 1.78 & 1.96  & \\%
\hline
\end{tabular}\\[2pt]
\end{table}

\setcounter{equation}{0} \setcounter{figure}{0} \setcounter{table}{0}\newpage
     \section{Electromagnetic structure of $1/2^+$ octet baryons} \label{IV}

 Experimentally we know eight baryons
$[p,n]$;$[\mit\Lambda]$;$[\mit{\Sigma^+,\Sigma^0,\Sigma^-}]$ and
$[\mit{\Xi^0,\Xi^-}]$ of the spin 1/2 and the positive parity,
which according to the SU(3) classification belong to the same
multiplet and are compound of 3 light (up, down and strange)
quarks.

   After more than half of the century of the discovery of
EM structure of hadrons, a knowledge about the EM structure of
octet $1/2^+$ baryons is unsatisfactory up to now. Almost all
experimental investigations are concentrated to the proton and
mainly in the space-like region. Less precise information exists
on the neutron. Concerning other members of the baryon $1/2^+$
octet, there is only one experimental point on the total cross
section of the $\Lambda\bar\Lambda$ production in $e^+e^-$
annihilation and an upper limit on the cross-section of the
$\Sigma^0\bar\Sigma^0$ production at $t=5.693 GeV^2$
\cite{Biselsigm}. There is no experimental information on the
electromagnetic structure of $\Sigma^{\pm}$ and $\Xi$-hyperons up
to now.

   The electromagnetic structure of each $1/2^+$ baryon is completely
described (see \ref{II1}) by two scalar functions of one variable,
commonly to be chosen in the form of the Sachs electric $G_{EB}(t)$
and magnetic $G_{MB}(t)$ FFs. They are defined by the relation
(\ref{sec27}), which have no more the asymptotic behavior of
VMD model (\ref{asymvmd}) and according to the quark structure of
baryons it takes the form

\begin{equation}
   G_{EB}(t)_{|t|\to\infty} = G_{MB}(t)_{|t|\to\infty} \sim
   \frac{1}{t^2}.\label{asymemff}
\end{equation}

   Inspired by a good experimental situation, many more or less
successful phenomenological models have been constructed for a
global description of the nucleon EM structure. However, missing
are analogous attempts to predict a behavior of EM FFs of other
baryons of the same octet. The main reason is that almost all
constructed phenomenological models depend on some number of free
parameters which, however, have to be fixed in a comparison with
experimental data.

   Further we show how it is possible to go round this problem to
some extent.

\medskip

 \subsection{Nucleon electromagnetic structure} \label{IV1}

\medskip

   The EM structure of the nucleons (isodublet compound of the proton
and neutron), as revealed first time in elastic electron-nucleon
scattering more than half century ago, is completely described by
four independent scalar functions of one variable called nucleon
EM FFs. They depend on the squared momentum transfer $t=-Q^2$ of
the virtual photon.

   Nucleon EM FFs can be chosen in a divers way, e.g. as the Dirac and
Pauli, $F^p_1(t)$, $F^n_1(t)$ and $F^p_2(t)$, $F^n_2(t)$, FFs, or
the Sachs electric and magnetic, $G_{Ep}(t)$, $G_{En}(t)$ and
$G_{Mp}(t)$, $G_{Mn}(t)$, FFs, or isoscalar and isovector Dirac
and Pauli, $F_{1N}^s(t)$, $F_{1N}^v(t)$ and $F_{N2}^s(t)$, $F^v_{2N}(t)$, FFs
and isoscalar and isovector electric and magnetic, $G_{EN}^s(t)$,
$G_{EN}^v(t)$ and $G_{MN}^s(t)$, $G_{MN}^v(t)$, FFs, respectively.

   All these always four independent sets of four nucleon EM FFs are related by

\begin{eqnarray} \label{d5}
\nonumber
G_{Ep}(t)&=&G_{EN}^s(t)+G_{EN}^v(t)=F_{1p}(t)+\frac{t}{4m^2_p}F_{2p}(t)=\\ &=& \nonumber
[F_{1N}^s(t)]+ F_{1N}^v(t)]+\frac{t}{4m_p^2}[F_{2N}^s(t)+F_{2N}^v(t)];\\
 G_{Mp}(t)&=&G_{MN}^s(t)+G_{MN}^v(t)=F_{1p}(t)+F_{2p}(t)=  \\ &=&\nonumber
[F_{1N}^s(t)+F_{1N}^v(t)]+
[F_{2N}^s(t)+F_{2N}^v(t)];\\
\nonumber
G_{En}(t)&=&G_{EN}^s(t)-G_{EN}^v(t)=F_{1n}(t)+\frac{t}{4m_n^2}F_{2n}(t)= \\ &=& \nonumber
[F_{1N}^s(t)-F_{1N}^v(t)]+\frac{t}{4m_n^2}[F_{2N}^s(t)-F_{2N}^v(t)];\\
\nonumber
G_{Mn}(t)&=&G^s_{MN}(t)-G_{MN}^v(t)=F_{1n}(t)+F_{2n}(t)=  \\ &=&\nonumber
[F_{1N}^s(t)-F_{1N}^v(t)]+
[F_{2N}^s(t)-F_{2N}^v(t)],
\end{eqnarray}
and at the value $t=0$ normalized as follows

\begin{eqnarray}
(i)&\nonumber&\hspace{-0.5cm}
\ G_{Ep}(0)=1;\ G_{Mp}(0)=1+\mu_p;\ G_{En}(0)=0;\ G_{Mn}(0)=\mu_n;\\
(ii)&\nonumber &\hspace{-0.5cm}  \ G_{EN}^s(0)=G_{EN}^v(0)=\frac{1}{2};\
G_{MN}^s(0)=\frac{1}{2}(1+\mu_p+\mu_n);\
 G_{MN}^v(0)=\frac{1}{2}(1+\mu_p-\mu_n);\\(iii)&\nonumber&\hspace{-0.5cm}
\ F_{1p}(0)=1;\ F_{2p}(0)=\mu_p;\ F_{1n}(0)=0;\ F_{2n}(0)=\mu_n;\\
(iv)&\label{NFFnorm} &\hspace{-0.5cm} \ F_{1N}^s(0)=F_{1N}^v(0)=\frac{1}{2};\
 F_{2N}^s(0)=\frac{1}{2}(\mu_p+\mu_n);\
 F_{2N}^v(0)=\frac{1}{2}(\mu_p-\mu_n),
\end{eqnarray}
where $\mu_p$ and $\mu_n$ are the proton and neutron anomalous
magnetic moments, respectively.

   The Dirac and Pauli FFs are naturally obtained (\ref{sec26})
in a decomposition of the nucleon matrix element of the EM current
into a maximum number of linearly independent covariants
constructed from the four-momenta, $\gamma$-matrices and Dirac
bispinors of nucleons

\begin{equation}
\langle N|J^{EM}_\mu|N\rangle=e\bar u(p')\{\gamma_\mu
F_{1N}(t)+\frac{i}{2m_N}\sigma_{\mu\nu}(p'-p)_{\nu}F_{2N}(t)\}u(p)\label{d1}
\end{equation}
with $m_N$ to be the nucleon mass.

   On the other hand, the electric and magnetic FFs are very suitable in
extracting  experimental information on the nucleon EM structure
from the measured cross sections

\begin{eqnarray}
\frac{d\sigma^{lab}(e^-N\to
e^-N)}{d\Omega}&=&\frac{\alpha^2}{4E^2}\frac{\cos^2(\theta/2)}{\sin^4(\theta/2)}
\frac{1}{1+(\frac{2E}{m_N})\sin^2(\theta/2)}\times\nonumber \\
&\times&\left[\frac{G^2_{EN}-\frac{t}{4m_N^2}G^2_{MN}}
{1-\frac{t}{4m_N^2}}-2\frac{t}{4m_N^2}G^2_{MN}\tan^2(\theta/2),\right]
\label{d2}
\end{eqnarray}
$\alpha=1/137$, $E$-the incident electron energy,
and

\begin{equation}
\sigma_{tot}^{c.m.}(e^+e^-\to N\bar
N)=\frac{4\pi\alpha^2\beta_N}{3t}
[|G_{MN}(t)|^2+\frac{2m_N^2}{t}|G_{EN}(t)|^2],\quad\quad
\beta_N=\sqrt{1-\frac{4m_N^2}{t}} \label{d3}
\end{equation}
or

\begin{equation}
\sigma_{tot}^{c.m.}(\bar p p\to
e^+e^-)=\frac{2\pi\alpha^2}{3p_{c.m.}\sqrt{t}}
[|G_{Mp}(t)|^2+\frac{2m_N^2}{t}|G_{Ep}(t)|^2], \label{d4}
\end{equation}
($p_{c.m.}$-antiproton momentum in the c.m. system)
as there are no interference terms between them.

   In the Breit frame, the Sachs FFs give the distribution of charge and
magnetization within the proton and neutron, respectively. From
all four Sachs FFs the neutron electric FF plays a particular
role. Though the total neutron charge is zero, there is a
nonvanishing distribution of charge inside of the neutron, which
leads to the nonvanishing neutron electric FF.

   The isoscalar and isovector Dirac and Pauli FFs are suitable for
a construction of various phenomenological models of the nucleon
EM structure, however with the correct asymptotic behavior
\begin{equation}
   F_1^s(t)_{|t|\to \infty} = F_1^v(t)_{|t|\to \infty} \sim
   \frac{1}{t^2}\label{Diracffasym}
\end{equation}
\begin{equation}
   F_2^s(t)_{|t|\to \infty} = F_2^v(t)_{|t|\to \infty} \sim
   \frac{1}{t^3}\label{Pauliffasym}
\end{equation}
in conformity with (\ref{sec27}) and (\ref{asymemff}).

   The most attractive of them is the Vector Meson Dominance (VMD)
picture  in the framework of which FFs are simply saturated (see
the paragraph \ref{II3}) by a set of isoscalar and isovector vector
meson poles on the positive real axis. However, this turns out to
be practically an insufficient approximation and in a more
realistic description of the data (especially in the time-like
region) instability of vector-mesons has to be taken into account
and the contributions of continua, to be created by n-particle
thresholds, like, e.g., $2\pi, 3\pi, K \bar K, N \bar N$ etc.,
together with the correct asymptotic behaviours
(\ref{Diracffasym}),(\ref{Pauliffasym}) and normalizations
(\ref{NFFnorm}) have to be included.

   In recent years, abundant and very accurate data on the nucleon
EM FFs appeared. Most of the references concerning the nucleon
space-like data can be found in \cite{Dubn88}. More recent precise
measurements are presented in \cite{Walker94}-\cite{Platsch90}.
Besides the latter, there are also new data on the neutron
electric FF from BATES \cite{Eden94}, MAMI
\cite{Ostrick99}-\cite{Rohe99} and NIKHEF \cite{Pass99}.

   For the time-like region data see \cite{Bassom83}-\cite{Voci97}.
There is, in particular, the FENICE experiment in Frascati (Roma,
Italy) measured, besides the proton EM FFs \cite{Antonelli94}, the
magnetic neutron FF in the time-like region \cite{Voci97} for the
first time. There are also valuable results on the magnetic proton
FF \cite{Armstrong93,Ambrog99} at higher energies measured in
FERMILAB (Batavia, USA).

   All this  stimulated  recent dispersion theoretical analysis
\cite{Mergel96,Furiuchi97} of the nucleon EM FF data in the
space-like region and in the time-like region \cite{Hammer96} as
well.

   The latter works are an update and extension of historically the
most competent nucleon FF analysis carried out by H\"ohler with
collaborators \cite{Hoehler76}. However, the model does not allow
one to describe all the time-like data consistently, while still
giving good description of the data in the space-like region.

   In the next paragraph, we construct a ten-resonance $U\&A$ model
of the nucleon EM structure \cite{DDW}, defined on the
four-sheeted Riemann surface with canonical normalizations and QCD
asymptotics, which provides a very effective framework for a
superposition of complex conjugate vector-meson pole pairs on
unphysical sheets and continua contributions in nucleon EM FFs .
The model contains, e.g., an explicit two-pion continuum
contribution given by the unitary cut starting from $t=4m_\pi^2$
and automatically predicts the strong enhancement of the left wing
of the $\rho(770)$ resonance in the isovector spectral functions
to be consistent with the results of \cite{Mergel96,Hoehler75}.

   Another result of the presented model is the prediction of
parameters of the fourth excited state of the  $\rho(770)$ meson
and the automatic prediction of isoscalar nucleon spectral
function behaviors. At the same time, a description of all
existing space-like and time-like nucleon EM FF data, including
also FENICE (Frascati) results on the neutron from the $e^+e^-\to
n\bar n$ process, is achieved.

\medskip

     \subsection{Unitary and analytic model of nucleon electromagnetic
     structure} \label{IV2}

\medskip

\begin{figure}[tb]
\begin{center}
\includegraphics[scale=.46]{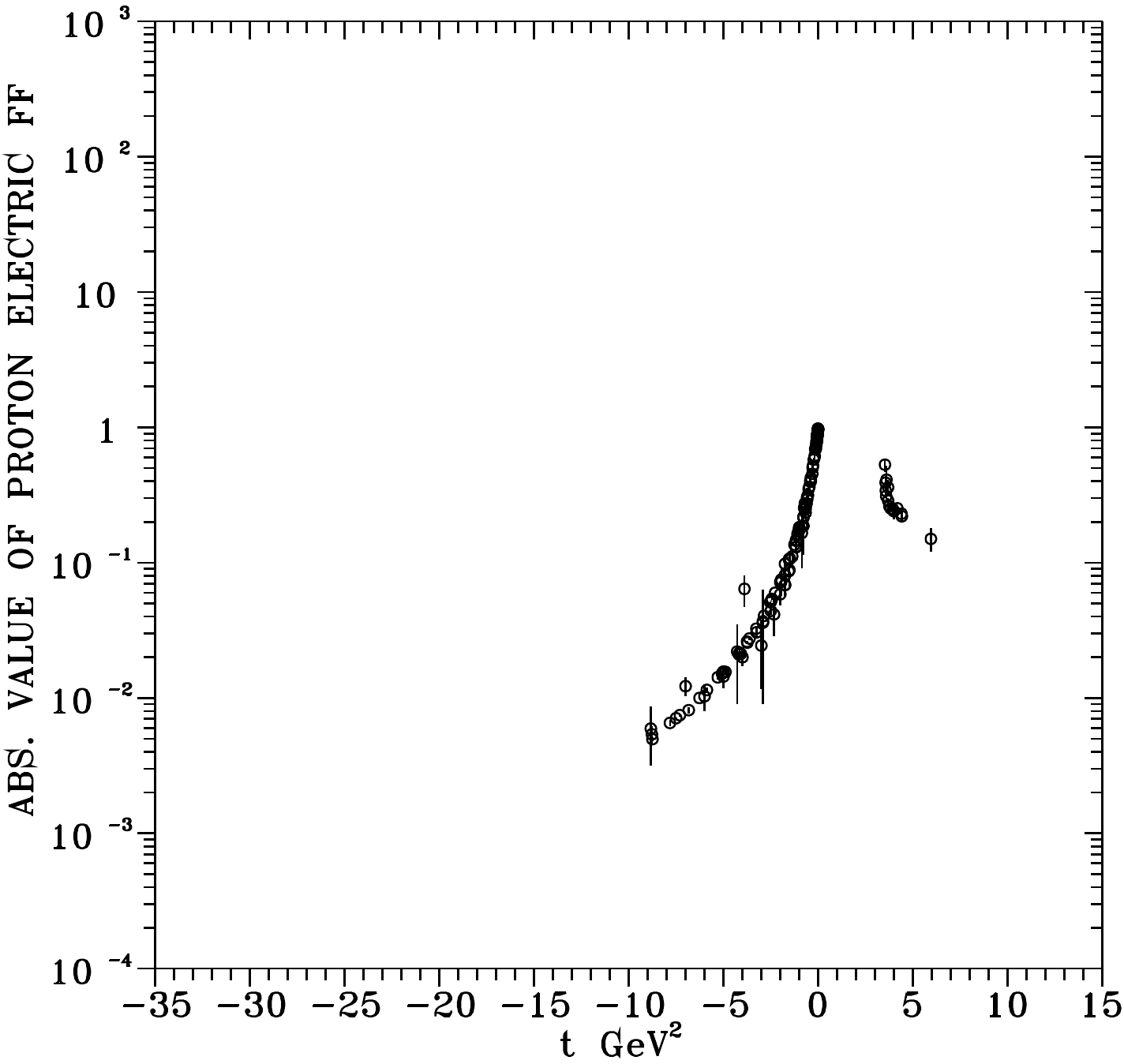}
\includegraphics[scale=.46]{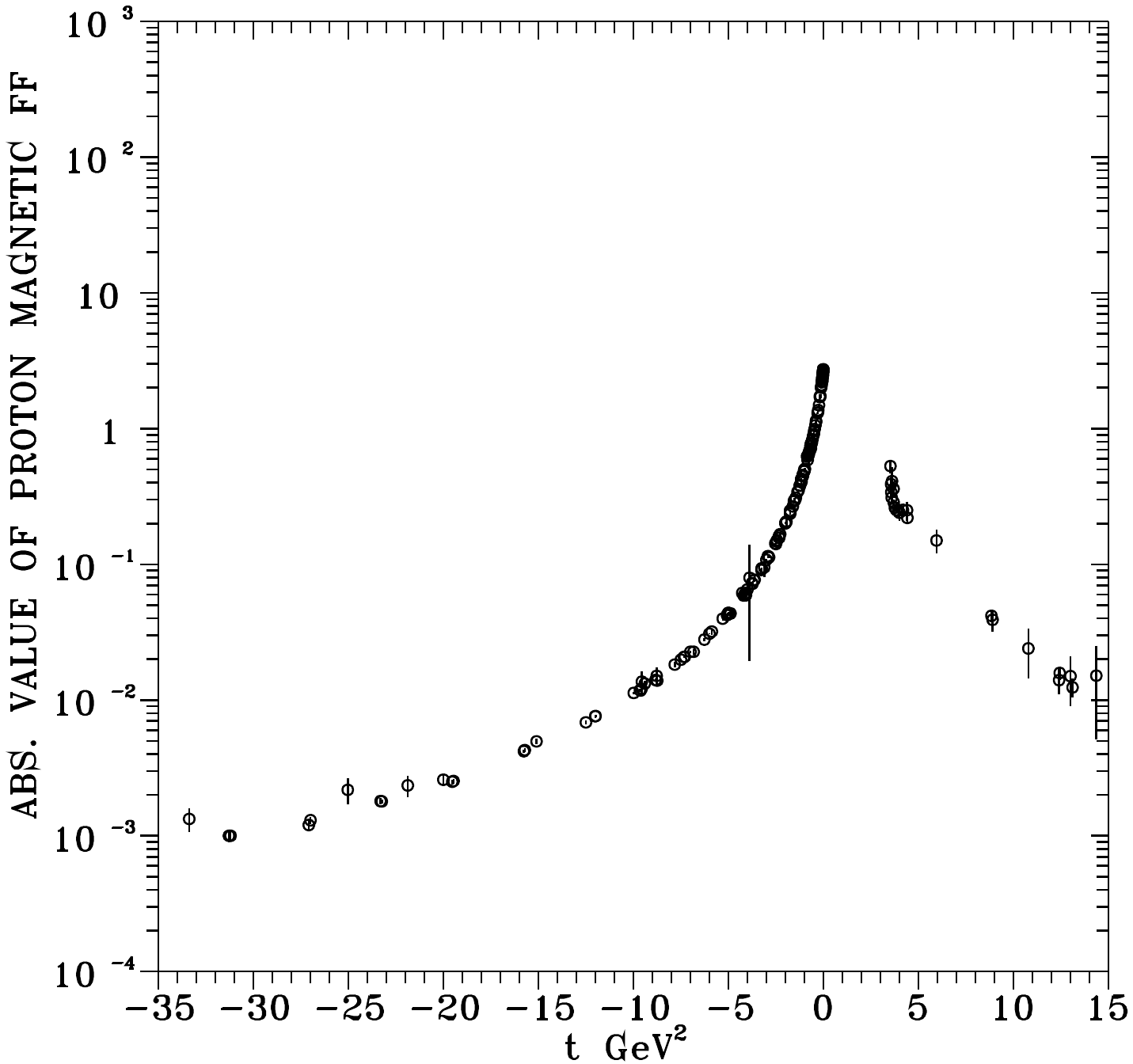}
\end{center}
  \caption{Compiled proton electric and magnetic FF data.}
  \label{fig18}
\end{figure}

\begin{figure}[tb]
\begin{center}
\includegraphics[scale=.46]{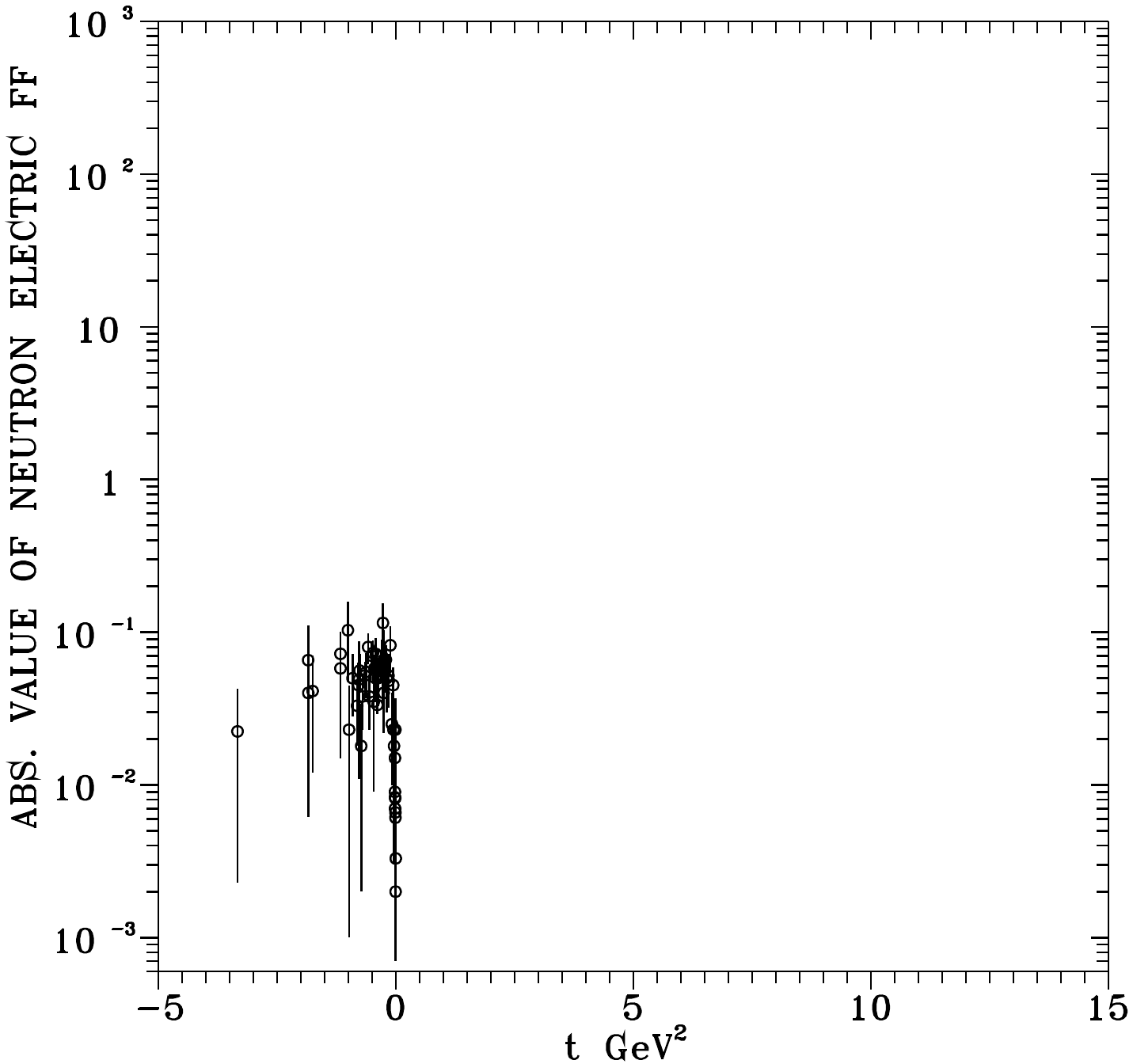}
\includegraphics[scale=.46]{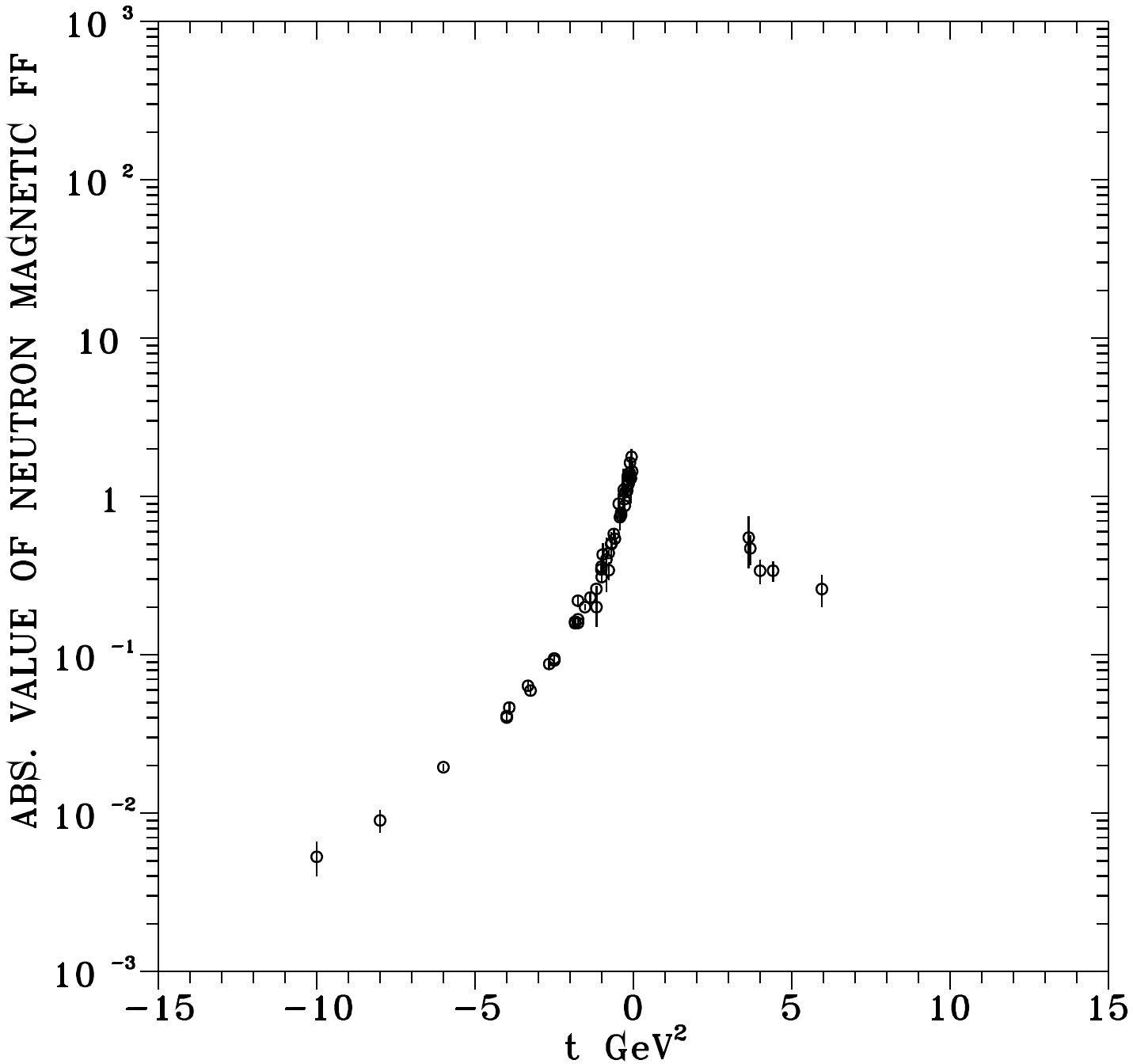}
\end{center}
  \caption{Compiled neutron electric and magnetic FF data.}
  \label{fig19}
\end{figure}

    There are more than 500 experimental points with errors
(see Figs.~\ref{fig18},  and ~\ref{fig19}) collected for qualified analyses
to be carried out by the 10 resonance $U\&A$ model of the nucleon
EM FFs.

   The model will represent, as we have mentioned previously, a consistent
unification of the following three fundamental features (besides
other properties) of the nucleon EM FFs
\begin{itemize}
\item[1.]
The experimental fact of  creation of unstable vector-meson
resonances in the $e^+e^-$-annihi\-lation processes into hadrons.
\item[2.]
The hypothetical analytic properties of the nucleon EM FFs on the
first (physical) sheet of the Riemann surface, by means of which
just the contributions of continua are taken into account.
\item[3.]
The asymptotic behavior of nucleon EM FFs (\ref{asymemff}) to be
proved also in the framework of the QCD \cite{Brodsky80}.
\end{itemize}

   Here we would like to note that a further procedure will not mean
any mathematically correct derivation of the $U\&A$ model, but
only an (noncommutative) algorithm of its construction which is,
however, generally valid also for any other strongly interacting
particles.

   It has been practically manifested to be optimal to saturate
the isoscalar and isovector Dirac and Pauli nucleon FFs by 5
isoscalars ($\omega ,\phi ,\omega ^{\prime },\omega ^{\prime
\prime },\phi ^{\prime }$) and 5 isovectors ($\rho ,\rho ^{\prime
},\rho ^{\prime\prime },\rho ^{\prime \prime \prime },\rho
^{\prime \prime \prime \prime }$), respectively, in the form of
the VMD parametrization (\ref{ff35}) to be automatically
normalized with required asymptotic behavior. For $F^s_{1N}(t)$ and
$F^v_{1N}(t)$ one takes the formula (\ref{ff35}) with $n = 5, m=2$, and
for $F^s_{2N}(t)$ and $F^v_{2N}(t)$ one takes the general solution
(\ref{ff35}) with $n = 5, m=3$. As a result the following
zero-width VMD expressions are obtained
\begin{eqnarray}
\nonumber F^s_1(t)&=&\frac{1}{2}\mmt\sc\sbb+\\
  \nonumber &+&\left\{\mmt\sc\sa\zll\mcsc\mcsa\mcsc\mcsb-
  \mmt\sbb\sa\zll\mcsb\mcsa\mcsc\mcsb-\right.\\
  &-&\left.\mmt\sc\sbb\right\}\fff 1\sa+\nonumber
\end{eqnarray}
\begin{eqnarray}
  \nonumber &+&\left\{\mmt\sc\sd\zll\mcsc\mcsd\mcsc\mcsb-
  \mmt\sbb\sd\zll\mcsb\mcsd\mcsc\mcsb-\right.\\
&-&\left.\mmt\sc\sbb\right\}\fff 1\sd-\label{d8}\\
  \nonumber &-&\left\{\mmt\se\sc\zll\mcse\mcsc\mcsc\mcsb-
  \mmt\se\sbb\zll\mcse\mcsb\mcsc\mcsb+\right.\\
  \nonumber &+&\left.\mmt\sc\sbb\right\}\fff 1\se,
\nonumber
\end{eqnarray}
\begin{eqnarray}
\nonumber F^v_1(t)&=&\frac{1}{2}\mmt\vc\vb+\\
   \nonumber &+&\left\{\mmt\vc\va\zll\mcvc\mcva\mcvc\mcvb-
\mmt\vb\va\zll\mcvb\mcva\mcvc\mcvb-\right.\\
   &-&\left.\mmt\vc\vb\right\}\fff 1\va+\label{d9}\\
   \nonumber &+&\left\{\mmt\vd\vb\zll\mcvd\mcvb\mcvc\mcvb-
\mmt\vd\vc\zll\mcvd\mcvc\mcvc\mcvb-\right.\\
   \nonumber &-&\left.\mmt\vc\vb\right\}\fff 1\vd-\\
   \nonumber &-&\left\{\mmt\ve\vc\zll\mcve\mcvc\mcvc\mcvb-
\mmt\ve\vb\zll\mcve\mcvb\mcvc\mcvb+\right.\\
   \nonumber &+&\left.\mmt\vc\vb\right\}\fff 1\ve,
\nonumber
\end{eqnarray}
\begin{eqnarray}
\nonumber F_2^s(t)&=&\frac{1}{2}(\mu_p+\mu_n)\mtt\sc\sbb\sa+\\
   \nonumber &+&\left\{\mtt\sc\sd\sa\zlll\mcsc\mcsd\mcsd\mcsa\mcsc\mcsb\mcsb\mcsa+\right.\\
   \nonumber &+&\mtt\sc\sbb\sd\zlll\mcsc\mcsd\mcsb\mcsd\mcsc\mcsa\mcsb\mcsa-\\
   \nonumber &-&\mtt\sbb\sd\sa\zlll\mcsb\mcsd\mcsd\mcsa\mcsc\mcsb\mcsc\mcsa-
\end{eqnarray}
\begin{eqnarray}
   &-&\left.\mtt\sc\sbb\sa\right\}\fff 2\sd+\label{d10} \\
   \nonumber &+&\left\{\mtt\se\sc\sbb\zlll\mcse\mcsc\mcse\mcsb\mcsc\mcsa\mcsb\mcsa-\right.\\
   \nonumber &-&\mtt\se\sc\sa\zlll\mcse\mcsc\mcse\mcsa\mcsc\mcsb\mcsb\mcsa+\\
   \nonumber &+&\mtt\se\sbb\sa\zlll\mcse\mcsb\mcse\mcsa\mcsc\mcsb\mcsc\mcsa-\\
   \nonumber &-&\left.\mtt\sc\sbb\sa\right\}\fff 2\se,
\nonumber
\end{eqnarray}
\begin{eqnarray}
\nonumber F_2^v(t)&=&\frac{1}{2}(\mu_p-\mu_n)\mtt\vc\vb\va+\\
  \nonumber &+&\left\{\mtt\vd\vb\va\zlll\mcvd\mcvb\mcvd\mcva\mcvc\mcvb\mcvc\mcva-\right.\\
  \nonumber &-&\mtt\vd\vc\va\zlll\mcvd\mcvc\mcvd\mcva\mcvc\mcvb\mcvb\mcva+\\
  \nonumber &+&\mtt\vd\vc\vb\zlll\mcvd\mcvc\mcvd\mcvb\mcvc\mcva\mcvb\mcva-\\
  &-&\left.\mtt\vc\vb\va\right\}\fff 2\vd+\label{d11}\\
  \nonumber &+&\left\{\mtt\ve\vb\va\zlll\mcve\mcvb\mcve\mcva\mcvc\mcvb\mcvc\mcva-\right.\\
  \nonumber &-&\mtt\ve\vc\va\zlll\mcve\mcvc\mcve\mcva\mcvc\mcvb\mcvb\mcva+\\
  \nonumber &+&\mtt\ve\vc\vb\zlll\mcve\mcvc\mcve\mcvb\mcvc\mcva\mcvb\mcva-\\
  \nonumber &-&\left.\mtt\vc\vb\va\right\}\fff 2\ve.
\end{eqnarray}
however, they are already automatically normalized (\ref{NFFnorm})
and they govern the asymptotics (\ref{Diracffasym}) and (\ref{Pauliffasym}), respectively,
 as predicted by QCD up to the logarithmic corrections.

   Despite of the latter properties the model is unable to reproduce the
existing experimental information properly and only its
unitarization (see paragraph \ref{II7}), i.e.,  inclusion of the
contributions of continua and instability of vector-meson
resonances, leads to a simultaneous description of the space-like
and time-like data.

   It is well known that the unitarity condition requires the imaginary part
of the nucleon EM FFs to be different from zero only above the
lowest branch point $t_0$ and, moreover, it just predicts its
smoothly varying behavior (see e.g. \cite{Mergel96,Hoehler75}).

   The unitarization of the model (\ref{d8})-(\ref{d11}) can be
achieved by application of the following special non-linear
transformations
\begin{eqnarray}
\nonumber t=t_0^s-\frac{4(t_{in}^{1s}-t_0^s)}{[1/V-V]^2}\\
t=t_0^v-\frac{4(t_{in}^{1v}-t_0^v)}{[1/W-W]^2}\label{d12}\\
\nonumber t=t_0^s-\frac{4(t_{in}^{2s}-t_0^s)}{[1/U-U]^2}\\
\nonumber t=t_0^v-\frac{4(t_{in}^{2v}-t_0^v)}{[1/X-X]^2},\nonumber
\end{eqnarray}
respectively, and a subsequent incorporation of the nonzero values
of vector meson widths.

   Here $t^s_0=9m_{\pi}^2, t^v_0=4m_{\pi}^2, t_{in}^{1s}, t_{in}^{1v}, t_{in}^{2s}, t_{in}^{2v}$
are square-root branch points, as it is transparent from the
inverse transformations to (\ref{d12}), e.g

\begin{equation}
V(t)=i\frac {\sqrt{\left (\frac{t_{in}^{1s}-t_0^s}{t_0^s}\right
)^{1/2}+\left (\frac{t-t_0^s}{t_0^s}\right )^{1/2}}-
 \sqrt{\left (\frac{t_{in}^{1s}-t_0^s}{t_0^s}\right )^{1/2}-\left (\frac{t-t_0^s}{t_0^s}\right )^{1/2}}}
{\sqrt{\left (\frac{t_{in}^{1s}-t_0^s}{t_0^s}\right )^{1/2}+\left
(\frac{t-t_0^s}{t_0^s}\right )^{1/2}}+
 \sqrt{\left (\frac{t_{in}^{1s}-t_0^s}{t_0^s}\right )^{1/2}-\left (\frac{t-t_0^s}{t_0^s}\right )^{1/2}}}
\label{d13}
\end{equation}
and similarly for $W(t), U(t)$ and $X(t)$.

   The interpretation of $t^s_0=9m_{\pi}^2$ and $t^v_0=4m_{\pi}^2$ is clear. They are the lowest
branch points of isoscalar and isovector Dirac and Pauli nucleon
FFs on the positive real axis, respectively, as in the isoscalar
case the 3-pion states and in the isovector case the 2-pion states
are the lowest intermediate mass states in the unitarity
conditions of the corresponding FFs.

   However, as it follows just from the unitarity conditions of FFs, there is an
infinite number of allowed higher mass intermediate states and as
a result there is an infinite number of the corresponding branch
points (and thus, an infinite number of branch cut contributions)
in every of the considered nucleon FFs.

    Since, in principle, an infinite number of cuts cannot be taken
    into account in any theoretical
scheme, we restrict ourselves in every isoscalar and isovector
Dirac and Pauli FF to the two-cut approximation. The second one,
an effective inelastic cut, in every isoscalar and isovector Dirac
and Pauli FF is generated just by the square-root branch points
$t_{in}^{1s}, t_{in}^{1v}, t_{in}^{2s}, t_{in}^{2v}$,
respectively. They are free parameters of the model and the data
themselves, by a fitting procedure, will choose for them such
numerical values  that the contributions of the corresponding
square-root cuts will be practically equivalent to the
contributions of an infinite number of unitary branch cuts in
every considered FF.

   Some experts privately have been suggesting us to fix these
square-root branch points at the $N \bar N$ threshold. However, it
has been demonstrated practically that this can be done only in
the case of isovector parts of Dirac and Pauli FFs, but in none of
the cases of the isoscalar parts of the Dirac and Pauli FFs. The
latter are found at  much lower values than the $N \bar N$
threshold. This result indicates that between the lowest
$t_0^s=9m_{\pi}^2$ branch point and the $N \bar N$ threshold there
is some allowed intermediate mass state in the unitarity condition
generating an important cut contribution which cannot be neglected
in a description of the nucleon e.m. structure. We know from other
considerations that it is just the $K \bar K$ threshold.

   So, by  the unitarization of (\ref{d8})-(\ref{d11}), one
gets one analytic function at the interval $-\infty < t +\infty$
for every FF

\begin{eqnarray}
F_{1}^{s}[V(t)] &=&\left( \frac{1-V^{2}}{1-V_{N}^{2}}\right)
^{4}\left\{
\frac{1}{2}H_{\omega ^{\prime \prime }}(V)\cdot L_{\omega ^{\prime }}(V)+%
\left[ H_{\omega ^{\prime \prime }}(V)\cdot L_{\omega }(V)\cdot \frac{%
C_{\omega ^{\prime \prime }}^{1s}-C_{\omega }^{1s}}{C_{\omega
^{\prime
\prime }}^{1s}-C_{\omega ^{\prime }}^{1s}}\right. \right. -  \nonumber \\
&-&L_{\omega ^{\prime }}(V)\cdot L_{\omega }(V)\frac{C_{{\omega^{^{\prime}}}%
}^{1s}-C_{\omega}^{1s}}{C_{\omega^{^{\prime\prime}}}^{1s}-C_{\omega^{^{%
\prime}}}^{1s}}-H_{\omega ^{\prime \prime }}(V)\cdot L_{\omega ^{\prime }}(V)%
\biggr ](f_{\omega {NN}}^{(1)}/f_{\omega })+  \nonumber \\
&+&\left[ H_{\omega ^{\prime \prime }}(V)\cdot L_{\phi }(V)\frac{%
C_{\omega^{^{\prime\prime}}}^{1s}-C_{\phi}^{1s}}{C_{\omega^{^{\prime%
\prime}}}^{1s}-C_{\omega^{^{\prime}}}^{1s}}-L_{\omega ^{\prime
}}(V)\cdot
L_{\phi }(V)\frac{C_{\omega^{^{\prime}}}^{1s}-C_{\phi}^{1s}}{%
C_{\omega^{^{\prime\prime}}}^{1s}-C_{\omega^{^{\prime}}}^{1s}}\right.
-
\label{d14} \\
&-&H_{\omega ^{\prime \prime }}(V)\cdot L_{\omega ^{\prime }}(V)\biggr ](f_{{%
\phi }{NN}}^{(1)}/f_{{\phi }})-\left[ H_{\phi ^{\prime }}(V)\cdot
H_{\omega
^{\prime \prime }}(V)\frac{C_{\phi^{^{\prime}}}^{1s}-C_{\omega^{^{\prime%
\prime}}}^{1s}}{C_{\omega^{^{\prime\prime}}}^{1s}-C_{\omega^{^{\prime}}}^{1s}%
}\right. -  \nonumber \\
&-&H_{\phi ^{\prime }}(V)\cdot L_{\omega ^{\prime }}(V)\frac{%
C_{\phi^{^{\prime}}}^{1s}-C_{\omega^{^{\prime}}}^{1s}}{C_{\omega^{^{\prime%
\prime}}}^{1s}-C_{\omega^{^{\prime}}}^{1s}}+H_{\omega ^{\prime
\prime
}}(V)\cdot L_{\omega ^{\prime }}(V)\biggr ](f_{{\phi ^{^{\prime }}}{NN}%
}^{(1)}/f_{{\phi ^{^{\prime }}}})\biggr\}  \nonumber
\end{eqnarray}

\begin{eqnarray}
F_{1}^{v}[W(t)] &=&\left( \frac{1-W^{2}}{1-W_{N}^{2}}\right)
^{4}\left\{ \frac{1}{2}L_{\rho ^{\prime \prime }}(W)\cdot L_{\rho
^{\prime }}(W)+\left[ L_{\rho ^{\prime \prime }}(W)\cdot L_{\rho
}(W)\frac{C_{\varrho ^{^{\prime \prime }}}^{1v}-C_{\varrho
}^{1v}}{C_{\varrho ^{^{\prime \prime
}}}^{1v}-C_{\varrho ^{^{\prime }}}^{1v}}-\right. \right.   \nonumber \\
&-&L_{\rho ^{\prime }}(W)\cdot L_{\rho }(W)\frac{C_{\varrho
^{^{\prime }}}^{1v}-C_{\varrho }^{1v}}{C_{\varrho ^{^{\prime
\prime }}}^{1v}-C_{\varrho ^{^{\prime }}}^{1v}}-L_{\rho ^{\prime
\prime }}(W)\cdot L_{\rho ^{\prime
}}(W)\biggr](f_{{\varrho }{NN}}^{(1)}/f_{{\varrho }})+ \label{d15} \\
&+&\left[ H_{\rho ^{\prime \prime \prime }}(W)\cdot L_{\rho ^{\prime }}(W)%
\frac{C_{\varrho ^{^{\prime \prime \prime }}}^{1v}-C_{\varrho
^{^{\prime }}}^{1v}}{C_{\varrho ^{^{\prime \prime
}}}^{1v}-C_{\varrho ^{^{\prime }}}^{1v}}\right. -H_{\rho ^{\prime
\prime \prime }}(W)\cdot L_{\rho ^{\prime \prime
}}(W)\frac{C_{\varrho ^{^{\prime \prime \prime }}}^{1v}-C_{\varrho
^{^{\prime \prime }}}^{1v}}{C_{\varrho ^{^{\prime \prime
}}}^{1v}-C_{\varrho
^{^{\prime }}}^{1v}}-  \nonumber\\
&-&L_{\rho ^{\prime \prime }}(W)\cdot L_{\rho ^{\prime }}(W)\biggr ](f_{{%
\varrho ^{^{\prime \prime \prime }}}{NN}}^{(1)}/f_{{\varrho
^{^{\prime \prime \prime }}}})-\left[ H_{\rho ^{\prime \prime
\prime \prime }}(W)\cdot L_{\rho ^{\prime \prime
}}(W)\frac{C_{\varrho ^{^{\prime \prime \prime \prime
}}}^{1v}-C_{\varrho ^{^{\prime \prime }}}^{1v}}{C_{\varrho
^{^{\prime
\prime }}}^{1v}-C_{\varrho ^{^{\prime }}}^{1v}}\right.   \nonumber \\
&-&H_{\rho ^{\prime \prime \prime \prime }}(W)\cdot L_{\rho ^{\prime }}(W)%
\frac{C_{\varrho ^{^{\prime \prime \prime \prime
}}}^{1v}-C_{\varrho ^{^{\prime }}}^{1v}}{C_{\varrho ^{^{\prime
\prime }}}^{1v}-C_{\varrho ^{^{\prime }}}^{1v}}+L_{\rho ^{\prime
\prime }}(W)\cdot L_{\rho ^{\prime
}}(W)\biggr](f_{{\varrho ^{^{\prime \prime \prime \prime }}}{NN}}^{(1)}/f_{{%
\varrho ^{^{\prime \prime \prime \prime }}}})\biggr\}  \nonumber
\end{eqnarray}

\begin{eqnarray}
F_{2}^{s}[U(t)] &=&\left( \frac{1-U^{2}}{1-U_{N}^{2}}\right)
^{6}\left\{ \frac{1}{2}(\mu _{p}+\mu _{n})H_{\omega ^{\prime
\prime }}(U)\cdot L_{\omega
^{\prime }}(U)\cdot L_{\omega }(U)\right. +  \nonumber
\end{eqnarray}

\begin{eqnarray}
\phantom{F_{2}^{s}[U(t)] }&+&\left[ H_{\omega ^{\prime \prime }}(U)\cdot L_{\phi }(U)\cdot
L_{\omega }(U)\frac{C_{\omega ^{^{\prime \prime }}}^{2s}-C_{\phi
}^{2s}}{C_{\omega ^{^{\prime \prime }}}^{2s}-C_{\omega ^{^{\prime
}}}^{2s}}.\frac{C_{\phi
}^{2s}-C_{\omega }^{2s}}{C_{\omega ^{^{\prime }}}^{2s}-C_{\omega }^{2s}}%
\right. +  \nonumber \\
&+&H_{\omega ^{\prime \prime }}(U)\cdot L_{\omega ^{\prime
}}(U)\cdot
L_{\phi }(U)\frac{C_{\omega ^{^{\prime \prime }}}^{2s}-C_{\phi }^{2s}}{%
C_{\omega ^{^{\prime \prime }}}^{2s}-C_{\omega
}^{2s}}.\frac{C_{\omega ^{^{\prime }}}^{2s}-C_{\phi
}^{2s}}{C_{\omega ^{^{\prime }}}^{2s}-C_{\omega
}^{2s}}-  \nonumber \\
&-&L_{\omega ^{\prime }}(U)\cdot L_{\phi }(U)\cdot L_{\omega }(U)\frac{%
C_{\omega ^{^{\prime }}}^{2s}-C_{\phi }^{2s}}{C_{\omega ^{^{\prime
\prime
}}}^{2s}-C_{\omega ^{^{\prime }}}^{2s}}.\frac{C_{\phi }^{2s}-C_{\omega }^{2s}%
}{C_{\omega ^{^{\prime \prime }}}^{2s}-C_{\omega }^{2s}}-  \label{d16} \\
&-&H_{\omega ^{\prime \prime }}(U)\cdot L_{\omega ^{\prime
}}(U)\cdot
L_{\omega }(U)\biggr](f_{{\phi }{NN}}^{(2)}/f_{{\phi }})+ \nonumber \\
&+&\left[ H_{\phi ^{\prime }}(U)\cdot H_{\omega ^{\prime \prime
}}(U)\cdot L_{\omega ^{\prime }}(U)\frac{C_{\phi ^{^{\prime
}}}^{2s}-C_{\omega ^{^{\prime \prime }}}^{2s}}{C_{\omega
^{^{\prime \prime }}}^{2s}-C_{\omega
}^{2s}}.\frac{C_{\phi ^{^{\prime }}}^{2s}-C_{\omega ^{^{\prime }}}^{2s}}{%
C_{\omega ^{^{\prime }}}^{2s}-C_{\omega }^{2s}}\right. -  \nonumber \\
&-&H_{\phi ^{\prime }}(U)\cdot H_{\omega ^{\prime \prime
}}(U)\cdot L_{\omega }(U)\frac{C_{\phi ^{^{\prime
}}}^{2s}-C_{\omega ^{^{\prime \prime
}}}^{2s}}{C_{\omega ^{^{\prime \prime }}}^{2s}-C_{\omega ^{^{\prime }}}^{2s}}%
.\frac{C_{\phi ^{^{\prime }}}^{2s}-C_{\omega }^{2s}}{C_{\omega
^{^{\prime
}}}^{2s}-C_{\omega }^{2s}}+  \nonumber \\
&+&H_{\phi ^{\prime }}(U)\cdot L_{\omega ^{\prime }}(U)\cdot L_{\omega }(U)%
\frac{C_{\phi ^{^{\prime }}}^{2s}-C_{\omega ^{^{\prime
}}}^{2s}}{C_{\omega ^{^{\prime \prime }}}^{2s}-C_{\omega
^{^{\prime }}}^{2s}}.\frac{C_{\phi ^{^{\prime }}}^{2s}-C_{\omega
}^{2s}}{C_{\omega ^{^{\prime \prime
}}}^{2s}-C_{\omega }^{2s}}-  \nonumber \\
&-&H_{\omega ^{\prime \prime }}(U)\cdot L_{\omega ^{\prime
}}(U)\cdot L_{\omega }(U)\biggr](f_{{\phi ^{^{\prime
}}}{NN}}^{(2)}/f_{{\phi ^{^{\prime }}}})\biggr\}  \nonumber
\end{eqnarray}

\begin{eqnarray}
F_{2}^{v}[X(t)] &=&\left( \frac{1-X^{2}}{1-X_{N}^{2}}\right)
^{6}\left\{ \frac{1}{2}(\mu _{p}-\mu _{n})L_{\rho ^{\prime \prime
}}(X)\cdot L_{\rho
^{\prime }}(X)\cdot L_{\rho }(X)\right. +  \nonumber \\
&+&\left[ H_{\rho ^{\prime \prime \prime }}(X)\cdot L_{\rho
^{\prime
}}(X)\cdot L_{\rho }(X)\frac{C_{\varrho^{^{\prime\prime\prime}}}^{2v}-C_{%
\varrho^{^{\prime}}}^{2v}}{C_{\varrho^{^{\prime\prime}}}^{2v}-C_{\varrho^{^{%
\prime}}}^{2v}}.\frac{C_{\varrho^{^{\prime\prime\prime}}}^{2v}-C_{%
\varrho}^{2v}}{C_{\varrho^{^{\prime\prime}}}^{2v}-C_{\varrho}^{2v}}\right.-
\nonumber \\
&-&H_{\rho ^{\prime \prime \prime }}(X)\cdot L_{\rho ^{\prime
\prime
}}(X)\cdot L_{\rho }(X)\frac{C_{\varrho^{^{\prime\prime\prime}}}^{2v}-C_{%
\varrho^{^{\prime\prime}}}^{2v}}{C_{\varrho^{^{\prime\prime}}}^{2v}-C_{%
\varrho^{^{\prime}}}^{2v}}.\frac{C_{\varrho^{^{\prime\prime%
\prime}}}^{2v}-C_{\varrho}^{2v}}{C_{\varrho^{^{\prime}}}^{2v}-C_{%
\varrho}^{2v}}+  \nonumber \\
&+&H_{\rho ^{\prime \prime \prime }}(X)\cdot L_{\rho ^{\prime
\prime
}}(X)\cdot L_{\rho ^{\prime }}(X)\frac{C_{\varrho^{^{\prime\prime%
\prime}}}^{2v}-C_{\varrho^{^{\prime\prime}}}^{2v}}{C_{\varrho^{^{\prime%
\prime}}}^{2v}-C_{\varrho}^{2v}}.\frac{C_{\varrho^{^{\prime\prime%
\prime}}}^{2v}-C_{\varrho^{^{\prime}}}^{2v}}{C_{\varrho^{^{%
\prime}}}^{2v}-C_{\varrho}^{2v}}-  \nonumber \\
&-&L_{\rho ^{\prime \prime }}(X)\cdot L_{\rho ^{\prime }}(X)\cdot
L_{\rho }(X)\biggr ](f_{{\varrho ^{^{\prime \prime \prime
}}}{NN}}^{(2)}/f_{{\varrho
^{^{\prime \prime \prime }}}})+  \label{d17} \\
&+&\left[ H_{\rho ^{\prime \prime \prime \prime }}(X)\cdot L_{\rho
^{\prime
}}(X)\cdot L_{\rho }(X)\frac{C_{\varrho^{^{\prime\prime\prime%
\prime}}}^{2v}-C_{\varrho^{^{\prime}}}^{2v}}{C_{\varrho^{^{\prime%
\prime}}}^{2v}-C_{\varrho^{^{\prime}}}^{2v}}.\frac{C_{\varrho^{^{\prime%
\prime\prime\prime}}}^{2v}-C_{\varrho}^{2v}}{C_{\varrho^{^{\prime%
\prime}}}^{2v}-C_{\varrho}^{2v}}\right. -  \nonumber \\
&-&H_{\rho ^{\prime \prime \prime \prime }}(X)\cdot L_{\rho
^{\prime \prime
}}(X)\cdot L_{\rho }(X)\frac{C_{\varrho^{^{\prime\prime\prime%
\prime}}}^{2v}-C_{\varrho^{^{\prime\prime}}}^{2v}}{C_{\varrho^{^{\prime%
\prime}}}^{2v}-C_{\varrho^{^{\prime}}}^{2v}}.\frac{C_{\varrho^{^{\prime%
\prime\prime\prime}}}^{2v}-C_{\varrho}^{2v}}{C_{\varrho^{^{%
\prime}}}^{2v}-C_{\varrho}^{2v}}+  \nonumber \\
&+&H_{\rho ^{\prime \prime \prime \prime }}(X)\cdot L_{\rho
^{\prime \prime
}}(X)\cdot L_{\rho ^{\prime }}(X)\frac{C_{\varrho^{^{\prime\prime\prime%
\prime}}}^{2v}-C_{\varrho^{^{\prime\prime}}}^{2v}}{C_{\varrho^{^{\prime%
\prime}}}^{2v}-C_{\varrho}^{2v}}.\frac{C_{\varrho^{^{\prime\prime\prime%
\prime}}}^{2v}-C_{\varrho^{^{\prime}}}^{2v}}{C_{\varrho^{^{%
\prime}}}^{2v}-C_{\varrho}^{2v}}-  \nonumber
\end{eqnarray}

\begin{eqnarray}
\phantom{F_{2}^{v}[X(t)]}&-&L_{\rho ^{\prime \prime }}(X)\cdot L_{\rho ^{\prime }}(X)\cdot
L_{\rho
}(X)\biggr ](f_{{\varrho ^{^{\prime \prime \prime \prime }}}{NN}}^{(2)}/f_{{%
\varrho ^{^{\prime \prime \prime \prime }}}})\biggr\}  \nonumber
\end{eqnarray}%
where
\begin{eqnarray}
&&L_{r}(V)=\frac{(V_{N}-V_{r})(V_{N}-V_{r}^{\ast
})(V_{N}-1/V_{r})(V_{N}-1/V_{r}^{\ast })}{(V-V_{r})(V-V_{r}^{\ast
})(V-1/V_{r})(V-1/V_{r}^{\ast })};  \nonumber \\
&&C_{r}^{1s}=\frac{(V_{N}-V_{r})(V_{N}-V_{r}^{\ast
})(V_{N}-1/V_{r})(V_{N}-1/V_{r}^{\ast })}{-(V-1/V_{r})(V^{\ast}-1/V_{r}^{\ast })}%
;\qquad r=\omega ,\phi ,\omega ^{^{\prime }},  \nonumber \\
&&H_{l}(V)=\frac{(V_{N}-V_{l})(V_{N}-V_{l}^{\ast
})(V_{N}+V_{l})(V_{N}+V_{l}^{\ast })}{(V-V_{l})(V-V_{l}^{\ast
})(V+V_{l})(V+V_{l}^{\ast })};  \nonumber \\
&&C_{l}^{1s}=\frac{(V_{N}-V_{l})(V_{N}-V_{l}^{\ast
})(V_{N}+V_{l})(V_{N}+V_{l}^{\ast })}{-(V_{l}-1/V_{l})(V_{l}^{\ast
}-1/V_{l}^{\ast })};\qquad l=\omega ^{^{\prime \prime }},\phi
^{^{\prime }}
\nonumber \\
&&L_{k}(W)=\frac{(W_{N}-W_{k})(W_{N}-W_{k}^{\ast
})(W_{N}-1/W_{k})(W_{N}-1/W_{k}^{\ast })}{(W-W_{k})(W-W_{k}^{\ast
})(W-1/W_{k})(W-1/W_{k}^{\ast })};  \nonumber \\
&&C_{k}^{1v}=\frac{(W_{N}-W_{k})(W_{N}-W_{k}^{\ast
})(W_{N}-1/W_{k})(W_{N}-1/W_{k}^{\ast
})}{-(W_{k}-1/W_{k})(W_{k}^{\ast }-1/W_{k}^{\ast })};\qquad k=\rho
,\rho ^{^{\prime }},\rho ^{^{\prime \prime
}},  \nonumber \\
&&H_{n}(W)=\frac{(W_{N}-W_{n})(W_{N}-W_{n}^{\ast
})(W_{N}+W_{n})(W_{N}+W_{n}^{\ast })}{(W-W_{n})(W-W_{n}^{\ast
})(W+W_{n})(W+W_{n}^{\ast })};  \nonumber \\
&&C_{n}^{1v}=\frac{(W_{N}-W_{n})(W_{N}-W_{n}^{\ast
})(W_{N}+W_{n})(W_{N}+W_{n}^{\ast })}{-(W_{n}-1/W_{n})(W_{n}^{\ast
}-1/W_{n}^{\ast })};\qquad n=\rho ^{^{\prime \prime \prime }},\rho
^{^{\prime \prime \prime \prime }}  \nonumber \\
&&L_{r}(U)=\frac{(U_{N}-U_{r})(U_{N}-U_{r}^{\ast
})(U_{N}-1/U_{r})(U_{N}-1/U_{r}^{\ast })}{(U-U_{r})(U-U_{r}^{\ast
})(U-1/U_{r})(U-1/U_{r}^{\ast })};  \nonumber \\
&&C_{r}^{2s}=\frac{(U_{N}-U_{r})(U_{N}-U_{r}^{\ast
})(U_{N}-1/U_{r})(U_{N}-1/U_{r}^{\ast })}{-(U-1/U_{r})(U^{\ast}-1/U_{r}^{\ast })}%
;\qquad r=\omega ,\phi ,\omega ^{^{\prime }},  \nonumber \\
&&H_{l}(U)=\frac{(U_{N}-U_{l})(U_{N}-U_{l}^{\ast
})(U_{N}+U_{l})(U_{N}+U_{l}^{\ast })}{(U-U_{l})(U-U_{l}^{\ast
})(U+U_{l})(U+U_{l}^{\ast })};  \nonumber \\
&&C_{l}^{2s}=\frac{(U_{N}-U_{l})(U_{N}-U_{l}^{\ast
})(U_{N}+U_{l})(U_{N}+U_{l}^{\ast })}{-(U_{l}-1/U_{l})(U_{l}^{\ast
}-1/U_{l}^{\ast })};\qquad l=\omega ^{^{\prime \prime }},\phi
^{^{\prime }}
\nonumber \\
&&L_{k}(X)=\frac{(X_{N}-X_{k})(X_{N}-X_{k}^{\ast
})(X_{N}-1/X_{k})(X_{N}-1/X_{k}^{\ast })}{(X-X_{k})(X-X_{k}^{\ast
})(X-1/X_{k})(X-1/X_{k}^{\ast })};  \nonumber \\
&&C_{k}^{2v}=\frac{(X_{N}-X_{k})(X_{N}-X_{k}^{\ast
})(X_{N}-1/X_{k})(X_{N}-1/X_{k}^{\ast
})}{-(X_{k}-1/X_{k})(X_{k}^{\ast }-1/X_{k}^{\ast })};\qquad k=\rho
,\rho ^{^{\prime }},\rho ^{^{\prime \prime
}},  \nonumber \\
&&H_{n}(X)=\frac{(X_{N}-X_{n})(X_{N}-X_{n}^{\ast
})(X_{N}+X_{n})(X_{N}+X_{n}^{\ast })}{(X-X_{n})(X-X_{n}^{\ast
})(X+X_{n})(X+X_{n}^{\ast })};  \nonumber \\
&&C_{n}^{2v}=\frac{(X_{N}-X_{n})(X_{N}-X_{n}^{\ast
})(X_{N}+X_{n})(X_{N}+X_{n}^{\ast })}{-(X_{n}-1/X_{n})(X_{n}^{\ast
}-1/X_{n}^{\ast })};\qquad n=\rho ^{^{\prime \prime \prime }},\rho
^{^{\prime \prime \prime \prime }}  \nonumber \\
&&  \nonumber
\end{eqnarray}
with $t_{0}^{s}=9m_{\pi }^{2},t_{0}^{v}=4m_{\pi
}^{2},t_{in}^{1v}=t_{in}^{2v}=4m_{N}^{2}$ and
$t_{in}^{1s},t_{in}^{2s}$ effective square root branch points,
which however can not be fixed at two-nucleon treshold as in the
isoscalar case there is a remarkable contribution of
$K\overline{K}$ intermediate state in the unitarity condition.

  There are no data in the unphysical region $0 < t < 4m_N^2$
in order to determine the parameters of $\rho ,\omega ,\phi ,\rho
^{\prime },\omega ^{\prime },\phi ^{\prime },\rho ^{\prime \prime
},\omega ^{\prime \prime }$ by the fitting procedure. Therefore,
they are taken from Review of Particle Physics \cite{Rpp}. The
parameters of $\rho ^{\prime \prime \prime }$ are taken from
\cite{rho4} and the parameters of $\rho ^{\prime \prime \prime
\prime }$ are left to be free. Then we are finally in
$F_{1}^{s}[V(t)]$ with 4 free parameters, in $F_{1}^{v}[W(t)]$
with 3 free parameters, in $F_{2}^{s}[U(t)]$ with 3 free
parameters and in $F_{2}^{v}[X(t)]$ with 4 free parameters. All of
them are determined in comparison of the model with the collected
512 experimental points on the nucleon EM FFs in space-like and
time-like region simultaneously \cite{DDW}. As a result, the
fourth excited state of $\rho $-meson with the slightly lower mass
$m_{\rho ^{\prime \prime \prime \prime }}=1455\pm 53$ MeV than in
\cite{Fredrho5} and $\Gamma _{\rho ^{\prime \prime \prime \prime
}}=728\pm 42$ MeV is determined. Its existence, however, has to be
confirmed also in other processes and not only in the $e^+e^-\to
N\bar N$ process.

   The quality of the achieved descriptions is graphically presented
in Figs.~\ref{fig20} and \ref{fig21} by dashed lines.

\begin{figure}[tb]
\centering
\scalebox{0.46}{\includegraphics{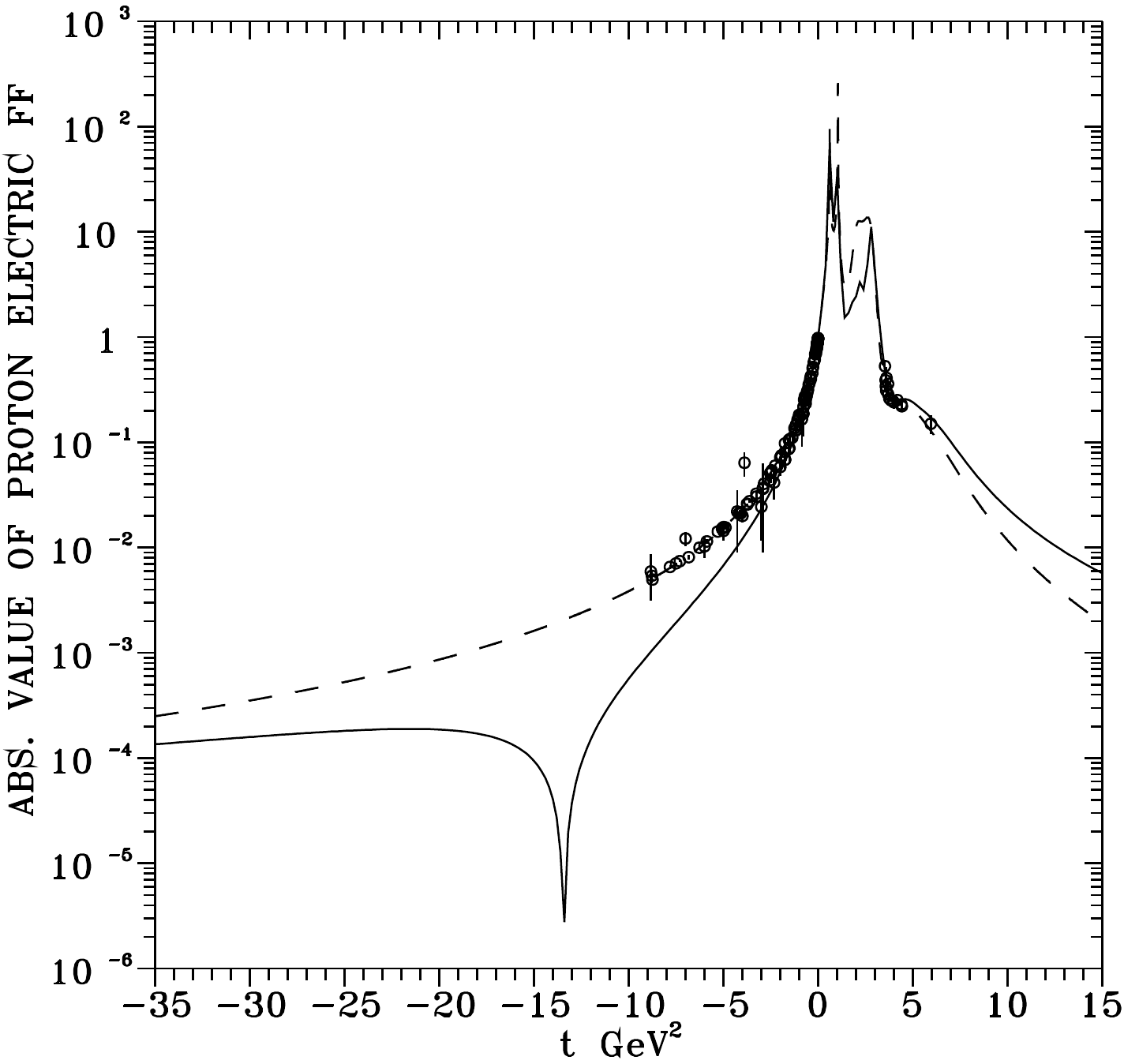}}\hspace{.1cm}
\scalebox{0.46}{\includegraphics{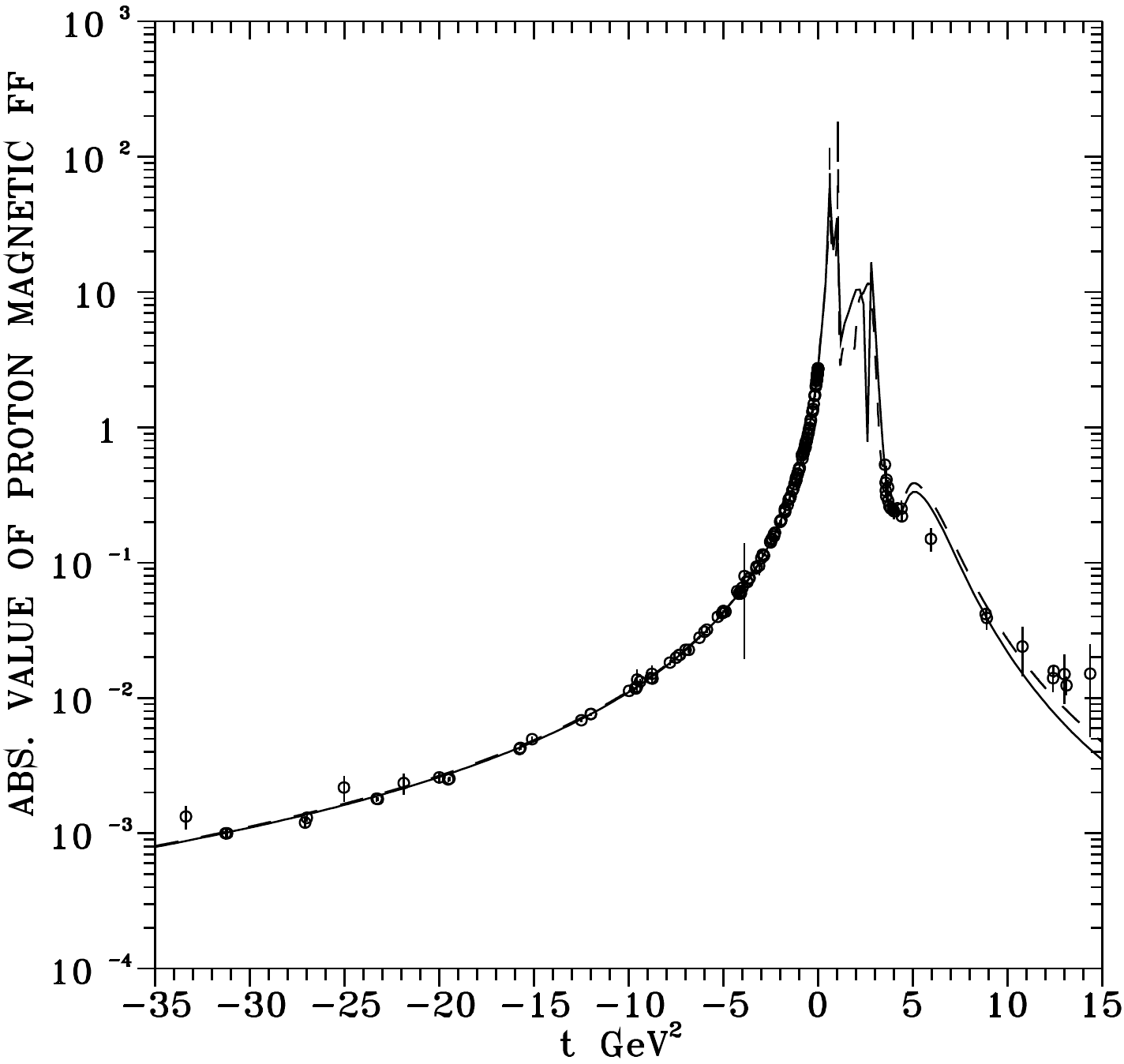}}
\caption{The predicted different behaviors of $G_{Ep}$ and $G_{Mp}$in $t<0$
region dependent on the fact if Rosenbluth technique data (dashed
line) or JLab proton polarization data (full line) are used in the
analysis}\label{fig20}
\end{figure}

\begin{figure}[tb]
\centering
\scalebox{0.45}{\includegraphics{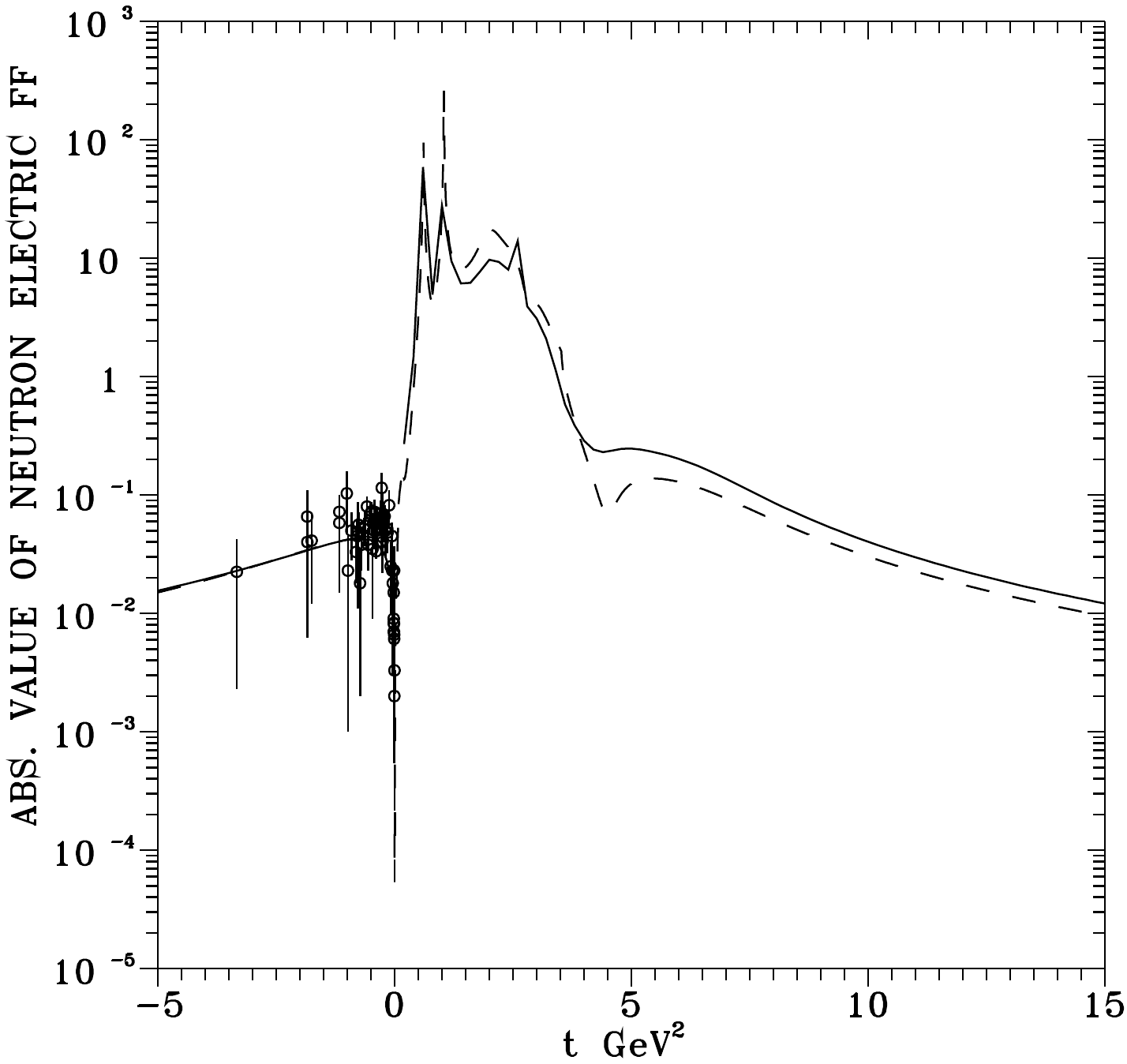}}\hspace{.1cm}
\scalebox{0.45}{\includegraphics{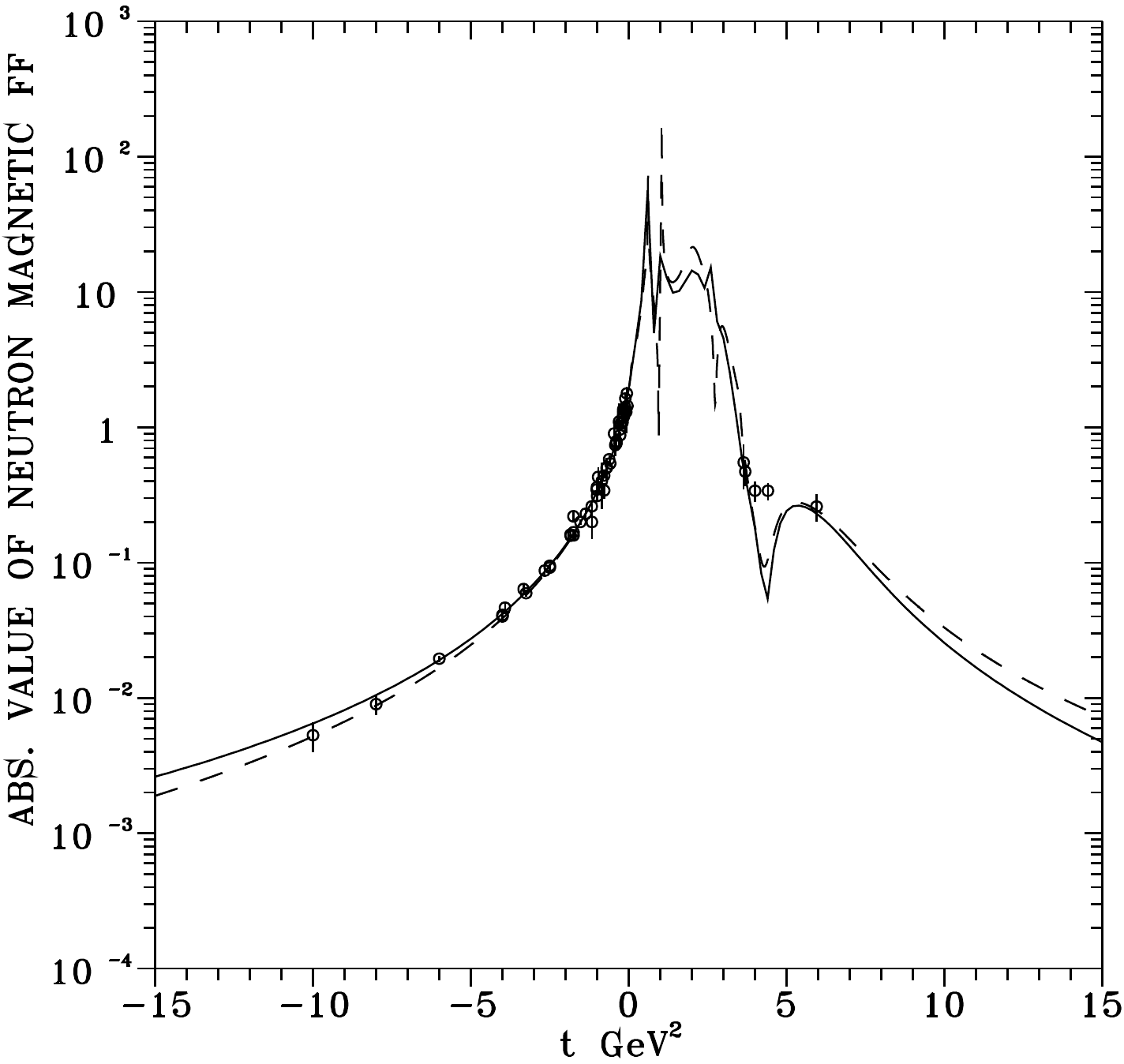}}
\caption{The predicted different behaviors of $G_{En}$ and  $G_{Mn}$ in $t<0$
region dependent on the fact if Rosenbluth technique data (dashed
line) or JLab proton polarization data (full line) are used in the
analysis}\label{fig21}
\end{figure}

   Such situation has existed up to the year 2000, when measurements
of the ratio of the electric to magnetic proton form factors were
carried out by a completely different method at Jefferson Lab. in
Newport News (Virginia).

\medskip

    \subsection{JLab proton polarization data puzzle} \label{IV3}

   \medskip

\begin{figure}[tb]
\centering
\includegraphics[scale=.6]{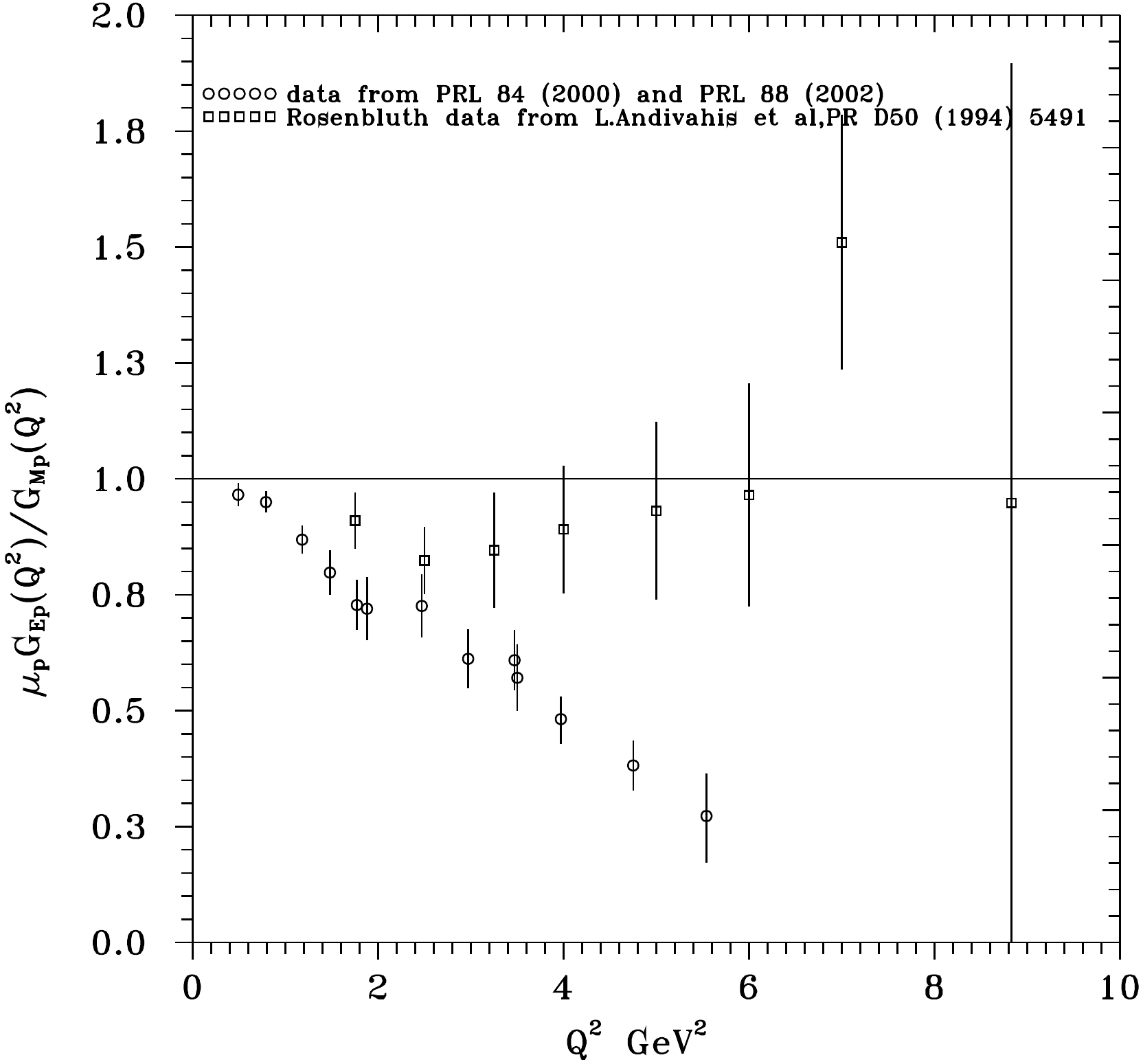}
\caption{The JLab proton polarization data on ratio
$\mu_pG_{Ep}/G_{Mp}$ (circles) and the same ratio calculated from
data on $G_{Ep}$ and $G_{Mp}$ obtained by Rosenbluth technique
(quadrangles).}
\label{fig22}
\end{figure}

   Starting from the year 2000 a large progress has been done \cite{Jones00,Gayon02,Punjabi05} in
obtaining of the ratio (see Fig.~\ref{fig22})
\begin{equation}
\frac{G_{Ep}}{G_{Mp}}=-\frac{P_t}{P_l}\frac{(E+E^{'})}{2m_p}\tan({\theta/2}).
\label{a5}
\end{equation}
in the space-like ($t=-Q^2<0$) region by measuring simultaneously
transverse
\begin{equation}
P_t=\frac{h}{I_0}(-2)\sqrt{\tau(1+\tau)}G_{Ep}G_{Mp}\tan({\theta/2})
\label{a6}
\end{equation}
and longitudinal
\begin{equation}
P_l=\frac{h(E+E^{'})}{I_0m_p}\sqrt{\tau(1+\tau)}G_{Mp}^2\tan^2({\theta/2}),
\label{a7}
\end{equation}
components of the recoil proton's polarization in the electron
scattering plane of the polarization transfer process $\vec{e}^-
p\to e^-\vec{p}$, where $h$ is the electron beam helicity, $I_0$
is the unpolarized cross-section excluding $\sigma_{Mott}$ and
$\tau=Q^2/4m_p^2$. As one can see from Fig.~\ref{fig22}, these ratio data
are in strong disagreement with the data obtained by Rosenbluth
technique.

It could be understandable as follows.

   Due to the fact that $G_{Mp}^2(t)$ in (\ref{d2}) is multiplied by
$-t/4m^2_p$ factor, the measured cross-section with increased $-t$
becomes dominant by $G_{Mp}^2(t)$ part contribution and the
extraction of $G_{Ep}^2(t)$ is more and more difficult. As a
result the extraction of $G_{Ep}$ at higher values of $-t$ by
Rosenbluth technique is not promising.

   We have carried out a test of this hypothesis \cite{Adamuscin05}
in the framework of the ten-resonance $U\&A$ model of nucleon EM
structure \cite{DDW}, which is formulated in the language of
isoscalar $F_{1,2}^s(t)$ and isovector $F_{1,2}^v(t)$ parts of the
Dirac and Pauli FF's and comprises all known nucleon FF
properties.

   First, we have carried out the analysis of all proton and neutron
data obtained by Rosenbluth technique together with all proton and
neutron data in time-like region.

Then all $|G_{Ep}(t)|$ space-like data obtained by Rosenbluth
technique were excluded and the new JLab proton polarization data on
$\mu_p G_{Ep}(Q^2)/G_{Mp}(Q^2)$ for $0.49$ $GeV^2$ $\leq Q^2 \leq
5.54 $ $GeV^2$ were analyzed together with all electric proton
time-like data and all space-like and time-like magnetic  proton, as
well as electric and magnetic neutron data.

The results of the analysis are presented in Figs.~\ref{fig20} and \ref{fig21} from
where it is seen that almost nothing is changed in a description
of $G_{Mp}(t)$, $G_{En}(t)$ and $G_{Mn}(t)$ in both space-like and
time-like regions, and also  $|G_{Ep}(t)|$ in the time-like
region. There is only a difference in behaviors of $G_{Ep}(t)$ in
$t<0$ region dependent on the fact if old data obtained by
Rosenbluth technique are used (dashed line) or the new JLab proton
polarization data are analysed (full line). The new JLab proton
polarization data require in $G_{Ep}(t)$ an existence of the zero
(diffraction minimum) around the value $t = -13 GeV^2$ of the
momentum transfer squared, which could change the charge
distribution behavior within proton.

   The proton charge distribution (assuming to be spherically symmetric) is an
inverse Fourier transform of the proton electric FF
\begin{equation}
\rho _{p}(r)=\frac{1}{(2\pi )^{3}}\int e^{-iQr}G_{Ep}(Q^{2})d^{3}Q
\end{equation}
from where
\begin{equation*}
\rho _{p}(r)=\frac{4\pi }{(2\pi )^{3}}\int_{0}^{\infty }G_{Ep}(Q^{2})\frac{
\sin (Qr)}{Qr}Q^{2}dQ.
\end{equation*}

\begin{figure}[tb]
\centering{\includegraphics[scale=.6]{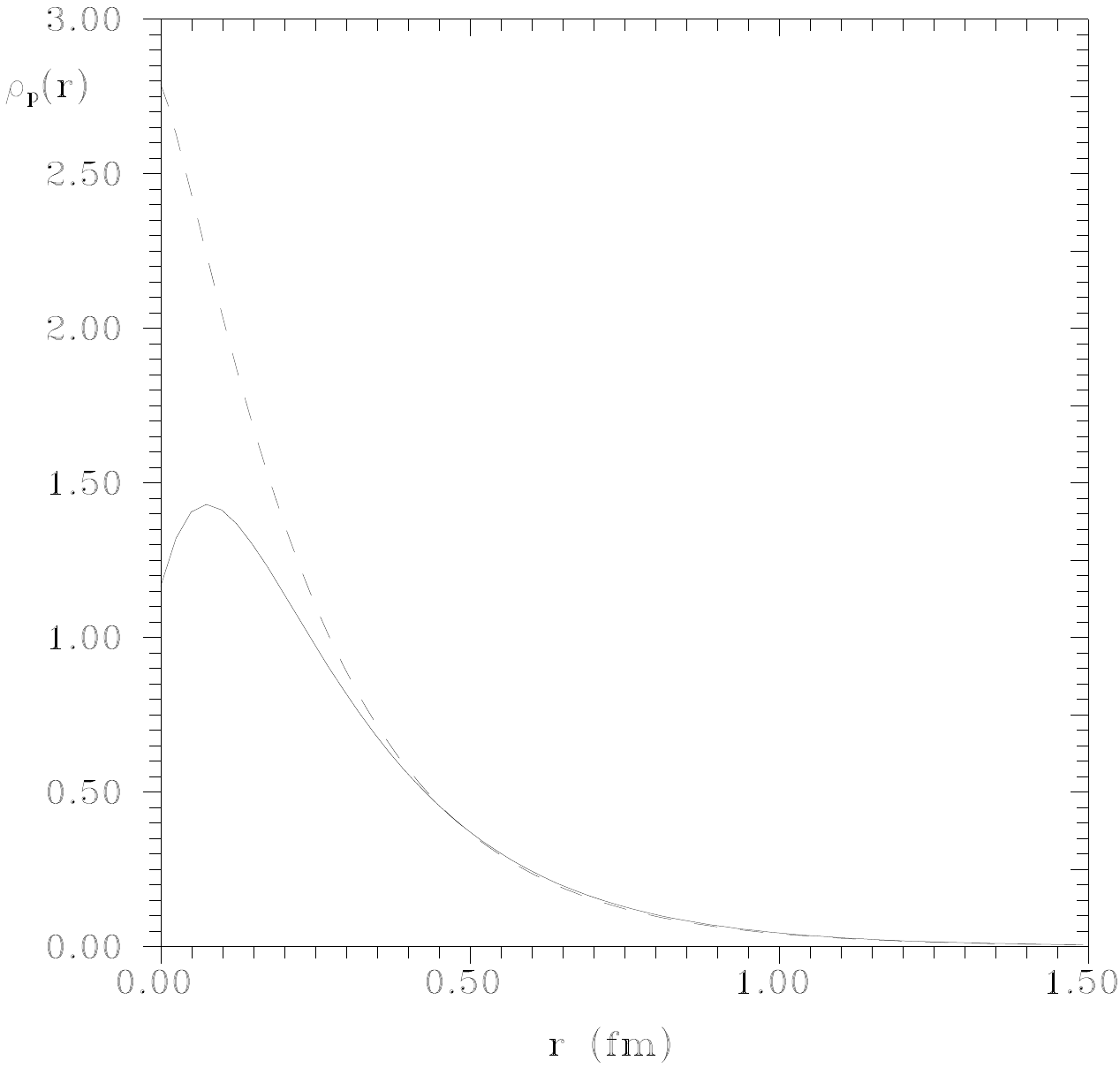}}\caption{Charge
distribution behavior within the proton}
\label{fig23}
\end{figure}

\begin{figure}[tb]
\centering
\includegraphics[scale=.6]{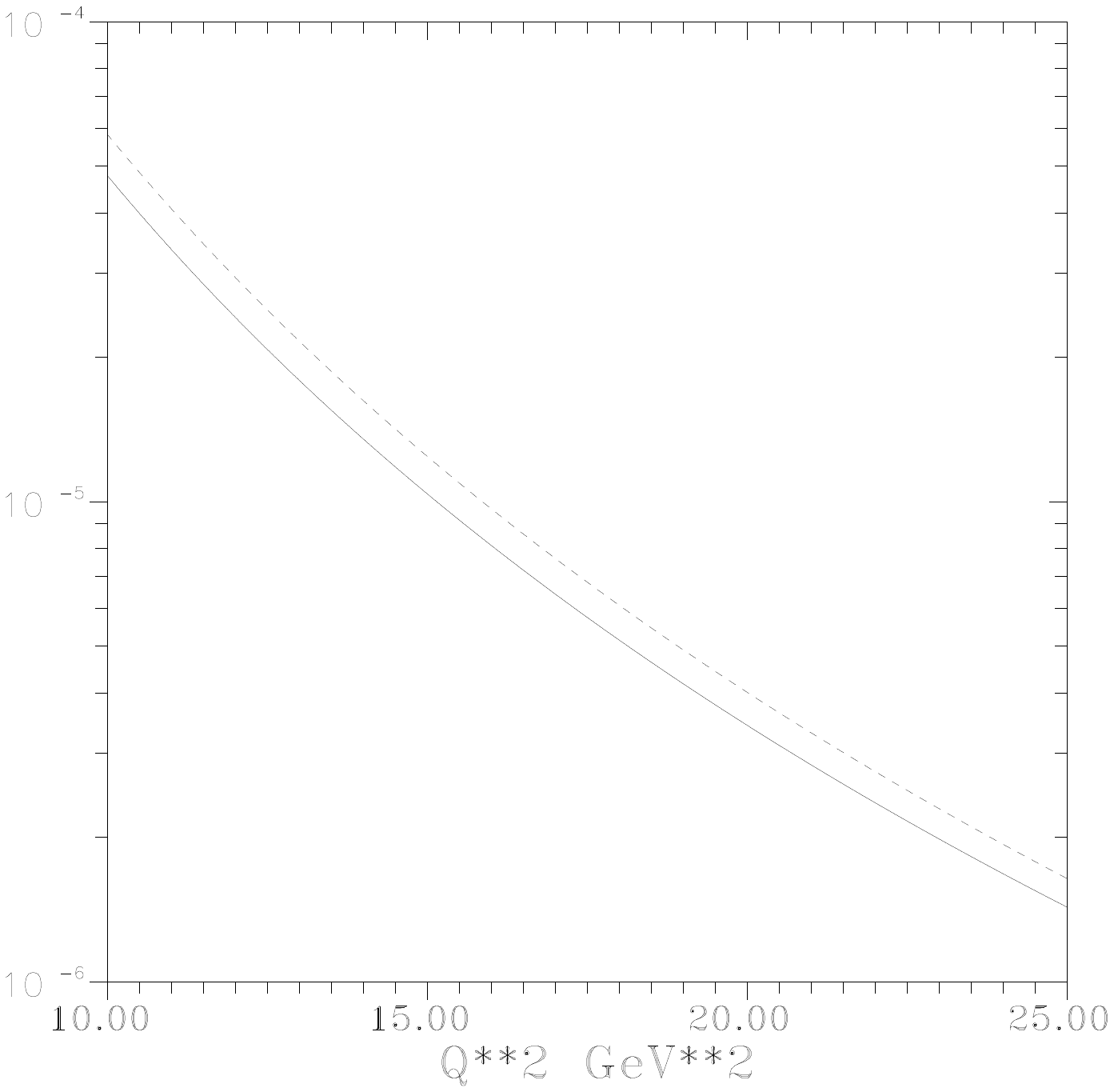}
\caption{A prediction of two different behaviors of the right-hand
side in (\ref{dynsumr}) following from two different behaviors of
$G_{Ep}$ in Fig.~\ref{fig20}.}\label{figdis}
\end{figure}
Substituting for the $G_{Ep}(Q^{2})$ under the integral either the
older behavior from Fig.~\ref{fig20} given by dashed line, or the new
behavior with the zero (see Fig.~\ref{fig20} given by full line) following
from the new JLab polarization data, one gets different charge
distributions within the proton given in Fig.~\ref{fig23} by dashed and full
lines, respectively. That all leads also to different mean square
proton charge radii. The old proton charge radius takes the value
$\left\langle r_{p}^{2}\right\rangle =0.68$fm$^{2}$ and the new
one $\left\langle r_{p}^{2}\right\rangle =0.72$fm$^{2}.$

\medskip

   In order to distinguish which of these behaviors of $G_{Ep}(t)$ in
the space-like ($t< 0$) region is correct, we suggest to employ
also the new sum rule \cite{BDKsr04} (to be derived later in this
review)
\begin{eqnarray}\label{dynsumr}
&&  F_{1p}^2(-{Q}^2)+\frac{{Q}^2}{4m_p^2} F_{2p}^2(-{Q}^2)-\\
&-& F_{1n}^2(-{Q}^2)-\frac{{Q}^2}{4m_n^2}
F_{2n}^2(-{Q}^2)=\nonumber\\
&=&1-2\frac{({Q}^2)^2}{\pi\alpha^2}
\left(\frac{d\sigma^{e^-p \to e^-X}}{d{Q}^2}-\frac{d\sigma^{e^-n \to
e^-X}}{d{Q}^2}\right),\nonumber
\end{eqnarray}
giving into a relation proton and neutron Dirac and Pauli FFs in
the space-like region with a difference of the differential proton
and neutron cross-sections describing $Q^2$ distribution in DIS.

Evaluating Dirac and Pauli FFs on the left-hand side corresponding
to the old (dashed line) and new (full line) space-like behavior
of $G_{Ep}(t)$ in Fig.~\ref{fig20} one predicts the corresponding behaviors
of the difference of deep inelastic cross sections in Fig.~\ref{figdis}. By
a measurement  of the latter the true $t< 0$ behavior of
$G_{Ep}(t)$ can be chosen.

\medskip

\subsection{Prediction of $\sigma_{tot}(e^+e^- \to Y \bar Y)$
     behaviors} \label{IV4}

\medskip

   It is well known, that according to SU(3) classification of
hadrons there is ${1/2}^+$ octet of baryons including nucleon
doublet $[n,p]$ together with 6 other hyperons $[\Lambda^0]$,
$[\Sigma^+,\Sigma^0,\Sigma^-]$ and $[\Xi^0,\Xi^-]$. Though there
is almost zero experimental information on the hyperon EM
structure, by using special nine-resonance $U\&A$ models of EM FFs
of all members of the ${1/2}^+$ octet of baryons, the experimental
information on the nucleon EM FFs and SU(3) symmetry, one can
predict behaviors of all hyperon EM FFs and as a consequence also
behaviors of
\begin{equation}
\sigma_{\rm tot}(e^+e^-\to Y\bar{Y}) =
\frac{4\pi\alpha^2\beta_Y}{3t} \left[|G_{MY}(t)|^2 +
\frac{2m^2_Y}{t}|G_{EY}(t)|^2\right]. \label{stothyp}
\end{equation}

   Practically, one has to start with a specific nine-resonance
$U\&A$ model of the EM FFs of ${1/2}^+$ octet of baryons, unifying
compatibly all known properties of FFs, which renders just
$\rho$-, $\omega$-~and $\phi$-meson coupling constant ratios as
free parameters.

   For nucleons, these free parameters are evaluated numerically by a
comparison of the nucleon $U\&A$ model with existing nucleon FF
data and then it is straightforward to find numerical values of
$f_{\rho NN}$, $f_{\omega NN}$ and $f_{\phi NN}$.

   On the other hand, the trace of SU(3) invariant Lagrangian for
vector-meson-baryon-antibaryon vertex
\begin{eqnarray}
{\rm Tr}(L_{VB\bar{B}}) &=&
\frac{i}{\sqrt{2}}f^F\left[\bar{B}^\alpha_\beta\gamma_\mu
B^\beta_\gamma -
\bar{B}^\beta_\gamma\gamma_\mu B^\alpha_\beta\right](V_\mu)^\gamma_\alpha + \nonumber\\
&+&\frac{i}{\sqrt{2}}f^D\left[\bar{B}^\beta_\gamma\gamma_\mu B^\alpha_\beta
+
\bar{B}^\alpha_\beta\gamma_\mu B^\beta_\gamma\right](V_\mu)^\gamma_\alpha + \nonumber\\
&+&\frac{i}{\sqrt{2}}f^S\bar{B}^\alpha_\beta\gamma_\mu
B^\beta_\alpha\omega^0_\mu
\end{eqnarray}
with $\omega-\phi$ mixing
\begin{eqnarray}
\phi^0 &=& \phi_8\cos\vartheta-\omega_1\sin\vartheta \nonumber \\
\omega^0 &=& \phi_8\sin\vartheta+\omega_1\cos\vartheta
\end{eqnarray}
$B$,$\bar{B}$ and $V$ baryon, antibaryon and vector-meson octuplet
matrices, $\omega^0_\mu$ omega-meson singlet, $f^F$, $f^D$ and
$f^S$ SU(3) coupling constants and $\vartheta$ mixing angle,
provides the following expressions for vector-meson-baryon
coupling constants
\begin{eqnarray}
f^{(1,2)}_{\rho NN} &=& \frac{1}{2}(f^D_{1,2}+f^F_{1,2}) \nonumber \\
f^{(1,2)}_{\omega NN} &=&\frac{1}{\sqrt{2}}\cos\vartheta
f^S_{1,2}-
      \frac{1}{2\sqrt{3}}\sin\vartheta(3f^F_{1,2}-f^D_{1,2}) \label{su3nuc}\\
f^{(1,2)}_{\phi NN} &=&\frac{1}{\sqrt{2}}\sin\vartheta f^S_{1,2} +
      \frac{1}{2\sqrt{3}}\cos\vartheta(3f^F_{1,2}-f^D_{1,2}) \nonumber
\end{eqnarray}

\begin{eqnarray}
f^{(1,2)}_{\omega\Lambda\Lambda} &=&
\frac{1}{\sqrt{2}}\cos\vartheta f^S_{1,2} +
          \frac{1}{\sqrt{3}}\sin\vartheta f^D_{1,2} \nonumber \\
 & & \label{su3lam}\\
f^{(1,2)}_{\phi\Lambda\Lambda} &=& \frac{1}{\sqrt{2}}\sin\vartheta
f^S_{1,2} -
          \frac{1}{\sqrt{3}}\cos\vartheta f^D_{1,2} \nonumber
\end{eqnarray}

\begin{eqnarray}
f^{(1,2)}_{\rho\Sigma\Sigma} &=& -f^F_{1,2} \nonumber \\
f^{(1,2)}_{\omega\Sigma\Sigma} &=& \frac{1}{\sqrt{2}}\cos\vartheta
f^S_{1,2} -
         \frac{1}{\sqrt{3}}\sin\vartheta f^D_{1,2} \label{su3sig}\\
f^{(1,2)}_{\phi\Sigma\Sigma} &=& \frac{1}{\sqrt{2}}\sin\vartheta
f^S_{1,2} +
         \frac{1}{\sqrt{3}}\cos\vartheta f^D_{1,2}\  \nonumber
\end{eqnarray}

\begin{eqnarray}
f^{(1,2)}_{\rho\Xi\Xi} &=& \frac{1}{2}(f^D_{1,2}-f^F_{1,2}) \nonumber \\
f^{(1,2)}_{\omega\Xi\Xi} &=&\frac{1}{\sqrt{2}}\cos\vartheta
f^S_{1,2}+
      \frac{1}{2\sqrt{3}}\sin\vartheta(3f^F_{1,2}+f^D_{1,2}) \label{su3xi}\\
f^{(1,2)}_{\phi\Xi\Xi} &=&\frac{1}{\sqrt{2}}\sin\vartheta
f^S_{1,2} -
      \frac{1}{2\sqrt{3}}\cos\vartheta(3f^F_{1,2}+f^D_{1,2}). \nonumber
\end{eqnarray}
Then the solutions of the system of algebraic eqs.~(\ref{su3nuc})
according to $f^D_{1,2}$, $f^F_{1,2}$, $f^S_{1,2}$ with numerical
values of $f_{\rho NN}$, $f_{\omega NN}$ and $f_{\phi NN}$, by
means of the expressions (\ref{su3lam})-(\ref{su3xi}), enable to
predict all free vector-meson-hyperon coupling constant ratios in
the EM FFs of hyperons $[\Lambda^0]$,
$[\Sigma^+,\Sigma^0,\Sigma^-]$ and $[\Xi^0,\Xi^-]$ and, as a
result, also behaviours of the total cross-sections
(\ref{stothyp}) (see Figs.~\ref{fig25}-\ref{fig27}).

\begin{figure}[tb]
\begin{center}
\includegraphics[scale=.47]{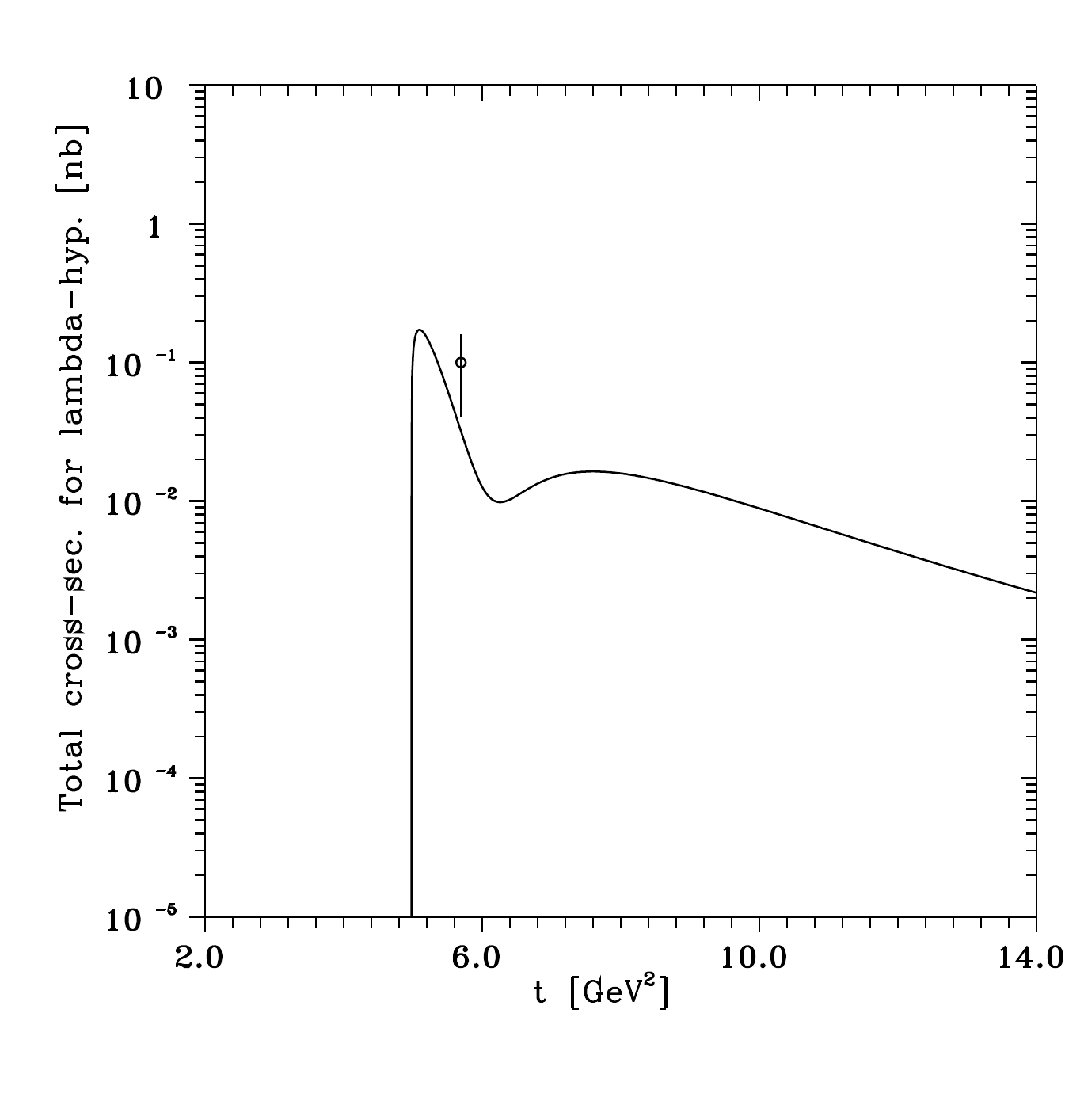}
\includegraphics[scale=.47]{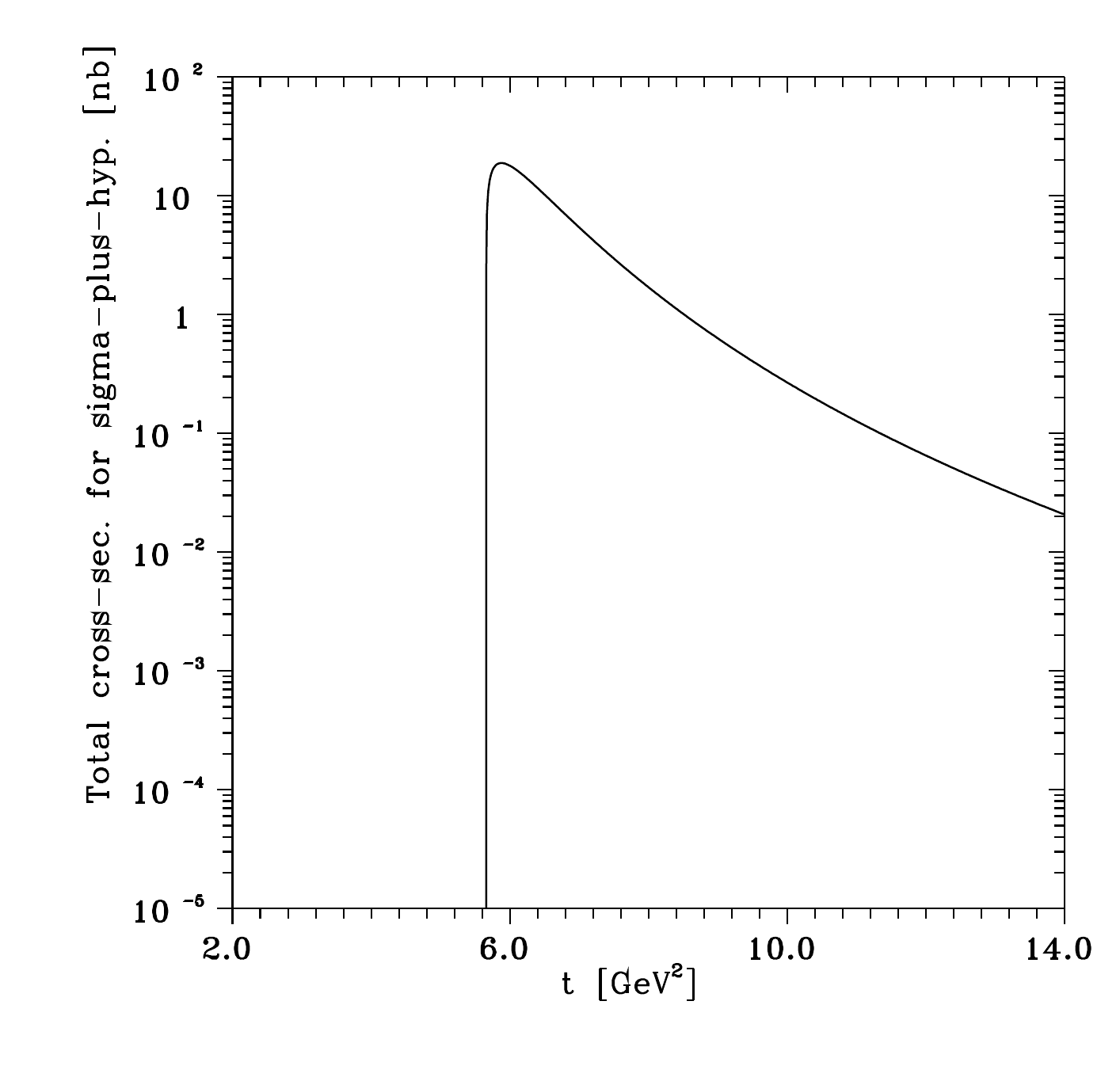}
\end{center}
  \caption{$e^+e^-$-annihilation cross-sections to $\Lambda$ and $\Sigma^+$ hyperons.}
  \label{fig25}
\end{figure}
\begin{figure}[tb!]
\begin{center}
\includegraphics[scale=.47]{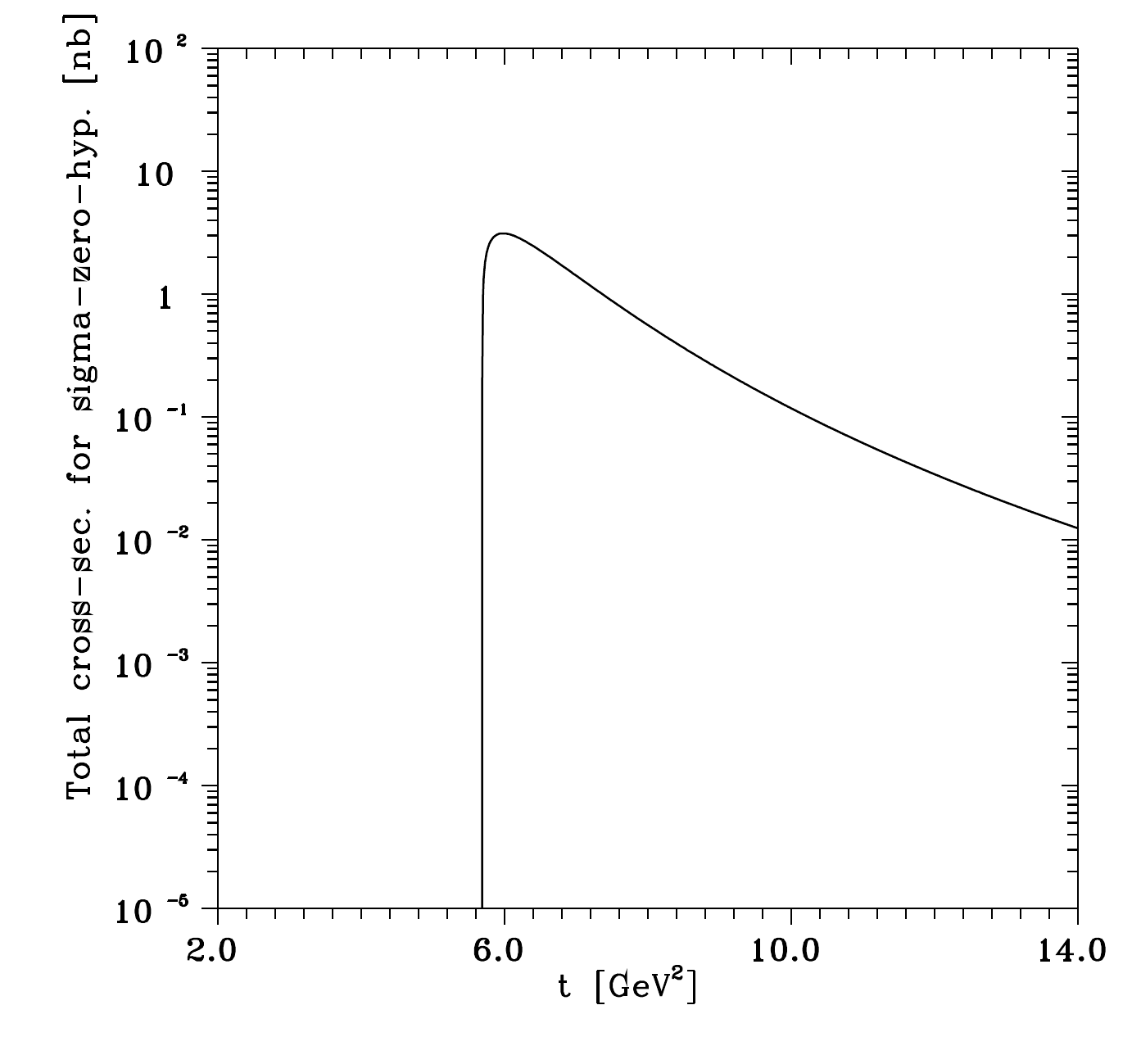}
\includegraphics[scale=.47]{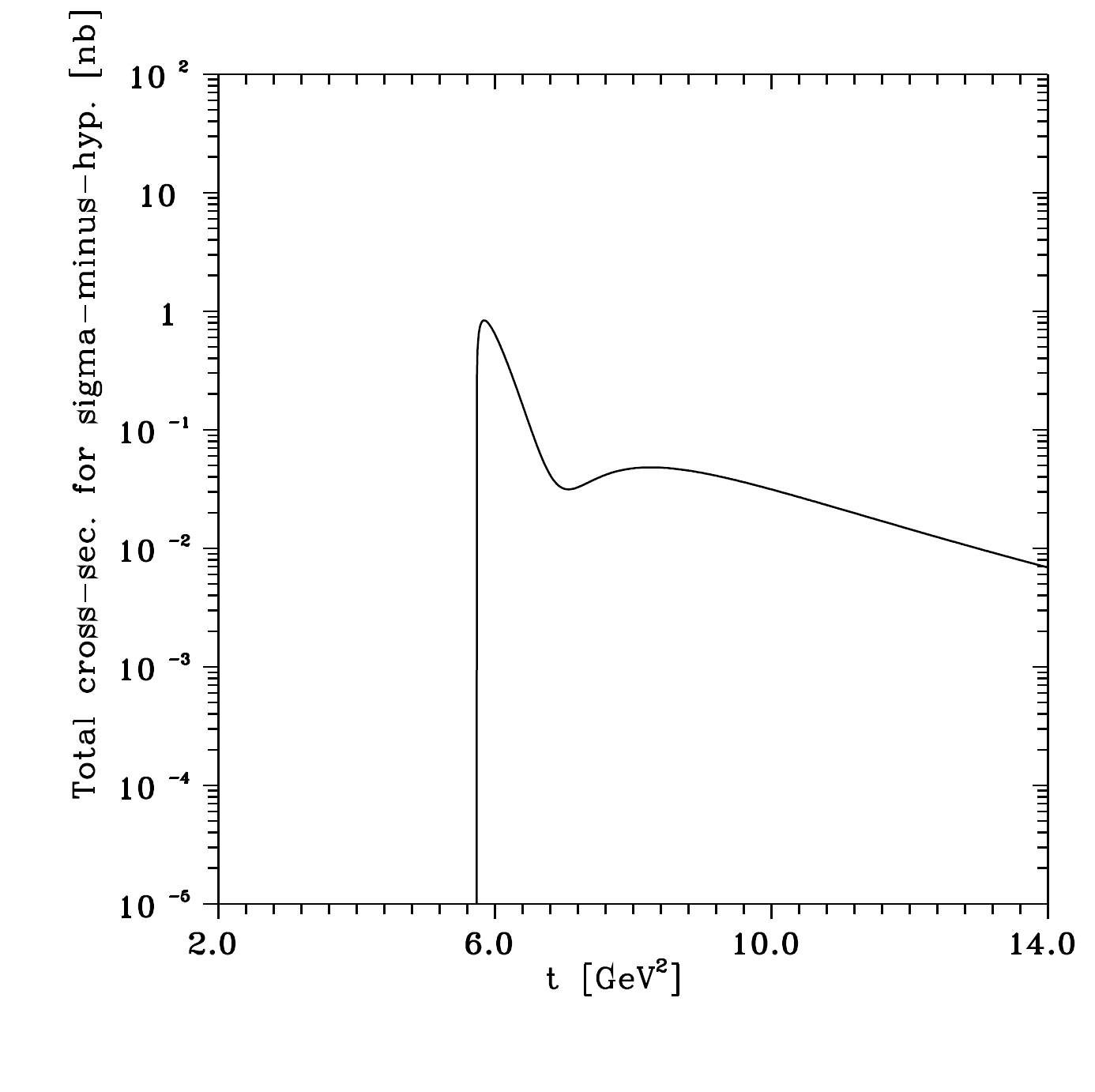}
\end{center}
  \caption{$e^+e^-$-annihilation cross-sections to $\Sigma^0$ and $\Sigma^-$ hyperons.}
  \label{fig26}
\end{figure}
\begin{figure}[tb!]
\begin{center}
\includegraphics[scale=.47]{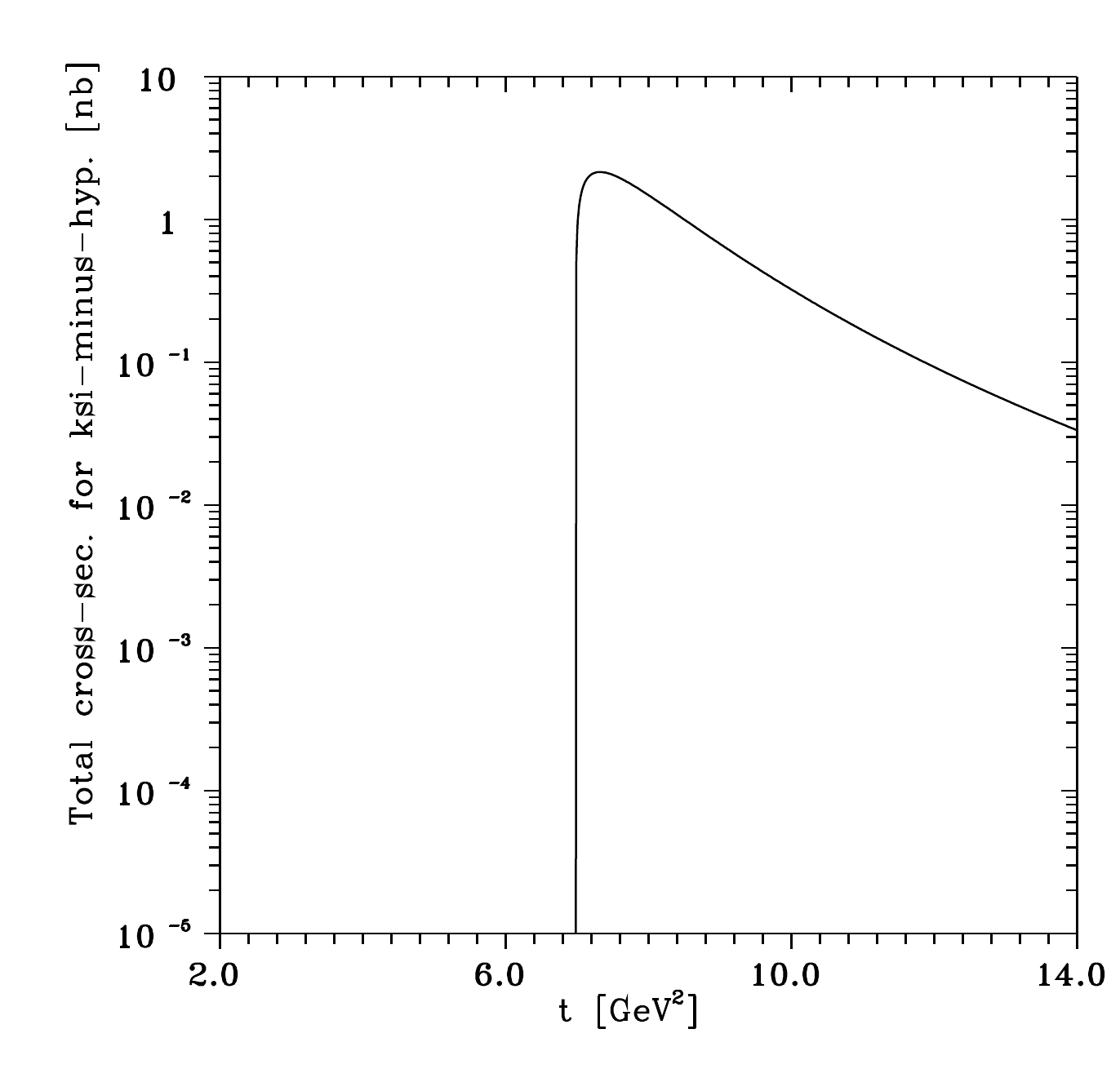}
\includegraphics[scale=.47]{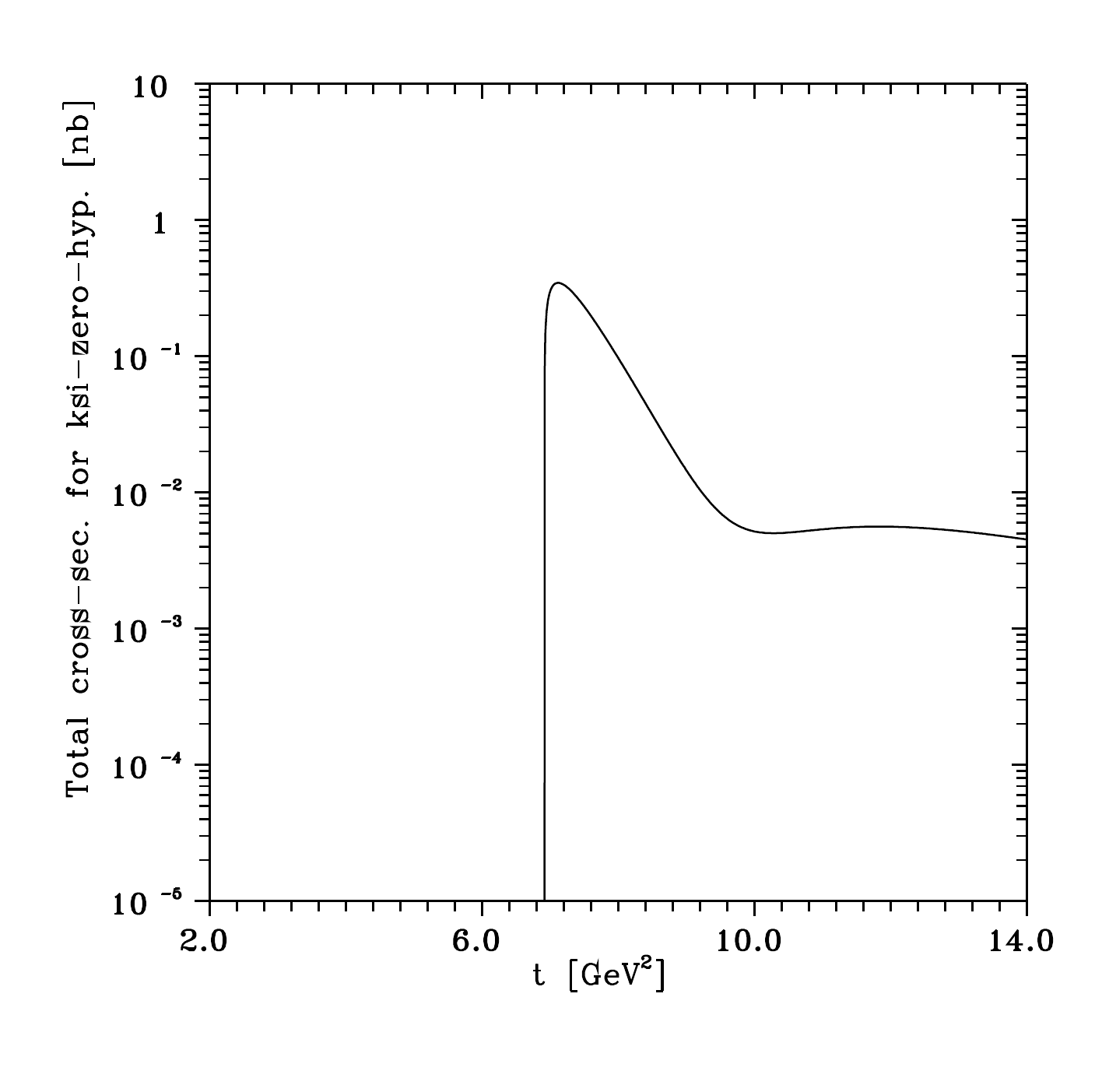}
\end{center}
  \caption{$e^+e^-$-annihilation cross-sections to $\Xi^-$ and $\Xi^0$ hyperons.}
  \label{fig27}
\end{figure}

\phantom{
Then the solutions of the system of algebraic eqs.~(\ref{su3nuc})
according to $f^D_{1,2}$, $f^F_{1,2}$, $f^S_{1,2}$ with numerical
values of $f_{\rho NN}$, $f_{\omega NN}$ and $f_{\phi NN}$, by
means of the expressions (\ref{su3lam})-(\ref{su3xi}), enable to
predict all free vector-meson-hyperon coupling constant ratios in
the EM FFs of hyperons $[\Lambda^0]$,
$[\Sigma^+,\Sigma^0,\Sigma^-]$ and $[\Xi^0,\Xi^-]$ and, as a
result, also behaviours of the total cross-sections
(\ref{stothyp}) (see Figs.~\ref{fig25}-\ref{fig27}).}

\setcounter{equation}{0} \setcounter{figure}{0} \setcounter{table}{0}\newpage
     \section{Deuteron electromagnetic structure}\label{V}

   The deuteron is the simplest object of all the existing nuclei and its EM
structure provides an illustration of continuity between nuclear
and particle physics at the microscopic level. The latter is
completely described by three scalar functions {\bf\ref{II1}} to
be normalized as follows
\begin{equation}
G_{Cd}(0)=1; \hspace{1cm} G_{Md}(0)=\frac{m_d}{m_p}\mu_d;
\hspace{1cm} G_{Qd}(0)=m^2_d Q \label{chap479}
\end{equation}
where  $\mu_d=0.857406\pm 0.000001 [\mu_N]$ is the magnetic moment
and $Q=0.2860\pm 0.0015 [fm^2]$ the quadrupole moment of the
deuteron $d$.

   Experimentally, the EM structure of the deuteron is practically
measured in the elastic scattering of electrons on deuterons
described by the cross-section (\ref{sigdeutlab}), from which
exploiting the Rosenbluth methods one obtains the data on the
elastic deuteron structure functions (\ref{deutelsf}) to be
compiled in \cite{Gilman02}.

   In principle,  by  analogy with mesons and baryons, there is
another source of experimental information on the EM structure of
the deuteron in the time-like region for $t>4m^2_d$ by means of
the $e^+e^-$ annihilation into deuteron-antideuteron pairs, which
is described by the following cross-section
\begin{eqnarray}
\sigma_{tot}(e^+e^- \to d\bar{d})  =
\frac{\pi\alpha^2\beta^3_d}{3t}
\left\{ 3|G_{Cd}(t)|^2+\frac{t}{m^2_d}\left[|G_{Md}(t)|^2+\frac{1}{6m^2_d}|G_{Qd}(t)|^2 \right]\right\}, \nonumber \\
\label{chap482}
\end{eqnarray}
where $\beta_d=[1-{4m^2_d\over t}]^{1/2}$ is a velocity of
produced deuterons or antideuterons. However, there are no
suitable $e^+e^-$ colliding beam accelerators giving such an
experimental information up to now.

   A lot of work has been done in theoretical attempts, like the
traditional nonrelativistic impulse approximation
\cite{Muzaf83,Gilman02}, sometimes also augmented by
meson-exchange currents and isobar contributions \cite{Khanna90},
as well as the relativistic impulse approximation
\cite{Frankf81,Gilman02}, the hybrid quark model \cite{Kissling87}
and the Skyrme model \cite{Carlson}, to describe the existing data
on $A(t)$ and $B(t)$ in the space-like region. But in the
framework of these models there is no concept of the time-like
region behavior of the nuclear EM FFs at all.

   In this section we demonstrate \cite{Adamuscin10} that the universal approach of the
$U\&A$ model of the EM structure of strongly interacting
particles, elaborated in the Chapter II.,  is suitable also for a
description of existing data on the elastic deuteron EM structure
functions $A(t)$, $B(t)$ in the space-like region and
simultaneously also for a prediction of the deuteron EM FFs
$G_{Cd}(t),G_{Md}(t)$ and $G_{Qd}(t)$ in the time-like region for
the first time.

\medskip

     \subsection{Unitary and analytic model of deuteron electromagnetic
     form factors} \label{V1}

\medskip

   Any model, describing the EM structure of deuteron, should satisfy
general properties of the deuteron FFs. As deuteron is a spin 1
particle, the ratios  of the deuteron EM FFs at large space-like
and time-like momentum transferred squared hold \cite{BrodHill}
the relations
\begin{equation}
G_{Cd}(t):G_{Md}(t):G_{Qd}(t)=(1-\tfrac{2}{3}\eta):2:-1,
\label{UAD:ratio}
\end{equation}
which together with the QCD \cite{Brodsky83}, \cite{Carlson87}
predictions for the asymptotic behavior of the deuteron EM
structure function $A(t)$
\begin{equation}
\left[ A(t)\right] ^{1/2}\sim t_{\left\vert t\right\vert
\rightarrow -\infty }^{-5},\label{deutasym}
\end{equation}
imply the asymptotic behaviors of all deuteron EM FFs in both
space-like and time-like regions  to be
\begin{eqnarray}\label{deutffasympt}
G_{Cd}(t) &\sim &t_{\left\vert t\right\vert \rightarrow -\infty }^{-5}  \nonumber \\
G_{Md}(t) &\sim &t_{\left\vert t\right\vert \rightarrow -\infty }^{-6}   \\
G_{Qd}(t) &\sim &t_{\left\vert t\right\vert \rightarrow -\infty
}^{-6}.\nonumber
\end{eqnarray}

   In the time-like region above the lowest branch point the deuteron
EM FFs are complex functions of t- variable, i.e. they have
imaginary parts different from zero to be given explicitly by the
unitarity conditions. In other words the deuteron EM FFs are
analytic functions in the whole complex t-plane besides cuts on
the positive real axis starting from the lowest threshold (see
paragraph \ref{II6})
\begin{equation}
t_0=4m_p^2 -\frac{(m_d^2-m_p^2-m_n^2)^2}{m_n^2}=1.7298 m_{\pi}^2
\label{t0}
\end{equation}
to be anomalous \cite{Dubnicka91} and generated by the Feynman
diagram in Fig.~\ref{fig28}. Its position is calculated from the dual
diagram presented in Fig.~\ref{fig29} by methods of elementary geometry.

\begin{figure}[tb]
    \centering
        \includegraphics[scale=.6]{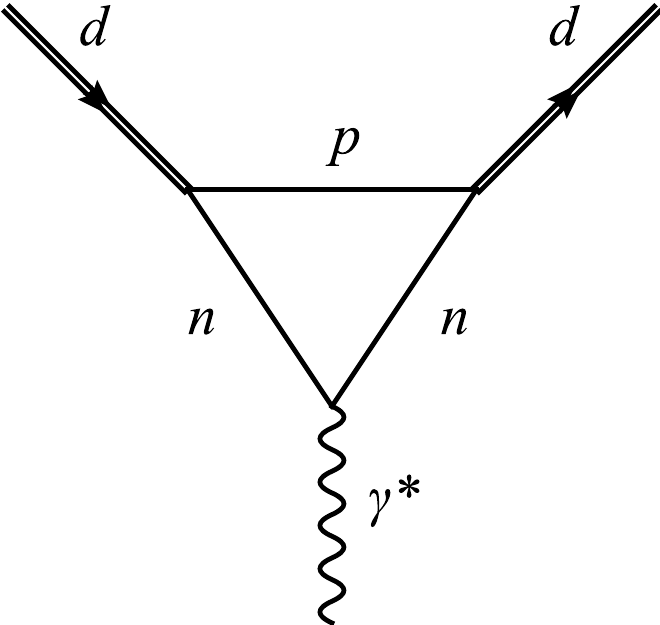}
    \caption{\small{The Feynman diagram of the deuteron
    EM vertex generating the lowest anomalous threshold
    in deuteron electromagnetic form factors.}}
    \label{fig28}
\end{figure}

\begin{figure}[tb]
    \centering
        \includegraphics[scale=.7]{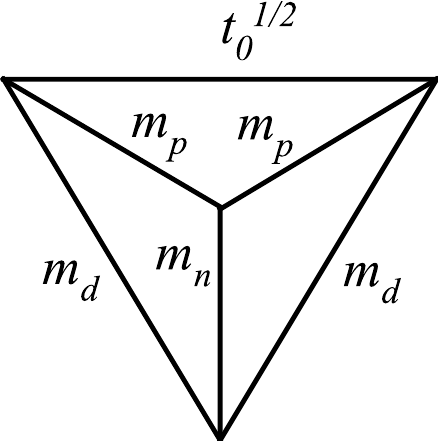}
    \caption{\small{The dual diagram from which by means
    of the methods of elementary geometry the position
    of the lowest anomalous threshold is calculated.}}
    \label{fig29}
\end{figure}

   In order to obtain $U\&A$ parametrization  of deuteron EM FFs
with the asymptotic behavior (\ref{deutffasympt}) one needs to utilize
the modified VMD parametrization with correct asymptotic behavior
and normalization
\begin{eqnarray}
F_h(t)& =&F_0\frac{\prod^m_{j=1}m_j^2}{\prod^m_{j=1}(m_j^2-t)}+ \label{ro35} \\
&+& \sum_{k=m+1}^n\biggl\{\sum_{i=1}^m\frac{m_k^2}{(m_k^2-t)}
\frac{\prod_{\substack{j=1\\j\neq
i}}^mm_j^2}{\prod_{\substack{j=1\\j\neq
i}}^m(m_j^2-t)}\frac{\prod^m_{\substack{j=1\\j\neq
i}}(m_j^2-m_k^2)}{\prod^m_{\substack{j=1\\j\neq
i}}(m_j^2-m_i^2)}-\frac{\prod_{j=1}^mm_j^2}{\prod_{j=1}^m(m_j^2-t)}\biggr
\}a_k \nonumber
\end{eqnarray}
discussed in the paragraph \ref{II5}, where $F(0)$ is normalization
of the FF, $m_x$'s are masses of the vector mesons, $a_x$'s are
ratios of coupling constants ($f_{xdd}/f_x$), $n$ is the number of
saturated vector mesons and $F_{h}(t)\big|_{|t|\rightarrow
{\infty}}=t^{-m}$ is demanded asymptotic behavior of the examined
FF.

   In what follows we apply the procedure of the unitarization
(see paragraph \ref{II7}) of every pole term in all three deuteron EM
FFs in the form of the modified VMD (\ref{ro35}) parametrization,
where $t_0$ is the lowest threshold (\ref{t0}) and $t_{in}$ is a
free parameter independent for every deuteron EM FF to be
simulating contributions of all higher branch points effectively.

   As a result one gets the following $U\&A$ parametrization of
the FF
\begin{eqnarray}
F_{h}(t) &=&\left( \frac{1-X(t)^{2}}{1-X_{N}^{2}}\right)
^{2m}\Bigg[\left(
F_{0}-\sum_{k=m+1}^{n}a_{k}\right) \prod_{v=1}^{m}LH(X_{v})+  \notag \\
&&+\sum_{k=m+1}^{n}LH(X_{k})\left[ \sum_{i=1}^{m}\prod_{v=1,v\neq
i}^{m}\left\{ LH(X_{v})\frac{C(X_{v})-C(X_{k})}{C(X_{v})-C(X_{i})}\right\} %
\right] a_{k}\Bigg],  \label{general UA}
\end{eqnarray}
where $LH(X_{v})$ equals
\begin{eqnarray*}
&&\mbox{ in case of:   } m_{x}^{2}-\Gamma
_{x}^{2}/4<t_{in}\\
&&LH(X_{x}) =\frac{(X_{N}-X_{x})(X_{N}-X_{x}^{\ast
})(X_{N}-1/X_{x})(X_{N}-1/X_{x}^{\ast
})}{(X(t)-X_{x})(X(t)-X_{x}^{\ast
})(X(t)-1/X_{x})(X(t)-1/X_{x}^{\ast })} \\
&&\mbox{ in case of:   } m_{x}^{2}-\Gamma
_{x}^{2}/4>t_{in}\\
&&LH(X_{x}) =\frac{(X_{N}-X_{x})(X_{N}-X_{x}^{\ast
})(X_{N}+X_{x})(X_{N}+X_{x}^{\ast
})}{(X(t)-X_{x})(X(t)-X_{x}^{\ast })(X(t)+X_{x})(X(t)+X_{x}^{\ast
})}
\end{eqnarray*}
and $C(X_{x})$ equals
\begin{eqnarray*}
&&\mbox{ in case of:   } m_{x}^{2}-\Gamma
_{x}^{2}/4<t_{in}\\
&&C(X_{x}) =\frac{(X_{N}-X_{x})(X_{N}-X_{x}^{\ast})
(X_{N}-1/X_{x})(X_{N}-1/X_{x}^{\ast })}{-(X_{x}-1/X_{x})(X_{x}^{\ast} -1/X_{x}^{\ast })} \\
&&\mbox{ in case of:   } m_{x}^{2}-\Gamma
_{x}^{2}/4>t_{in}\\
&&C(X_{x}) =\frac{(X_{N}-X_{x})(X_{N}-X_{x}^{\ast})
(X_{N}+X_{x})(X_{N}+X_{x}^{\ast })}{-(X_{x}-1/X_{x})(X_{x}^{\ast}
-1/X_{x}^{\ast })}.
\end{eqnarray*}

   The minimal number $n$ of isoscalar vector mesons needed to describe deuteron
EM FFs depends on the asymptotic behavior, as $n\geq m$ should be
valid for every deuteron EM FF. However, it also depends on the
existence of a node in the FF behavior.  It can be shown, that
special solution of Eq. (\ref{general UA}) for $n=m$
\begin{equation}
F_{h}(t)=F_{0}\left( \frac{1-X(t)^{2}}{1-X_{N}^{2}}\right)
^{2m}\prod_{v=1}^{m}LH(X_{v}),  \label{special UA}
\end{equation}
is nonzero for any real value of $t$. As there are nodes in the
behaviors of deuteron EM FFs $G_C(t)$ and $G_M(t)$, according to
(\ref{deutffasympt}) the minimal number of vector mesons has to be
7(=6+1). Moreover, the positions \cite{Garcon01} of the nodes
$t_{0C}\simeq - 0.7$\, GeV$^2$ and $t_{0M}\simeq - 2.0$ \,GeV$^2$ give
us additional constraints for the fitting procedure.

   The masses $m_x$ and corresponding widths $\Gamma_x$ of the first
5 vector mesons can be fixed to the world average values of the
'light' vector mesons $\omega, \phi, \omega', \omega'', \phi'$.
Another ones can be fixed to the parameters of the charmed vector
meson $J/\Psi$ and $m_x$,$\Gamma_x$ of the last vector meson will
remain to be free, as we expect it to correspond to yet unknown
'light' vector meson resonance $\phi''$ following from the SU(3)
classification to be valid also for excited states of vector
mesons.

   Finally, this procedure leads us to the $U\&A$ parametrizations
of the deuteron EM FFs
\begin{eqnarray}
G_{Cd}(t) &=&\left( \frac{1-W(t)^{2}}{1-W_{N}^{2}}\right) ^{10}
\Bigg[ (1-a_{C:J/\Psi} - a_{C:x}) \prod_{
v=\omega ,\phi ,\omega^{\prime}, \atop \omega ^{\prime \prime }, \phi ^{\prime }}
LH(W_{v})\nonumber \\
&+& \mathcal{LH}(W_{J/\Psi})a_{C:J/\Psi} +
\mathcal{LH}(W_{x})a_{C:x}\Bigg]\\
G_{Md}(t) &=&\left( \frac{1-V(t)^{2}}{1-V_{N}^{2}} \right) ^{12}
\left[ \left(\frac{m_{d}}{m_{p}}\mu _{d}-a_{M:x}\right)
\prod_{v=\omega ,\phi ,\omega ^{\prime
}, \atop \omega ^{\prime \prime },\phi ^{\prime }, J/\Psi} LH(V_{v}) + \mathcal{LH}'(V_x) a_{M:x} \right]\nonumber\\
G_{Qd}(t) &=&\left( \frac{1-U(t)^{2}}{1-U_{N}^{2}}\right)
^{12}\left[ \left(m_{d}^{2}Q-a_{Q:x}\right) \prod_{v=\omega ,\phi
,\omega ^{\prime }, \atop \omega ^{\prime \prime },\phi ^{\prime },
J/\Psi} LH(U_{v}) + \mathcal{LH}'(U_x) a_{Q:x} \right],\nonumber
\end{eqnarray}
where
\begin{eqnarray}
\mathcal{LH}(X_w)&=&LH(X_w)\sum_{i=\omega ,\phi ,\omega ^{\prime
}, \atop \omega ^{\prime \prime },\phi ^{\prime }}\prod_{v=\omega ,\phi
,\omega ^{\prime }, \atop \omega ^{\prime \prime },\phi ^{\prime } ,v\neq
i}\left\{ LH(X_{v})\frac{C(X_{v})-C(X_w)} {C(X_{v})-C(X_{i})}\right\} \nonumber \\
\mathcal{LH}'(X_w)&=&LH(X_w)\sum_{i=\omega ,\phi ,\omega ^{\prime
}, \atop \omega ^{\prime \prime },\phi ^{\prime },J/\Psi}\prod_{v=\omega
,\phi ,\omega ^{\prime },\omega ^{\prime \prime }, \atop \phi ^{\prime
},J/\Psi ,v\neq i}\left\{ LH(X_{v})\frac{C(X_{v})-C(X_w)}
{C(X_{v})-C(X_{i})}\right\}
\end{eqnarray}
with two conditions for ratios of coupling constants arising from
the existence of nodes
\begin{eqnarray}
a_{C:J/\Psi}&=& \frac{(1-a_{C:x}) \prod_{v=\omega ,\phi ,\omega
^{\prime },\omega ^{\prime \prime },\phi ^{\prime
}}LH(W_{v}(t_{0C}))+
\mathcal{LH}(W_x(t_{0C}))a_{C:x}}{\prod_{v=\omega ,\phi ,\omega
^{\prime
},\omega ^{\prime \prime },\phi ^{\prime }}LH(W_{v}(t_{0C}))-\mathcal{LH}(W_{J/\Psi}(t_{0C}))}\nonumber \\
a_{M:x}&=& \frac{\frac{m_{d}}{m_{p}}\mu _{d} \prod_{v=\omega ,\phi
,\omega ^{\prime },\omega ^{\prime \prime },\phi ^{\prime },
J/\Psi} LH(V_{v}(t_{0M}))} {\prod_{v=\omega ,\phi ,\omega ^{\prime
},\omega ^{\prime \prime },\phi ^{\prime }, J/\Psi}^{m}LH(V_{v}(t_{0M}))-\mathcal{LH}(V_{x}(t_{0M}))}. \label{ratioCON}
\end{eqnarray}

The function $X(t)$ takes different form for each deuteron FF,
$W(t)$ for $G_{C}(t)$, $V(t)$ for $G_{M}(t)$ and $U(t)$ for
$G_{Q}(t)$
\begin{eqnarray*}
W(t)&=&i\frac{\sqrt{\left( \frac{t_{inC}-t_{0}}{t_{0}}\right)
^{1/2}+\left(
\frac{t-t_{0}}{t_{0}}\right) ^{1/2}}-\sqrt{\left( \frac{t_{inC}-t_{0}}{t_{0}}
\right) ^{1/2}-\left( \frac{t-t_{0}}{t_{0}}\right)
^{1/2}}}{\sqrt{\left(
\frac{t_{inC}-t_{0}}{t_{0}}\right) ^{1/2}+\left( \frac{t-t_{0}}{t_{0}}
\right) ^{1/2}}+\sqrt{\left( \frac{t_{inC}-t_{0}}{t_{0}}\right)
^{1/2}-\left( \frac{t-t_{0}}{t_{0}}\right) ^{1/2}}} \\
V(t) &=&i\frac{\sqrt{\left( \frac{t_{inM}-t_{0}}{t_{0}}\right)
^{1/2}+\left(
\frac{t-t_{0}}{t_{0}}\right) ^{1/2}}-\sqrt{\left( \frac{t_{inM}-t_{0}}{t_{0}}
\right) ^{1/2}-\left( \frac{t-t_{0}}{t_{0}}\right)
^{1/2}}}{\sqrt{\left(
\frac{t_{inM}-t_{0}}{t_{0}}\right) ^{1/2}+\left( \frac{t-t_{0}}{t_{0}}
\right) ^{1/2}}+\sqrt{\left( \frac{t_{inM}-t_{0}}{t_{0}}\right)
^{1/2}-\left( \frac{t-t_{0}}{t_{0}}\right) ^{1/2}}} \\
U(t) &=&i\frac{\sqrt{\left( \frac{t_{inQ}-t_{0}}{t_{0}}\right)
^{1/2}+\left(
\frac{t-t_{0}}{t_{0}}\right) ^{1/2}}-\sqrt{\left( \frac{t_{inQ}-t_{0}}{t_{0}}
\right) ^{1/2}-\left( \frac{t-t_{0}}{t_{0}}\right)
^{1/2}}}{\sqrt{\left(
\frac{t_{inQ}-t_{0}}{t_{0}}\right) ^{1/2}+\left( \frac{t-t_{0}}{t_{0}}
\right) ^{1/2}}+\sqrt{\left( \frac{t_{inQ}-t_{0}}{t_{0}}\right)
^{1/2}-\left( \frac{t-t_{0}}{t_{0}}\right) ^{1/2}}}.
\end{eqnarray*}

\medskip

     \subsection{Analysis of data on deuteron electromagnetic structure}\label{V2}

\medskip

   In order to have experimental information on all three deuteron
FFs, one needs another observable, usually the component $t_{20}$
of the tensor polarization \cite{Gilman02} of the recoil deuteron,
which contains the following combination of all three FFs
\begin{eqnarray}
 t_{20}&=&-\frac{1}{\sqrt{2}R(t)} \left[\frac{8}{3}\eta G_{Cd} G_{Qd} \right.+ \label{t20}
\\
&&\left. \frac{8}{9}\eta^2 G_{Qd}^2+\frac{1}{3}\left(1+2(1+\eta)\tan^2 \tfrac{\theta}{2}\right)G_{Md}^2\right],\nonumber
\end{eqnarray}
where $R(t)=A(t)+B(t)\tan^2 \tfrac{\theta}{2}$.

   There are experimental data on the structure function $A(t)$
\cite{Grosset66}-\cite{Abbot99} in the range
$-5.9536$  $GeV^2$ $<t<-0.0016 \, GeV^2$, on the structure function $B(t)$
\cite{Buckanan65}-\cite{Bosted90} in the range $ -2.7556$ $GeV^2<t<
-0.01 \, GeV^2$ and on the $t_{20}$
\cite{Schulze84}-\cite{Nikolenko01} in the range $-1.7161
\, GeV^2<t<-0.0289\, GeV^2$.

   The constructed $U\&A$ model of deuteron EM FFs depends on physical parameters like
$m_{v},\Gamma _{v}$ for $v=\omega ,\phi ,\omega ^{\prime},\omega
^{\prime \prime },\phi ^{\prime },J/\Psi,x$,

\begin{figure}[tb]
\includegraphics[scale=.35]{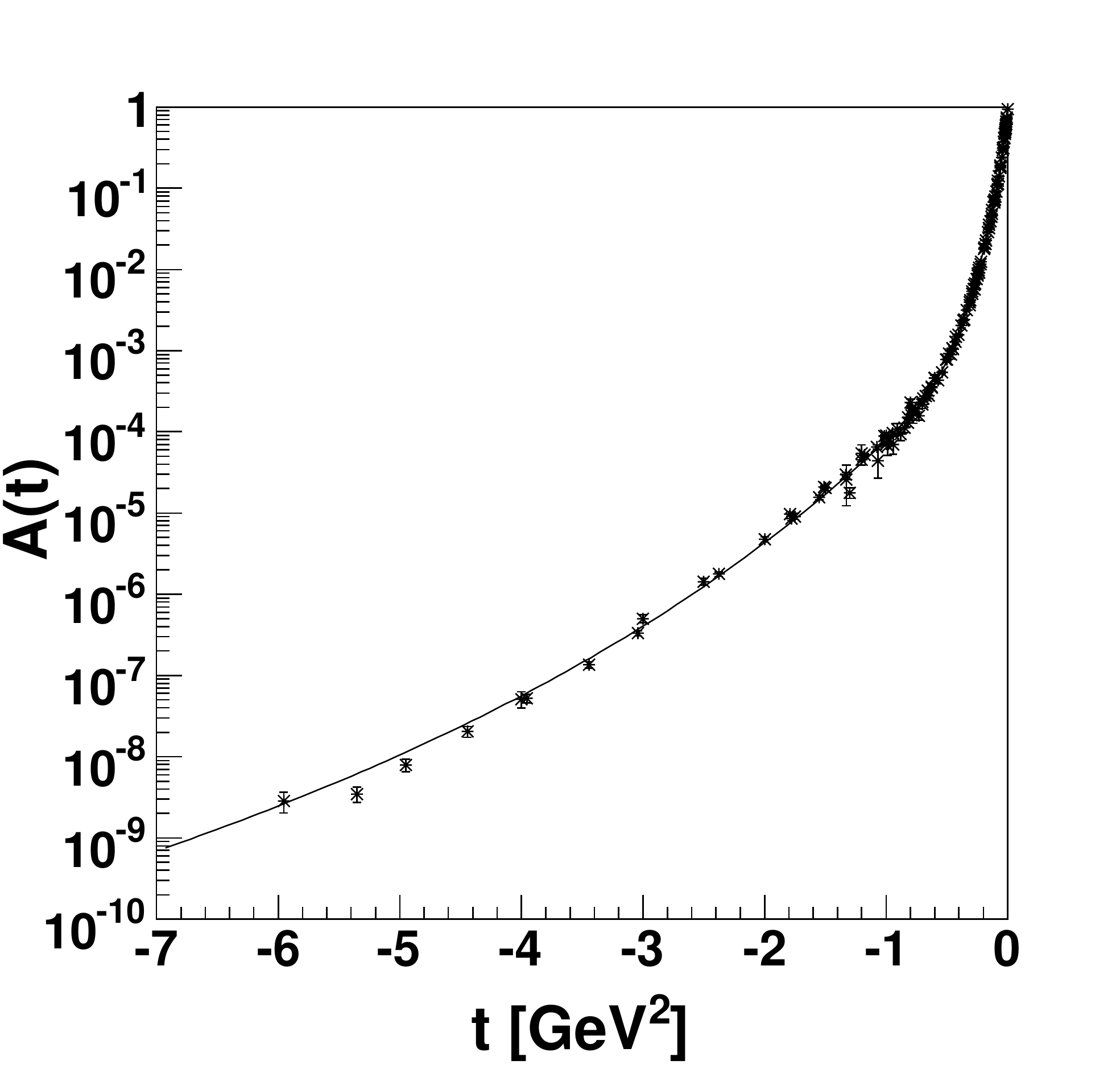}
   \hspace{.4cm}
\includegraphics[scale=.35]{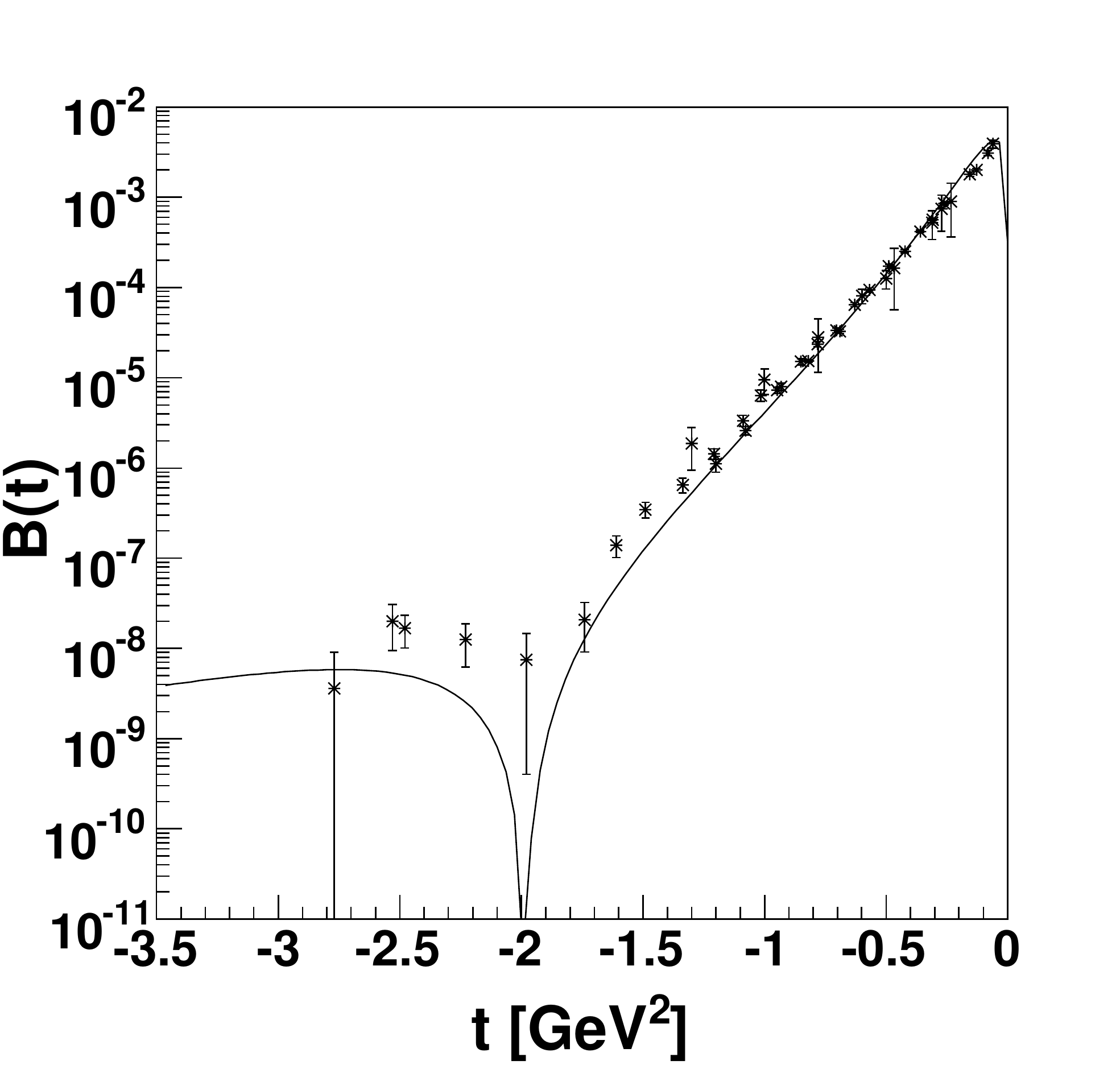}
  \caption{\small{Behavior of the deuteron elastic structure functions $A(t)$ and $B(t)$ obtained by a simultaneous
    comparison of our  $U\&A$ model of the deuteron EM structure  to existing  experimental data on $A(t)$,
    $B(t)$ and polarization observable $t_{20}$}}
  \label{fig30}
\end{figure}

\begin{figure}[tb]
    \centering
        \includegraphics[scale=.35]{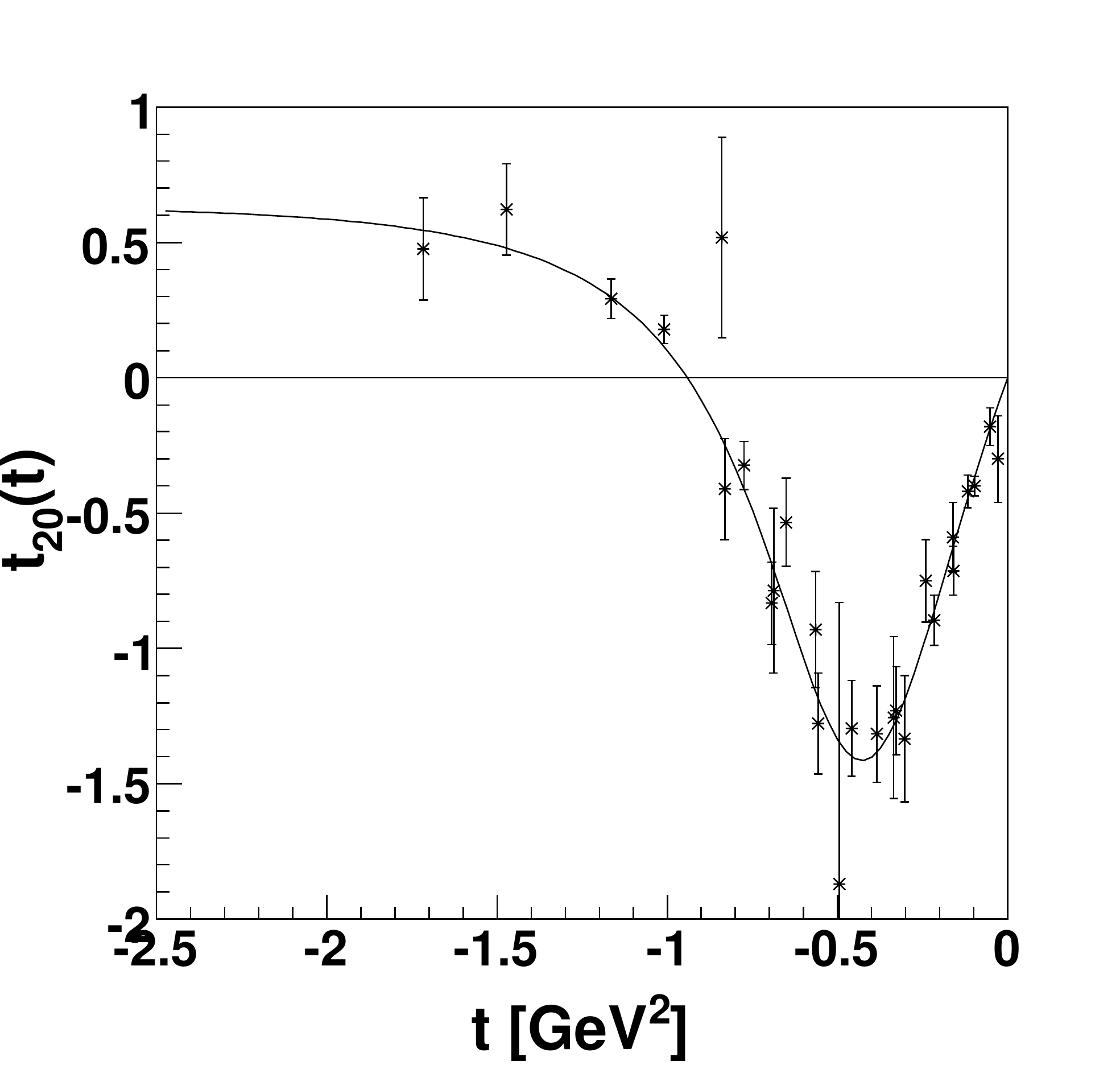}
    \caption{\small{Behavior of the deuteron tensor polarization observable ${t}_{20}(t)$
    obtained by a simultaneous comparison of our  $U\&A$ model of the deuteron EM structure
     to existing  experimental data on $A(t)$, $B(t)$ and $t_{20}$}}
    \label{fig31}
\end{figure}
\begin{figure}[tb!]
\includegraphics[scale=.35]{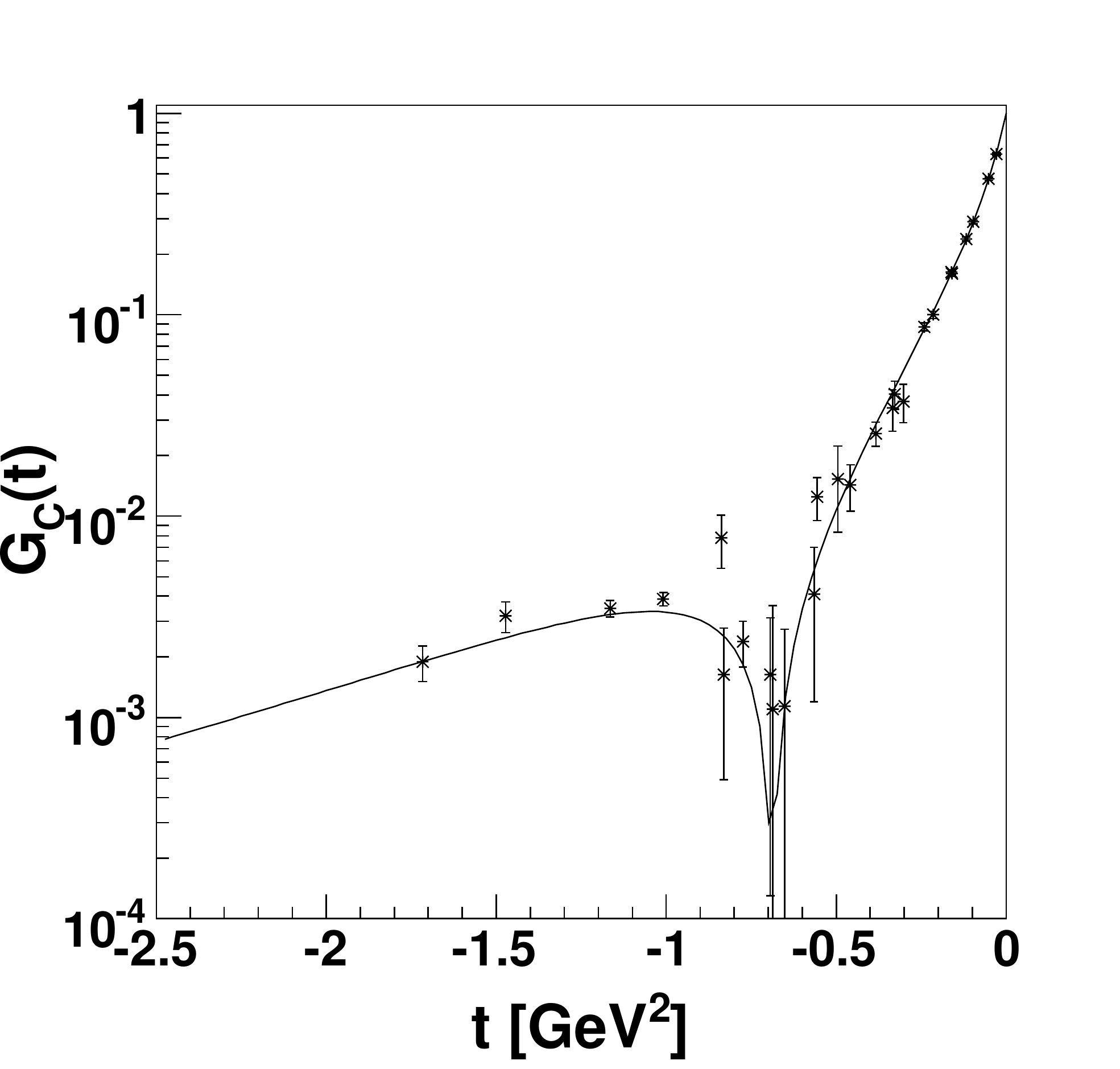}
   \hspace{.4cm}
\includegraphics[scale=.35]{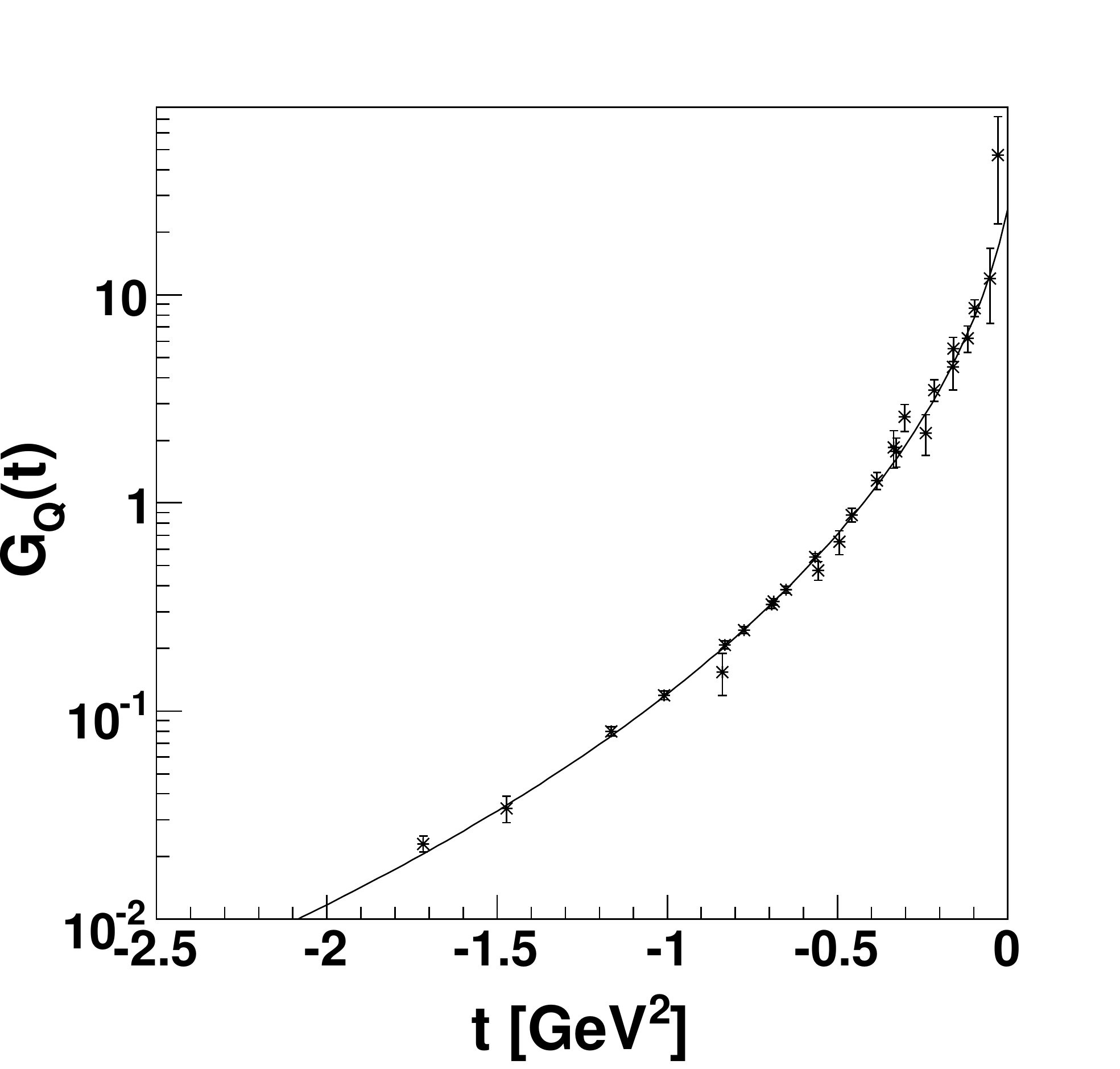}
  \caption{\small{The predicted behavior of the space-like deuteron charge form factor
    $G_C(t)$ and quadrupole form factor $G_Q(t)$
    by our U\&A model of the deuteron electromagnetic structure and their comparison to existing experimental data.}}
    \label{fig:UAdGC}
\end{figure}

on the effective thresholds $t_{inC},t_{inM}, t_{inQ}$ and on
unknown ratios $a_{C:x}$ and $a_{Q:x}$.

\begin{figure}[tb]
    \centering
        \includegraphics[scale=0.5]{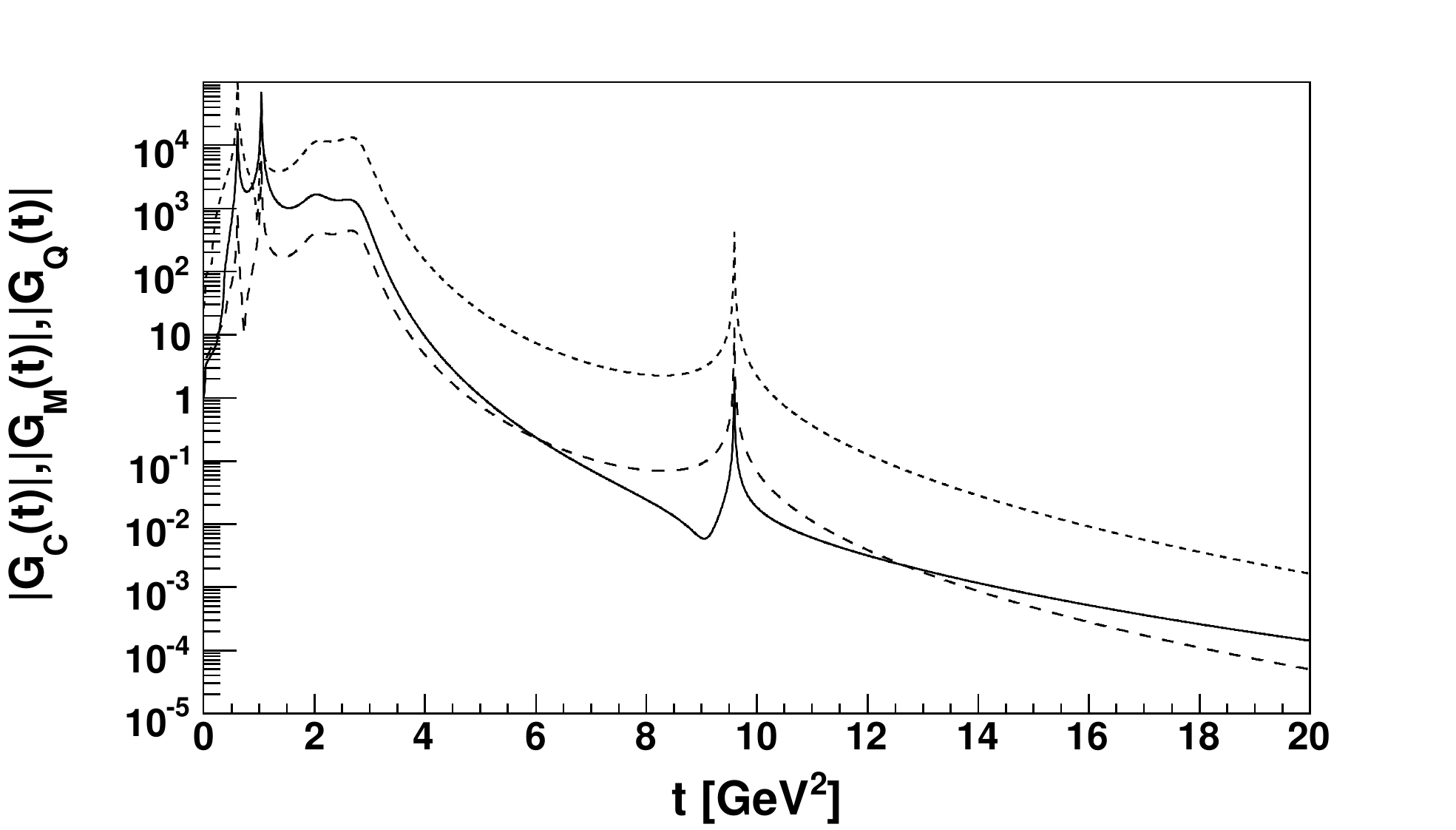}
    \caption{\small{The predicted time-like behavior of the absolute values of the deuteron
    complex EM form factors $G_C(t),G_M(t),G_Q(t)$  by our U\&A model of the deuteron
    electromagnetic structure. Full line corresponds to the charge FF $|G_C(t)|$, dashed
    line corresponds to the magnetic FF $|G_M(t)|$ and dotted line corresponds to the
    quadrupole FF $|G_Q(t)|$.}}
    \label{fig:UAffTL}
\end{figure}
\begin{figure}[tbh!]
    \centering
        \includegraphics[scale=0.5]{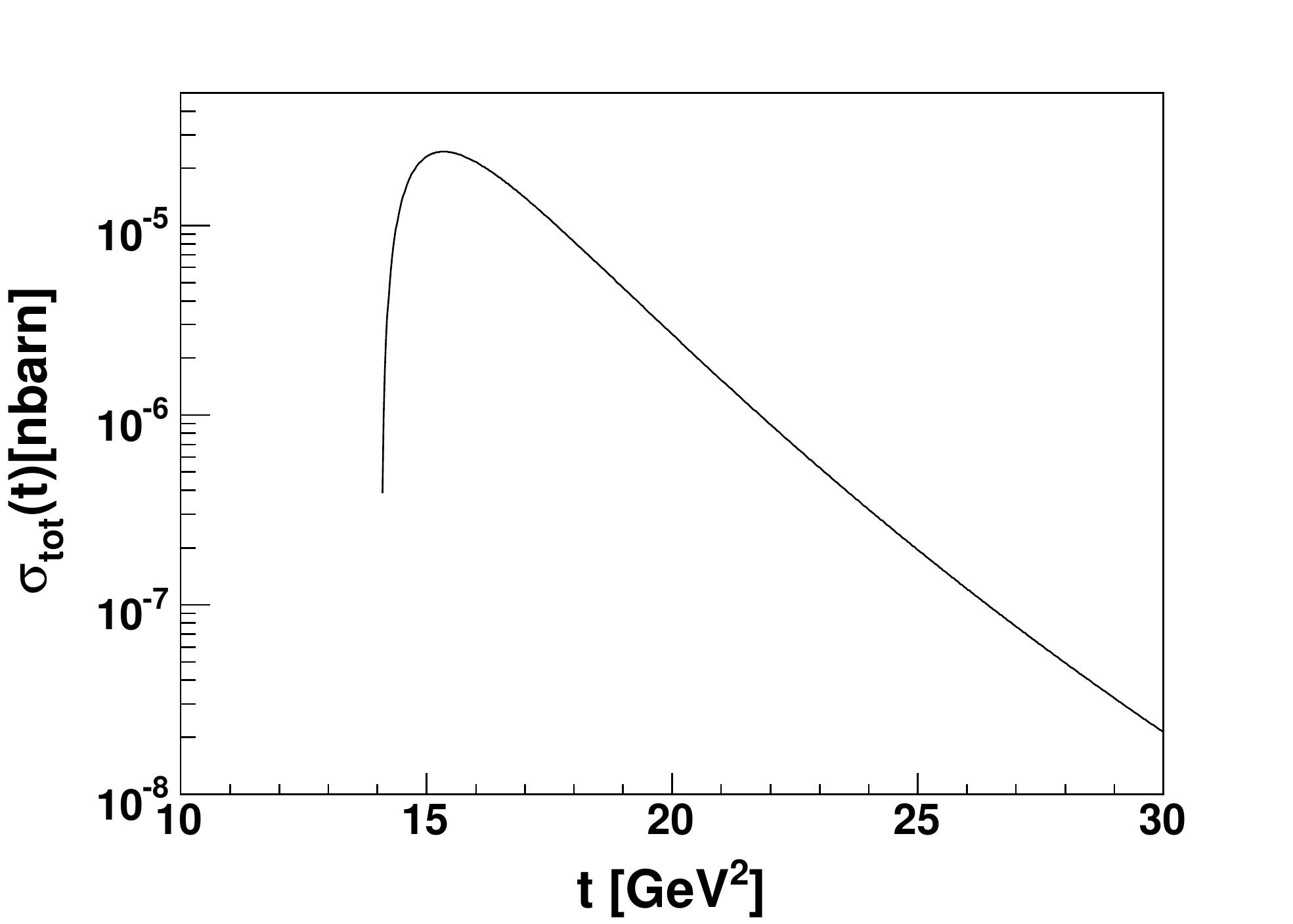}
    \caption{\small{The total cross section behavior  of the $e^- e^+ \rightarrow d \bar{d}$
    annihilation process to be  estimated by our U\&A model of the deuteron EM structure.}}
    \label{fig:UAdSIGtot}
\end{figure}

 Naturally, we can fix masses
$m_{v}$ and widths $\Gamma_{v}$ of all known vector mesons
($\omega ,\phi ,\omega ^{\prime },\omega ^{\prime\prime },\phi
^{\prime },$ $J/\Psi$), what means, that there will remain seven free
parameters, which are numerically evaluated in the optimal
description of all available experimental data on the deuteron
elastic structure functions $A(t)$, $B(t)$ and the polarization
observable ${t}_{20}(t)$. The results were obtained by using CERN
program ROOT \cite{rootcern} and corresponding behaviors are shown
on Figs.~\ref{fig30}, \ref{fig31} and \ref{fig:UAdGC}, together with data. The values of free
parameters are given in Table~\ref{uadeuteronTAB}.
\begin{table}[!b]
\tabcolsep=2.5pt
\caption{Fitted parameters of the U\&A model of the deuteron EM structure.}
\medskip
    \centering
        \small
        \begin{tabular}{|c|c|c|c|c|c|c|}
        \hline
        ${m_x}{\mathrm{MeV}}${\phantom{$1^{1^1}$}}&${\Gamma_x}{\mathrm{MeV}}$&${t_{inC}}
        {\mathrm{GeV}^2}$&${t_{inM}}{\mathrm{GeV}^2}$&${t_{inQ}}{\mathrm{GeV}^2}$&$a_{C:x}$&$a_{Q:x}$\\
        \hline
        $504.9\pm0.1$&$677.6\pm0.2$&$18.2\pm0.1$&$20.2\pm0.2$&$7.8\pm0.1$&$3.43\pm0.01$&$28.14\pm0.03$\\
        \hline
        \end{tabular}

    \label{uadeuteronTAB}
\end{table}

   As one can see, we have obtained a quite reasonable description of
the deuteron structure functions $A(t)$, $B(t)$ and polarization
observable ${t}_{20}(t)$, however with $\chi^2/n.d.f.=4.61$
indicating an inconsistency among independent sets of data.

   Moreover, the constructed model with the same values of free
parameters can be used for the description of experimental data on
the additional deuteron EM FFs $G_{Cd}(t)$ and $G_{Qd}(t)$, as
well as for estimation of all three deuteron EM FFs time-like
region behavior ( see Fig.~\ref{fig:UAffTL}). They allow us to estimate the total cross section
of the electron-positron annihilation into deuteron-antideuteron
(\ref{chap482}), which is planned to be measured on BES3 in Peking
for the first time.

   Its predicted behavior is given in Fig.~\ref{fig:UAdSIGtot}.

\medskip

     \subsection{Study of two-photon contribution relevance into
     deuteron electromagnetic structure by impulse approximation}\label{V3}

\medskip

   As a result of a consideration of one-photon exchange
approximation in elastic scattering of electrons on any hadron
with nonzero spin, one obtains the same expression in the form for
the differential cross-section in the laboratory system to be
expressed through the elastic structure functions $A(t)$ and
$B(t)$. For comparison see (\ref{sec213}) and (\ref{sigdeutlab}).

   If also two-photon contributions are considered, no more simple
form of the differential cross-section of elastic scattering of
electrons on hadrons with nonzero spin is obtained.

   In the previous chapter we have analysed the new JLab
proton polarization data \cite{Jones00,Gayon02,Punjabi05} in the
framework of the ten-resonance $U\&A$ model \cite{DDW} of nucleon
EM structure. The parameters, in comparison with those in
\cite{DDW} have been found to be changed very little. But a
reasonable description of the new JLab data was achieved, whereby
almost nothing has been changed in the description of $G_{Mp}(t)$,
$G_{En}(t)$, $G_{Mn}(t)$ in both the space-like and time-like
regions and $|G_{Ep}(t)|$ in the time-like region. However, the
existence of the zero (see the full line in Fig.~\ref{fig20}), i.e. a
diffraction minimum in the space-like region of $|G_{Ep}(t)|$ at
$t=-Q^2\approx -13~ GeV^2$ is predicted by such ten-resonance
unitary and analytic model and so, $|G_{Ep}(t)|$ has no more
dipole behavior in the space-like region.

   Recently it was suggested \cite{Guichon03}-\cite{Afanasiev05} that the two-photon
corrections could be responsible for the found non-dipole behavior
of $|G_{Ep}(t)|$ in the space-like region. Then there is a natural
question arisen, if two-photon corrections play so important role
in the elastic electron-proton scattering, what about a size of
two-photon corrections in elastic electron-deuteron scattering.

   In this paragraph \cite{Adamuscin08} the non-relativistic impulse approximation
(NIA), which requires a knowledge only of the deuteron wave
functions and the nucleon EM FFs, is used to study the previous
question.

   As deuteron can be found in S- ($\approx 96\%$) and D-state ($\approx 4\%$), then
NN non-relativistic full wave function of the deuteron can be
written in terms of two scalar wave functions
\begin{eqnarray}
\nonumber \Psi _{abm}
&=&\sum_{l}\sum_{m_{s}}\frac{z_{l}(r)}{r}Y_{l,m-m_{s}}(\widehat{
\mathbf{r}})\chi _{ab}^{1m_{s}}\\
&&\left\langle l,1,m-m_{s},m_{s}|1,m\right\rangle \nonumber  \\
&=&\frac{u(r)}{r}Y_{0,0}(\widehat{\mathbf{r}})\chi_{ab}^{1m}+\label{nrel}
\\
&+&\frac{w(r)}{r} \sum_{m_{s}}Y_{2,m-m_{s}}(\widehat{
\mathbf{r}})\chi _{ab}^{1m_{s}}\left\langle
2,1,m-m_{s},m_{s}|1,m\right\rangle \nonumber,
\end{eqnarray}
where $\left\langle l,1,m-m_{s},m_{s}|1,m\right\rangle $ are
Clebsh-Gordan coefficients, $Y_{l,m_{l}}$ are spherical harmonics
normalized to unity on the unit sphere, $\chi _{ab}^{1m_{s}}$ is
the spin part of the wave function and $z_{0}=u$, $z_{2}=w$ are
reduced $S-$ and $D$-state wave functions, respectively.

The normalization condition
\begin{equation}
\int d^{3}r\Psi _{abm^{\prime }}^{\dag }\Psi _{abm}=\delta
_{m^{\prime }m} \nonumber
\end{equation}
implies normalization
\begin{equation}
\int_{0}^{\infty }dr\left[ u^{2}(r)+w^{2}(r)\right] =1,
\end{equation}
which could be understood as the probability of finding deuteron
in $S-$ or $D$-state. The $D$-state probability
\begin{equation}
P_{D}=\int_{0}^{\infty }drw^{2}(r) \nonumber
\end{equation}
is an interesting measurement of the strength of the tensor
component of the $NN$ force.

   The non-relativistic wave functions are calculated from the
Schr\"{o}dinger equation using potentials adjusted to fit NN
scattering data for laboratory energies from 0 to 350 MeV. In this
paper we will use one of the most common potentials called
\textit{Paris potential} \cite{Lacombe80}, which depends on the
minimal number of free parameters and it was among the first
potentials to be determined from such realistic fit. The $S-$ and
$D-$state wave functions determined from this model are presented
in Fig.~\ref{fig:wavefunctions}.

   The deuteron is a pure iso-scalar target, therefore within NIA its
FFs depend only on the iso-scalar nucleon form factors $G_{EN}^s$
and $G_{MN}^s$
\begin{eqnarray}
G_{EN}^s&=&G_{Ep}+G_{En} \nonumber \\
G_{MN}^s&=&G_{Mp}+G_{Mn} \label{nuciso}
\end{eqnarray}
in the following way
\begin{eqnarray}
G_{Cd} &=&G_{EN}^{s}D_{C}  \nonumber \\
G_{Md} &=&\frac{m_d}{2m_p}\left[G_{MN}^{s}D_{M}+G_{EN}^{s}D_{E}\right]\label{deutffisos}\\
G_{Qd} &=&G_{EN}^{s}D_{Q},  \nonumber
\end{eqnarray}
\begin{figure}[tb]
    \centering
        \scalebox{0.8}{\includegraphics{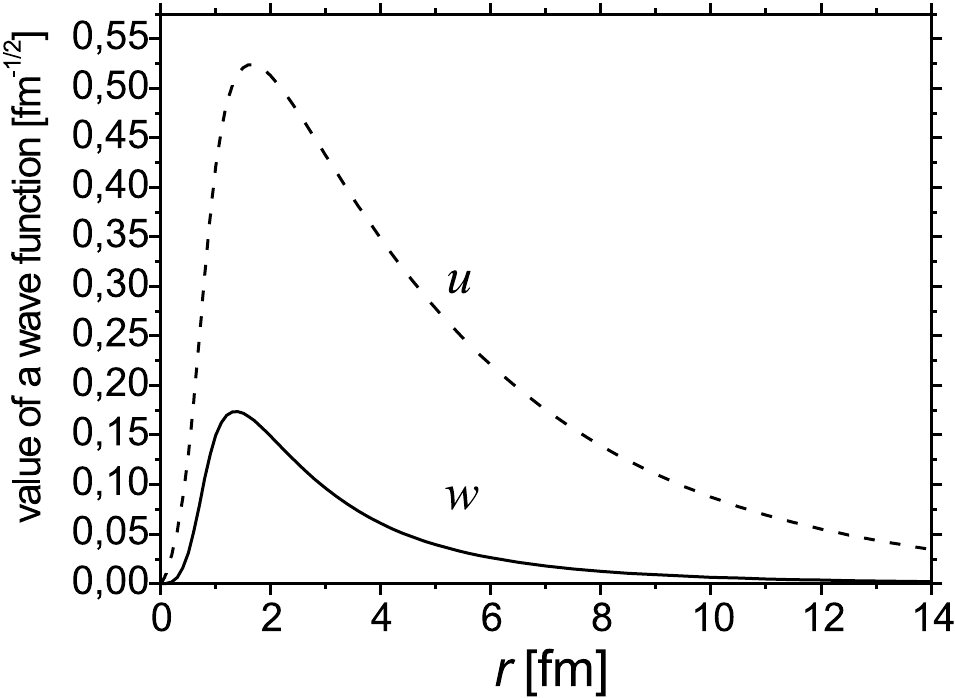}}
    \caption{\small{The $S$ and $D$ state deuteron wave functions ($u$,$w$)
    behaviors for Paris potential \cite{Lacombe80}.}}
    \label{fig:wavefunctions}
\end{figure}
where the body form factors $D_{C}$, $D_{M}$, $D_{E}$ and $D_{Q}$
are functions of the momentum transfer squared $t$. The
non-relativistic formulas for the body form factors $D$ involve
overlaps of the wave functions $u(r),w(r)$, weighted by spherical
Bessel functions
\begin{eqnarray}
D_{C}(q^{2}) &=&\int_{0}^{\infty }dr\left[
u^{2}(r)+w^{2}(r)\right]
j_{0}(\kappa )  \nonumber \\
D_{M}(q^{2}) &=&\int_{0}^{\infty }dr\left[
2u^{2}(r)-w^{2}(r)\right]
j_{0}(\kappa)+\label{bodyffs} \\
&+&\left[ \sqrt{2}u(r)w(r)+w^{2}(r)\right] j_{2}(\kappa)  \nonumber \\
D_{E}(q^{2}) &=&\frac{3}{2}\int_{0}^{\infty }drw^{2}(r)\left[
j_{0}(\kappa)+j_{2}(\kappa)\right]\nonumber \\
D_{Q}(q^{2}) &=&\frac{3}{\sqrt{2}\eta }\int_{0}^{\infty }dr
w(r)\left[ u(r)- \frac{w(r)}{\sqrt{8}}\right] j_{2}(\kappa ),
\nonumber
\end{eqnarray}
where $\kappa =qr/2$. At $q^{2}=0,$ the body form factors become
\begin{eqnarray}
D_{C}(0) &=&\int_{0}^{\infty }dr\left[ u^{2}(r)+w^{2}(r)\right] =1  \nonumber \\
D_{M}(0) &=&\int_{0}^{\infty }dr\left[ 2u^{2}(r)-w^{2}(r)\right]
=2-3P_{D}
\nonumber \\
D_{E}(0) &=&\frac{3}{2}\int_{0}^{\infty
}drw^{2}(r)=\frac{3}{2}P_{D}\\
D_{Q}(0) &=&\frac{m_{d}^{2}}{\sqrt{50}}\int_{0}^{\infty }dr w(r)\left[ u(r)-
\frac{w(r)}{\sqrt{8}}\right]  \nonumber
\end{eqnarray}
giving the non-relativistic predictions
\begin{eqnarray}
Q_{d} &=&D_{Q}(0)  \label{quad} \\
\mu _{d} &=&\mu _{N}^s D_{M}(0)+D_{E}(0)=\mu
_{N}^s(2-3P_{D})+1.5P_{D},\nonumber
\end{eqnarray}%
where $Q_{d}$ is the quadrupole moment of the deuteron, $\mu _{d}$
is the magnetic moment of the deuteron and $\mu
_{N}^s=\frac{1}{2}(\mu_p+\mu_n-1)$ is the isoscalar nucleon
magnetic moment. The experimental value of the deuteron magnetic
moment $\mu _{d}=1.7139$, leads to the probability of $D$-state
$P_{D}=4.0\%.$ But this is only approximate value, because the
magnetic moment is very sensitive to relativistic corrections.

\begin{figure}[tb]
    \centering\vspace{-0.3cm}
        \scalebox{0.35}{\includegraphics{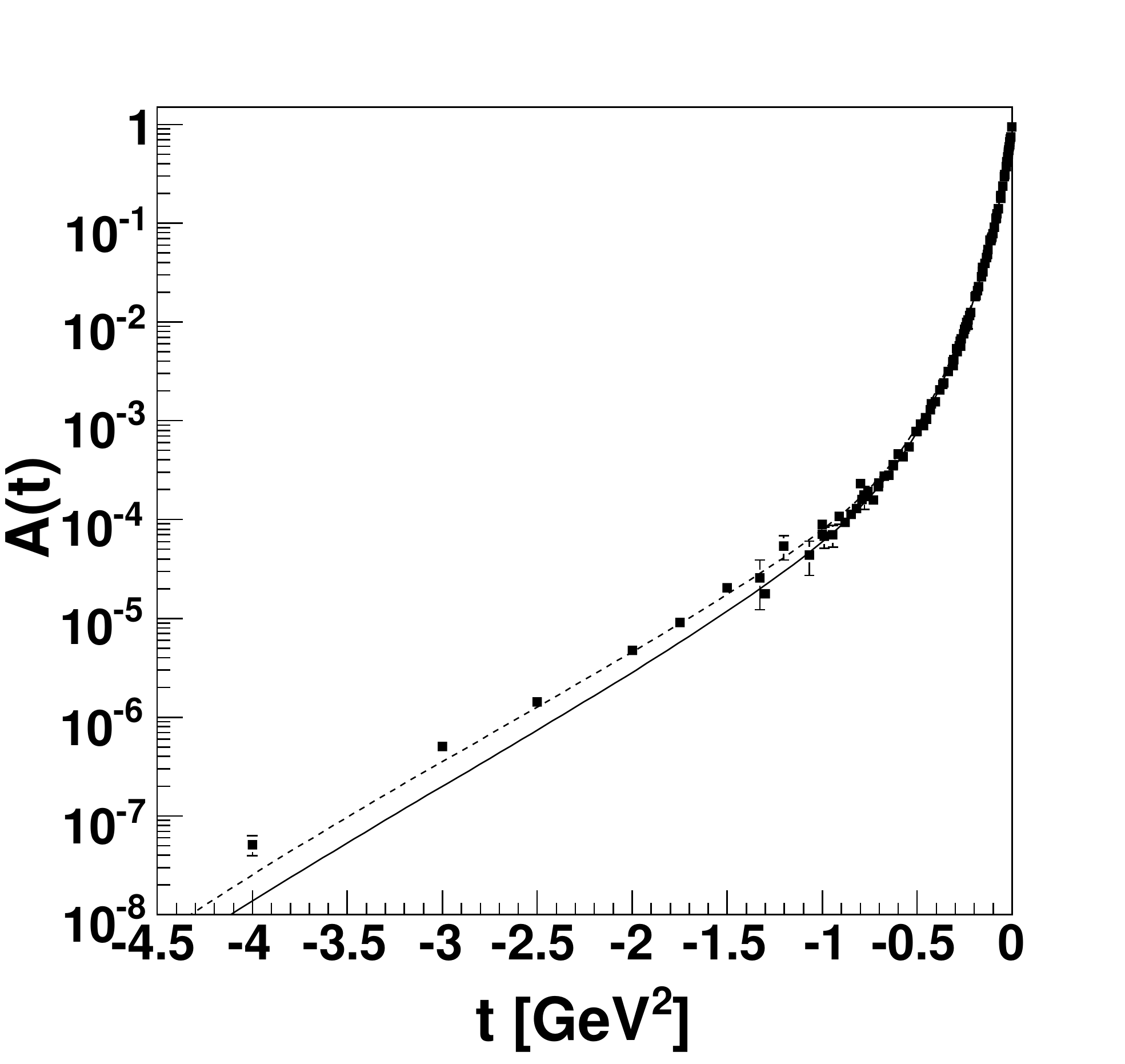}}
    \caption{\small{Deuteron structure function $A(t)$ data and their
    comparison with NIA predictions using non-dipole (full line) and dipole
    (dashed line) $G_{Ep}(t)$ behavior.}}
    \label{fig:A}
\end{figure}
\begin{figure}[tb!h]
    \centering\vspace{-0.3cm}
        \scalebox{0.35}{\includegraphics{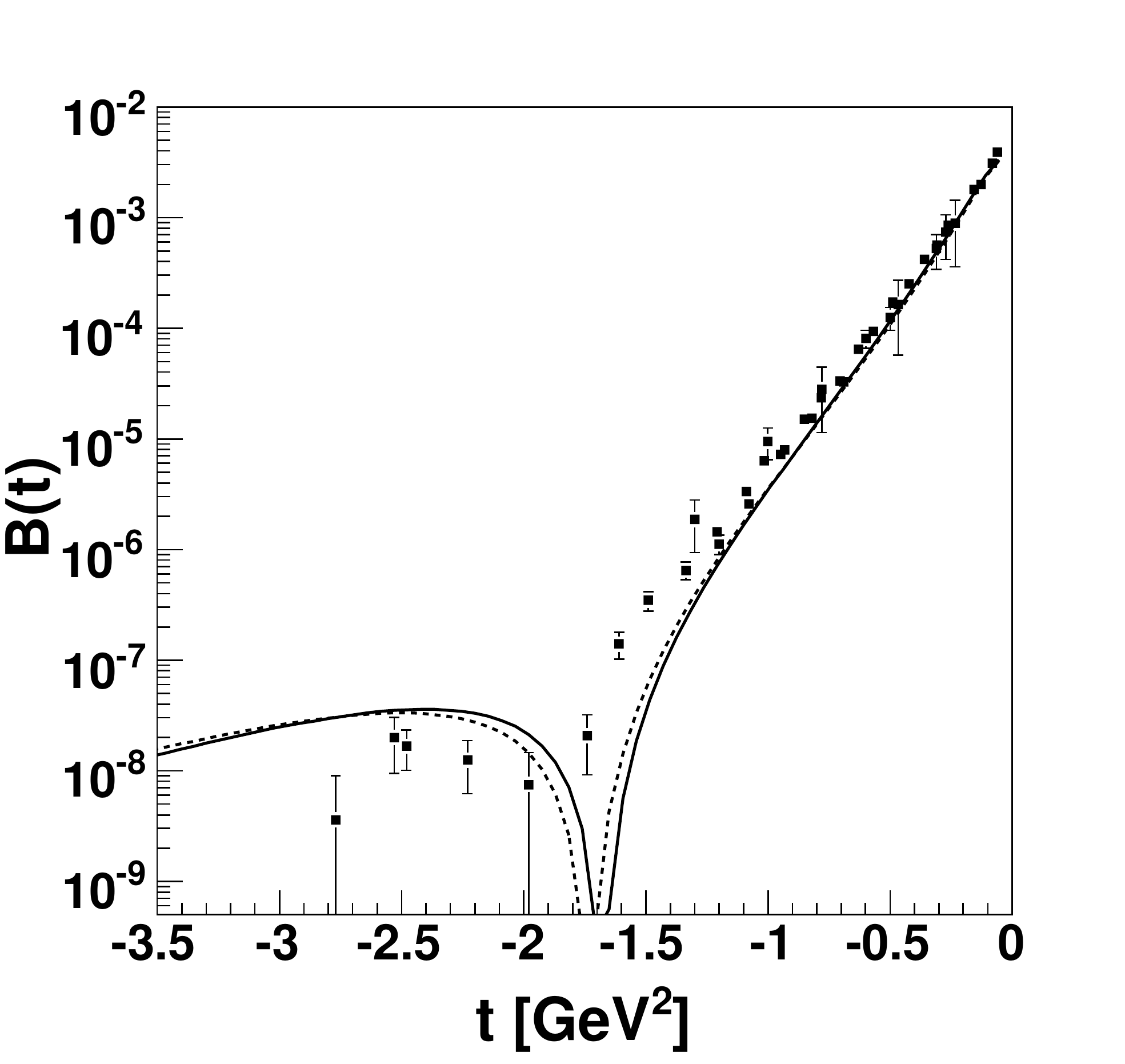}}
    \caption{\small{Deuteron structure function $B(t)$ data and their
    comparison with NIA predictions using non-dipole (full line) and dipole
    (dashed line) $G_{Ep}(t)$ behavior.}}
    \label{fig:B}
\end{figure}

 Experimentally the EM structure of the deuteron is
measured in the elastic scattering of electrons on deuterons,
described by the differential cross-section (\ref{sigdeutlab}) to be
calculated in the one-photon-exchange approximation with the
deuteron elastic structure functions. In order to see predicted
behaviors of the deuteron elastic structure functions $A(t)$ and
$B(t)$ to be caused by the non-dipole behavior of the proton
electric FF with the zero around $t=-Q^2=-13 GeV^2$, we use the
$G_{Ep}(t)$ (see full line in Fig.~\ref{fig20}) together with all other
nucleon FF behaviors in the calculation of the nucleon EM FF
isoscalar parts by means of the relation (\ref{nuciso}). Then
through the relations (\ref{deutelsf}) and (\ref{deutffisos}) one
comes to the behavior of $A(t)$ and $B(t)$ as presented in Fig.~\ref{fig:A}
and Fig.~\ref{fig:B}, respectively, by full lines.

    For comparison the deuteron elastic structure functions
$A(t),B(t)$ are predicted also by means of the nucleon EM FFs
\cite{DDW} obtained in the elastic scattering of unpolarized
electrons on unpolarized protons by Rosenbluth technique, i.e. by
using the proton electric FF in $t<0$ region having more or less
dipole behavior (see dashed line in Fig.~\ref{fig20}). The results are
presented in Fig.~\ref{fig:A} and Fig.~\ref{fig:B} by dashed lines. As one can see
from figures the examined difference is negligible, indicating
that the two-photon-exchange contribution is unimportant for the
unpolarized electron-deuteron elastic scattering in comparison
with the unpolarized electron-proton elastic scattering.

\setcounter{equation}{0} \setcounter{figure}{0} \setcounter{table}{0}\newpage
  \section{Sea-strange quark contributions to electromagnetic
     structure of hadrons} \label{VI}

   Hadrons are classified according to the $SU(3)$ symmetries into
multiplets, the most of them are known nonet of pseudoscalar
mesons and $1/2^+$ octet of baryons to be already treated in this
review. The mesons in the quark model are $q\bar q$ bound states
and members of octuplet are 3-quark $qqq$ configurations plus any
number of quark-antiquark pairs of light quarks u, d and s.

   At the nonet of pseudoscalar mesons the $\pi$ and $\eta$ are
compound from only up and down quarks, whereas $K$-mesons and
$\eta'$ to some extent are compound also from strange quark.

   At the octet of baryons only proton and nucleon are compound
from up and down quarks. The hyperons contain one or two strange
quarks.

   This section we start with experimental indications on sea
$s \bar s$ pairs contributions to the nucleon structure.

   They are as follows

   \begin{itemize}
\item[i)] {\it Pion-nucleon sigma term $\sigma_{\pi N}$}

   The $\sigma$-term of $\pi N$ scattering is defined by the relation
\begin{equation}
\sigma_{\pi N}(t=0)=\langle p|(\bar u u + \bar d d)|p\rangle
\label{a1}
\end{equation}
and it can be determined from extrapolation of scattering data on
the isospin even $\pi N$ on shell amplitude $\Sigma= f_\pi^2
\bar{D}(\nu, t)$ to the unphysical Cheng-Dashen point $\nu=0$,
$t=2\mu^2$ in the chiral limit $m_{\pi}=m_u=m_d=0$.

The results obtained from the analysis of $\pi N$ data are now
\begin{equation}
\Sigma(2\mu^2)=\sigma_{\pi N}(2\mu^2)=60 MeV \label{a2}
\end{equation}
and consequently $\sigma_{\pi N}(0)=45 MeV$.

On the other hand, the SU(3) mass splittings are due to the quark
mass term $H_m$ (the isospin symmetry is assumed) in the QCD
Hamiltonian
\begin{equation}
H_m=\hat{m}(\bar u u + \bar d d) + m_s\bar s s \label{a3}
\end{equation}
in which $m_s$ is the strange quark mass. The latter can be
written in terms of {\it singlet} and {\it octet} contributions:
\begin{equation}
H_m=1/3(m_s+2\hat{m})(\bar u u +\bar d d +\bar s
s)-1/3(m_s-\hat{m})((\bar u u +\bar d d -2\bar s s).\label{a4}
\end{equation}

The octet matrix element
\begin{equation}
\delta=\langle N|(\bar u u +\bar d d -2\bar s s)|N \rangle=\langle
N|(\bar u u +\bar d d)|N \rangle-2 \langle N|\bar s s|N\rangle
\label{a5}
\end{equation}
is determined from the octet masses by  current algebra and SU(3)
symmetry. As a result one finds
\begin{equation}
\delta \simeq 26 MeV \label{a6}
\end{equation}
Since $\sigma_{\pi N} \neq \delta$, the scalar strange quark
density $\langle N|\bar s s|N\rangle$ is non-vanishing, leading to
the following fraction of strange quarks in the proton
\begin{equation}
Y=\frac{\langle N|\bar s s |N \rangle }{\langle N|\bar u u +\bar d
d |N \rangle }=0.21 \label{a7}
\end{equation}
\item[ii)] {\it Proton spin crisis}

Deep inelastic scattering of polarized charged leptons ($e^-$ or
$\mu^-$) by polarized  protons has been investigated at CERN and
SLAC and it was found so-called EMC effect, i.e. quarks appeared
to contribute very little to the proton spin.

If $\triangle\Sigma$ is the fraction of  proton spin
contributed by light quarks, then
\begin{equation}
\triangle\Sigma=\triangle u +\triangle d + \triangle s, \label{a8}
\end{equation}
where $\triangle q's$ are integrals
\begin{equation}
\triangle q \equiv \int_0^1dx \{[q^{\uparrow}(x)-\bar
q^{\uparrow}(x)]- [q^{\downarrow}(x)-\bar q^{\downarrow}(x)]\}
\label{a9}
\end{equation}
of the difference of differences between parallel and
antiparallel probability quark distributions,
$q^{\uparrow\downarrow}(x)$ and $\bar q^{\uparrow\downarrow}(x)$
(q=u,d,s), for quarks and antiquarks with spin parallel and
antiparallel to the proton spin and $x$ is the Bjorken scaling
variable.

In the non-relativistic constituent quark model
\begin{equation}
\triangle u=4/3, \quad\quad\quad \triangle d=-1/3, \quad\quad\quad  \triangle s=0
\label{a10}
\end{equation}
and
\begin{equation}
\triangle\Sigma =1.\label{a11}
\end{equation}

However, the $\triangle u$,  $\triangle d$ and  $\triangle s$ can
be determined from DIS spin-dependent proton structure function
$g_1^p(x)$ and baryon $\beta$-decays.

Really
\begin{enumerate}
\item
The integral of $g_1^p(x)$ is a sum over $\triangle q's$, each
weighted by the square of that flavor quark's electric charge
\begin{equation}
\Gamma_1^p\equiv \int g_1^p(x)dx=1/2 \{4/9\triangle u
+1/9\triangle d +1/9\triangle s \}; \label{a12}
\end{equation}
\item
The matrix elements $F$ and $D$ measured in hyperon decays are
related to $\triangle u$ - $\triangle d$ and $\triangle u$ +
$\triangle d$ - 2$\triangle s$ as follows
\begin{eqnarray}
g_A\equiv &F+D&=\triangle u -\triangle d \label{a13}\\
 &3F-D&=\triangle u+\triangle d-2\triangle s. \label{a14}
\end{eqnarray}
A recent analysis of $\Gamma_1^p$, $F$ and $D$ finds
\begin{equation}
\triangle u=0.83\pm 0.03; \quad\quad\quad \triangle d=-0.43\pm 0.03; \quad\quad\quad
\triangle s=-0.10\pm 0.03 \label{a15}
\end{equation}
and
\begin{equation}
\triangle\Sigma =0.30\pm 0.05 \label{a16}
\end{equation}
indicating that very little of the proton spin is carried by the
light quarks and that the strange sea is polarized
antiparallel to the proton spin.
\end{enumerate}

\item[iii)] {\it  OZI rule violation.}
The OZI rule states that quark diagrams with disconnected quark
lines are strongly suppressed. As a result the process $\bar p
p\to \phi X$ has to be remarkably suppressed in comparison with
the process $\bar p p \to \omega X$.

Recent experiments at LEAR at CERN have found violations of the
OZI rule of up to two orders of the magnitude in $\bar p N\to \phi
X$ annihilations.

Such data could be explained if there are $\bar s s$ pairs present
in the nucleon (and antinucleon).
\item[iv)]{\it Neutrino experiments}

Elastic $\nu$ and $\bar\nu$ scattering on protons is sensitive to
strange axial-vector form factor, indicating on the presence of
$\bar s s$ pairs in the proton.

On the other hand, in deep inelastic scattering of $\nu$ and
$\bar\nu$  on protons the  charmed particles can be
produced in $d\to c$ and $s\to c$ transitions. The probability
\begin{eqnarray*}
d\to c &\sim& \sin^2{\theta_c}\\
s\to c &\sim& \cos^2{\theta_c}
\end{eqnarray*}
where $\theta_c$=0.23 is the Cabibbo angle.

Due to the smallness of $\theta_c$ the $s\to c$ transition is
dominant and gives the probability of studying the  strange
sea in $N$.
\end{itemize}
All these evidences for strangeness in $N$ mean that various
elements of
\begin{itemize}
\item
 scalar $\bar s s$
\item
  pseudoscalar $\bar s\gamma_5 s$
\item
 vector $\bar s\gamma_{\mu} s$
 \item
axial-vector $\bar s\gamma_{\mu}\gamma_5 s$
\item
tensor $\bar s\sigma_{\mu\nu} s$
\end{itemize}

currents, appearing in various processes with nucleons $N$, are
non vanishing and it has sense to investigate them theoretically.

   Further we concentrate \cite{Dubnicka08} only to the nucleon matrix element of
the strange-quark vector current $J^s_\mu=\bar s \gamma_\mu s$,
which is experimentally accessible in parity-violating elastic and
quasi-elastic electron scattering from the proton and light atomic
nuclei, where the strange electric and the strange magnetic
nucleon form factors (FFs) (or their combinations) are measured.

\medskip

     \subsection{Prediction of strange nucleon form factors behaviors}\label{VI1}

\medskip

   The momentum dependence of the nucleon matrix element of the strange-quark
vector current $J^{s}_{\mu}$=$\bar{s}\gamma_{\mu} s$ is contained
in the strange Dirac $F^{s}_{1} (t)$ and Pauli $F^{s}_{2} (t)$
nucleon FFs
\begin{equation}
\langle N|\bar{s}\gamma_{\mu} s|N\rangle=\bar{u}(p^{'})
\left[\gamma_{\mu} F^{s}_{1}(t)+
i{{\sigma_{\mu\nu}q^\nu}\over{2m_N}} F^{s}_{2}(t)\right]u(p),
\label{FFS}
\end{equation}
by means of which the strange electric and strange magnetic
nucleon FFs are defined
\begin{equation}
G_{E}^s(t)=F_{1}^s(t)+\frac{t}{4m_N^2}F_{2}^s(t),\quad
G_{M}^s(t)=F_{1}^s(t)+F_{2}^s(t).\label{ffs1}
\end{equation}
The latter can be measured in parity-violating elastic and
quasi-elastic scattering of electrons on protons and light atomic
nuclei.

   Since the strange-quark vector current $J_{\mu}^s = \bar s
\gamma _{\mu} s$  carries the quantum numbers of the isoscalar
part of the EM current $J_{\mu}^{I=0}$ and thus both currents
couple to the nucleon through the same intermediate states, it is
natural to expect \cite{Jaffe89} that one can extract the behavior
of strange Dirac and Pauli nucleon FFs just from the isoscalar
parts of Dirac and Pauli nucleon EM FFs.

   The main idea of a prediction of strange nucleon FFs behaviors
from the known isoscalar parts of the Dirac and Pauli nucleon EM
FFs is based on two assumptions
\begin{itemize}
\item
the $\omega-\phi$ mixing is valid also for coupling constants
between EM current (the strong strange quark current as well) and
vector-meson
\begin{equation}
\frac{1}{f_\omega}= \frac{1}{f_{\omega_0}} \cos{\epsilon} -
\frac{1}{f_{\phi_0}} \sin{\epsilon};\quad
\frac{1}{f_\phi}=\frac{1}{f_{\omega_0}} \sin{\epsilon} +
\frac{1}{f_{\phi_0}} \cos{\epsilon},\label{const}
\end{equation}
where $\epsilon=3.7^{\circ}$ is a deviation from the ideally mixing
angle $\theta_{0}=35.3^{\circ}$;
\item
the quark current of some flavor couples with universal strength
$\kappa$ exclusively to the vector-meson wave function component
of the same flavor
\begin{equation}
\langle0|\bar q_r\gamma q_r|(\bar q_t q_t)_V\rangle = \kappa m^2_V
\delta_{rt}\varepsilon_\mu,\label{curr}
\end{equation}
\end{itemize}
where $m_V$ and $\varepsilon_\mu$ are the mass and the
polarization vector of the considered vector-meson.

   Starting from a definition of the virtual-photon vector-meson
transition coupling constants $1/f^e_V$ by the relation
\begin{equation}
<0|J^e_\mu|V>=\frac{m_V^2}{f^e_V}\epsilon_\mu \label{elmag}
\end{equation}
and the second assumption for the isoscalar EM current
$J^{I=0}_\mu$ to be expressed by quark fields, one comes to the
equations
\begin{eqnarray}
<0|J^{I=0}_\mu|\omega_0>&=&<0|\frac{1}{6}(\bar{u}\gamma_\mu
u+\bar{d}\gamma_\mu d) -\frac{1}{3}\bar{s}\gamma_\mu
s|\frac{1}{\sqrt{2}}(|\bar{u}u>+|\bar{d}d>)= \nonumber \\
&=&\frac{1}{6}(\frac{1}{\sqrt{2}}+\frac{1}{\sqrt{2}})\kappa
m^2_{\omega_0}\varepsilon_\mu\equiv\frac{m^2_{\omega_0}}{f^e_{\omega_0}}\varepsilon_\mu
\label{isocurr}
\end{eqnarray}

\begin{eqnarray}
<0|J^{I=0}_\mu|\phi_0>&=&<0|\frac{1}{6}(\bar{u}\gamma_\mu
u+\bar{d}\gamma_\mu d)-\frac{1}{3}\bar{s}\gamma_\mu
s|\bar{s}s>)= \nonumber\\
&=&-\frac{1}{3}\kappa
m^2_{\phi_0}\varepsilon_\mu\equiv\frac{m^2_{\phi_0}}{f^e_{\phi_0}}\varepsilon_\mu
\label{isocurr1}
\end{eqnarray}
from where expressions for EM coupling constants follow
\begin{equation}
\frac{1}{f^e_{\omega_0}}=\frac{1}{6}(\frac{1}{\sqrt{2}}+\frac{1}{\sqrt{2}})\kappa=
\frac{1}{\sqrt{6}}\frac{1}{\sqrt{3}}\kappa;\quad
\frac{1}{f^e_{\phi_0}}=-\frac{1}{3}\kappa=
-\frac{1}{\sqrt{6}}\sqrt\frac{2}{3}\kappa. \label{frac}
\end{equation}
Substituting the latter into the first assumption, together with
identities $\frac{1}{\sqrt{3}}=\sin{\theta_0}$ and
$\sqrt\frac{2}{3}=\cos{\theta_0}$, one obtains coupling constants
of real $\omega$ and $\phi$
\begin{equation}
\frac{1}{f^e_{\omega}}=\frac{\kappa}{\sqrt{6}}\sin(\varepsilon+\theta_0;\quad
\frac{1}{f^e_{\phi}}=-\frac{\kappa}{\sqrt{6}}\cos(\varepsilon+\theta_0).\label{emcplcons}
\end{equation}

These relations, together with
$\frac{1}{f^e_\rho}=\frac{1}{\sqrt{2}}\kappa$ following from
\begin{eqnarray}
<0|J^{I=1}_\mu|\rho>&=&<0|\frac{1}{2}(\bar{u}\gamma_\mu
u-\bar{d}\gamma_\mu d)|\frac{1}{\sqrt{2}}(|\bar{u}u>-|\bar{d}d>)=\nonumber \\
&=&\frac{1}{2}(\frac{1}{\sqrt{2}}+\frac{1}{\sqrt{2}})\kappa
m^2_{\rho}\varepsilon_\mu\equiv\frac{m^2_{\rho}}{f^e_{\rho}}\varepsilon_\mu,\label{isovekc}
\end{eqnarray}
give for the ratios of the universal vector-meson coupling
constants the values

$\frac{1}{f^e_\rho}:\frac{1}{f^e_\omega}:\frac{1}{f^e_\phi}=0.71:0.25:(-0.32)$

in a very good agreement with experimental values

$\frac{1}{f^e_\rho}:\frac{1}{f^e_\omega}:\frac{1}{f^e_\phi}=0.79:0.23:(-0.31)$

obtained from leptonic widths $\Gamma(V\to e^+e^-)$ of considered
vector-mesons. Just this agreement demonstrates the previous two
assumptions to be compatible with physical reality and one can
extend their validity also for strong strange-quark current
vector-meson transition coupling constants $1/f^s_V$.

   Then analogically one can write for the strong strange-quark
current the equations
\begin{equation}
<0|J^s_\mu|\omega_0>=<0|(\bar{s}\gamma_\mu
s|\frac{1}{\sqrt{2}}(|\bar{u}u>+|\bar{d}d>)=0
\equiv\frac{m^2_{\omega_0}}{f^s_{\omega_0}}\varepsilon_\mu
\label{js}
\end{equation}
 \begin{equation} <0|J^s_\mu|\phi_0>=<0|(\bar{s}\gamma_\mu
s||\bar{s}s>=1.\kappa m^2_{\phi_0}\varepsilon_\mu\equiv
\frac{m^2_{\phi_0}}{f^s_{\phi_0}}\varepsilon_\mu, \label{jsmu}
\end{equation}
from where one gets $\frac{1}{f^s_{\omega_0}}=0$ and
$\frac{1}{f^s_{\phi_0}}=1.\kappa$. Substituting them into
$\omega-\phi$ mixing relations one comes to the strange coupling
constants of the real $\omega$ and $\phi$
\begin{equation}
\frac{1}{f^s_\omega}=-\kappa \sin{\varepsilon};\quad
\frac{1}{f^s_\phi}=+\kappa \cos{\varepsilon}
\end{equation}
Bringing these expressions for $\omega$ and $\phi$ vector mesons
into ratios with EM coupling constants (\ref{emcplcons}),
respectively, one gets rid of the unknown parameter $\kappa$ and
comes to the relations
\begin{eqnarray}
({f^{(i)}_{\omega
NN}}/{f^{s}_{\omega}})&=&-\sqrt{6}\frac{\sin{\epsilon}}
{\sin(\epsilon+\theta_{0})}({f^{(i)}_{\omega
NN}}/{f^{e}_{\omega}})\nonumber \\
({f^{(i)}_{\phi NN}}/{f^{s}_{\phi}})
&=&-\sqrt{6}\frac{\cos{\epsilon}}
{\cos(\epsilon+\theta_{0})}({f^{(i)}_{\phi NN}}/{f^{e}_{\phi}})
\quad(i=1,2)\label{CC}
\end{eqnarray}
giving a possibility to calculate the unknown strange coupling
constant ratios (parameters of strange nucleon FFs) from the known
EM coupling constant ratios (parameters of EM nucleon FFs) to be
determined in a description of all existing nucleon EM FF data by
a suitable model of the EM structure of nucleons.

   The derived relations (\ref{CC}) are valid also for any pairs of excited
states $\omega'$,$\phi'$, $\omega''$, $\phi''$, etc. of the ground
state of $\omega$ and $\phi$ isoscalar vector-mesons.

   Then for a prediction of strange nucleon form factors the $U\&A$ model of the nucleon EM
structure \cite{DDW} will be used, which comprises all known
nucleon FF properties to be contained in the following models of
isoscalar parts of Dirac and Pauli nucleon EM FFs
\begin{eqnarray}
F^{I=0}_{1}[V(t)]&=&\left(\frac{1-V^{2}}{1-V^{2}_{N}}\right)^{4}\bigg\{\frac{1}{2}
 L(V_{\omega''})L(V_{\omega'})+
[L(V_{\omega''})L(V_{\omega})
 \frac{(C_{\omega''}-C_{\omega})}
 {(C_{\omega''}-C_{\omega'})} -\nonumber \\
 &-&L(V_{\omega'})L(V_{\omega})
 \frac{(C_{\omega'}-C_{\omega})}
 {(C_{\omega''}-C_{\omega'})}-
L(V_{\omega''})L(V_{\omega'})]
 (f^{(1)}_{\omega NN}/f^{e}_{\omega})+\nonumber \\
&+&[L(V_{\omega''})L(V_{\phi})
 \frac{(C_{\omega''}-C_{\phi})}
 {(C_{\omega''}-C_{\omega'})}-
 L(V_{\omega'})L(V_{\phi})
 \frac{(C_{\omega'}-C_{\phi})}
 {(C_{\omega''}-C_{\omega'})}-\nonumber \\
 &-&L(V_{\omega''})L(V_{\omega'})]
 (f^{(1)}_{\phi NN}/f^{e}_{\phi})\bigg \}\label{F01}
\end{eqnarray}
\begin{eqnarray}\nonumber
F^{I=0}_{2}[V(t)]&=&\left(\frac{1-V^2}{1-V^2_N}\right)^6\bigg \{L(V_{\omega''})
L(V_{\omega'})L(V_{\omega})\left [1-{\frac{C_{\omega}}{(C_{\omega''}-C_{\omega'})}}\right.\times\\
&\times&\left.\left(\frac{(C_{\omega''}-C_{\omega})}{C_{\omega'}} -
\frac{(C_{\omega'}-C_{\omega})}{C_{\omega''}}\right)\right ] (f^{(2)}_{\omega
NN}/f^{e}_{\omega})+\nonumber \\
&+&L(V_{\omega''})L(V_{\omega'})L(V_{\phi})
\left[1-{\frac{C_{\phi}}{(C_{\omega''}-C_{\omega'})}}
\left(\frac{(C_{\omega''}-C_{\phi})}{C_{\omega'}}\right.\right.-\nonumber \\
&-&\left.\left.\left.\frac{(C_{\omega'}-C_{\phi})}{C_{\omega''}}\right)\right](f^{(2)}_{\phi
NN}/f^{e}_{\phi})\right\}\label{F02}
\end{eqnarray}
and of the Dirac and Pauli strange nucleon FFs
\begin{eqnarray} \label{FS1}
F^{s}_{1}[V(t)]&=&\left(\frac{1-V^{2}}{1-V^{2}_{N}}\right)^{4}
\bigg\{\left[L(V_{\omega''})L(V_{\omega})\frac{(C_{\omega''}-C_{\omega})}
{(C_{\omega''}-C_{\omega'})}\right.-\nonumber\\
&-&\left.L(V_{\omega'})L(V_{\omega})
\frac{(C_{\omega'}-C_{\omega})}{(C_{\omega''}-C_{\omega'})}-
L(V_{\omega''})L(V_{\omega'})\right](f^{(1)}_{\omega
NN}/f^{s}_{\omega})\nonumber+\\
&+& \bigg[L(V_{\omega''})L(V_{\phi})\frac{(C_{\omega''}-C_{\phi})}{(C_{\omega''}-C_{\omega'})}-
L(V_{\omega'})L(V_{\phi})\frac{(C_{\omega'}-C_{\phi})}
{(C_{\omega''}-C_{\omega'})}- \nonumber\\
&-&L(V_{\omega''})L(V_{\omega'})\bigg]
(f^{(1)}_{\phi NN}/f^{s}_{\phi})\bigg \}
\end{eqnarray}
\begin{eqnarray} \label{FS2}
F^{s}_{2}[V(t)]&=&\left(\frac{1-V^2}{1-V^2_N}\right)^6\bigg\{L(V_{\omega''})
L(V_{\omega'})L(V_{\omega})\bigg[1-{\frac{C_{\omega}}{(C_{\omega''}-C_{\omega'})}}\times\\
&\times&\left(\frac{(C_{\omega''}-C_{\omega})}{C_{\omega'}}
-\frac{(C_{\omega'}-C_{\omega})}{C_{\omega''}}\right)\bigg](f^{(2)}_{\omega
NN}/f^{s}_{\omega})+L(V_{\omega''})L(V_{\omega'})L(V_{\phi})\times\nonumber \\
&\times&\left[1-\frac{C_{\phi}}{(C_{\omega''}-C_{\omega'})}
\left(\frac{(C_{\omega''}-C_{\phi})}{C_{\omega'}}-
\frac{(C_{\omega'}-C_{\phi})}{C_{\omega''}}\right)\right] (f^{(2)}_{\phi
NN}/f^{s}_{\phi})\bigg\},\nonumber
\end{eqnarray}
where
\begin{eqnarray*}
\nonumber
L(V_r)=\frac{(V_N-V_r)(V_N-V^{\ast}_r)(V_N-1/V_r)(V_N-1/V^{\ast}_r)}
 {(V-V_r)(V-V^{\ast}_r)(V-1/V_r)(V-1/V^{\ast}_r)},\\
C_r=\frac{(V_N-V_r)(V_N-V^{\ast}_r)(V_N-1/V_r)(V_N-1/V^{\ast}_r)}
 {-(V_r-1/V_r)(V^{\ast}_r-1/V^{\ast}_r)},\\
V_N=V(t)_{|t=0}; V_r=V(t)_{|t=(m_r-i\Gamma_r/2)^2};
 (r=\omega,\phi,\omega',\omega''),\nonumber
\end{eqnarray*}

\begin{equation}
 V(t)=i\frac
 {\sqrt{[\frac{t_{N\bar N}-t^{I=0}_0}{t^{I=0}_0}]^{1/2}+
 [\frac{t-t^{I=0}_0}{t^{I=0}_0}]^{1/2}}-
 \sqrt{[\frac{t_{N\bar N}-t^{I=0}_0}{t^{I=0}_0}]^{1/2}-
 [\frac{t-t^{I=0}_0}{t^{I=0}_0}]^{1/2}}}
 {\sqrt{[\frac{t_{N\bar N}-t^{I=0}_0}{t^{I=0}_0}]^{1/2}+
 [\frac{t-t^{I=0}_0}{t^{I=0}_0}]^{1/2}}+
 \sqrt{[\frac{t_{N\bar N}-t^{I=0}_0}{t^{I=0}_0}]^{1/2}-
 [\frac{t-t^{I=0}_0}{t^{I=0}_0}]^{1/2}}}
\label{ITR}
\end{equation}
and $t_{N\bar N}=4m_N^2$ is a square-root branch point
corresponding to $N \bar N$ threshold.

\begin{figure}[tb]
\centering
\psfig{file=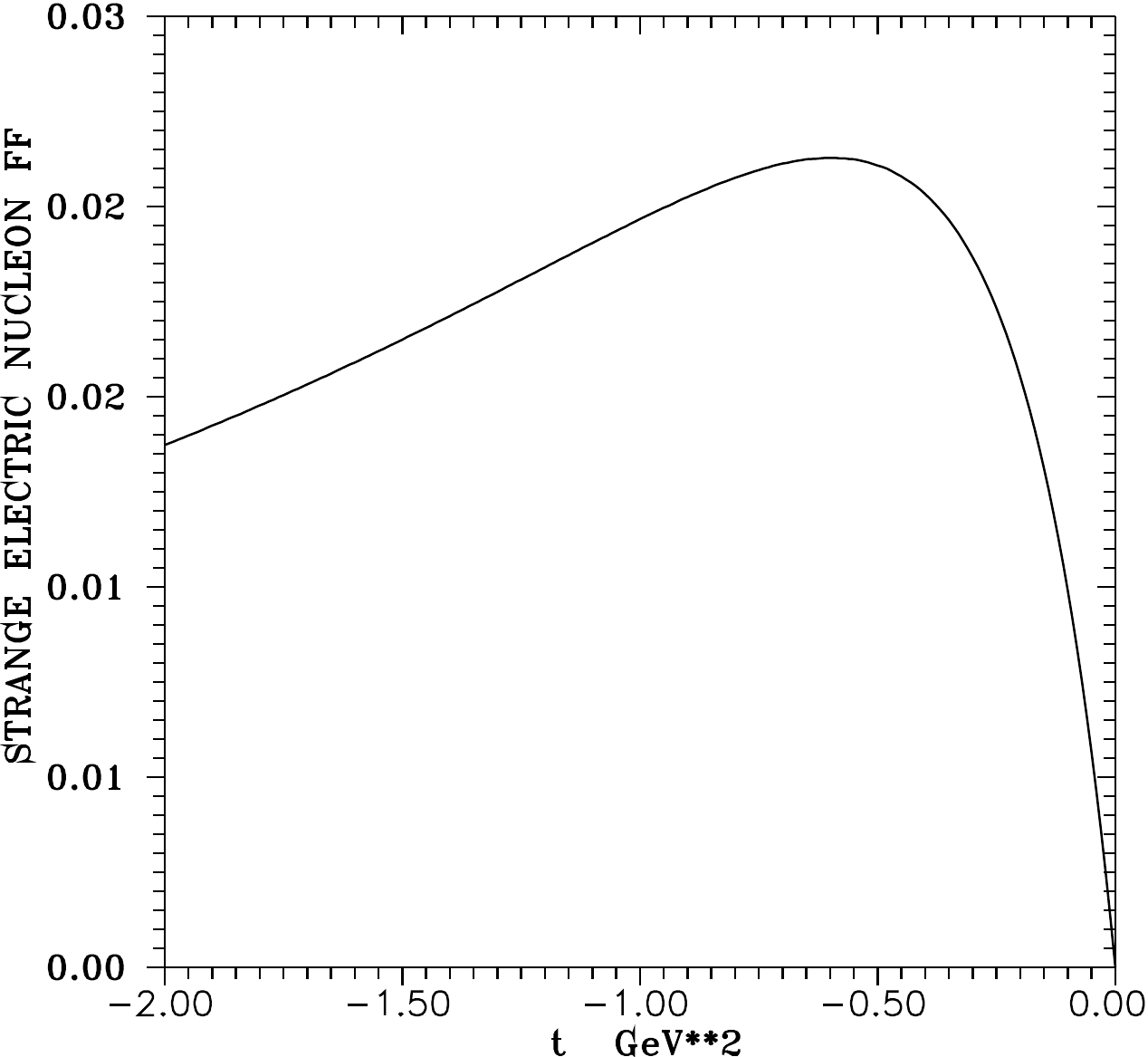,width=6.48cm}\ \
\psfig{file=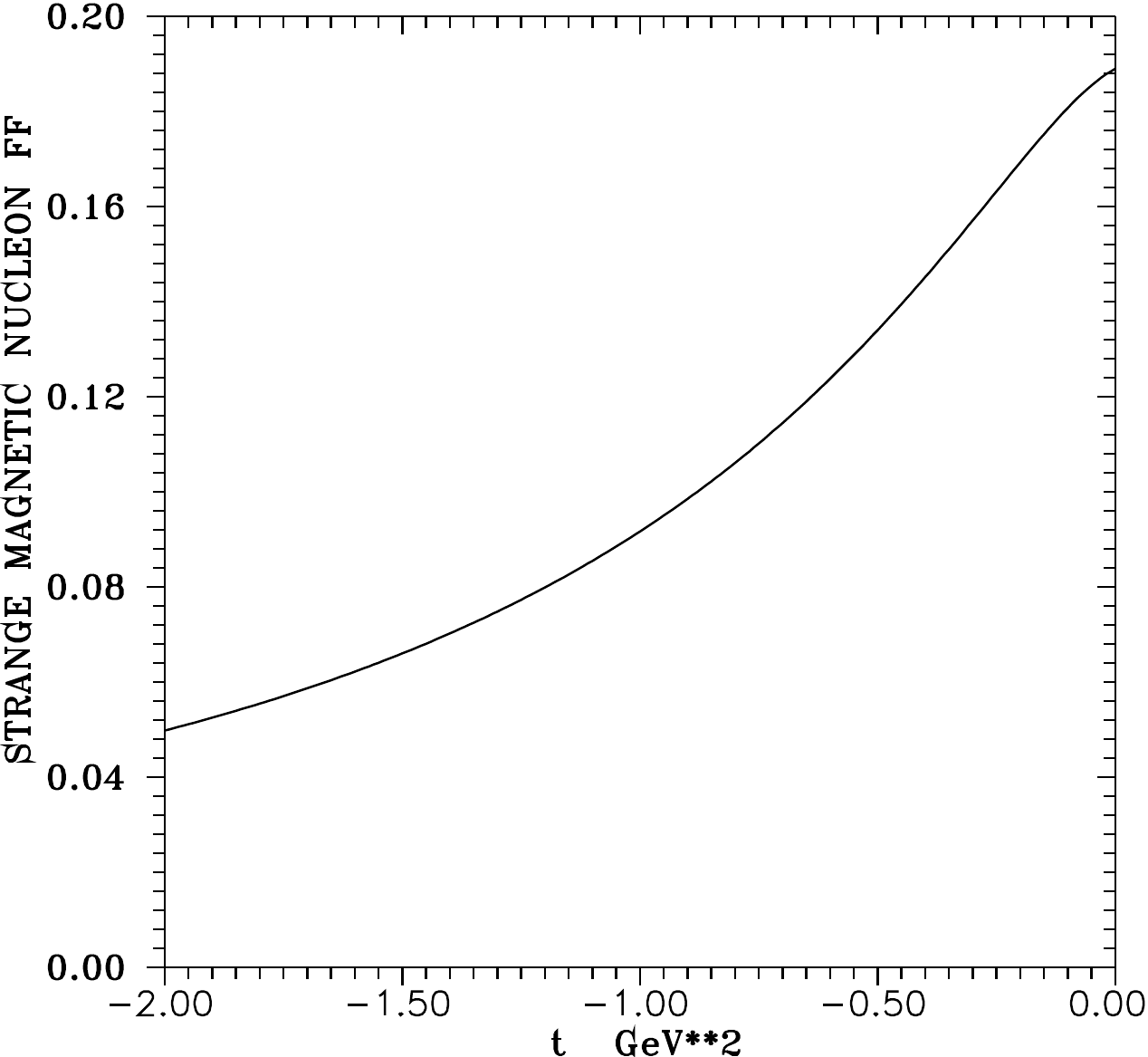,width=6.48cm}  \caption{Predicted strange
electric and magnetic nucleon FFs behaviors \label{figstrsl}}
\end{figure}
\begin{figure}[tb!h]
\centering \psfig{file=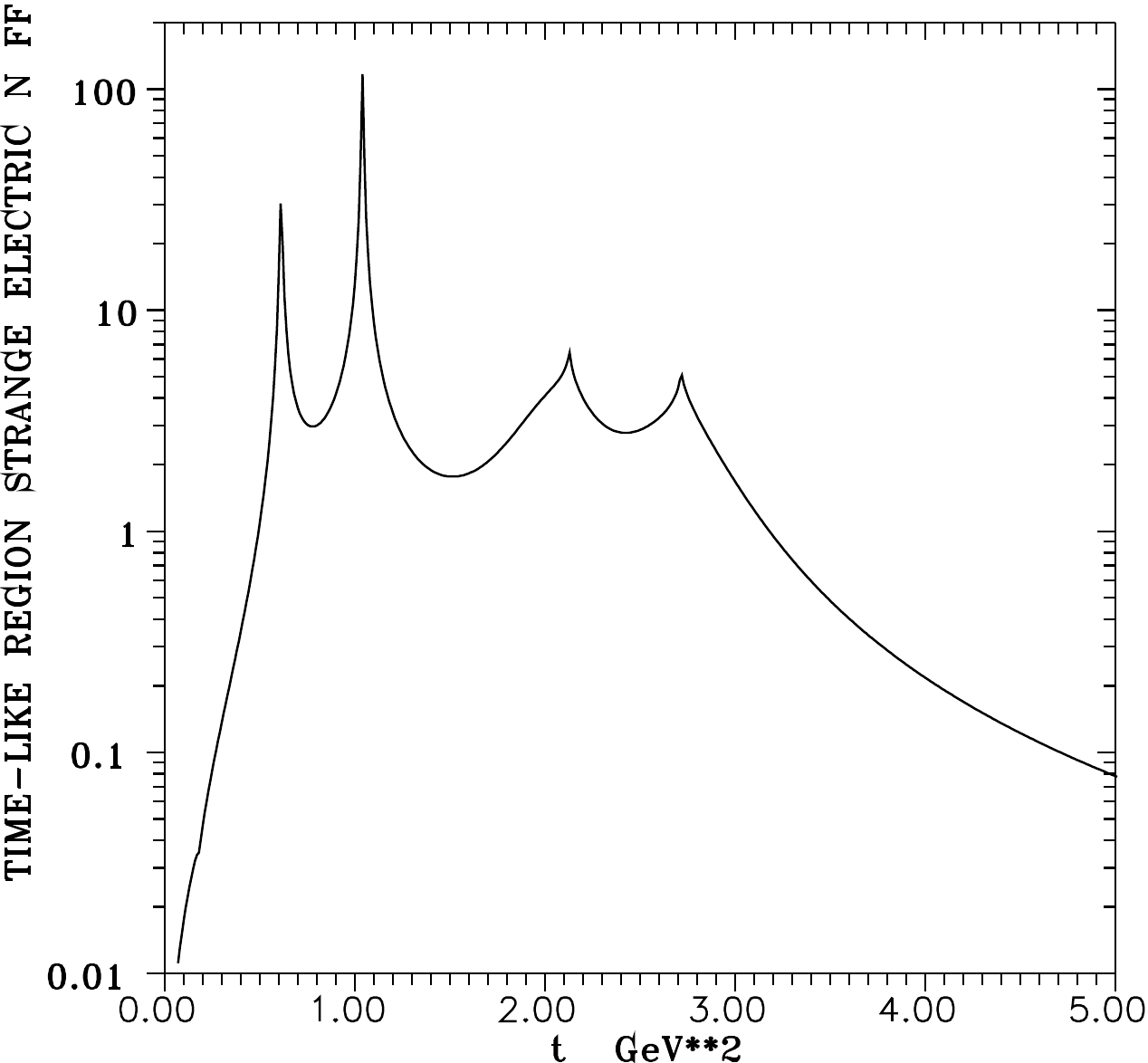,width=6.48cm}\ \
\psfig{file=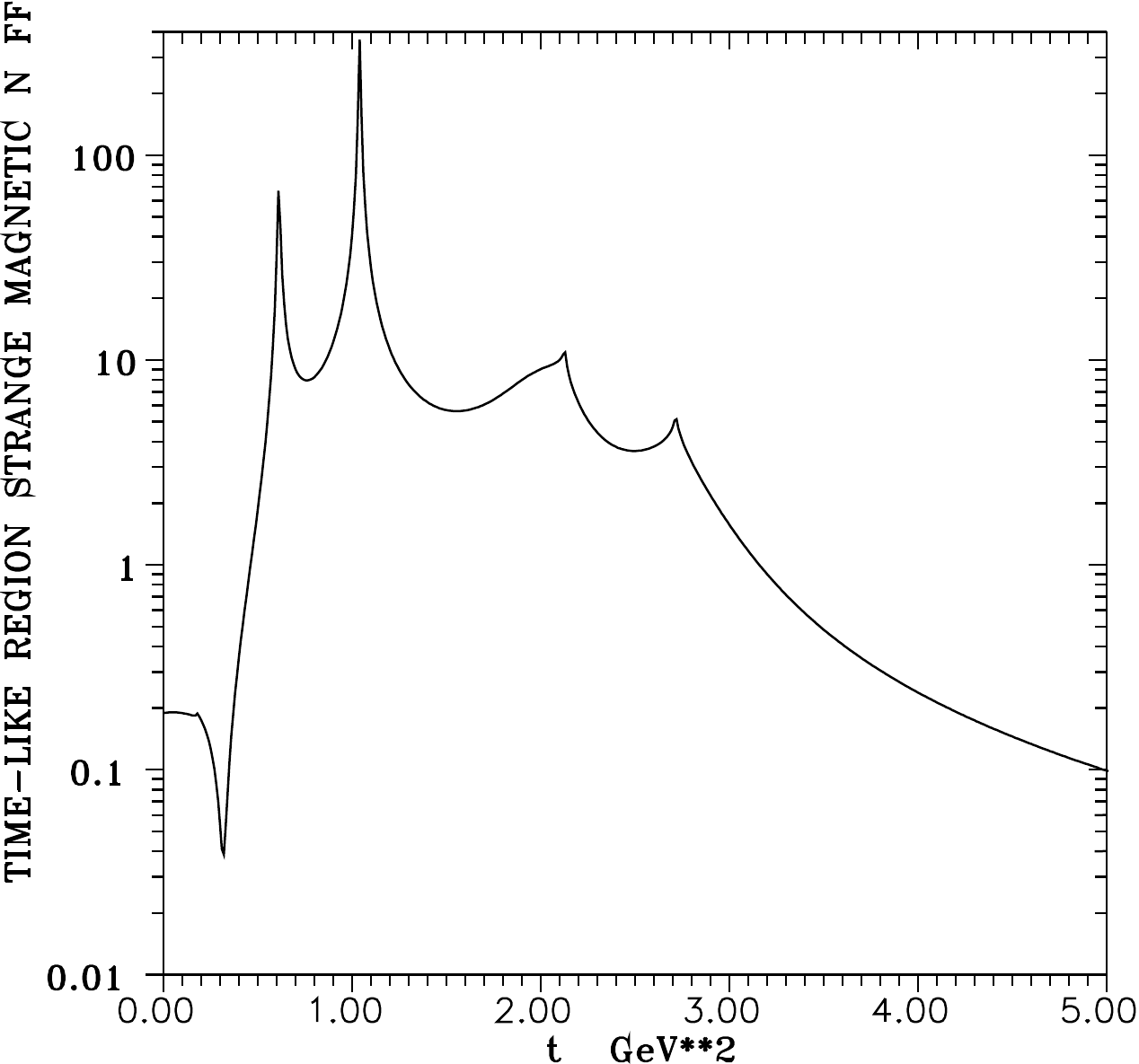,width=6.48cm} \caption{Predicted behaviors
of the strange nucleon FFs in the time-like region \label{figstretl}}
\end{figure}
    The expressions (\ref{F01}) and (\ref{F02})
for $F_1^{I=0}, F_2^{I=0}$, respectively, together with similar
expressions  for $F_1^{I=1}, F_2^{I=1}$ \cite{DDW} were used for a
description of all solid nucleon EM FF data and a reasonable
description of them has been achieved. The found numerical values
of $(f^{(i)}_{\omega N N}/f^e_\omega)$, $(f^{(i)}_{\phi N
N}/f^e_\phi)\quad i=1,2$ in the relations (\ref{F01}), (\ref{F02})
are used to calculate the unknown strange coupling constant ratios
$(f^{(i)}_{\omega N N}/f^s_\omega), (f^{(i)}_{\phi N N}/f^s_\phi)\quad
i=1,2$ in (\ref{FS1}), (\ref{FS2}) by means of the relations
(\ref{CC}). As a result the behaviors of the strange nucleon FFs
are predicted (see Fig.~\ref{figstrsl}) without the use of any experimental
point obtained in parity-violating elastic and quasi-elastic
scattering of electrons on protons and light atomic nuclei.

   Due to the fact, that the expressions (\ref{FS1}) and
(\ref{FS2}) for the strange nucleon FFs are parame\-trized also by
the $U\&A$ model and so, both FFs are analytic functions for
$-\infty<t<+\infty$, naturally we are predicting the time-like
behavior of strange nucleon FFs (see Fig.~\ref{figstretl}), though there is no
method to be known for their experimental determination until now.

   Moreover, as the $U\&A$ model represents compatible unification
of the pole and continuum (given by cuts on the positive real axis
of the analytic strange nucleon FFs) contributions, one can
predict even the imaginary parts behaviors of the strange nucleon
FFs to be given by unitarity conditions of FFs under
consideration.

   Since the first results reported by the SAMPLE Collaboration in 1997
\cite{Mueller97}, approximatelly ten independent measurements of the
parity-violating contribution to the elastic EM FFs of the nucleons
have now been completed. But only four of them
\cite{Maas04}-\cite{Maas05} declare clearly nonzero experimental
values of the strangeness within the proton and our theoretical
predictions are compatible with them.

   The A4 Collaboration \cite{Maas04} result at $Q^2=0.230 GeV^2$ is
$G^s_E+0.225G^s_M=0.039\pm 0.034$ to be in accordance with our
prediction $0.055$; the SAMPLE Collaboration \cite{Spyde04} result at
$Q^2=0.1 GeV^2$ is $G^s_M=0.37\pm 0.34$ to be also in accordance
with our result $0.18$; another A4 Collaboration \cite{Maas05} result at
$Q^2=0.108 GeV^2$ is $G^s_E+0.225 G^s_M=0.039\pm 0.034$ to be
again in accordance with our theoretical prediction $0.030$.
\begin{figure}[tb]
\begin{center}
\epsfig{file=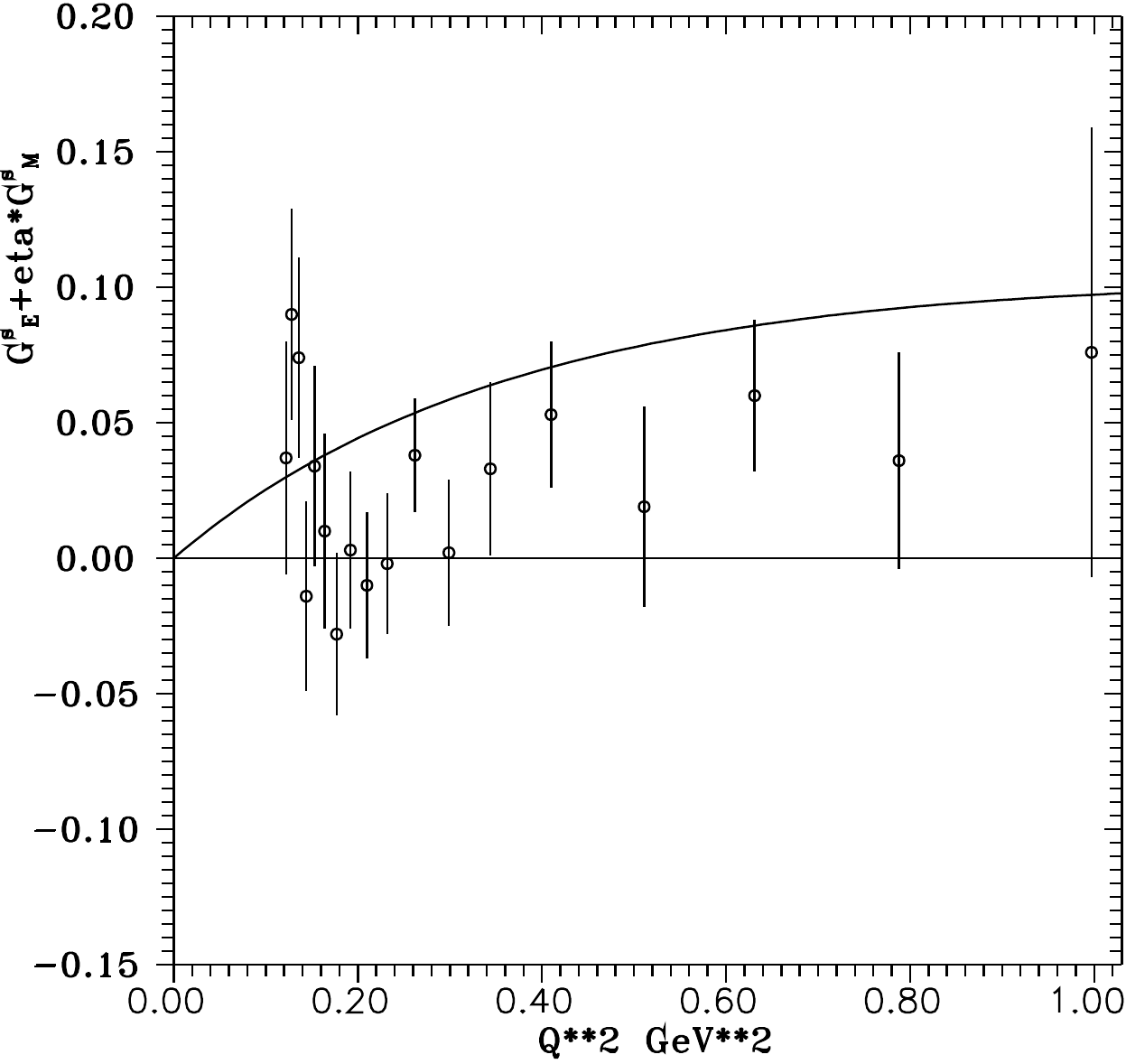,scale=0.7}
\caption{Compatibility of our theoretically predicted curve with
G0 Collab. data.\label{figstrange}}
\end{center}
\end{figure}

   May be the most impressive is a compatibility of our
theoretically predicted curve (see Fig.~\ref{figstrange}) with the recent data
\cite{Armstrong05} obtained by G0 Collab. on the combination
$G^s_E(Q^2)+\eta(Q^2) G^s_M(Q^2)$ for the interval $0.12
GeV^2<Q^2<1.0 GeV^2$ of momentum transfer squared values, which
strengthen our belief in the strangeness in the nucleon.

\medskip

     \subsection{Strange vector form factor of $K$-mesons}\label{VI2}

\medskip

   We have extended the method elaborated for prediction of the
strange nucleon form factors behaviors to $K$-mesons.

   Taking into account the splitting of FFs for charge kaon (\ref{Kffch}) and
for neutral kaon (\ref{Kff0}) into isoscalar and isovector FFs and
having the $U\&A$ models of the kaon EM structure
\begin{eqnarray}
F^{I=0}_K[V(t)]&=& \left(\frac{1-V^2}{1-V^2_N}\right)^2
     \left\{\frac{1}{2}H(V_{\omega'})+\left[L(V_\omega)-H(V_{\omega'})\right](f_{\omega KK}/f^e_\omega) + \right. \nonumber \\
 & +&
\biggl.   [L(V_\phi) - H(V_{\omega'})](f_{\phi KK}/f^e_\phi)\biggr\} \label{ka1}\\
F^{I=1}_K[W(t)]&=& \left(\frac{1-W^2}{1-W^2_N}\right)^2
     \left\{\frac{1}{2}H(W_{\rho'''})+[L(W_\rho)-H(W_{\rho'''})](f_{\rho KK}/f^e_\rho) + \right. \nonumber \\
 & + &
\biggl.  [L(W_{\rho'}) - H(W_{\rho'''})](f_{\rho'
KK}/f^e_{\rho'})\biggr\} \label{ka2}
\end{eqnarray}
and also for the strange FF of kaon with the inner structure of
$F^{I=0}_K[V(t)]$
\begin{eqnarray}
F^{S}_K[V(t)]&=& \left(\frac{1-V^2}{1-V^2_N}\right)^6
     \left\{-H(V_{\omega'})+[L(V_\omega)-H(V_{\omega'})](f_{\omega KK}/f^s_\omega) + \right. \nonumber \\
 & +&
\biggl.   [L(V_\phi) - H(V_{\omega'})](f_{\phi
KK}/f^s_\phi)\biggr\} \label{ffstrange}
\end{eqnarray}
set up, one predicts \cite{Dubnicka02} the behavior (see Fig.~\ref{figstrekaon})
 of the strange FF of $K$-mesons (\ref{ffstrange}) by means
of an evaluation of $(f_{\omega KK}/f^s_\omega)$, $(f_{\phi
KK}/f^s_\phi)$ from $(f_{\omega KK}/f^e_\omega)$, $(f_{\phi
KK}/f^e_\phi)$ determined in a comparison of (\ref{Kffch}) and
(\ref{Kff0}) with data on charge and neutral kaon EM FFs.
\begin{figure}[tb]
\centering
\psfig{file=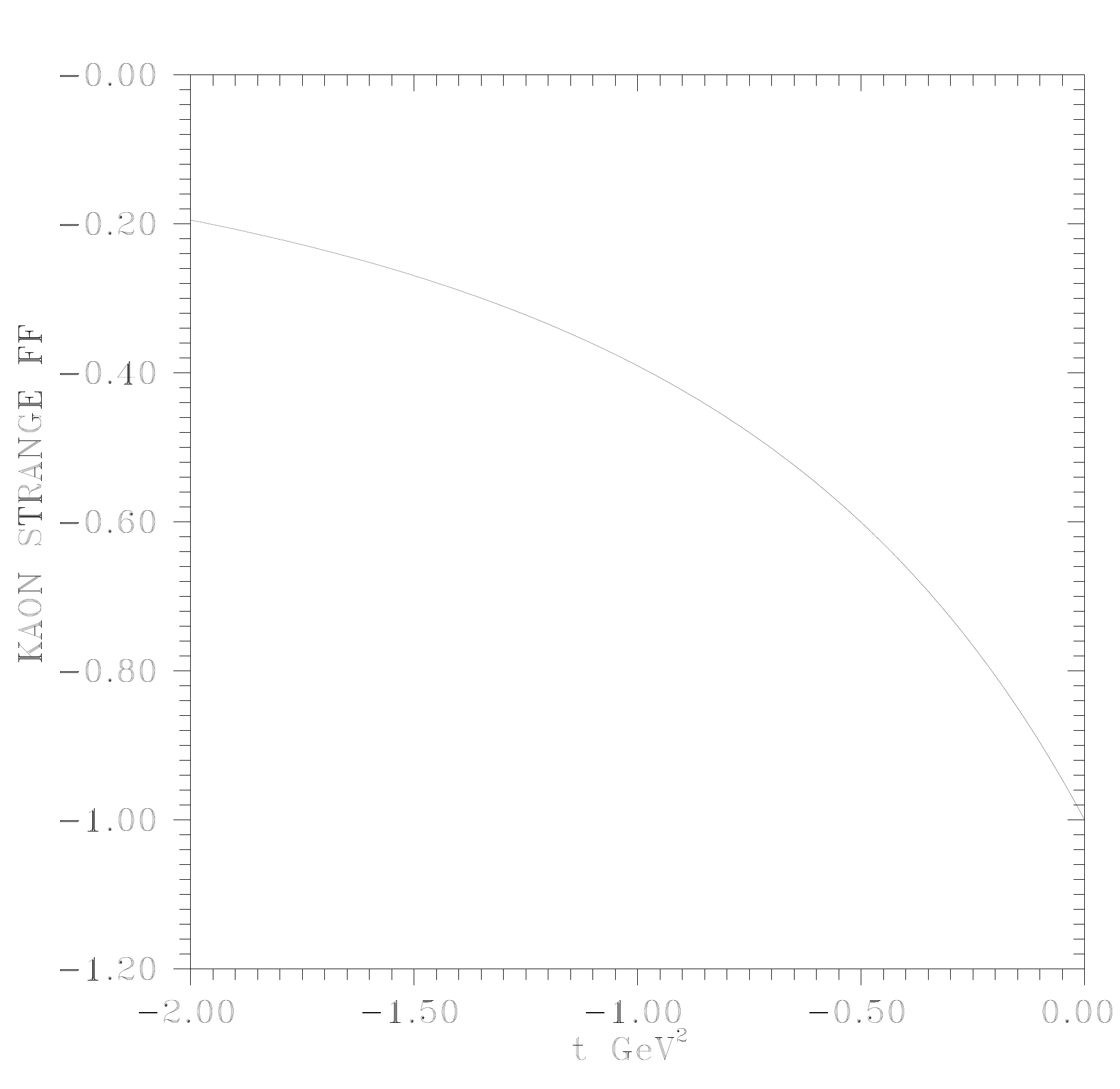,width=6.48cm}\ \
\psfig{file=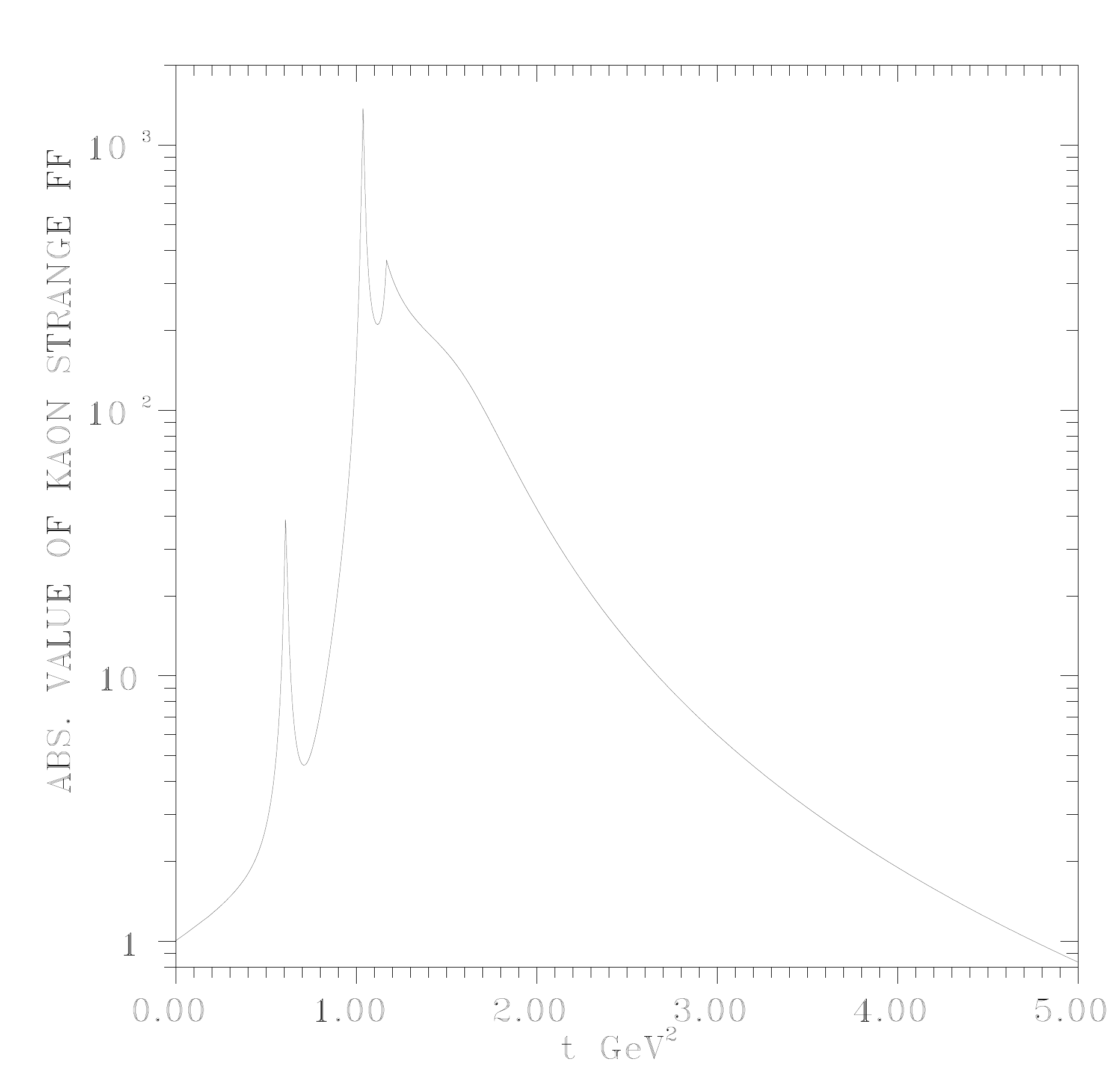,width=6.48cm} \caption{Predicted behaviors
of the strange kaon FFs in the space-like and the time-like region \label{figstrekaon}}
\end{figure}

   The obtained result is interesting from the point of view that
in the case of the $K$-meson, which is naturally compound from one
non-strange (up ore down) and one strange valence quark, the
nonzero behavior of the strange vector FF of kaons presented in
Fig.~\ref{figstrekaon} means that besides explicit contribution of the
strange valence quark into the kaon structure there are nonzero
contributions to the $K$-meson structure also from sea strange
quark $s \bar s$ pairs. However, we have no idea how to measure it
experimentally.

\setcounter{equation}{0} \setcounter{figure}{0} \setcounter{table}{0}\newpage
     \section{Polarization phenomena in electromagnetic interactions
     of hadrons}\label{VII}

    The polarization phenomena appear if beam particles or target
particles, or both are considered to be polarized. If colliding
particles are unpolarized, nevertheless final particles can be found
in polarized states.

    An investigation of the polarization phenomena is very important
in particle physics phenomenology as they provide more wealthy
information on the EM interactions of hadrons. Sometimes they reveal
new unexpected results. Just typical example is JLab proton
polarization data puzzle, discussed at  \ref{IV3}

   Prior to the year 2000 all data on proton EM FFs $G_{Ep}(t)$
and $G_{Mp}(t)$ in the space-like ($t<0$) region were obtained by
measuring  the differential cross-section of elastic scattering of
unpolarized electrons on unpolarized protons in the laboratory
system, utilizing the Rosenbluth technique. Both FFs manifested more
or less dipole behaviors and their ratio in error bars is
approximately equal one.

   Though polarization techniques have been suggested long time ago
\cite{Rekalo68}, only at the beginning of 21st century they were
used \cite{Jones00,Gayon02,Punjabi05} in obtaining of the ratio (see
Fig.~\ref{fig22})
\begin{equation}
\frac{G_{Ep}}{G_{Mp}}=-\frac{P_t}{P_l}\frac{(E+E^{'})}{2m_p}\tan({\theta/2}).
\label{a5}
\end{equation}
which revealed a non-dipole behavior of the proton electric FF
$G_{Ep}(t)$ leading to the JLab proton polarization data puzzle.

   Further we present some other useful exploiting of polarization
effects in particle physics phenomenology.

\medskip

     \subsection{Prediction of polarization observables in $e^+e^- \to p
     \bar p$ process}\label{VII1}

\medskip

   This paragraph is devoted to the analysis of polarization
effects \cite{Dubnickova96} in the process $e^+e^-\to p\bar p$
calculated in the framework of the one-photon exchange
approximation.

   The above-mentioned process is interesting as it has noticeable
polarization effects even if there are no polarized particles in the
initial state. The appearance of such polarization effects is due to
$G_{Ep}(t)$ and $G_{Mp}(t)$ being complex with non-zero relative
phase. On that account there are also nontrivial polarization
effects in the scattering of longitudinally polarized electrons on
unpolarized target.
\begin{figure}[tb]
\begin{center}
\epsfig{file=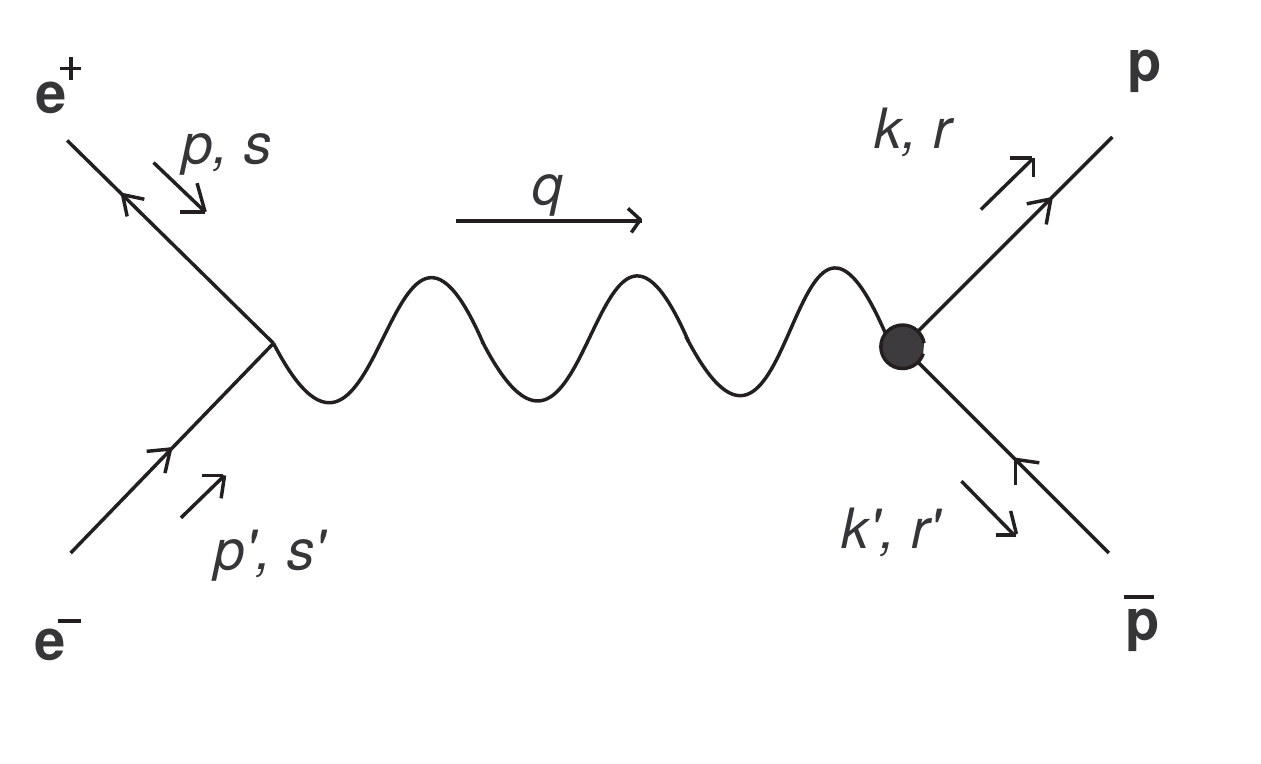,scale=0.5}
\end{center}\vspace{-0.5cm}
\caption{The one photon exchange diagram of the $e^+e^-\to p\bar p$
process.\label{diagram}}
\end{figure}

   The matrix element of the process $e^+e^-\to p\bar{p}$ in the
framework of the one-photon exchange approximation to be presented
in Fig.~\ref{diagram} is defined by the formulae
\begin{eqnarray}
M &=& \frac{e^2}{k^2} j_{\mu}J_{\mu},\nonumber \\
j_{\mu} &=& \bar u(-p)\gamma_{\mu} u(p'),\label{eqn:1}\\
J_{\mu} &=& \bar u(k)\bigl[F_{1p}(t)\gamma_{\mu} -
F_{2p}(t)\frac{\sigma_{\mu\nu}k_{\nu}}{2m_p}\bigr]u(-k'),\nonumber
\end{eqnarray}
where $t=q^2\ge 4m_p^2$. We note, that the c.m. system of the
reaction $e^+ e^-\to p\bar{p}$ is the most suitable for the analysis
of polarization effects.

   The EM currents $j_{\mu}$ and $J_{\mu}$ are conserved $q\cdot j=
q\cdot J=0$ and the matrix element $M$ is completely determined by
the product of spatial components of the currents $\vec{j}$ and
$\vec{J}$.

The electromagnetic current $\vec{J}$ can be expressed through
two-component spinors $\varphi_1$ and $\varphi_2$
\begin{equation}
\vec{J}=\sqrt{t}\varphi^+_1\left[G_{Mp}(t)(\vec{\sigma}-\vec{n}\vec{\sigma}\cdot\vec{n})+
\frac{2m_p}{\sqrt{t}}G_{Ep}(t)\vec{n}\vec{\sigma}\cdot\vec{n}\right]\varphi_2,
\label{eq:1}
\end{equation}
 where we denote
\begin{equation}
 \vec{F} =
\sqrt{t}\,[\,G_{Mp}(t)(\vec{\sigma} - \vec{n}\,
\vec{\sigma}.\vec{n}) + \frac{2m_p}{\sqrt{t}}\, G_{Ep}(t)\vec{n}\,
\vec{\sigma}.\vec{n}\,], \label{eq:2}
\end{equation}
 $\vec{\sigma}$  are  Pauli matrices,
 $\vec{n} = (0,0,1)$ is the unit vector along the three momentum $\vec{q}$ of
the proton, $\vec{m} = (-\sin\vartheta,0,\cos\vartheta)$ is the unit
vector of incoming electron (see Fig.~\ref{surad}) and $G_{Ep}(t)$ and
$G_{Mp}(t)$ are defined by the relations (\ref{d5}).

\begin{figure}[tb]
\begin{center}
\epsfig{file=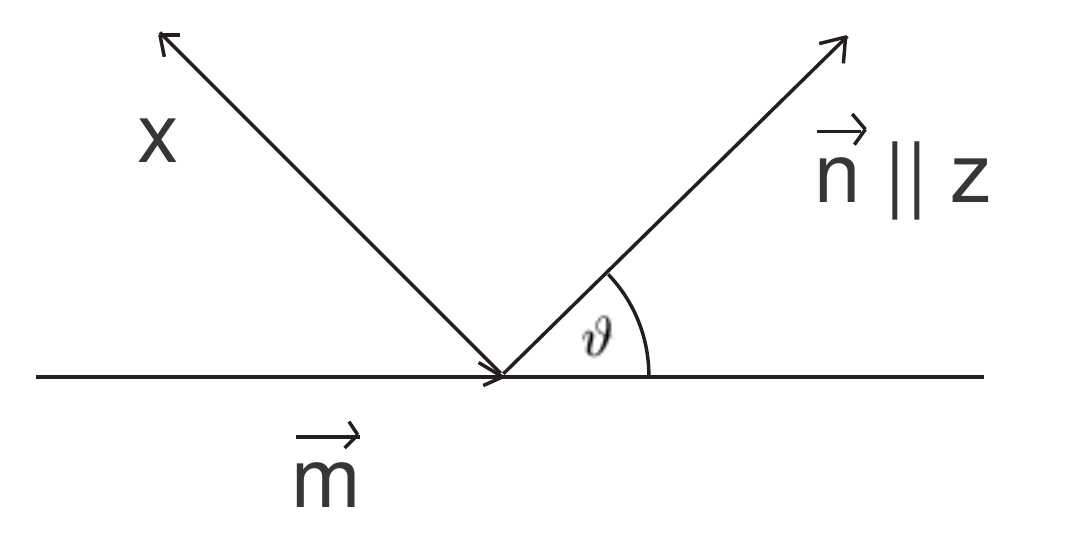,scale=0.5}
\end{center}
\caption{Definition of the scattering plane of the $e^+e^-\to p\bar
p$ process\label{surad}}
\end{figure}

  In order to find the corresponding cross-section, one has to
calculate $|M|^2$. The differential cross section of the reaction
$e^+e^-\to p\bar p$ in terms of EM FFs, for the case of unpolarized
particles, has the  form
\begin{equation}
\frac{d\sigma}{d\Omega}=\frac{\alpha^2}{4s}\left(\frac{1}{\tau}|G_{Ep}|^2\sin^2{\vartheta}
+ |G_{Mp}|^2[1+\cos^2{\vartheta}]\right),\;\quad
\alpha=\frac{e^2}{4\pi}; \quad \tau=\frac{t}{4m_p^2}.
\end{equation}

   Further single and double spin polarization observables are
calculated explicitly.

   First the single spin polarization observables (either for proton or for
antiproton) in the case of
\begin{itemize}
\item
 unpolarized incoming leptons
\item
 incoming electron to be longitudinally polarized.
\end{itemize}
are calculated.

   In the case of unpolarized initial leptons the corresponding
lepton tensor takes the form
\begin{equation}
j_{ij} = 2t( \delta_{ij} - m_i m_j)\label{eq:cur1}
\end{equation}
and the vector polarization is
\begin{equation}
\vec{P} =\frac{j_{ij} Tr[F_i F_j^\dag\vec{\sigma}]}{j_{ij} Tr[F_i
F_j^\dag]}. \label{eq:z2}
\end{equation}

   Calculating the corresponding trace in the numerator, and
similarly in the denominator, one obtains only $P_y$ component of
the vector polarization to be nonzero
\begin{eqnarray}
P_x &=& 0 \nonumber \\
 P_y &=& \frac{\sin2\vartheta \cdot Im[G^*_{Mp}(t)G_{Ep}(t)]}{\sqrt{\tau}[1/\tau|G_{Ep}|^2\sin^2\vartheta
+|G_{Mp}|^2(1+\cos^2{\vartheta})]}\label{eqn:2pol} \\
P_z &=& 0. \nonumber
\end{eqnarray}

 The y-axis is orthogonal to the scattering plane (see Fig.~\ref{surad}) defined by the
unit vectors $\vec{m}$ and  $\vec{n}$, along the three-momentum of
the electron and along the three-momentum of the created proton,
respectively.

The  contributions of $P_x$ and $P_z$ in proton polarization are
different from zero only if the electron (or positron) is
longitudinally polarized, i.e. the lepton tensor  takes the
following form
\begin{equation}
j_{ij} = 2t(\, \delta_{ij} - m_i m_j + \lambda
i\varepsilon_{ijl}m_l\, ). \label{eq:cur2}
\end{equation}

   Then the components of the vector polarization $\vec{P}$  of the
created proton (or antiproton) in the  reaction $e^+e^-\to p
\bar{p}$ are

\begin{eqnarray}
P_x&=&-\frac{2\sin\vartheta\cdot Re[G_{Ep}(t) G_{Mp}^*(t)]}
{\sqrt{\tau}[1/\tau |G_{Ep}(t)|^2\sin^2\theta + |G_{Mp}(t)|^2(1+\cos^2\theta)]} \label{eqn:pol3}\\
P_y &=& \frac{\sin{2\vartheta}\cdot Im[G^*_{Mp}(t)G_{Ep}(t)]}
{\sqrt\tau[1/\tau |G_{Ep}(t)|^2\sin^2\vartheta
+|G_{Mp}(t)|^2(1+\cos^2\vartheta)]}\nonumber \\
 P_z&=&\frac{2\cos\vartheta
|G_{Mp}(t)|^2}{[1/\tau |G_{Ep}(t)|^2\sin^2\theta +
|G_{Mp}(t)|^2(1+\cos^2\theta)]},\nonumber
\end{eqnarray}
assuming $100\%$ longitudinal polarization of one of the leptons.

\medskip

   In a similar procedure one can find explicit forms of double spin polarization observables
in the $e^+e^-\to p\bar{p}$ process, where we are interested for
polarizations of created proton and antiproton simultaneously. The
corresponding polarization tensor  is
\begin{eqnarray}
P_{kl} =\frac{j_{ij} Tr[F_i \sigma_k F_j^\dag \sigma_l]}{j_{ij}
Tr[F_i F_j^\dag]},\label{eq:poll1}
\end{eqnarray}
where $k,l= x,y,z$.
 Calculating the trace in the numerator and similarly the trace in the denominator,
considering unpolarized incoming leptons, one finds
\begin{eqnarray}
P_{xx}&=& \sin^2 \vartheta\cdot\frac{\tau|G_{Mp}(t)|^2
+|G_{Ep}(t)|^2}{{\tau}[1/\tau |G_{Ep}(t)|^2\sin^2\vartheta
+|G_{Mp}(t)|^2(1+\cos^2\vartheta)]};\nonumber \\
P_{yy}&=&\sin^2\vartheta\cdot\frac{ |G_{Ep}(t)|^2
-{\tau}|G_{Mp}(t)|^2}{\tau[1/\tau |G_{Ep}(t)|^2\sin^2\vartheta
+|G_{Mp}(t)|^2(1+\cos^2\vartheta)]};\label{eqn:pyy} \\
P_{zz}&=& \frac{\tau(1+\cos^2{\vartheta})|G_{Mp}(t)|^2
-sin^2{\vartheta}|G_{Ep}(t)|^2}{{\tau}[1/\tau
|G_{Ep}(t)|^2\sin^2\vartheta
+|G_{Mp}(t)|^2(1+\cos^2\vartheta)]};\nonumber \\
P_{xy}&=& P_{yx} = 0; \nonumber \\
P_{xz}&=& P_{zx} =-\frac{\sin2{\vartheta}
Re[G^*_{Mp}(t)G_{Ep}(t)]}{\sqrt{\tau}[1/\tau
|G_{Ep}(t)|^2\sin^2\vartheta
+|G_{Mp}(t)|^2(1+\cos^2\vartheta)]};\nonumber \\
P_{yz} &=& P_{zy} = 0.\nonumber
\end{eqnarray}

Now, if the expression (\ref{eq:cur2}) for the lepton tensor is used
with the longitudinally polarized electron or positron, for
components $P_{xx}$, $P_{yy}$, $P_{zz}$, $P_{xy}$, $P_{yx}$,
$P_{xz}$, $P_{zx}$ one obtains identical expressions with
(\ref{eqn:pyy}), but the last two components are now nonzero as well
\begin{equation}
P_{yz}= P_{zy} = \frac{\sin \vartheta
Im[G^*_{Mp}(t)G_{Ep}(t)]}{\sqrt{\tau}[1/\tau
|G_{Ep}(t)|^2\sin^2\vartheta
+|G_{Mp}(t)|^2(1+\cos^2\vartheta)]}.\label{eq:pyz}
\end{equation}

   Every of components $P_{kl}$ characterize a polarization of the proton $p$
in the direction  $k$, if antiproton $\bar{p}$ is polarized at the
direction  $l$ and they all are calculated for $100\%$ polarization
of one of the initial leptons.

   As one can see from explicit formulae the vector polarization
component $P_y$ and the tensor polarization components $P_{xz}
$=$P_{zx}$ depend on the imaginary parts of electric and magnetic
proton FFs. So, only our $U\&A$ models of the nucleon EM structure
can give a sophisticate prediction of behaviors of $P_y$ and $P_{xz}
$=$P_{zx}$. The eight and ten resonance U\&A models represent
compatible unification of pole and continua (in the language of the
analyticity  given by cuts on positive real axis) contributions,
which give just imaginary parts of FFs different from zero starting
always from the lowest branch point representing the lowest opened
physical threshold.

   Now, exploiting behaviors of proton electric and magnetic FFs in the
time-like region as predicted by our eight and ten-resonance $U\&A$
models one finds \cite{Dubnickova08} behaviors of single and double
spin polarization observables of the $e^+e^-\to p\bar p$ process as
they are presented in figures Fig.~\ref{vecpo}, \ref{tenpo}.

\begin{figure}[tb]
\centering
\psfig{file=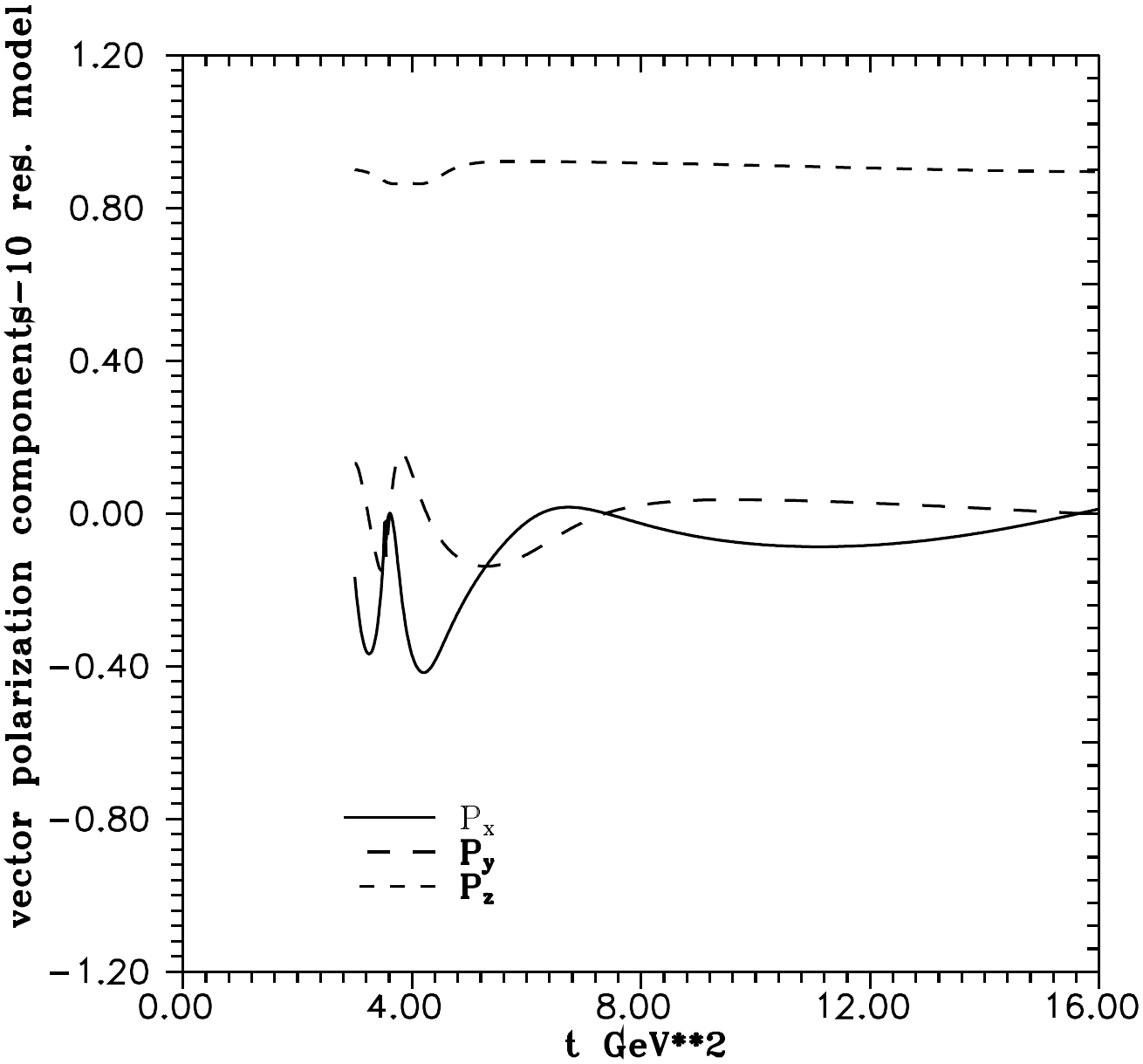,width=6.48cm}\ \
\psfig{file=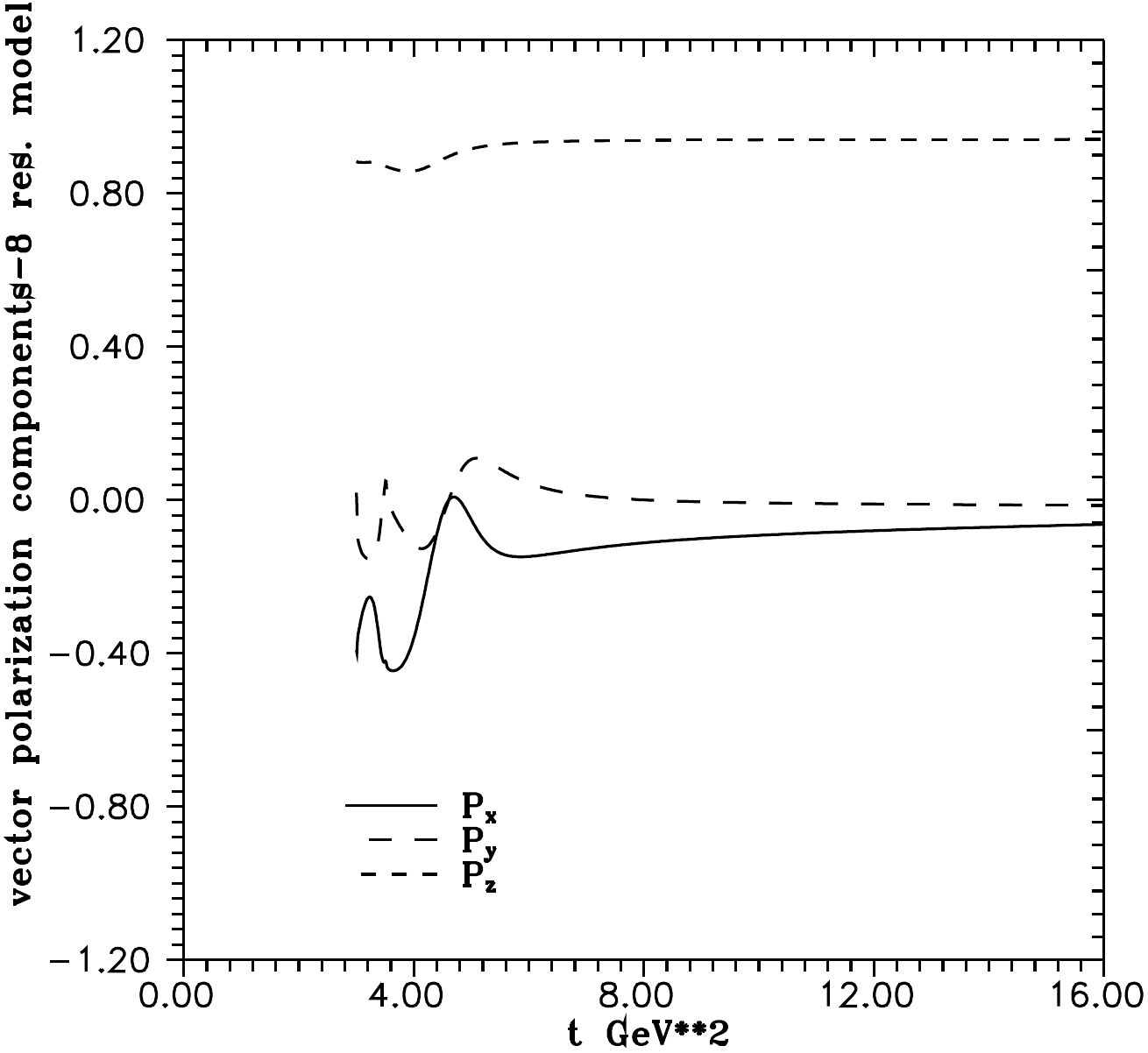,width=6.48cm}  \caption{Prediction of the
single polarizations observables by ten-resonance (left-hand) and
eight-resonance (right-hand) U\&A model\label{vecpo}}
\end{figure}
\begin{figure}[tb!]
\centering
\psfig{file=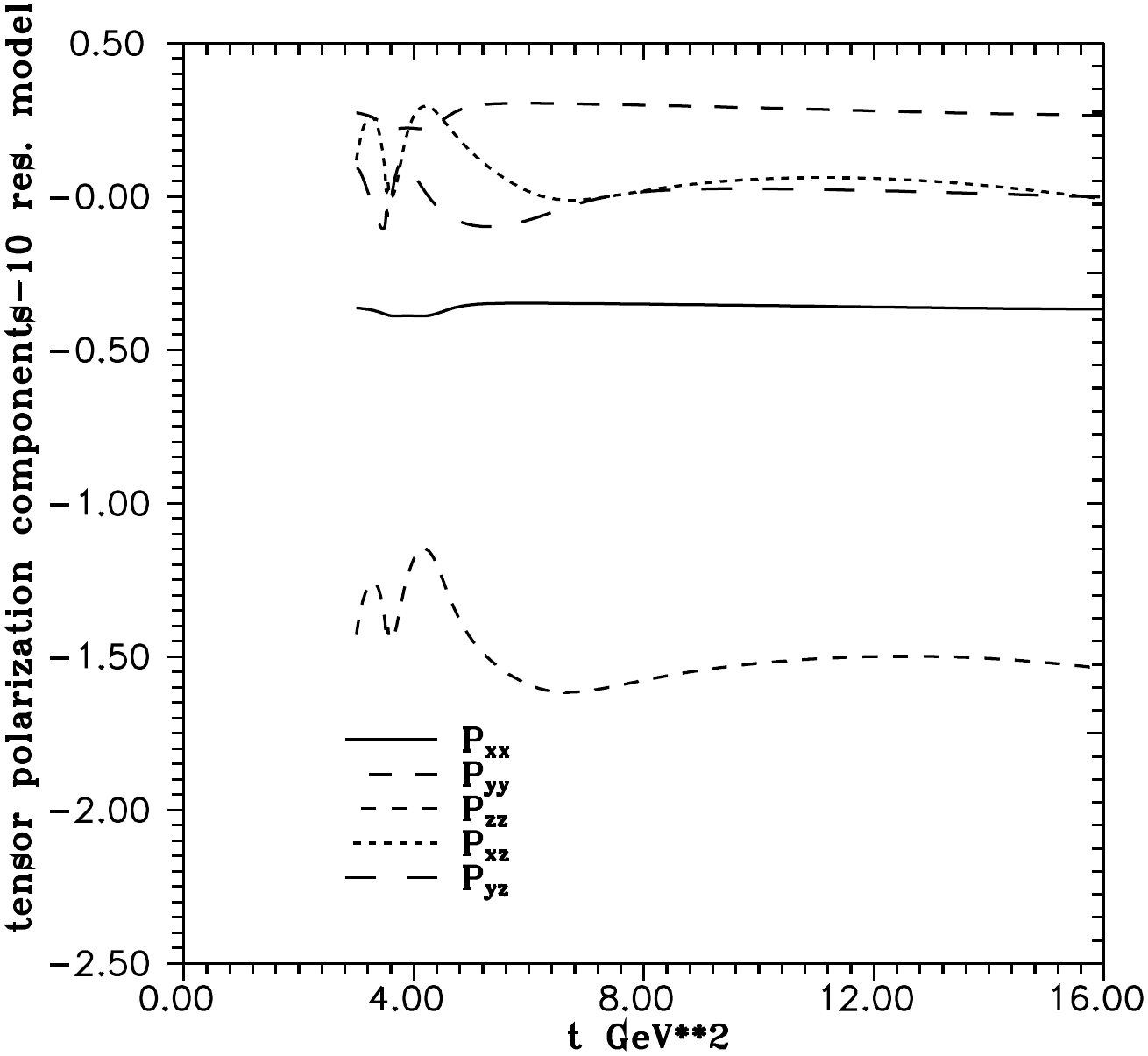,width=6.48cm}\ \
\psfig{file=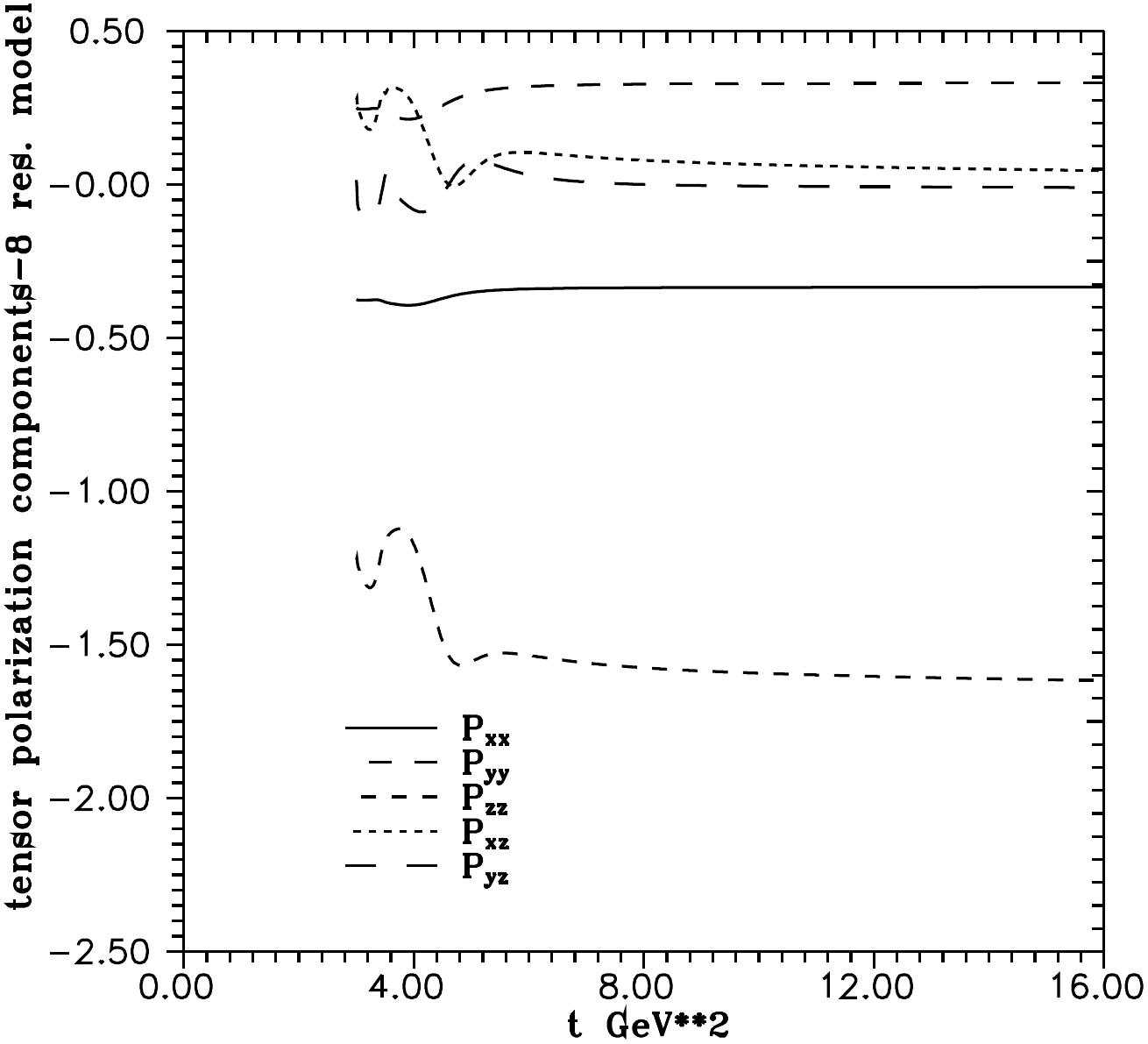,width=6.48cm}
\caption{Prediction of the double polarizations observables by
ten-resonance (left-hand) and eight-resonance (right-hand) U\&A
model\label{tenpo}}
\end{figure}

\medskip

     \subsection{Polarization effects in $e^+ e^-\rightarrow  d \bar d$
     and experimental determination of time like deuteron form
     factors}\label{VII2}

\medskip

   In the present paragraph the polarization observables in the
reaction
\begin{equation}\label{eq:eq1}
e^-(k_1)+e^+(k_2)\rightarrow d(p_1) + \bar d(p_2),
\end{equation}
where the momenta of the particles are indicated in brackets, are
calculated \cite{Gakh06} explicitly.

   With this aim one considers the case of unpolarized and longitudinally-polarized
electron beam with production of vector- and tensor-polarized
deuterons. The expressions of polarization observables are given in
terms of the deuteron EM FFs in time-like region, where they are
complex functions of the momentum transfer squared.

   In the one-photon approximation, the differential cross section
of the reaction (\ref{eq:eq1}) in terms of the leptonic $L_{\mu\nu}$
and hadronic $W_{\mu\nu}$ tensors contraction (in the Born
approximation one can neglect the electron mass) is written as
\begin{equation}\label{eq:eq2}
\frac{d\sigma}{d\Omega } = \frac{\alpha^2\beta }{4q^2}\frac{
L_{\mu\nu}W_{\mu\nu}}{q^4},
\end{equation}
where $\alpha=1/137$ is the electromagnetic constant, $\beta
=\sqrt{1-4m_d^2/q^2}$ is the deuteron velocity in the reaction
center of mass system (CMS), $m_d$ is the deuteron mass and $q$ is
the four momentum of the virtual photon, $q=k_1+k_2=p_1+p_2$ (note
that the cross section is not averaged over the spins of the initial
beams).

   The leptonic tensor (for the case of longitudinally polarized
electron beam) is
\begin{equation}\label{eq:eq3}
L_{\mu\nu}=-q^2g_{\mu\nu}+2(k_{1\mu}k_{2\nu}+k_{2\mu}k_{1\nu}) +
2i\lambda \varepsilon_{\mu\nu\sigma\rho}k_{1\sigma}k_{2\rho}\ ,
\end{equation}
where $\lambda$ is the degree of the beam polarization (further we
assume that the electron beam is completely polarized and
consequently $\lambda=1$).

   The hadronic tensor can be expressed via the deuteron
electromagnetic current $J_{\mu}$, describing the transition $\gamma
^*\rightarrow \bar dd$, as
\begin{equation}\label{eq:eq4}
W_{\mu\nu} = J_{\mu}J^*_{\nu}\ .
\end{equation}
As the deuteron is a spin-one nucleus, its electromagnetic current
is completely described by three FFs. Assuming the P and C
invariance of the hadron EM interaction this current can be
decomposed (\ref{sec215}) into three deuteron EM FFs $F_1$, $F_2$
and $F_3$, by means of which analogically to the nucleon Sachs EM
FFs $G_{Cd}$, $G_{Md}$ and $G_{Qd}$ (\ref{sec216}) are defined .

   When calculating the expression for the hadron tensor $W_{\mu\nu}$
in terms of the deuteron electromagnetic FFs, the spin-density
matrices of the deuteron and antideuteron to be defined by means of
deuteron polarization vectors $\xi$ and $\xi*$ are
\begin{eqnarray}
\xi_{1\mu}\xi^*_{1\nu}&=&-\left(g_{\mu\nu}-\frac{p_{1\mu}
p_{1\nu}}{m_d^2}\right) +\frac{3i}{2m_d}
 \varepsilon_{\mu\nu\rho\sigma}s_{\rho} p_{1\sigma}
+3Q_{\mu\nu} \label{eq:eq8} \\
\xi_{2\mu}\xi^*_{2\nu}&=&-\left(g_{\mu\nu}-\frac{p_{2\mu}
p_{2\nu}}{m_d^2}\right)\nonumber
\end{eqnarray}
if the deuteron polarization is measured and the antideuteron
polarization is not measured. Here $s_{\mu}$ and $Q_{\mu\nu}$ are
the deuteron polarization four vector and quadrupole tensor,
respectively. The four vector of the deuteron vector polarization
$s_{\mu}$ and the deuteron quadrupole-polarization tensor $Q_{\mu\nu}$
satisfy the following conditions
\begin{equation*}
 s^2=-1,\; sp_1=0,\; Q_{\mu\nu}=Q_{\nu\mu}, \quad Q_{\mu\mu}=0,\;
p_{1\mu}Q_{\mu\nu}=0.
\end{equation*}

   Taking into account Eqs. (\ref{eq:eq4}) and
(\ref{eq:eq8}), the hadronic tensor in the general case can be
written as the sum of three terms
\begin{equation}\label{eq:eq9}
W_{\mu\nu}=W_{\mu\nu}(0)+W_{\mu\nu}(V)+W_{\mu\nu}(T),
\end{equation}
where $W_{\mu\nu}(0)$ corresponds to the case of unpolarized
deuteron and $W_{\mu\nu}(V)$ $(W_{\mu\nu}(T))$ corresponds to the case
of the vector (tensor) polarized deuteron. The explicit form of
these terms is

\noindent\underline{ - the unpolarized term $W_{\mu\nu}(0)$:}
\begin{eqnarray}
W_{\mu\nu}(0)&=&W_1(q^2)\tilde{g}_{\mu\nu}+\frac{W_2(q^2)}{m_d^2}
\tilde{p}_{1\mu}\tilde{p}_{1\nu} \nonumber \\
\tilde{g}_{\mu\nu}&=&g_{\mu\nu}-\frac{q_{\mu}q_{\nu}}{q^2}\quad
\tilde{p}_{1\mu}=p_{1\mu}-\frac{p_1q}{q^2}q_{\mu} \nonumber\\
W_1(q^2)&=&8m_d^2\tau (1-\tau
)|G_{Md}|^2, \nonumber\\
W_2(q^2)&=&12m_d^2(|G_{Cd}|^2-\frac{2}{3}\tau
|G_{Md}|^2+\frac{8}{9}\tau ^2|G_{Qd}|^2)\label{eq:eq10, structure functions}
\end{eqnarray}
and $\tau = -q^2/{4m_d^2}$.

\noindent\underline{ - the term for vector polarization
$W_{\mu\nu}(V)$:}

\begin{eqnarray}\nonumber
W_{\mu\nu}(V)&=&\frac{i}{m_d}S_1(q^2)\varepsilon_{\mu\nu\sigma\rho}
s_{\sigma}q_{\rho}+ \frac{i}{m_d^3}S_2(q^2)[\tilde{p}_{1\mu}
\varepsilon_{\nu\alpha\sigma\rho}
s_{\alpha}q_{\sigma}p_{1\rho}-\tilde{p}_{1\nu}
\varepsilon_{\mu\alpha\sigma\rho} s_{\alpha}q_{\sigma}p_{1\rho}]+\\
&+&\frac{1}{m_d^3}S_3(q^2)[\tilde{p}_{1\mu}
\varepsilon_{\nu\alpha\sigma\rho}
s_{\alpha}q_{\sigma}p_{1\rho}+\tilde{p}_{1\nu}\varepsilon_{\mu\alpha\sigma\rho}
s_{\alpha}q_{\sigma}p_{1\rho}], \label{eq:eq11, structure functions} \\
S_1(q^2)&=&-3m_d^2(\tau -1)|G_{Md}|^2, \nonumber \\
S_2(q^2)&=&3m_d^2[|G_{Md}|^2-2Re(G_{Cd}-\frac{\tau
}{3}G_{Qd})G_{Md}^*], \nonumber \\
S_3(q^2)&=&6m_d^2Im(G_{Cd}-\frac{\tau }{3}G_{Qd})G_{Md}^*.\nonumber
\end{eqnarray}

\noindent\underline{ - the term for tensor polarization
$W_{\mu\nu}(T)$:}

\begin{eqnarray}\label{eq:eq12}
W_{\mu\nu}(T)&=&V_1(q^2)\bar Q\tilde{g}_{\mu\nu}+V_2(q^2)\frac{\bar
Q}{m_d^2} \tilde{p}_{1\mu}\tilde{p}_{1\nu}+\\
&+&V_3(q^2)(\tilde{p}_{1\mu}\widetilde{Q}_{
\nu}+\tilde{p}_{1\nu}\widetilde{Q}_{\mu})+
V_4(q^2)\widetilde{Q}_{\mu\nu}\ +
iV_5(q^2)(\tilde{p}_{1\mu}\widetilde{Q}_{
\nu}-\tilde{p}_{1\nu}\widetilde{Q}_{\mu}),\nonumber
\end{eqnarray}
 where
\begin{eqnarray}\nonumber
\widetilde{Q}_{\mu}&=&Q_{\mu\nu}q_{\nu}-\frac{q_{\mu}}{q^2}\bar
{Q} \, \widetilde{Q}_{\mu}q_{\mu}=0  \nonumber \\
\widetilde{Q}_{\mu\nu}&=& Q_{\mu\nu}+\frac{q_{\mu}q_{\nu}}{q^4}\bar Q-
\frac{q_{\nu}q_{\alpha}}{q^2}Q_{\mu\alpha}-
\frac{q_{\mu}q_{\alpha}}{q^2}Q_{\nu\alpha}\,
\widetilde{Q}_{\mu\nu}q_{\nu} = 0, \nonumber\\
\bar{Q}&=&Q_{\alpha\beta}q_{\alpha}q_{\beta}.\label{eq:eq13}
\end{eqnarray}

The tensor structure functions $V_i(q^2)$ are combinations of
deuteron FFs as follows
\begin{eqnarray}
V_1(q^2)&=&-3|G_{Md}|^2, \nonumber\\
V_2(q^2)&=&3\left [|G_{Md}|^2+\frac{4}{1-\tau
}Re(G_{Cd}-\frac{\tau }{3}G_{Qd} -\tau G_{Md})G_{Qd}^*\right ], \nonumber\\
V_3(q^2)&=&-6\tau \left [|G_{Md}|^2+2ReG_{Qd}G_{Md}^* \right ], \label{eq:eq14}\\
V_4(q^2)&=&-12m_d^2\tau (1-\tau )|G_{Md}|^2, \quad V_5(q^2)=-12\tau Im
(G_{Qd}G_{Md}^*).\nonumber
\end{eqnarray}
   Using the definitions of the cross--section (\ref{eq:eq2}),
leptonic (\ref{eq:eq3}) and hadronic (\ref{eq:eq9}) tensors, one can
easily derive the expression for the unpolarized differential cross
section in terms of the structure functions $W_{1,2}$ (after
averaging over the spins of the initial particles)
\begin{equation}\label{eq:eq15}
\frac{d\sigma^{un}}{d\Omega }=\frac{\alpha^2\beta }{4q^4} \left \{
-W_1(q^2)+\frac{1}{2}W_2(q^2) \left [\tau
-1-\frac{(u-t)^2}{4m_d^2q^2}\right ]\right \}\ ,
\end{equation}
where $t=(k_1-p_1)^2$, $ u=(k_1-p_2)^2.$

   In the reaction CMS this expression can be written as
\begin{equation}\label{eq:eq16}
\frac{d\sigma^{un}}{d\Omega }=\frac{\alpha^2\beta ^3}{4q^2}D, \
D=\tau (1+\cos^2\theta )|G_{Md}|^2+\frac{3}{2}\sin^2\theta \left
(|G_{Cd}|^2+ \frac{8}{9}\tau ^2|G_{Qd}|^2 \right ),
\end{equation}
where $\theta $ is the angle between the momenta of the deuteron
(${\vec p}$) and the electron beam (${\vec k}$). Integrating the
expression (\ref{eq:eq16}) with respect to the deuteron angular
variables one obtains the following formula for the total cross
section of the reaction (\ref{eq:eq1})
\begin{equation}\label{eq:i1}
\sigma_{tot}(e^+e^-\to \bar dd)= \frac{\pi\alpha
^2\beta^3}{3q^2}\biggl [3|G_{Cd}|^2+4\tau
(|G_{Md}|^2+\frac{2}{3}\tau |G_{Qd}|^2)\biggr ].
\end{equation}

   One can define also an angular asymmetry, $R$, with respect to the
differential cross section measured at $\theta =\pi /2, \sigma_0$
\begin{equation}\label{eq:i2}
\frac{d\sigma^{un}}{d\Omega }=\sigma_0(1+Rcos^2\theta ),
\end{equation}
where $R$ can be expressed as a function of the deuteron FFs
\begin{equation}\label{eq:i3}
R=\frac{2\tau (|G_{M}|^2-\frac{4}{3}\tau |G_{Q}|^2)-3|G_{C}|^2}
{2\tau (|G_{M}|^2+\frac{4}{3}\tau |G_{Q}|^2)+3|G_{C}|^2}.
\end{equation}
This observable should be sensitive to the different underlying
assumptions on deuteron FFs; therefore, a precise measurement of
this quantity, which does not require polarized particles, would be
very interesting.

   One can see from (\ref{eq:eq16}) that, as in the space-like
region, the measurement of the angular distribution of the outgoing
deuteron determines the modulus of the magnetic FF. The separation
of the charge and quadrupole FFs requires the measurement of
polarization observables to be more convenient derived in  CMS.
When considering the polarization of the final
particle, we choose a reference system with the  $z$ axis along the
momentum of this particle (in our case it is ${\vec p}$). The $y$
axis is normal to the reaction plane in the direction of ${\vec
k}\times {\vec p}$, $x$, $y$ and $z$ form a right-handed coordinate
system.

   The cross section can be written, in the general case, as the sum
of unpolarized and polarized terms, corresponding to the different
polarization states and polarization directions of the incident and
scattered particles
\begin{equation}\label{eq:eq21}
\displaystyle\frac{d\sigma}{d\Omega}=
\displaystyle\frac{d\sigma^{un}}{d\Omega} \left [1+P_y+\lambda
P_x+\lambda P_z+ P_{zz}R_{zz}+ P_{xz}R_{xz}+P_{xx}(R_{xx}-R_{yy})
+\lambda P_{yz}R_{yz}\right ],
\end{equation}
where $P_i$ ($P_{ij}$), $i,j=x,y,z$ are the components of the
polarization vector (tensor) of the outgoing deuteron, $R_{ij}$,
$i,j=x,y,z$ the components of the quadrupole polarization tensor of
the outgoing deuteron $Q_{\mu\nu}$, in its rest system and
$\displaystyle\frac{d\sigma^{un}}{d\Omega}$ is the differential
cross-section for the unpolarized case.

   The degree of longitudinal polarization of the electron beam,
$\lambda$, is explicitly indicated, in order to stress the origin of
the specific polarization observables.

  Now, let us consider the different polarization observables and give
their expression in terms of the deuteron EM FFs.

\begin{itemize}
\item

The vector polarization of the outgoing deuteron, $P_y$, which does
not require polarization in the initial state is
\begin{equation}\label{eq:eq17}
P_y=-\frac{3}{2}\sqrt{\tau }\sin(2\theta )Im \left [\left
(G_{Cd}-\frac{\tau }{3}G_{Qd} \right ) G_{Md}^*\right ]
\end{equation}

\item
The part of the differential cross section that depends on the
tensor polarization can be written as follows
\begin{eqnarray}
&\displaystyle\frac{d\sigma_T}{d\Omega}&=
\frac{d\sigma_{zz}}{d\Omega}R_{zz}+\frac{d\sigma_{xz}}{d\Omega}R_{xz}+
\frac{d\sigma_{xx}}{d\Omega}(R_{xx}-R_{yy}),
\nonumber \\
&\displaystyle\frac{d\sigma_{zz}}{d\Omega}&=\frac{\alpha ^2\beta
^3}{4q^2} \frac{3\tau }{4}\left [(1+\cos^2\theta
)|G_{Md}|^2+8\sin^2\theta \left (
\frac{\tau }{3}|G_{Qd}|^2-Re(G_{Cd}.G_{Qd}^*)\right ) \right], \nonumber \\
&\displaystyle\frac{d\sigma_{xz}}{d\Omega}&=-\frac{\alpha ^2\beta
^3}{4q^2} 3\tau ^{3/2}\sin(2\theta )Re(G_{Qd}.G_{Md}^*),
\nonumber \\
&\displaystyle\frac{d\sigma_{xx}}{d\Omega}&=-\frac{\alpha ^2\beta
^3}{4q^2} \frac{3\tau }{4}\sin^2\theta |G_{Md}|^2,\label{crosssec}
\end{eqnarray}

\item
Let us consider now the case of a longitudinally polarized electron
beam. The other two components of the deuteron vector polarization
($P_x$, $P_z$) require the initial particle polarization and are
\begin{equation}\label{eq:eq20}
P_x=-3\frac{\sqrt{\tau }}{D}\sin\theta Re \left (G_{Cd}-\frac{\tau
}{3}G_{Qd} \right )G_{Md}^*, \ P_z=\frac{3\tau }{2D}\cos\theta
|G_{Md}|^2.
\end{equation}
\end{itemize}

   From angular momentum and helicity conservations it follows that
the sign of the deuteron polarization component $P_z$ in the forward
direction ($\theta =~0$) must coincide with the sign of the electron
beam polarization. This requirement is satisfied by Eq.
(\ref{eq:eq20}).

   A possible nonzero phase difference between the deuteron FFs leads
to another T-odd polarization observable proportional to the
$R_{yz}$ component of the tensor polarization of the deuteron. The
part of the differential cross section that depends on the
correlation between the longitudinal polarization of the electron
beam and the deuteron tensor polarization can be written as follows
\begin{equation}\label{eq:eq21a}
\frac{d\sigma_{\lambda T}}{d\Omega}=\frac{\alpha ^2\beta ^3}{4q^2}
6\tau ^{3/2}\sin\theta Im(G_{Md}.G_{Qd}^*)R_{yz}.
\end{equation}

   The deuteron FFs in the time-like region are complex functions. In the
case of unpolarized initial and final particles, the differential
cross section depends only on the squared modulus $|G_{Md}|^2$ and
on the combination $G = |G_{Cd}|^2+\frac{8}{9}\tau ^2|G_{Qd}|^2.$
So, the measurement of the angular distribution allows one to
determine $|G_{Md}|$ and the quantity $G$, as in the elastic
electron-deuteron scattering.

   Let us discuss, which information can be obtained by measuring the
polarization observables derived above. Three relative phases exist
for three deuteron EM FFs, which we note as follows:
$\alpha_1=\alpha_M-\alpha_Q,$ $\alpha_2=\alpha_M-\alpha_C,$ and
$\alpha_3=\alpha_Q-\alpha_C,$ where $\alpha_M=ArgG_{Md},$
$\alpha_C=ArgG_{Cd},$ and $\alpha_Q=ArgG_{Qd}.$ These phases are
important characteristics of FFs in the time-like region since they
result from the strong interaction between final particles.

   Let us consider the ratio of the polarizations $P_{yz}$ (we
would like to note that it requires a longitudinally polarized
electron beam) and $P_{xz}$ (when the electron beam is unpolarized).
One finds
\begin{equation}\label{eq:eq22}
R_1=\frac{P_{xz}}{P_{yz}}=-\cos\theta \cot\alpha_1.
\end{equation}
So, the measurement of this ratio gives us information about the
relative phase $\alpha_1$. The measurement of another ratio of
polarizations, $R_2=P_{xz}/P_{xx}$ gives us information about the
quantity $|G_{Qd}|$
\begin{equation}\label{eq:eq23}
R_2=\frac{P_{xz}}{P_{xx}}=8\sqrt{\tau }\cot\theta
\cos\alpha_1\frac{|G_{Qd}|}{|G_{Md}|}.
\end{equation}
This allows one to obtain the modulus of the charge FF, $|G_{Cd}|$,
from the quantity $G$, known from the measurement of the
differential cross section. The measurement of a third ratio
\begin{equation}\label{eq:eq24}
R_3=\frac{P_{y}}{P_{x}}=-\cos\theta
\frac{\sin\alpha_2-r\sin\alpha_1} {\cos\alpha_2-r\cos\alpha_1}, \ \
r=\frac{\tau }{3}\frac{|G_{Qd}|}{|G_{Cd}|}
\end{equation}
allows to determine the phase difference $\alpha_2$. And at last, if
we measure the ratio of the polarizations $P_{zz}$ and $P_{xx}$
\begin{equation}\label{eq:eq25}
R_4=\frac{P_{zz}}{P_{xx}}=-\frac{1}{\sin^2\theta} \left
[1+\cos^2\theta +8\sin^2\theta
\frac{|G_{Cd}||G_{Qd}|}{|G_{Md}|^2}(r-\cos\alpha_3) \right ]
\end{equation}
we can obtain information about the third phase difference
$\alpha_3$. Moreover, one can verify the relation:
$$\alpha_3=\alpha_2-\alpha_1.$$

   Thus, the measurement of the polarization observables in the
process (\ref{eq:eq1}) allows to determine the deuteron EM FFs in the
time-like region for the firs time.

\medskip

     \subsection{Alternative method of experimental determination of deuteron
     electromagnetic form factors}\label{VII3}

\medskip

   In \ref{IV3} we have demonstrated that alternative
method of a measurement of the same physical quantity, in this case
the proton electric FF $G_{Ep}(t)$, can lead to very surprising
results. To be motivated by means of this example, here we propose
\cite{Adamuscin09} to experimentalists a new method of experimental
measurement of all three deuteron EM FFs in the space-like region
for the first time.

   Knowing experimentally the FFs  (\ref{sec215}) $F_1(t)$, $F_2(t)$ and
$F_3(t)$, one can specify experimental behaviors of $G_C(t)$,
$G_M(t)$ and $G_Q(t)$  by (\ref{sec216}).

   The FF $F_2(t)$ can be found by a measurement of the unpolarized
differential cross-section in the laboratory system
\begin{equation}
\frac{d\sigma_{unp}^{Lab} }{d\Omega }=\frac{1}{64 \pi^2 m_d^2} \bigg
[\frac{1}{1+(2E_e/m_d) \sin^2(\theta/2)}\bigg ]
^2|\mathcal{M}|_{\text{unp}}^2 \label{difcs}
\end{equation}
with

\begin{equation}
|\mathcal{M}|_{\text{unp}}^2=\frac{e^4 m_d^2}{E^{2}_{e}}\frac{\cos
^{2}(\theta /2)}{\sin^{4}(\theta /2)} \bigg (1+(2E_e/m_d)\sin
^{2}(\theta /2)\bigg )\bigg [A(t)+B(t)\tan ^{2}(\theta /2)\bigg ],
\label{ampln}
\end{equation}
and deuteron elastic structure functions (\ref{deutelsf}).

   In order to find $F_1(t)$, we suggest to measure elastic
electron-deuteron scattering with only vector polarized deuterons,
where the incoming (target) deuteron is polarized in the direction
$\vec{n}$ of the three momenta of outgoing deuteron and the vector
polarization of the outgoing deuteron has the same direction. Then
in the laboratory system (the rest frame of the incoming deuteron)
the four vectors of electron and deuteron momenta ($k_1$, $p_1$,
$p_2$) and the deuteron vector polarizations $\xi_1$, $\xi_2$ can be
expressed in components as
\begin{eqnarray}
k_1=(E_e,0,0,E_e) ; \quad \quad  p_1&=&(m_d,0,0,0); \quad  \quad p_2=(E_2, |\vec p_2| \vec n);\nonumber \\
\xi_1&=& (0,\vec n) ; \quad \quad \xi_2=\frac{1}{m_d}(|\vec p_2|,E_2 \vec n) ;\nonumber \\
|\vec p_2|&=&2m_d \sqrt{\eta(1+\eta)} ; \quad \quad
E_2=m_d(1+2\eta), \label{cmpn}
\end{eqnarray}
where the outgoing deuteron vector polarization was obtained by
Lorentz boost to be the kinematic frame of the recoil deuteron. As a
result one obtains the following relations
\begin{eqnarray}
(\xi_1 \cdot \xi_2)&=& -1-2\eta ; \quad \quad (\xi_1 \cdot p_2) = -|\vec p_2| ; \quad \quad (\xi_2\cdot p_1) = |\vec p_2|; \nonumber\\
(\xi_1 \cdot k_1) (\xi_2 \cdot p_1)&=&-2 m_d (E_e+m_d)\eta;\nonumber \\
(\xi_1\cdot k_1)(\xi_2\cdot k_1)&=&\frac{\eta}{1+\eta} (E_e+m_d)[m_d(1+2\eta)-E_e]; \nonumber\\
(p_1\cdot k_1)&=&\quad E_e m_d ; \quad \quad (p_1\cdot p_2) =
m_d^2(1+2\eta) \label{kinmt}
\end{eqnarray}
to be useful in further calculations.

   The absolute value squared for the amplitude of the polarized
elastic electron-deuteron scattering is
\begin{eqnarray}
 |\mathcal{M}| _{\uparrow \uparrow}^2&=&\frac{e^4}{q^4}(2k_{1\mu}k_{2\nu}+2k_{1\nu}k_{2\mu}+
 g_{\mu\nu}q^2)\sum_{spins} \bigg\{ \Big [-F_1 (\xi_2^* \cdot \xi_1)+ \nonumber \\
 &+& F_3 \frac{(\xi_2^*\cdot q)( \xi_1 \cdot q)}{2 m_d^2}\Big ](p_1^{\mu}+p_2^{\mu})
  -F_2\Big [\xi_1^{\mu} (\xi_2^* \cdot q) - \xi_2^{*\mu}(\xi_1 \cdot q)\Big ]\bigg\}\times \nonumber \\
 &\times& \bigg \{\Big [-F_1 (\xi_2 \cdot \xi_1^*) +
 F_3 \frac{(\xi_2 \cdot q)( \xi_1^* \cdot q)}{2 m_d^2}\Big ](p_1^{\nu}+p_2^{\nu})- \nonumber \\
&-& G_2\Big [\xi_1^{*\nu} (\xi_2 \cdot q) - \xi_2^{\mu}(\xi_1^*
\cdot q)\Big ]\bigg\},\label{amplit}
\end{eqnarray}
then by using properties of the deuteron polarization vectors
$\xi_1$ , $\xi_2$, (\ref{cmpn}) and (\ref{kinmt}), it can be
arranged by the computer program FORM and simplified by Maxima to
the form
\begin{equation}
|\mathcal{M}|_{\uparrow \uparrow}^2
=\frac{1}{3}\frac{e^4m^2_d}{E_e^2}\frac{\cos ^{2}(\theta
/2)}{\sin^{4}(\theta /2)} \bigg (1+(2E_e/m_d)\sin ^{2}(\theta
/2)\bigg )\bigg [A(t)+B(t)\tan ^{2}(\theta /2)+ \frac{3}{2}F^2_1(t)
\bigg ]. \label{amplit}
\end{equation}

   In the latter equation one can identify the absolute value squared
of the unpolarized matrix element (\ref{ampln}) and as a result one
obtains
\begin{equation}
|\mathcal{M}|_{\uparrow \uparrow}^2 =
\frac{1}{3}|\mathcal{M}|_{\text{unp}}^2 +
\frac{e^4m_d^2}{2E^2_e}\frac{\cos^2(\theta/2)}{\sin^4(\theta/2)}\bigg
(1+(2E_e/{m_d})\sin^2(\theta/2)\bigg )F_1^2(t).\label{amppol}
\end{equation}
Multiplying the relation (\ref{amppol}) by
\begin{equation*}
\frac{1}{64\pi^2m_d^2} \bigg
(\frac{1}{1+(2{E_e}/{m_d})\sin^2(\theta/2)}\bigg )^2
\end{equation*}
one comes to the equation
\begin{equation}\label{difcsect}
 \frac{d\sigma_{\uparrow
\uparrow}^{Lab}}{d\Omega} -
\frac{1}{3}\frac{d\sigma^{Lab}_{unp}}{d\Omega} =
\frac{\alpha^2}{8E^2_e}\frac{ \cos^2(\theta/2)}{\sin^4 (\theta/2)}
\frac{1}{1+(2{E_e}/{m_d})\sin^2(\theta/2)} F_1^2(t).
\end{equation}
from which the EM FF $F_1(t)$ can be extracted.

 The third FF $F_3(t)$ is given by a solution of the quadratic equation
\begin{eqnarray}
&&\bigg [\frac{12}{9}\eta^2(1+ \eta^2\bigg ]F_3^2(t) +
\label{eq:3ffdeut} \\
&+& \bigg [\frac{4}{3}\eta(1+\eta)F_1(t) +
\frac{16}{9}\eta^2(1+\eta)(F_1(t) - F_2(t))\bigg ]\cdot F_3(t) +
\nonumber
\\
&+&\bigg\{\Big [F_1(t) + \frac{2}{3}\eta (F_1(t)- F_2(t))\Big ]^2
+\frac{2}{3}\eta F_2^2(t) + \frac{8}{9}\eta^2(F_1(t)-F_2(t))^2-
A(t)\bigg\}=0 \nonumber
\end{eqnarray}
(with $\eta=t/{4m_d^2}$) to be obtained by a substitution of
(\ref{sec216}) into the first relation of (\ref{deutelsf}).

Ones the data on $F_1(t)$, $F_2(t)$ and $F_3(t)$ are known, by using
(\ref{sec216}) one can obtain the experimental behavior of $G_C(t)$,
$G_M(t)$ and $G_Q(t)$.

\setcounter{equation}{0} \setcounter{figure}{0} \setcounter{table}{0}\newpage
     \section{Muon anomalous magnetic moment}\label{VIII}

   The muon is described by the Dirac equation and its magnetic
moment is related to the spin by means of the expression
\begin{equation}
  \vec{\mu}=g\left({e\over 2m_\mu}\right)\vec{s} \label{z1}
\end{equation}
where the value of gyromagnetic ratio $g$ is predicted (in the
absence of the Pauli term) to be exactly 2.

   However, interactions existing in nature modify $g$
to be exceeding the \mbox{value 2} because of the emission and
absorption of

\begin{itemize}
\item
 virtual photons (electromagnetic effects),
\item
 intermediate vector and Higgs bosons (weak interaction effects)
\item
 vacuum polarization into virtual hadronic states (strong interaction
effects).
\end{itemize}

In order to describe this modification of $g$ theoretically, the
magnetic anomaly was introduced by the relation

\begin{eqnarray}
  a_\mu\equiv \frac{g-2}{2}&=&a_\mu^{(1)}\left({\alpha\over\pi}
  \right) + \left(a_\mu^{(2)QED}+a_\mu^{(2)had}\right)
  \left({\alpha\over\pi}\right)^2 +  \nonumber \\
  &+& a_\mu^{(2)weak} + O\left({\alpha\over\pi}\right)^3 \label{z2}
\end{eqnarray}
where to every order some Feynman diagrams (see Figs.
\ref{fig:1}-\ref{fig:3}) correspond and $$\alpha=1/137.03599976(50)$$
is the fine structure constant.

\begin{figure}[b]
\begin{center}
\includegraphics[scale=0.6]{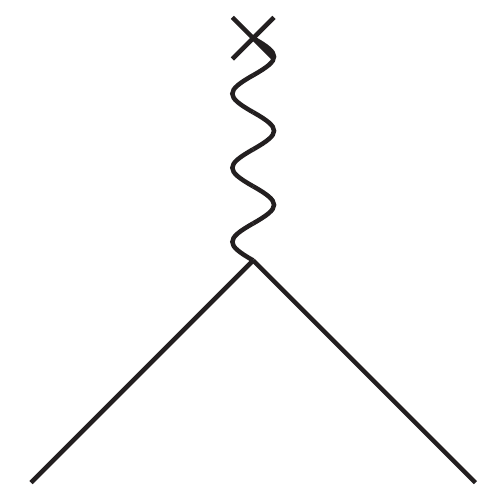}
\caption{The simplest Feynman diagram of an interaction of the muon
with an external magnetic field.} \label{fig:1}
\end{center}
\end{figure}

\begin{figure}[tb]
\begin{center}
\includegraphics[scale=0.6,clip]{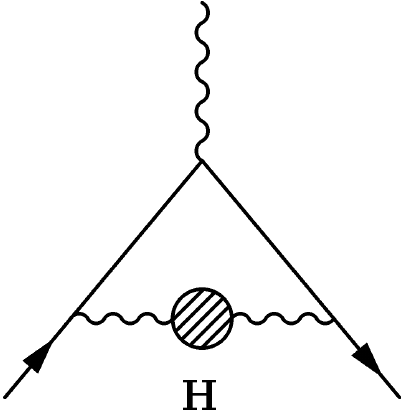}
\caption{The lowest-order hadronic vacuum-polarization contribution
to the anomalous magnetic moment of the muon.} \label{fig:2}
\end{center}
\end{figure}
\begin{figure}[tb!h]
\begin{center}
\includegraphics[scale=0.6,clip]{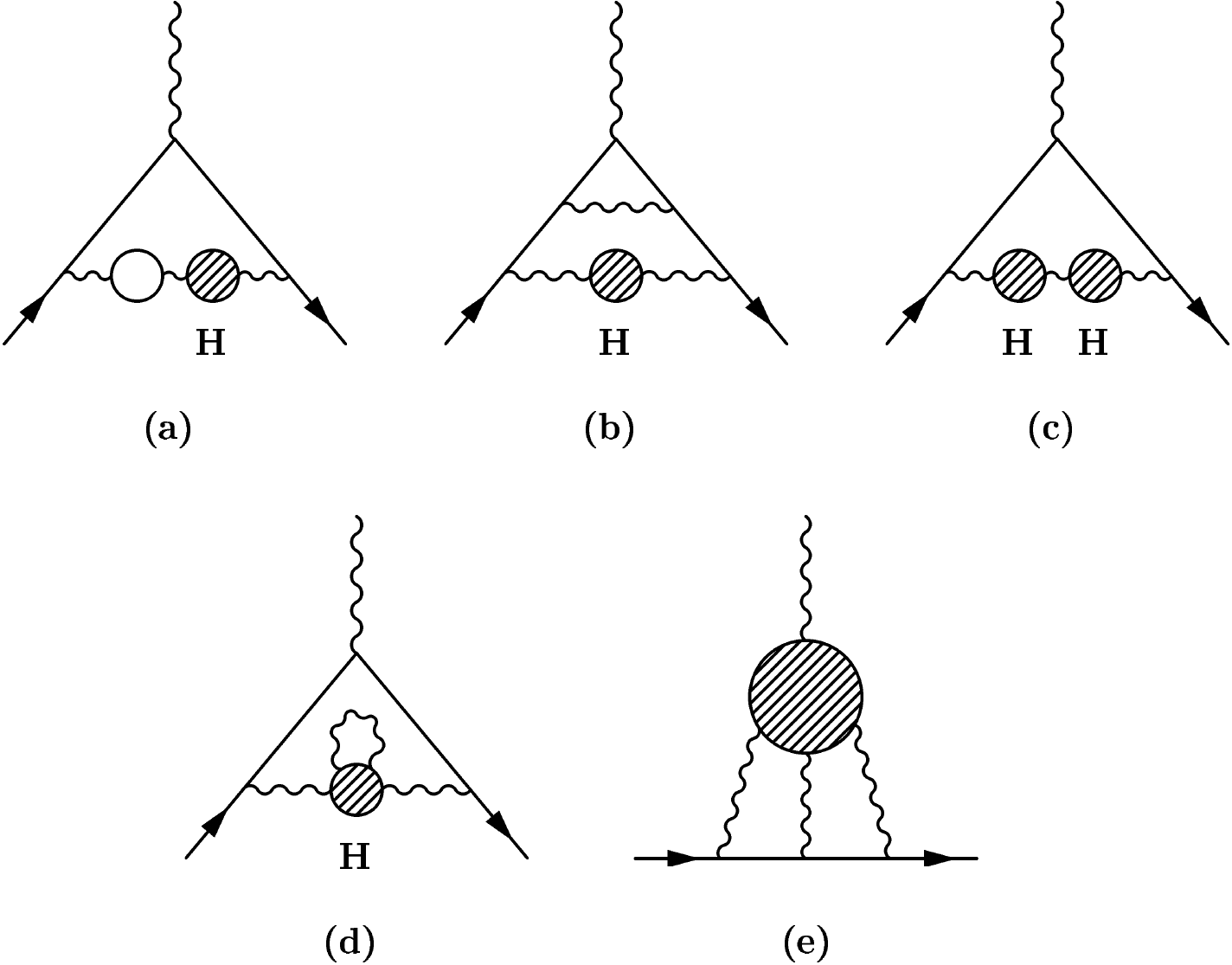}
\caption{The third-order hadronic vacuum-polarization contributions
to the anomalous magnetic moment of the muon.} \label{fig:3}
\end{center}
\end{figure}

   The muon anomalous magnetic moment $a_{\mu}$ is
very interesting object for theoretical investigations due to the
following reasons

\begin{enumerate}
  \item[$\left.i\right)$] it is one of the best measured
   quantities (BNL E-821 experiment) in physics \cite{Bennet02}
\begin{equation}
  a_\mu^{exp}=(116 592 040 \pm 86)\times 10^{-11} \label{z3}
\end{equation}

  \item[$\left.ii\right)$] its accurate theoretical evaluation
  provides an extremely clean test of "Electroweak theory" and may
  give hints on possible deviations from Standard Model (SM)

  \item[$\left.iii\right)$] moreover, in near future the
  measurement in BNL is expected to be performed yet with an improved
  accuracy
\begin{equation}
  \Delta a_\mu^{exp}=\pm 40\times 10^{-11}\label{z4}
\end{equation}

i.e. it is aimed at obtaining a factor 2 in a precision above that
of the last E-821 measurements.
\end{enumerate}

   At the aimed level of the precision (\ref{z4}) a sensibility will
already exist to contributions
\begin{equation}
  a_\mu^{(2,3)weak}=(152 \pm 4)\times 10^{-11},\label{z5}
\end{equation}
arising from single- and two-loop weak interaction diagrams.
   And so, if we compare theoretical evaluations of
 QED contributions up to 8th order
\begin{center}
   $a_{\mu}^{QED}=(116 584 705.7 \pm 2.9)\times10^{-11}$\cite{Hughes99}
\end{center}
the single- and two-loop weak contributions
\begin{center}
  $a_\mu^{(2,3)weak}=(151 \pm 4)\times 10^{-11}$\cite{Czarnecki99}\\

  $a_\mu^{(2,3)weak}=(153 \pm 3)\times 10^{-11}$\cite{Degrassi98}\\

  $a_\mu^{(2,3)weak}=(152 \pm 1)\times 10^{-11}$\cite{Knecht02}\\
\end{center}
strong interaction contributions

\begin{center}
  $a_\mu^{had}=(7068 \pm 172)\times 10^{-11}$\cite{Kinoshita85}\\

  $a_\mu^{had}=(7100 \pm 116)\times 10^{-11}$\cite{Casas85}\\

  $a_\mu^{had}=(7052 \pm 76)\times 10^{-11}$\cite{Martinovic90}\\

  $a_\mu^{had}=(7024 \pm 152)\times 10^{-11}$\cite{Eidelman95}\\

  $a_\mu^{had}=(7021 \pm 76)\times 10^{-11}$\cite{Narison01}\\
\end{center}
it is straightforward to see that the largest uncertainty is in
$a_{\mu}^{had}$.

   Error is comparable, or in the best case two times
smaller than the weak interaction contributions.

   So, in order to test the SM predictions for
$a_{\mu}$ and to look for new physics in comparison with BNL E-821
experiment, one has still to improve an evaluation of
$a_{\mu}^{had}$ and its error.

   The most critical from all hadronic contributions are the
light-by-light (LBL) contributions  (see Fig.~\ref{fig:4}) to be
approximated by a sum of meson pole terms (see Fig.~\ref{fig:5}).

\begin{figure}[tb]
\begin{center}
\includegraphics[scale=0.6]{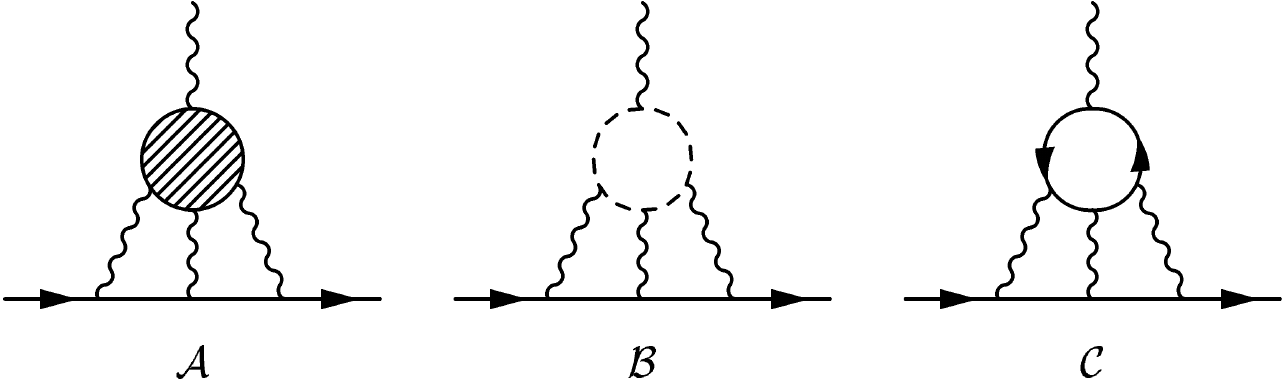}
\caption{Third order hadronic light-by-light scattering contribution
to $a_\mu^{had}$ ($\mathcal{A}$) and class of pseudoscalar meson
square loop diagrams ($\mathcal{B}$) and quark square loop diagrams
($\mathcal{C}$) contributing to ($\mathcal{A}$).} \label{fig:4}
\end{center}
\end{figure}
\begin{figure}[tb!h]
\begin{center}
\includegraphics[scale=0.85]{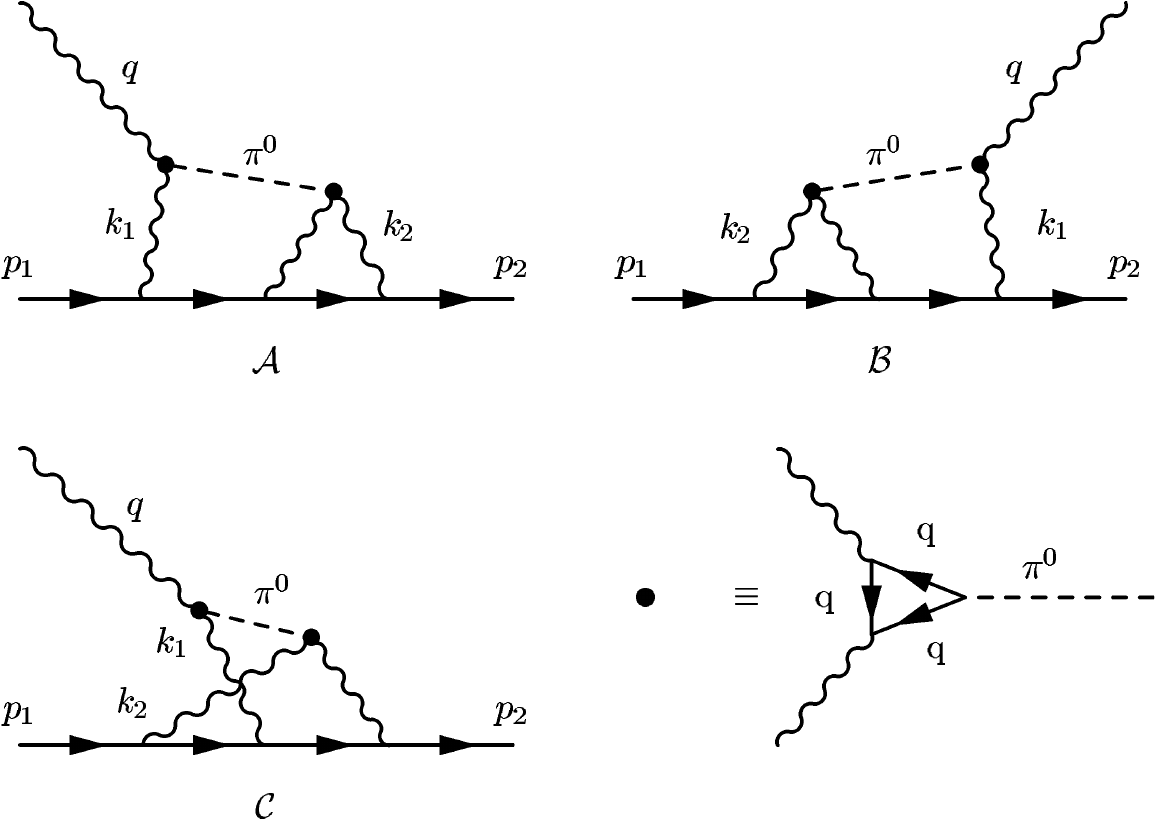}
\caption{Meson pole terms contribution to LBL diagram} \label{fig:5}
\end{center}
\end{figure}

   Therefore, we have recalculated \cite{BDADKZ} the third-order hadronic LBL
contributions to the anomalous magnetic moment of the muon
$a_\mu^{had}$ from the pole terms of the scalar $\sigma$, $a_0$ and
pseudoscalar $\pi^0$, $\eta$, $\eta'$ mesons $(M)$ in the framework
of the linearized extended Nambu-Jona-Lasinio model

\begin{center}
$\mathcal{L}_{q\bar qM}= g_M \bar q(x)\left[\sigma(x)+i \pi(x)
\gamma_5\right]q(x).$
\end{center}

   The reason for the latter are predictions of series of recent
papers

\begin{center}
   $a_\mu^{LBL}=(+52 \pm 18) \times 10^{-11}$\cite{Haykawa01}\\

   $a_\mu^{LBL}=(+92 \pm 32) \times 10^{-11}$\cite{Bijnens01}\\

   $a_\mu^{LBL}=(+83 \pm 12) \times 10^{-11}$\cite{Knecht01}\\

   $a_\mu^{LBL}(\pi_0)=(+58 \pm 10)\times 10^{-11}$\cite{Blokland01}\\
\end{center}
which differ in the magnitude.

   Moreover, in these papers only the pseudoscalar meson pole contributions
were considered. We include the scalar meson $(\sigma, a_0)$ pole
contributions as well.

   Current methods in a description of the  $\gamma^*\rightarrow M\gamma^*$
transition form factors are ChPT and the vector-meson-dominance
(VMD) model.

   Here the corresponding transition form factors by the constituent
quark triangle loops (see Fig.~\ref{fig:5}) with colorless and
flavorless quarks with charge equal to the electron one are
represented.

   An application of a similar modified constituent quark triangle
loop model for a prediction of the pion electromagnetic form factor
behavior was carried out in \cite{Dubnicka85} where also a
comparison with the naive VMD model prediction was demonstrated.

   The mass of the quark in the triangle loop is taken to be

\begin{center}
$m_u =m_d=m_q= (280 \pm 20)$ MeV\\
\end{center}
determined \cite{Nagy01} in the framework of the chiral quark model
of the Nambu-Jona-Lasinio type by exploiting the experimental values
of the pion decay constant, the $\rho$-meson decay into two-pions
constant, the masses of pion and kaon and the mass difference of
$\eta$ and $\eta '$ mesons.

   The unknown strong coupling constants of $\pi^0,\eta,\eta '$ and
$a_0$ mesons with quarks are evaluated in a comparison of the
corresponding theoretical two-photon widths with experimental ones.

   The $\sigma$-meson coupling constant is taken to be equal to
$\pi^0$-meson coupling constant as it follows from the corresponding
Lagrangian.

   The $\sigma$-meson mass is taken to be $m_\sigma$=$(496 \pm 47)$
MeV as an average of the values recently obtained experimentally
from the decay $D^+ \rightarrow \pi^-\pi^+\pi^+$ \cite{Aitala01} and
excited $\Upsilon$ decay \cite{Komada01} processes.

   As a result we present explicit formulas for $a_\mu^{LBL}(M)$
$(M=\pi^0, \eta, \eta ', \sigma, a_0)$ in terms of Feynman
parametric integrals of 10-dimensional order, which subsequently are
calculated by MIKOR method.

   Finally, one finds
\begin{eqnarray}
     a_\mu^{LBL}(\pi^0) &=& (81.83 \pm 16.50) \times 10^{-11}\label{AMULBL1} \\
     a_\mu^{LBL}(\eta) &=& (5.62 \pm 1.25) \times 10^{-11}\nonumber\\
     a_\mu^{LBL}(\eta ') &=& (8.00 \pm 1.74) \times 10^{-11}\nonumber \\
     a_\mu^{LBL}(\sigma) &=& (11.67 \pm 2.38) \times 10^{-11}\nonumber\\
     a_\mu^{LBL}(a_0) &=& (0.62 \pm 0.24) \times
     10^{-11}.\nonumber
\end{eqnarray}

   So, the total contribution of meson poles in LBL is

\begin{equation}
             a_\mu^{LBL}(M) = (107.74 \pm 16.81) \times 10^{-11},
\end{equation}
where the resultant error is the addition in quadrature of all
partial errors of (\ref{AMULBL1}).

   Together with the contributions of the pseudoscalar meson $(\pi^\pm,
K^\pm)$ square loops and constituent quark square loops
(Fig.~\ref{fig:4}) it gives

\begin{equation}\label{amulbl}
            a_\mu^{LBL}(total) = (111.20 \pm 16.81) \times 10^{-11}.
\end{equation}

   The others 3-loop hadronic contributions derived from the hadronic
vacuum polarizations $(VP)$ were most recently evaluated by
Krause\cite{Krause97}

\begin{equation*}
              a_\mu^{(3)VP} = (-101 \pm 6) \times 10^{-11}.
\end{equation*}
Then the total 3-loop hadronic correction is
\begin{equation}
       a_\mu^{(3)had} = a_\mu^{LBL}(total) + a_\mu^{(3)VP} =
                        (10.20 \pm 17.28) \times 10^{-11}
                        \label{amu3}
\end{equation}
where the errors have been again added in quadratures.

   If we take into account the most recent evaluation \cite{Narison01}
of the lowest-order hadronic vacuum-polarization contribution to the
anomalous magnetic moment of the muon

\begin{equation*}
              a_\mu^{(2)had} = (7021 \pm 76) \times 10^{-11}
\end{equation*}
together with the pure QED contribution up to 8th order and the
single- and two-loop weak interaction contribution, finally one gets
the SM theoretical prediction of the muon anomalous magnetic moment
value to be
\begin{equation}\label{amuth}
              a_\mu^{th} = (116 591 888.9 \pm 78.1) \times10^{-11}.
\end{equation}

   Comparing this theoretical result with experimental, one finds

\begin{equation}\label{amuexp}
              a_\mu^{exp} - a_\mu^{th} = (151 \pm 116) \times 10^{-11}
\end{equation}
which implies a reasonable consistency of the SM prediction for the
anomalous magnetic moment of the muon with experiment.

   However, one expects in near future a two times  lowering of the error in BNL E-821
experiment and then there can appear still a room for a new physics
beyond the SM.

\medskip

   Therefore, one has to think on further improvement of the
central value of the muon anomalous magnetic moment and especially
of the lowering of its total theoretical error.

\medskip

     \subsection{Remarkable suppression of the $e^{+}e^{-}\rightarrow\pi^{+}\pi^{-}$
contribution error into muon $g-2$}\label{VIII1}

 \medskip

   The first improvements we see still in the lowest-order hadronic
vacuum-polarization diagram contributions (Fig.~\ref{fig:2}) to be
dominant among all other hadronic contributions, which can be
expressed by the integral
\begin{equation}
a_\mu^{(2) had}=\frac{1}{4\pi^3}\int_{4m_\pi^2}^{\infty} \sigma^h(s)
K_\mu(s) ds;\label{LOhad}
\end{equation}
where  $\sigma^h(s)$ stands for the total cross section
$\sigma(e^+e^-\to had)$ and
\begin{equation*}
K_\mu(s)=\int_0^1\frac{x^2(1-x)}{x^2+(1-x)s/m_\mu^2} dx.
\end{equation*}

   Moreover, the two-pion channel $a_{\mu}^{had,LO}(e^{+}e^{-}\rightarrow\pi^{+}\pi^{-})$,
is dominant among all the lowest hadronic vacuum-polarization
contributions. The data are used to evaluate the integral up to few
GeV, above a perturbative calculation is possible.

   The expression (\ref{LOhad}) was always evaluated
\cite{Davier07,Hagiwara07} only by the direct integration of the
$\sigma_{LO}$ $\left(\pi^{+}\pi^{-}\right)$ data (with small
exceptions, see \cite{Yndurain05}).

   In this paragraph we demonstrate \cite{BelickaHS09} on
$a_{\mu}^{had,LO}(e^{+}e^{-}\rightarrow\pi^{+}\pi^{-})$, that by
using the $U\&A$ model of the pion EM structure (see \ref{III3})
one can achieve remarkable suppression of the contribution error in
comparison with a direct integration over existing experimental
data. In order to compare our results with results of other authors
obtained by integrating over the existing experimental data, we
chose three different limits for the upper integration limit
($3.24\:\mathrm{GeV^{2}}$, $2.0449\:\mathrm{GeV^{2}}$ and
$0.8\:\mathrm{GeV^{2}}$).

   Two approaches were used for the error evaluation.

   The first one was a Monte Carlo method based on a random number generator with
the assumption of the Gaussian distribution for the uncertainties of
the published data points. For each point a new one was randomly
generated using the Gaussian probability density function with the
mean identical to the original point and $\sigma$ equal to the
published error. Doing this for each point, new random data set was
obtained. This data set was then fitted and the values of the
parameters of the model $p_{i}$ as well as the value of
$a_{\mu}^{had,LO}\left(\pi^{+}\pi^{-}\right)$ were extracted.
Repeating the whole procedure 4000 times, we reached statistics high
enough to allow us for a reliable error calculation. The mean
$\overline{a_{\mu}^{had,LO}}\left(\pi^{+}\pi^{-}\right)$ and the
$\sigma$ were calculated from the 4000 values and since the mean is
not, in general, identical with the optimal-fit value
$a_{\mu,OPT}^{had,LO}\left(\pi^{+}\pi^{-}\right)$ we present
asymmetric uncertainties
\begin{equation*}
a_{\mu}^{had,LO}\left(\pi^{+}\pi^{-}\right)=a_{\mu,OPT}^{had,LO}\left(\pi^{+}\pi^{-}\right)_{+B}^{-A},
\end{equation*}
where \begin{equation*}
A=\sigma+a_{\mu,OPT}^{had,LO}\left(\pi^{+}\pi^{-}\right)-\overline{a_{\mu}^{had,LO}}\left(\pi^{+}\pi^{-}\right)
\end{equation*}
and
\begin{equation*}
B=\sigma+\overline{a_{\mu}^{had,LO}}\left(\pi^{+}\pi^{-}\right)-a_{\mu,OPT}^{had,LO}\left(\pi^{+}\pi^{-}\right).
\end{equation*}

   In the second approach the program MINUIT was used to establish
the uncertainties of the model parameters. Then, taking the
numerical derivatives for $\frac{\partial}{\partial
p_{i}}a_{\mu}^{had,LO}\left(\pi^{+}\pi^{-}\right)$, the uncertainty
was propagated to $a_{\mu}^{had,LO}\left(\pi^{+}\pi^{-}\right)$
using the covariance matrix. In this method the errors are
symmetric.

In addition to the integration of the model, we also performed a
direct integration of the data points based on the trapezoidal rule,
so as to cross-check our compatibility with other authors. Our
results and some results from other authors
\cite{Davier07,Hagiwara07,Yndurain05} are summarized in Table~\ref{tab:Results}.

\begin{table}[b!]
\caption{\label{tab:Results}Our results and results of other authors
\cite{Davier07,Hagiwara07,Yndurain05}.}
\medskip
\begin{centering}
\begin{tabular}{llll}
\hline
 & \multicolumn{3}{c}{$a_{\mu}^{had,LO}(e^{+}e^{-}\rightarrow\pi^{+}\pi^{-})\times10^{11}$}\tabularnewline
\cline{2-4} \noalign{\vskip\doublerulesep} Interval
$\left[GeV^{2}\right]$ & $4m_{\pi}^{2}<t<3.24$ &
$4m_{\pi}^{2}<t<2.0449$ & $4m_{\pi}^{2}<t<0.8$\tabularnewline \hline
\noalign{\vskip\doublerulesep} $U\&A$ Model, Meth.1 &
$5132.36_{+0.83}^{-0.83}$ & $5128.22_{+0.73}^{-0.67}$ &
$4870.24_{+0.20}^{-0.20}$\tabularnewline
\noalign{\vskip\doublerulesep}$U\&A$ Model, Meth.2 &
$5132.37\pm3.00$ & $5128.25\pm2.86$ &
$4870.44\pm2.64$\tabularnewline \noalign{\vskip\doublerulesep}
Integration over Data & $5035.33_{+28.32}^{-17.22}$ &
$5031.22_{+28.94}^{-16.43}$ &
$4756.77_{+27.55}^{-18.14}$\tabularnewline
\noalign{\vskip\doublerulesep} \noalign{\vskip\doublerulesep}
\emph{Davier} & $5040.00\pm31.05$ &  &
\tabularnewline[\doublerulesep] \noalign{\vskip\doublerulesep}
\emph{Hagiwara et al.} &  & $5008.2\pm28.70$ & \tabularnewline
\noalign{\vskip\doublerulesep} \emph{Yndur\'{a}in et al.} &  &  &
$4715\pm33.53$\tabularnewline \hline \noalign{\vskip\doublerulesep}
\end{tabular}
\par\end{centering}
\end{table}

   The use of the U\&A model dramatically reduces the error on
   $a_{\mu}^{had,LO}\left(\pi^{+}\pi^{-}\right)$.
This is not an arbitrary feature of the model but originates from
model-independent information which is additional to the data in the
integration region and which can be taken into account when the
model is used.

   The most important sources contributing to error reduction are
\begin{itemize}
\item
Expected smoothness of the $F_{\pi}(t)$ at small scale $\Delta t$:
The model provides a function behaving smoothly at small $\Delta t$.
\item
Experimental data outside the integration region: The fit is done
not only to the data inside, but also to the data outside the
integration region.
\item
Theoretical knowledge on $F_{\pi}(t)$: The model respects all known
properties of the pion EM FF.
\end{itemize}

   Especially the first point plays an important role. The new
precise data tend to lie above older, less precise data and, in some
regions, the vertical spread of the data is very important, at the
limit of inconsistency. If the calculation of the integral is based
directly on data, then less precise data shift the mean value of the
integral and enlarge the uncertainty.

   When the $U\&A$ model is used, the predicted behavior of $F_{\pi}(t)$ as
given by the result of the fit is mostly determined by precisely
measured points and is only little influenced by data with important
uncertainties. This leads to more appropriate mean value and smaller
errors.

   As a result, one arrives to the mean value of $a_{\mu}^{had,LO} \left(\pi^{+}\pi^{-}\right)$
which is higher than what is obtained by the direct data integration
and to much reduced uncertainty. The shift in the mean value goes in
the right direction and brings the theoretical value closer to the
experimental one.

   The error estimates from the two used methods are not fully
compatible, the first method gives smaller errors. This might be
related to statistical fluctuations (1st method) and to
approximations as numerical derivatives and linearization (2nd
method).

   Here we have presented the calculation of
$a_{\mu}^{had,LO}\left(\pi^{+}\pi^{-}\right)$ based on the $U\&A$
model of the pion EM structure. This approach allows for important
error reduction. It can be extended also to an evaluation of the
contributions of other channels to $a_{\mu}^{had,LO}$ with
two-particles in final states. As the important multi-particle
states can be mostly reduced to two-resonance final states, the
method exploiting the $U\&A$ models can be also applied to them.

\setcounter{equation}{0} \setcounter{figure}{0} \setcounter{table}{0}\newpage
     \section{Sum rules for hadron photo-production on hadrons and
     photon}\label{IX}

   Under the sum rules commonly one understands the expressions
bringing into relation some physical quantities with other physical
quantities. There exist various approaches for a derivation of a
complex of sum rules as it is fully demonstrated e.g. in
\cite{Alfaro73}. In this paper we are concerned of the sum rules for
hadron photo-production on hadrons and photon.

   Historically there was only one attempt, by Kurt Gottfried
\cite{Gottfried67}, to derive sum rule for hadron photo-production,
specially on proton target, relating the proton mean squared charge
radius $<r^2_{Ep}>$ and the proton magnetic moment
$\mu_p=1+\kappa_p$ to the integral over the total hadron
photo-production cross-section on proton, considering very
high-energy electron-proton scattering and the non-relativistic
quark model of hadrons. However, the corresponding integral is
divergent and the Gottfried sum rule practically cannot be
satisfied.

   Here we are interested in a derivation of hadron photo-production
sum rules, bringing into relation static properties of pseudoscalar
meson nonet, $1/2^+$ ground state octet baryons and quarks with the
convergent integral over the total cross-sections of hadron
photo-production on pseudoscalar mesons, on octet baryons and over
$\gamma\gamma \to 2 jets$ cross-section, respectively, either by
exploiting analytic properties of "the retarded forward Compton
scattering amplitude on hadron" \cite{Kuraev06} $\tilde
A(s_1,\bf{q})$ in $s_1$-plane  (such amplitude represents only a
class of diagrams in which the initial state photon is first
absorbed by a hadron line and then emitted by the scattered hadron)
for meson \cite{DDK07} and baryon \cite{DDK06} targets, or by its
explicit calculation in quark loops approximation for the case of
photon \cite{BDDK07} target. The variable $s_1$ is the c.m. energy
squared of the virtual Compton scattering process and $\bf q$ is the
third component of the transversal part $q_{\bot}=(0,0,\bf q)$ of
the virtual photon four-momentum $q$.

   This new approach gives sum rules with convergent integrals and
specially the proton-neutron sum rule \cite{BDKsr04} was tested by
using existing data and  it is fulfilled with high precision. The
latter fact gives us confidence also to all other derived sum rules
by a similar approach.

\medskip

      \subsection{$q^2$-dependent meson sum rules}\label{IX1}

\medskip

\begin{figure}[tb]
\begin{center}
\includegraphics[clip,scale=.6]{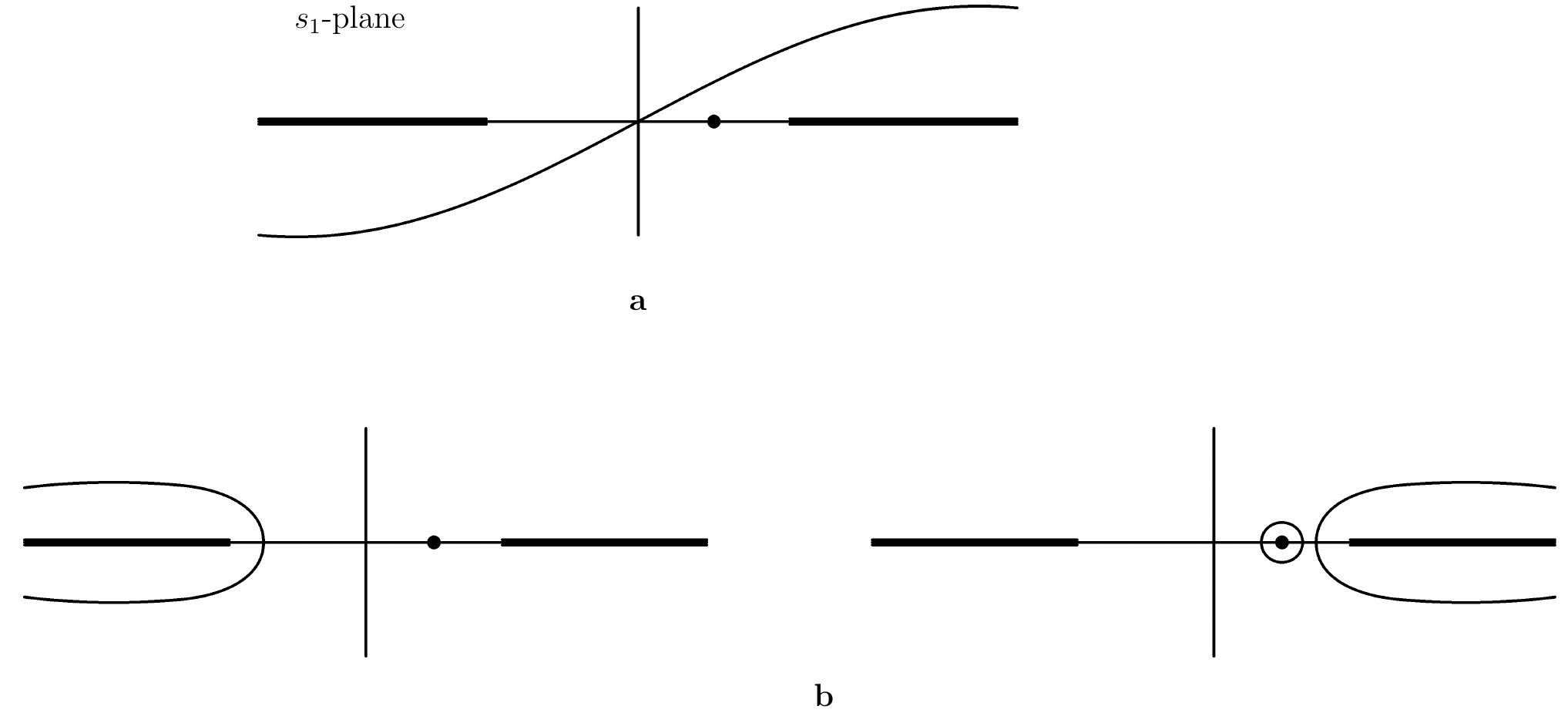}
\caption {Sum rule interpretation in $s_1$ plane.}
\end{center}\label{figsumes}
\end{figure}

The $q^2$-dependent meson sum rules are derived by investigating the
analytic properties of the retarded Compton scattering amplitude
$\tilde{A}^h(s_1,{\bf{q}})$ in $s_1$- plane as presented in
Fig.~\ref{figsumes}a, then defining the integral $I$ over the path $C$
(for more detail see \cite{Baier81}) in the $s_1$-plane
\begin{equation}\label{a1}
I=\int\limits_C d s_1 \frac{p_1^\mu p_1^\nu}{s^2}
\left(\tilde{A}^{h}_{\mu\nu}(s_1,{\bf{q}})-
\tilde{A}^{h'}_{\mu\nu}(s_1,{\bf{q}})\right )
\end{equation}
from the gauge invariant light-cone projection
$p_1^{\mu}p_1^{\nu}\tilde{A}_{\mu\nu}^h(s_1,{\bf{q}})$ of the
amplitude $\tilde{A}^h(s_1,{\bf{q}})$ and once closing the contour
$C$ to upper half-plane, another one to lower half-plane (see
Fig.~\ref{figsumes}b).

As a result the following sum rule appears
\begin{eqnarray}
& &\pi (Res^{h'}-Res^{h})
={\bf{q}}^2\int\limits_{r.h.}^\infty\frac{ds_1} {s_1^2} [Im
\tilde{A}^{h}(s_1,{\bf{q}})-Im \tilde{A}^{h'}(s_1,{\bf{q}})].
\label{a2}
\end{eqnarray}

The left-hand cut contributions expressed by an integral over the
difference $[Im \tilde{A}^{h}(s_1,{\bf{q}}) - Im
\tilde{A}^{h'}(s_1,{\bf{q}})]$ are assumed to be mutually annulated.

Now, one has to take into account the corresponding residuum of the
intermediate state pole (see Fig.~\ref{figsumes}).

As the electromagnetic structure of mesons is described by one
charge form factor, the residuum takes the form
\begin{equation}
\label{a3} Res^{(M)}=2\pi \alpha F^2_M(\bf{-q^2}),
\end{equation}
where an averaging over the initial photon spin is performed.

Then, substituting (\ref{a3}) into (\ref{a2}) and taking into
account \cite{DDK07}
\begin{eqnarray} \nonumber
&&\Big(\frac{d\sigma^{e^-h\to e^-X}(s,{\bf q})}{d^2{\bf{q}}} -
\frac{d\sigma^{e^-h'\to
e^-X'}(s,{\bf q})}{d^2{\bf{q}}}\Big)=\\
\label{a4} &=&\frac{\alpha}
{4\pi^2{\bf{q}^2}}\int\limits_{s_1^{thr}}^\infty\frac{d
s_1}{s_1^2}[Im \tilde{A^h}(s_1,{\bf{q}})- Im
\tilde{A^{h'}}(s_1,{\bf{q}})].
\end{eqnarray}
with $d^2{\bf q}=\pi d{\bf q^2}$, one comes to the $q^2$- dependent
meson sum rule
\begin{eqnarray}
& &[F^2_{P'}({\bf{-q^2}})-F^2_{P'}(0)] -
[F^2_P({\bf{-q^2}})-F^2_P(0)] = \nonumber\\
&=&\frac{2}{\pi \alpha^2}({\bf{q^2}})^2\Big(\frac{d\sigma^{e^- P\to
e^-X}}{d{\bf{q}}^2} - \frac{d\sigma^{e^- P'\to
e^-X'}}{d{\bf{q}}^2}\Big)\label{a5},
\end{eqnarray}
where the left-hand side was re-normalized in order to separate the
pure strong interactions from electromagnetic ones.

\medskip

     \subsection{Universal static sum rules for total hadron
     photo-production cross-sections on mesons}\label{IX2}

\medskip

   Now, employing the Weics\"acker-Williams like relation \cite{DDK07}
\begin{eqnarray} \label{a6}
&&{\bf{q}}^2\Big(\frac{d\sigma^{e^- P\to e^-X}}{d{\bf{q}}^2} -
\frac{d\sigma^{e^- P'\to e^-X'}}{d{\bf{q}}^2}\Big)_
{|_{{\bf{q}}^2\to 0}}=\\
\nonumber &=&\frac{\alpha} {\pi}\int\limits_{s_1^{th}}^\infty\frac{d
s_1}{s_1} [\sigma_{tot}^{\gamma P\to X}(s_1)-\sigma_{tot}^{\gamma
P'\to X'}(s_1)]
\end{eqnarray}
for mesons, taking a derivative according to ${\bf q^2}$ of both
sides in $q^2$-dependent meson sum rule for ${\bf q^2} \to 0$ and
using the laboratory reference frame by $s_1=2m_B\omega$, one comes
to the new universal meson sum rule relating meson mean square
charge radii to the integral over a difference of the corresponding
total photo-production cross-sections on mesons
\begin{eqnarray}\label{a7}
&&\frac{1}{3}(\langle r_{P'}^2 \rangle-\langle r_P^2 \rangle)= \\
&=&\frac{2}{\pi^2\alpha}\int\limits_{\omega_P}^\infty \frac{d\omega}
{\omega}\big[\sigma_{tot}^{\gamma P\to X}(\omega)-
\sigma_{tot}^{\gamma P'\to X}(\omega)\big ],\nonumber
\end{eqnarray}
 in which just a mutual cancelation of the
rise of the latter cross sections for $\omega \to \infty$ is
achieved.

\medskip

   According to the SU(3) classification of existing hadrons the
following ground state pseudoscalar meson nonet $\pi^-$, $\pi^0$,
$\pi^+$, $K^-$, $\bar K^0$, $K^0$, $K^+$, $\eta$, $\eta'$ exists.
However, in consequence of CPT invariance the meson electromagnetic
form factors $F_P(-{\bf q^2})$ hold the following relation
\begin{equation}
F_P({\bf -q^2})= - F_{\bar P}({\bf -q^2}),\label{a8}
\end{equation}
where $\bar P$ means antiparticle.

Since $\pi_0$, $\eta$ and $\eta'$ are true neutral particles, their
electromagnetic form factors are owing to the (\ref{a8}) zero in the
whole region of a definition and therefore we exclude them from
further considerations.

If one considers couples of particle-antiparticle like $\pi^{\pm}$,
$K^{\pm}$ and $K^0$, $\bar K^0$, the left hand side of (\ref{a5}) is
owing to the relation (\ref{a8}) equal zero and we exclude couples
$\pi^{\pm}$, $K^{\pm}$ and $K^0$, $\bar K^0$ from further
considerations as well.

Considering a couple of the iso-doublet of kaons $K^+, K^0$ and
$K^-, \bar K^0$, the following Cabibbo-Radicati like sum rules
\cite{Cabibbo66} for kaons can be written
\begin{eqnarray}\label{a9}
&&\frac{1}{6}{\pi^2\alpha}\langle r_{K^+}^2\rangle= \\
&=&\int_{\omega_{th}}^{\infty} \frac{d\omega}
{\omega}\left[\sigma_{tot}^{\gamma K^+\to X}(\omega)-
\sigma_{tot}^{\gamma K^0\to X}(\omega)\right] \nonumber
\end{eqnarray}

\begin{eqnarray}\label{a10}
&&\frac{1}{6}{\pi^2\alpha}(-1)\langle r_{K^-}^2\rangle= \\
&=&\int_{\omega_{th}}^{\infty} \frac{d\omega}
{\omega}\left[\sigma_{tot}^{\gamma K^-\to X}(\omega)-
\sigma_{tot}^{\gamma \bar K^0\to X}(\omega)\right ],\nonumber
\end{eqnarray}

in which the relation $\langle r_{K^+}^2\rangle=-\langle
r_{K^-}^2\rangle$ for kaon mean squared charge radii, following
directly from (\ref{a8}), holds and divergence of the integrals, due
to an increase of the total cross-sections $\sigma_{tot}^{\gamma
K^\pm\to X}(\omega)$ for large values of $\omega$, is taken off by
the increase of total cross-sections $\sigma_{tot}^{\gamma K^0\to
X}(\omega)$ and $\sigma_{tot}^{\gamma \bar K^0\to X}(\omega)$,
respectively.
   If besides the latter, also the relations
\begin{eqnarray}\label{a11}
&&\sigma_{tot}^{\gamma K^0\to X}(\omega)\equiv\sigma_{tot}^{\gamma
\bar K^0\to X}(\omega) \nonumber \\
&& \sigma_{tot}^{\gamma K^+\to
X}(\omega)\equiv\sigma_{tot}^{\gamma K^-\to X}(\omega),\nonumber
\end{eqnarray}
following from $C$ invariance of the electromagnetic interactions,
are taken into account, one can see the sum rule (\ref{a10}), as
well as all other possible sum rules obtained by combinations
$K^+\bar K^0, K^-K^0,$ to be contained already in (\ref{a9}).

The last possibility is a consideration of a couple of mesons taken
from the isomultiplet of pions and the isomultiplet of kaons leading
to the following sum rules
\begin{eqnarray}\label{a12}
&&\frac{1}{6}{\pi^2\alpha}[(\pm1)\langle r_{\pi^{\pm}}^2\rangle-
(\pm1)\langle r_{K^{\pm}}^2\rangle] = \\
&=&\int_{\omega_{th}}^{\infty}
\frac{d\omega} {\omega}\left[\sigma_{tot}^{\gamma \pi^{\pm}\to
X}(\omega)- \sigma_{tot}^{\gamma K^{\pm}\to X}(\omega)\right]\nonumber
\end{eqnarray}

\begin{eqnarray}\label{a13}
 &&\frac{1}{6}{\pi^2\alpha}(\pm1)\langle r_{\pi^{\pm}}^2\rangle = \\
&=&\int_{\omega_{th}}^{\infty} \frac{d\omega}
{\omega}\left[\sigma_{tot}^{\gamma \pi^{\pm}\to X}(\omega)-
\sigma_{tot}^{\gamma K^0\to X}(\omega)\right].\nonumber
\end{eqnarray}

   Now taking the experimental values \cite{Rpp}
\begin{equation*}
(\pm1)\langle r_{\pi^{\pm}}^2\rangle=+0.4516\pm0.0108 \quad [fm^2]
\end{equation*}
\begin{equation*}
(\pm1)\langle r_{K^{\pm}}^2\rangle=+0.3136\pm0.0347 \quad [fm^2]
\end{equation*}
one comes to the conclusion that in average

\begin{eqnarray}\label{a14}
&&[\bar\sigma_{tot}^{\gamma \pi^{\pm}\to X}(\omega)- \bar\sigma_{tot}^{\gamma
K^{\pm}\to X}(\omega)] > 0 \nonumber \\
&& [\bar\sigma_{tot}^{\gamma K^-\to
X}(\omega)- \bar\sigma_{tot}^{\gamma \bar K^0\to X}(\omega)] > 0,
\end{eqnarray}
from where the following chain of inequalities for finite values of
$\omega$ in average follow
\begin{equation}\label{a15}
\bar\sigma_{tot}^{\gamma \pi^{\pm}\to X}(\omega)> \bar\sigma_{tot}^{\gamma
K^{\pm}\to X}(\omega)
 > \bar\sigma_{tot}^{\gamma \bar K^0\to X}(\omega)
> 0.
\end{equation}

Subtracting up (\ref{a9}) or (\ref{a10}) from the relation
(\ref{a13}), the sum rule (\ref{a12}) is obtained, what demonstrates
a mutual consistency of all considered sum rules.

\medskip

     \subsection{$q^2$-dependent octet baryon sum rules}\label{IX3}

\medskip

   Similarly to the meson case, as the electromagnetic structure of
octet baryons is described by Dirac and Pauli form factors, the
residuum takes the form
\begin{equation}\label{b1}
Res^{B}=2\pi\alpha( F^2_{1B}+ \frac{{\bf{q}^2}}{4m_B^2}F_{2B}^2),
\end{equation}
where an averaging over the initial baryon and photon spins is
performed. Finally one obtains the $q^2$-dependent baryon sum rule
in the form \cite{DDK06}
\begin{eqnarray}
& &[F^2_{1B'}({-\bf{q^2}})-F^2_{1B'}(0)] -
[F^2_{1B}({\bf{-q^2}})-F^2_{1B}(0)] + \nonumber\\
&+&{\bf{q^2}}\big [\frac{F^2_{2B'}({\bf-{q^2}})}{4m_{B'}^2}-
\frac{F^2_{2B}({\bf-{q^2}})}{4m_{B}^2}\big ]=\nonumber \\
 &=&\frac{2}{\pi
\alpha^2}({\bf{q^2}})^2\Big(\frac{d\sigma^{e^- B\to
e^-X}}{d{\bf{q}}^2} - \frac{d\sigma^{e^- B'\to
e^-X}}{d{\bf{q}}^2}\Big ),\label{b2}
\end{eqnarray}
where the left-hand side was again re-normalized in order to
separate the pure strong interactions from electromagnetic ones.

\medskip

     \subsection{Universal static sum rules for total hadron
     photo-production cross-sections on baryons}\label{IX4}

\medskip

   Employing in (\ref{b2}) the Weics\"acker-Williams relation for
baryons, taking a derivative  according to ${\bf{q}}^2$ of both
sides in $q^2$-dependent baryon sum rule for ${\bf{q}^2}\to 0$ and
using the laboratory reference frame by $s_1=2m_B\omega$ , one comes
to the new universal static baryon sum rule
\begin{eqnarray}\nonumber
&&\frac{1}{3}\big [F_{1B}(0)\langle r_{1B}^2
\rangle-F_{1B'}(0)\langle r_{1B'}^2 \rangle\big ]-
\big [\frac{\kappa_B^2}{4m_B^2}-\frac{\kappa_{B'}^2}{4m^2_{B'}}\big ]=
\\
 &=&\frac{2}{\pi^2\alpha}\int\limits_{{\omega_B}}^{\infty}
\frac{d\omega} {\omega}\big[\sigma_{tot}^{\gamma B\to X}(\omega)-
\sigma_{tot}^{\gamma B'\to X}(\omega)\big]\label{b3}
\end{eqnarray}
relating Dirac baryon mean square radii $\langle r_{1B}^2\rangle$
and baryon anomalous magnetic moments $\kappa_B$ to the convergent
integral, in which a mutual cancelation of the rise of the
corresponding total cross-sections for $\omega\to\infty$ is
achieved.

\medskip

   According to the SU(3) classification of existing hadrons, there
are known the following members of the ground state $1/2^+$ baryon
octet ($p$, $n$, $\Lambda^0$, $\Sigma^+$,  $\Sigma^0$, $\Sigma^-$,
$\Xi^0$, $\Xi^-$). As a result, by using the universal expression
(\ref{b3}) one can write down $28$ different sum rules for total
cross-sections of hadron photo-production on ground state $1/2^+$
octet baryons.

In order to evaluate their left hand sides and to draw out some
phenomenological consequences, one needs the reliable values of
Dirac baryon mean square radii $\langle r_{1B}^2 \rangle $ and baryon
anomalous magnetic moments $\kappa_B$.

The latter are known (besides $\Sigma^0$, which is found from the
well known relation $\kappa_{\Sigma^+}+
\kappa_{\Sigma^-}$=$2\kappa_{\Sigma^0}$) experimentally \cite{Rpp}.
However, to calculate $\langle r_{1B}^2 \rangle$ by means of the
difference of the baryon electric mean square radius $\langle
r_{EB}^2 \rangle$ and Foldy term, well known for all ground state
octet baryons from the experimental information on the magnetic
moments given by Review of Particle Physics
\begin{equation}
\langle r_{1B}^2 \rangle =\langle r_{EB}^2
\rangle-\frac{3\kappa_B}{2 m_B^2}, \label{b4}
\end{equation}
we are in need of the reliable values of $\langle r_{EB}^2 \rangle$.

They are known experimentally only for the proton, neutron and
$\Sigma^-$-hyperon.

Fortunately there are recent results \cite{Kubis01} to fourth order
in relativistic baryon chiral perturbation theory (giving
predictions for the $\Sigma^-$ charge radius and the
$\Lambda$-$\Sigma^0$ transition moment in excellent agreement with
the available experimental information), which solve our problem
completely.

 Calculating the left-hand sides of all sum rules one finds

\begin{equation}
\label{a57} \frac{2}{\pi^2\alpha}\int\limits_{\omega_{p}}^{\infty}
\frac{d\omega} {\omega}\big[\sigma_{tot}^{\gamma p\to X}(\omega)-
\sigma_{tot}^{\gamma n\to X}(\omega)\big ]=2.0415
  {\textrm mb},\quad \Rightarrow \quad \bar\sigma_{tot}^{\gamma p \to X}(\omega)> \bar\sigma_{tot}^{\gamma
n\to X}(\omega)
\end{equation}

\begin{equation}
\label{a58}
\frac{2}{\pi^2\alpha}\int\limits_{\omega_{\Sigma^+}}^{\infty}
\frac{d\omega} {\omega}\big[\sigma_{tot}^{\gamma \Sigma^+\to
X}(\omega)- \sigma_{tot}^{\gamma \Sigma^0\to X}(\omega)\big
]=2.0825
  {\textrm mb},\quad \Rightarrow \quad \bar\sigma_{tot}^{\gamma \Sigma^+\to X}(\omega)>
\bar\sigma_{tot}^{\gamma \Sigma^0\to X}(\omega)
\end{equation}

\begin{equation}
\label{a59}
\frac{2}{\pi^2\alpha}\int\limits_{\omega_{\Sigma^+}}^{\infty}
\frac{d\omega} {\omega}\big[\sigma_{tot}^{\gamma \Sigma^+\to
X}(\omega)- \sigma_{tot}^{\gamma \Sigma^-\to X}(\omega) \big
]=4.2654  {\textrm mb}, \quad \Rightarrow\quad
\bar\sigma_{tot}^{\gamma \Sigma^+\to X}(\omega)>\bar \sigma_{tot}^{\gamma
\Sigma^-\to X}(\omega)
\end{equation}

\begin{equation}
\label{a60}
\frac{2}{\pi^2\alpha}\int\limits_{\omega_{\Sigma^0}}^{\infty}
\frac{d\omega} {\omega}\big[\sigma_{tot}^{\gamma \Sigma^0\to
X}(\omega)- \sigma_{tot}^{\gamma \Sigma^-\to X}(\omega)\big ]=
2.1829  {\textrm mb}, \quad \Rightarrow\quad \bar\sigma_{tot}^{\gamma
\Sigma^0\to X}(\omega)> \bar\sigma_{tot}^{\gamma \Sigma^-\to
X}(\omega)
\end{equation}

\begin{equation}
\label{a61}
\frac{2}{\pi^2\alpha}\int\limits_{\omega_{\Xi^0}}^{\infty}
\frac{d\omega} {\omega}\big[\sigma_{tot}^{\gamma \Xi^0\to
X}(\omega)- \sigma_{tot}^{\gamma \Xi^-\to X}(\omega)\big ]=1.5921
    {\textrm mb}, \quad \Rightarrow\quad \bar\sigma_{tot}^{\gamma
\Xi^0\to X}(\omega)> \bar\sigma_{tot}^{\gamma \Xi^-\to X}(\omega)
\end{equation}

\begin{equation}
\label{a62} \frac{2}{\pi^2\alpha}\int\limits_{\omega_{p}}^{\infty}
\frac{d\omega} {\omega}\big[\sigma_{tot}^{\gamma p\to X}(\omega)-
\sigma_{tot}^{\gamma \Lambda^0\to X}(\omega)\big ]=1.6673 {\textrm
mb},\quad \Rightarrow\quad \bar\sigma_{tot}^{\gamma p\to X}(\omega)>
\bar\sigma_{tot}^{\gamma \Lambda^0\to X}(\omega)
\end{equation}

\begin{equation}
\label{a63} \frac{2}{\pi^2\alpha}\int\limits_{\omega_{p}}^{\infty}
\frac{d\omega} {\omega}\big[\sigma_{tot}^{\gamma p \to X}(\omega)-
\sigma_{tot}^{\gamma \Sigma^+\to X}(\omega)\big ]=-0.4158 {\textrm
mb},\quad \Rightarrow\quad \bar\sigma_{tot}^{\gamma p \to
X}(\omega)<\bar\sigma_{tot}^{\gamma \Sigma^+\to X}(\omega)
\end{equation}

\begin{equation}
\label{a64} \frac{2}{\pi^2\alpha}\int\limits_{\omega_{p}}^{\infty}
\frac{d\omega} {\omega}\big[\sigma_{tot}^{\gamma p\to X}(\omega)-
\sigma_{tot}^{\gamma \Sigma^0\to X}(\omega)\big ]=1.6667  {\textrm
mb},\quad \Rightarrow\quad \bar\sigma_{tot}^{\gamma p\to X}(\omega)>
\bar\sigma_{tot}^{\gamma \Sigma^0\to X}(\omega)
\end{equation}

\begin{equation}
\label{a65} \frac{2}{\pi^2\alpha}\int\limits_{\omega_{p}}^{\infty}
\frac{d\omega} {\omega}\big[\sigma_{tot}^{\gamma p \to X}(\omega)-
\sigma_{tot}^{\gamma \Sigma^-\to X}(\omega)\big ]= 3.8496 {\textrm
mb},\quad \Rightarrow\quad \bar\sigma_{tot}^{\gamma p \to X}(\omega)>
\bar\sigma_{tot}^{\gamma \Sigma^-\to X}(\omega)
\end{equation}

\begin{equation}
\label{a66} \frac{2}{\pi^2\alpha}\int\limits_{\omega_{p}}^{\infty}
\frac{d\omega} {\omega}\big[\sigma_{tot}^{\gamma p\to
X}(\omega)-\Rightarrow \bar\sigma_{tot}^{\gamma \Xi^0\to
X}(\omega)\big ]=1.7259  {\textrm mb},\quad \quad
\bar\sigma_{tot}^{\gamma p\to X}(\omega)> \sigma_{tot}^{\gamma
\Xi^0\to X}(\omega)
\end{equation}

\begin{equation}
\label{a67} \frac{2}{\pi^2\alpha}\int\limits_{\omega_{p}}^{\infty}
\frac{d\omega} {\omega}\big[\sigma_{tot}^{\gamma p \to X}(\omega)-
\sigma_{tot}^{\gamma \Xi^-\to X}(\omega)\big ]=3.3180   {\textrm
mb},\quad \Rightarrow\quad \bar\sigma_{tot}^{\gamma p \to X}(\omega)>
\bar\sigma_{tot}^{\gamma \Xi^-\to X}(\omega)
\end{equation}

\begin{equation}
\label{a68} \frac{2}{\pi^2\alpha}\int\limits_{\omega_{n}}^{\infty}
\frac{d\omega} {\omega}\big[\sigma_{tot}^{\gamma n\to X}(\omega)-
\sigma_{tot}^{\gamma \Lambda^0\to X}(\omega)\big ]=-0.3260
{\textrm mb},\quad \Rightarrow\quad \bar\sigma_{tot}^{\gamma n\to
X}(\omega)< \bar\sigma_{tot}^{\gamma \Lambda^0\to X}(\omega)
\end{equation}

\begin{equation}
\label{a69} \frac{2}{\pi^2\alpha}\int\limits_{\omega_{n}}^{\infty}
\frac{d\omega} {\omega}\big[\sigma_{tot}^{\gamma n\to X}(\omega)-
\sigma_{tot}^{\gamma \Sigma^+\to X}(\omega)\big ]=-2.4573 {\textrm
mb},\quad \Rightarrow\quad \bar\sigma_{tot}^{\gamma n\to X}(\omega)<
\bar\sigma_{tot}^{\gamma \Sigma^+\to X}(\omega)
\end{equation}

\begin{equation}
\label{a70} \frac{2}{\pi^2\alpha}\int\limits_{\omega_{n}}^{\infty}
\frac{d\omega} {\omega}\big[\sigma_{tot}^{\gamma n\to X}(\omega)-
\sigma_{tot}^{\gamma \Sigma^0\to X}(\omega)\big ]=-0.3747 {\textrm
mb},\quad \Rightarrow\quad \bar\sigma_{tot}^{\gamma n\to X}(\omega)<
\bar\sigma_{tot}^{\gamma \Sigma^0\to X}(\omega)
\end{equation}

\begin{equation}
\label{a71} \frac{2}{\pi^2\alpha}\int\limits_{\omega_{n}}^{\infty}
\frac{d\omega} {\omega}\big[\sigma_{tot}^{\gamma n\to X}(\omega)-
\sigma_{tot}^{\gamma \Sigma^-\to X}(\omega)\big ]= 1.8082 {\textrm
mb},\quad \Rightarrow\quad \bar\sigma_{tot}^{\gamma n\to X}(\omega)>
\bar\sigma_{tot}^{\gamma \Sigma^-\to X}(\omega)
\end{equation}

\begin{equation}
\label{a72} \frac{2}{\pi^2\alpha}\int\limits_{\omega_{n}}^{\infty}
\frac{d\omega} {\omega}\big[\sigma_{tot}^{\gamma n\to X}(\omega)-
\sigma_{tot}^{\gamma \Xi^0\to X}(\omega)\big ]= -0.3156  {\textrm
mb},\quad \Rightarrow\quad \bar \sigma_{tot}^{\gamma n\to X}(\omega)<
\bar\sigma_{tot}^{\gamma \Xi^0\to X}(\omega)
\end{equation}

\begin{equation}
\label{a73} \frac{2}{\pi^2\alpha}\int\limits_{\omega_{n}}^{\infty}
\frac{d\omega} {\omega}\big[\sigma_{tot}^{\gamma n\to X}(\omega)-
\sigma_{tot}^{\gamma \Xi^-\to X}(\omega)\big ]=1.2766  {\textrm
mb},\quad \Rightarrow\quad \bar\sigma_{tot}^{\gamma n\to X}(\omega)>
\bar\sigma_{tot}^{\gamma \Xi^-\to X}(\omega)
\end{equation}

\begin{equation}
\label{a74}
\frac{2}{\pi^2\alpha}\int\limits_{\omega_{\Lambda^0}}^{\infty}
\frac{d\omega} {\omega}\big[\sigma_{tot}^{\gamma \Lambda^0\to
X}(\omega)- \sigma_{tot}^{\gamma \Sigma^+\to X}(\omega)\big ]=
-2.0831 {\textrm mb},\quad \Rightarrow\quad \bar\sigma_{tot}^{\gamma
\Lambda^0\to X}(\omega)< \bar\sigma_{tot}^{\gamma \Sigma^+\to
X}(\omega)
\end{equation}

\begin{equation}
\label{a75}
\frac{2}{\pi^2\alpha}\int\limits_{\omega_{\Lambda^0}}^{\infty}
\frac{d\omega} {\omega}\big[\sigma_{tot}^{\gamma \Lambda^0\to
X}(\omega)- \sigma_{tot}^{\gamma \Sigma^0\to X}(\omega)\big ]=
-0.0006   {\textrm mb},\quad \Rightarrow\quad \bar\sigma_{tot}^{\gamma
\Lambda^0\to X}(\omega)\approx \bar\sigma_{tot}^{\gamma \Sigma^0\to
X}(\omega)
\end{equation}

\begin{equation}
\label{a76}
\frac{2}{\pi^2\alpha}\int\limits_{\omega_{\Lambda^0}}^{\infty}
\frac{d\omega} {\omega}\big[\sigma_{tot}^{\gamma \Lambda^0\to
X}(\omega)- \sigma_{tot}^{\gamma \Sigma^-\to X}(\omega)\big
]=2.1823
  {\textrm{mb}},\quad \Rightarrow\quad \bar\sigma_{tot}^{\gamma
\Lambda^0\to X}(\omega)> \bar\sigma_{tot}^{\gamma \Sigma^-\to X}(\omega)
\end{equation}

\begin{equation}
\label{a77}
\frac{2}{\pi^2\alpha}\int\limits_{\omega_{\Lambda^0}}^{\infty}
\frac{d\omega} {\omega}\big[\sigma_{tot}^{\gamma \Lambda^0\to
X}(\omega)- \sigma_{tot}^{\gamma \Xi^0\to X}(\omega)\big ]=0.0586
{\textrm{mb}},\quad \Rightarrow\quad \bar\sigma_{tot}^{\gamma
\Lambda^0\to X}(\omega)> \bar\sigma_{tot}^{\gamma \Xi^0\to X}(\omega)
\end{equation}

\begin{equation}
\label{a78}
\frac{2}{\pi^2\alpha}\int\limits_{\omega_{\Lambda^0}}^{\infty}
\frac{d\omega} {\omega}\big[\sigma_{tot}^{\gamma \Lambda^0\to
X}(\omega)- \sigma_{tot}^{\gamma \Xi^-\to X}(\omega)\big ]=2.1823
{\textrm{mb}} ,\quad \Rightarrow\quad \bar\sigma_{tot}^{\gamma
\Lambda^0\to X}(\omega)> \bar\sigma_{tot}^{\gamma \Xi^-\to X}(\omega)
\end{equation}

\begin{equation}
\label{a79}
\frac{2}{\pi^2\alpha}\int\limits_{\omega_{\Sigma^+}}^{\infty}
\frac{d\omega} {\omega}\big[\sigma_{tot}^{\gamma \Sigma^+\to
X}(\omega)- \sigma_{tot}^{\gamma \Xi^0\to X}(\omega)\big ]=2.1417
\textrm{mb},\quad \Rightarrow\quad \bar\sigma_{tot}^{\gamma
\Sigma^+\to X}(\omega)> \bar\sigma_{tot}^{\gamma \Xi^0\to X}(\omega)
\end{equation}

\begin{equation}
\label{a80}
\frac{2}{\pi^2\alpha}\int\limits_{\omega_{\Sigma^+}}^{\infty}
\frac{d\omega} {\omega}\big[\sigma_{tot}^{\gamma \Sigma^+\to
X}(\omega)- \sigma_{tot}^{\gamma \Xi^-\to X}(\omega)\big ]=3.7338
\textrm{mb},\quad \Rightarrow\quad \bar\sigma_{tot}^{\gamma
\Sigma^+\to X}(\omega)> \bar\sigma_{tot}^{\gamma \Xi^-\to X}(\omega)
\end{equation}

\begin{equation}
\label{a81}
\frac{2}{\pi^2\alpha}\int\limits_{\omega_{\Sigma^0}}^{\infty}
\frac{d\omega} {\omega}\big[\sigma_{tot}^{\gamma \Sigma^0\to
X}(\omega)- \sigma_{tot}^{\gamma \Xi^0\to X}(\omega)\big ]=0.1168
\textrm{mb},\quad \Rightarrow\quad \bar\sigma_{tot}^{\gamma
\Sigma^0\to X}(\omega)> \bar\sigma_{tot}^{\gamma \Xi^0\to X}(\omega)
\end{equation}

\begin{equation}
\label{a82}
\frac{2}{\pi^2\alpha}\int\limits_{\omega_{\Sigma^0}}^{\infty}
\frac{d\omega} {\omega}\big[\sigma_{tot}^{\gamma \Sigma^0\to
X}(\omega)- \sigma_{tot}^{\gamma \Xi^-\to X}(\omega)\big ]=1.5732
\textrm{mb},\quad \Rightarrow\quad \bar\sigma_{tot}^{\gamma
\Sigma^0\to X}(\omega)> \bar\sigma_{tot}^{\gamma \Xi^\to X}(\omega)
\end{equation}

\begin{equation}
\label{a83}
\frac{2}{\pi^2\alpha}\int\limits_{\omega_{\Sigma^-}}^{\infty}
\frac{d\omega} {\omega}\big[\sigma_{tot}^{\gamma \Sigma^-\to
X}(\omega)- \sigma_{tot}^{\gamma \Xi^0\to X}(\omega)\big ]=-2.1238
\textrm{mb},\quad \Rightarrow\quad \bar\sigma_{tot}^{\gamma
\Sigma^-\to X}(\omega)< \bar\sigma_{tot}^{\gamma \Xi^0\to X}(\omega)
\end{equation}

\begin{equation}
\label{a84}
\frac{2}{\pi^2\alpha}\int\limits_{\omega_{\Sigma^-}}^{\infty}
\frac{d\omega} {\omega}\big[\sigma_{tot}^{\gamma \Sigma^-\to
X}(\omega)- \sigma_{tot}^{\gamma \Xi^-\to X}(\omega)\big ]=-0.5316
\textrm{mb},\quad \Rightarrow\quad \bar\sigma_{tot}^{\gamma
\Sigma^-\to X}(\omega)< \bar\sigma_{tot}^{\gamma \Xi^-\to X}(\omega),
\end{equation}

from where one gets the following  chain of
inequalities
\begin{eqnarray*}
&&\bar\sigma_{tot}^{\gamma \Sigma^+\to X}(\omega)> \bar\sigma_{tot}^{\gamma
p\to X}(\omega)> \bar\sigma_{tot}^{\gamma \Lambda^0\to X}(\omega)
\approx   \bar\sigma_{tot}^{\gamma \Sigma^0\to X}(\omega)>\\
&>&\bar\sigma_{tot}^{\gamma \Xi^0\to X}(\omega)>
 \bar\sigma_{tot}^{\gamma n\to X}(\omega)>
\bar\sigma_{tot}^{\gamma \Xi^-\to X}(\omega)>\bar\sigma_{tot}^{\gamma
\Sigma^-\to X}(\omega)
\end{eqnarray*}

for total cross-sections of hadron photo-production on ground state
$1/2^+$ octet baryons to be valid in average for finite values of
$\omega$.

\medskip

     \subsection{Sum rule for photon target}\label{IX5}

\medskip

Let us consider the two photon exchange electron-photon zero angle
scattering amplitude in the process
\begin{equation} \label{eq:proc}
e(p,\lambda)+\gamma(k,\varepsilon)\to
e(p,\lambda)+\gamma(k,\varepsilon),
\end{equation}
in two-loop ($\alpha^3$) approximation as presented in Fig.~\ref{Fphsumr},
with $p^2=m_{e}^2$, $k^2=0$ and assuming that
$s=2p.k\gg m_{e}^2$.

   \begin{figure}[tb]
\centerline{\includegraphics[scale=0.8]{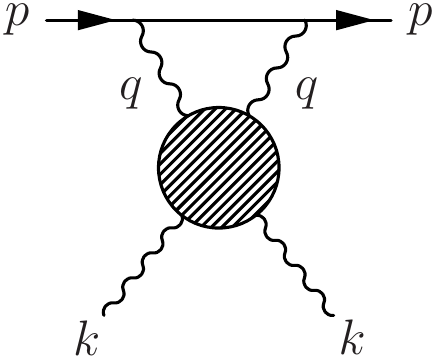} \qquad
\includegraphics[scale=0.8]{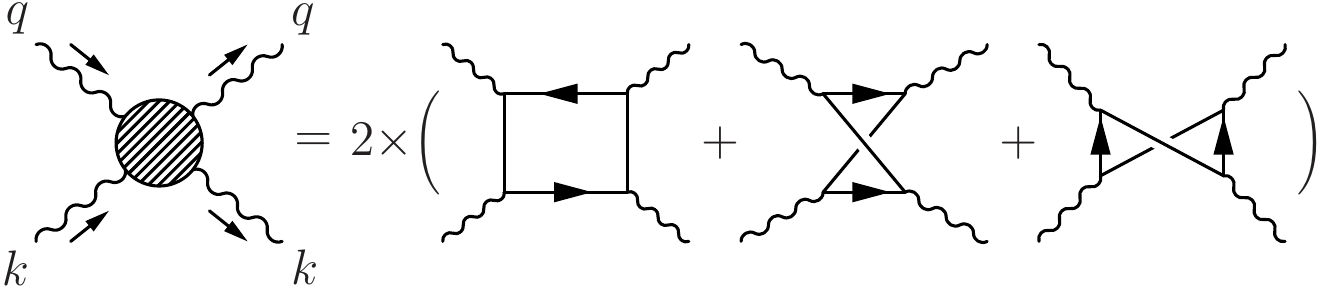}}
\caption{Feynman diagram of $e\gamma\to e\gamma$ scattering with LBL
mechanism to be realized by quark-loops}\label{Fphsumr}
\end{figure}

Averaging over the initial electron and photon spin states (initial
and final spin states are supposed to coincide) one can write  down
the amplitude of the process (\ref{eq:proc}) in the following form
\begin{eqnarray} \label{eq:ampl}
A^{e\gamma\to e\gamma}(s,t=0)=
s\frac{\alpha}{4\pi^2}\int\frac{d^2\bf{q}}{(q^2)^2}d s_1
\sum_\varepsilon
A^{\gamma\gamma\to\gamma\gamma}_{\mu\nu\alpha\beta}\frac{p^\mu p^\nu
\varepsilon^{\alpha} \varepsilon^{*\beta}}{s^2},
\end{eqnarray}
where the light-cone projection of the light-by-light (LBL)
scattering tensor takes the form
\begin{eqnarray}\label{c1}
A^{\gamma\gamma\to\gamma\gamma}(s_1,{\bf{q}})=A^{\gamma\gamma\to\gamma\gamma}_{\mu\nu\alpha\beta}
\frac{p^\mu p^\nu \varepsilon^{\alpha}
\varepsilon^{*\beta}}{s^2}=-\frac{8\alpha^2}{\pi^2}N_cQ^4_q\int
d^4q_{-}\Big[\frac{S_1}{D_1}+\frac{S_2}{D_2}+\frac{S_3}{D_3}\Big ]
\end{eqnarray}
with
\begin{eqnarray}
\frac{S_1}{D_1}\hspace{-0.1cm}=\hspace{-0.1cm}\frac{(1/4)Tr\hat{p}(\hat{q}_-+m_q)\hat{p}(\hat{q}_--\hat{q}+m_q)\hat{\varepsilon}^*
(\hat{q}_--\hat{q}+\hat{k}+m_q)\hat{\varepsilon}(\hat{q}_--\hat{q}+m_q)}{(q_-^2-m^2_q)((q_--q)^2-m^2_q)^2
((q_--q+k)^2-m^2_q)},\\
\frac{S_2}{D_2}\hspace{-0.1cm}=\hspace{-0.1cm}\frac{(1/4)Tr\hat{p}(\hat{q}_-+m_q)\hat{p}(\hat{q}_--\hat{q}+m_q)\hat{\varepsilon}^
(\hat{q}_--\hat{q}-\hat{k}+m_q)\hat{\varepsilon}^*(\hat{q}_--\hat{q}+m_q)}{(q_-^2-m^2_q)((q_--q)^2-m^2_q)^2
((q_--q-k)^2-m_q^2)},\\
\frac{S_3}{D_3}\hspace{-0.1cm}=\hspace{-0.1cm}\frac{(1/4)Tr\hat{p}(\hat{q}_-+m_q)\hat{\varepsilon}(\hat{q}_--\hat{k}+m_q)\hat{p}
(\hat{q}_-+\hat{q}-\hat{k}+m_q)\hat{\varepsilon}^*(\hat{q}_-+\hat{q}+m_q)}{(q_-^2-m^2_q)((q_-+q)^2-m^2_q)
((q_-+q+k)^2-m^2_q)((q_--k)^2-m^2_q)}.
\end{eqnarray}
where $q_{-}$ means the quark four-momentum in the quark loop of the
process $\gamma\gamma\to\gamma\gamma$, $N_c$ is the number of colors
in QCD and $Q_q$ is the charge of the quark $q$ in electron charge
units.
\begin{figure}[tb]
\centerline{\includegraphics[width=0.5\textwidth]{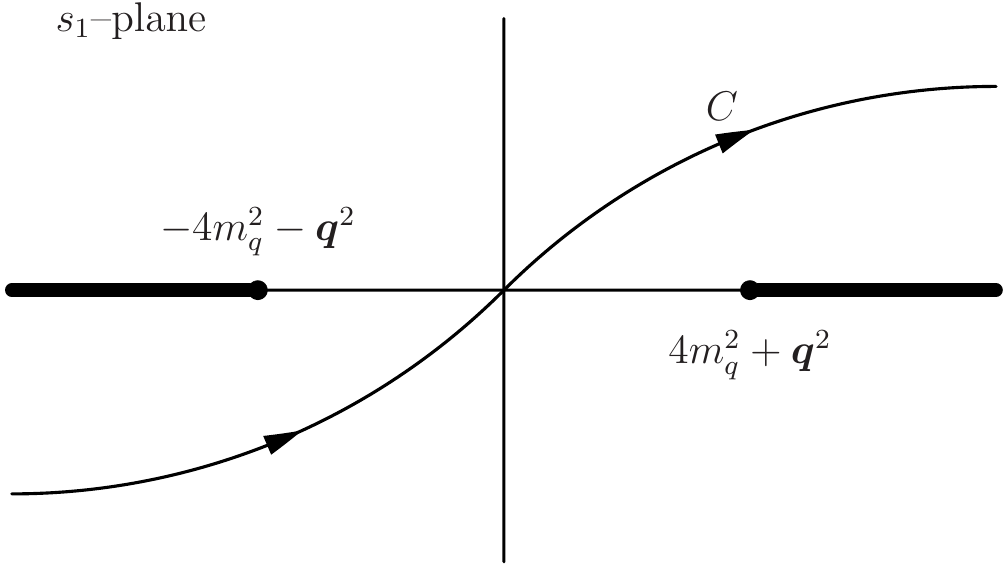}}\caption{The
path $C$ of an integration in (\ref{c2})}\label{confphsumr}
\end{figure}

Now, taking a derivative of the relation (\ref{eq:ampl}) according
to $d^2{\bf{q}}$ and investigating the analytic properties of the
obtained expression in the $s_1$-plane one gets the configuration as
presented in Fig.~\ref{confphsumr}, where also the path $C$ of the
integral expression

\begin{equation}\label{c2}
I=\int_Cds_1\frac{dA^{e\gamma\to
e\gamma}(s_1,{\bf{q}})}{d^2{\bf{q}}}
\end{equation}
is drawn. When the integration contour is closed to the right (on
s-channel cut) and to the left (on the $u$-channel cut) one comes to
the relation
\begin{equation}\label{c3}
\int^{-\infty}_{-4m^2_q-{\bf{q}^2}}ds_1\Delta_u\frac{dA^{e\gamma\to
e\gamma}(s_1,{\bf{q}})}{d^2{\bf{q}}}{|_{left}}=\int^\infty_{4m^2_q+{\bf{q}}^2}
ds_1\Delta_s\frac{dA^{e\gamma\to
e\gamma}(s_1,{\bf{q}})}{d^2{\bf{q}}}{|_{right}},
\end{equation}
where the right s-cannel discontinuity by means of
Eq.(\ref{eq:ampl}) is related, due to optical theorem in a
differential form
\begin{equation}\label{c4}
\Delta_s\frac{dA^{e\gamma\to e\gamma}(s,0)}{d^2{\bf{q}}}=
2s\frac{d\sigma^{e\gamma\to e q\bar{q}}}{d^2{\bf{q}}},
\end{equation}
to the $Q^2$=${\bf{q}}^2$=$-q^2$ dependent differential
cross-section of $q\bar q$ pair creation by electron on photon, to
be well known in the framework of QED \cite{Kelner67} for $l^+l^-$
pair creation
\begin{eqnarray}\label{c6}
&&\frac{4\alpha^3}{3(q^2)^2}f(\frac{{\bf{q}}^2}{m^2_q})N_cQ^4_q=\frac{d\sigma^{e\gamma\to
e q\bar{q}}}{d{\bf{q}}^2},\\
&&f(\frac{{\bf{q}}^2}{m^2_q})=({\bf{q}}^2-m^2_q)J+1,\nonumber \\
&&J=\frac{4}{\sqrt{{\bf{q}}^2({\bf{q}}^2+4m^2_q)}}
\ln[\sqrt{{\bf{q}}^2/(4m^2_q)}+\sqrt{1+{\bf{q}}^2/(4m^2_q)}].\nonumber
\end{eqnarray}
But the right hand cut concerns of two real quark production for
$s_1> 4 m_q^2$, which is associated wit 2 jets production.

The left-hand cut contribution has the same form as in QED case with
constituent quark masses and as a result one obtains
\begin{equation} \label{eq:sum}
\frac{4\alpha^3}{3({\bf{q}}^2)^2}N_c\sum_q
Q_q^4f(\frac{{\bf{q}}^2}{m_q^2})=\frac{d\sigma^{e\gamma\to e2
jets}}{d{\bf{q}}^2}.
\end{equation}
Finally, for the case of small ${\bf{q}}^2$ and applying the
Weizs\"acker-Williams relation, one comes to the sum rule for photon
target \cite{BDDK07} as follows
\begin{equation}\label{c8}
\frac{14}{3}\sum_q\frac{Q_q^4}{m_q^2}=
\frac{1}{\pi\alpha^2}\int\limits_{4m^2_q}^{\infty} \frac{d s_1}{s_1}
\sigma_{tot}^{\gamma\gamma\to 2{jet}}(s_1).
\end{equation}

The quantity $\sigma_{tot}^{\gamma\gamma\to 2 jets}(s_1)$ is assumed
to degrease for large values of $s_1$. It corresponds to the events
in $\gamma\gamma$ collisions with creation of two jets, which are
not separated by rapidity gaps and for which until the present days
there is no experimental information. The latter complicates a
verification of the obtained sum rule for photon target.

\setcounter{equation}{0} \setcounter{figure}{0} \setcounter{table}{0}\newpage
        \section{Conclusions}

   In this review we have formulated the main principles of a
construction of the universal Unitary and Analytic model of the
hadron electro-weak structure. Before the problem of inconsistency
of the asymptotic behavior of VMD model with the asymptotic
behavior of form factors of baryons and nuclei is solved, leading
to a formulation of the asymptotic conditions and their general
solutions giving explicitly all three possible forms of VMD form
factor representations to be automatically normalized and
possessing the right asymptotic behavior as required by the quark
model of strongly interacting particles. Also a general approach
for determination of the lowest normal and anomalous singularities
of form factors from the corresponding Feynman diagrams is
formulated. Finally, a number of concrete results to be obtained
by making use of the analytic properties of the dynamical physical
quantities, like electro-weak form factors of hadrons or
amplitudes of various electromagnetic processes of hadrons is
presented. Some of them are already successfully verified in
practice, others are expecting for experimental confirmation. All
this together demonstrates the analyticity to be an unavoidable
powerful tool of the present particle physics phenomenology.

\addcontentsline{toc}{section}{Acknowledgment}
\section*{Acknowledgments}
   The authors are indebted to C.Adamuscin for a technical help
in the preparation of this paper.
   The work was in part supported by the Slovak Grant Agency for
Sciences VEGA, Grant No. 2/0009/10.

\newpage
\fancyhead[LO]{References}

\end{document}